%% file: main_with_appendix.tex
\documentclass[12pt]{article}

\usepackage{makecell}
\usepackage{bm, commath}
\usepackage{natbib}
\usepackage{caption}
\usepackage{graphicx}
\usepackage{subfig}
\usepackage{amsmath, amsfonts, amsthm}
\usepackage{float}
\usepackage{booktabs,siunitx}
\usepackage{hyperref}
\usepackage{url}
\usepackage{multirow}
\usepackage{amssymb}
\usepackage{blkarray}
\usepackage{bbm}
\usepackage{multicol}
\usepackage{authblk}
\usepackage{natbib}
\usepackage[shortlabels]{enumitem}
\usepackage{longtable}
\usepackage{adjustbox}
\usepackage{threeparttable}
\usepackage{hhline}
\usepackage{colortbl}
\usepackage{lscape}
\usepackage{mathabx}

\usepackage{listings}
\usepackage{bbm}
\usepackage{textcomp}
\usepackage{authblk}
\usepackage[ruled]{algorithm2e}
\SetKwInput{KwParam}{Parameter}
\SetAlgoCaptionLayout{centerline}

\usepackage{sectsty}
\usepackage[doublespacing]{setspace}

\usepackage[utf8]{inputenc}
\usepackage{float}
\usepackage{tikz}
\usetikzlibrary{shapes,decorations,arrows,calc,arrows.meta,fit,positioning}

\usepackage{mwe}
\usepackage{graphicx}

\usepackage[textsize=scriptsize, textwidth = 2cm, shadow]{todonotes}

\usepackage[textsize=scriptsize, textwidth = 2cm, shadow]{todonotes}

\newcommand{\indep}{\mathrel{\perp\mspace{-10mu}\perp}}

\newtheorem{theorem}{Theorem}
\newtheorem{definition}{Definition}

\newtheorem{remark}{Remark}
\newtheorem{example}{Example}
\newtheorem{assumption}{Assumption}
\newtheorem*{assumption*}{Assumption}

\newtheorem{proposition}{Proposition}

\newtheorem{lemma}{Lemma}

\newtheorem{innercustomgeneric}{\customgenericname}
\providecommand{\customgenericname}{}
\newcommand{\newcustomtheorem}[2]{%
  \newenvironment{#1}[1]
  {%
   \renewcommand\customgenericname{#2}%
   \renewcommand\theinnercustomgeneric{##1}%
   \innercustomgeneric
  }
  {\endinnercustomgeneric}
}

\newcustomtheorem{customthm}{Theorem}
\newcustomtheorem{customassumption}{Assumption}

\usepackage{mathtools}

\usepackage{xcolor}

\makeatletter
\renewcommand{\algocf@captiontext}[2]{#1\algocf@typo. \AlCapFnt{}#2} 
\def\@algocf@capt@plain{top}
\renewcommand{\algocf@makecaption}[2]{%
  \addtolength{\hsize}{\algomargin}%
  \sbox\@tempboxa{\algocf@captiontext{#1}{#2}}%
  \ifdim\wd\@tempboxa >\hsize
  \hskip .5\algomargin%
  \parbox[t]{\hsize}{\algocf@captiontext{#1}{#2}}
  \else%
  \global\@minipagefalse%
  \hbox to\hsize{\box\@tempboxa}
  \fi%
  \addtolength{\hsize}{-\algomargin}%
}
\makeatother

\begin{document}

\def\spacingset#1{\renewcommand{\baselinestretch}%
{#1}\small\normalsize} \spacingset{1}

\sectionfont{\bfseries\large\sffamily}%

\subsectionfont{\bfseries\sffamily\normalsize}%

\title{Nested Instrumental Variables Analysis: Switcher Average Treatment Effect, Identification, Efficient Estimation and Generalizability}

\author[1]{Rui Wang }
\author[2]{Ying-Qi Zhao }
\author[3]{Oliver Dukes }
\author[4]{Bo Zhang\thanks{Assistant Professor of Biostatistics, Vaccine and Infectious Disease Division, Fred Hutchinson Cancer Center. Email: {\tt bzhang3@fredhutch.org}. }}

\affil[1]{Department of Biostatistics, University of Washington}
\affil[2]{Public Health Sciences Division, Fred Hutchinson Cancer Center}
\affil[3]{Department of Applied Mathematics, Computer Science and Statistics, Ghent University}
\affil[4]{Vaccine and Infectious Disease Division, Fred Hutchinson Cancer Center}
\date{}
\maketitle

\begin{abstract}
\input{abstract}
\end{abstract}
\vspace{0.3 cm}

\newpage
\spacingset{1.6}

\setlength\abovedisplayskip{2pt}
\setlength\belowdisplayskip{1pt}%

\input{1_Introduction}
\input{2.Estimand_and_identification}

\input{3.Estimation}
\input{4.Testing_homogeneity}
\input{5.Simulation}
\input{6.Case_study}
\input{7.Discussion}
\input{10_Acknowledgement}

\singlespacing
\bibliographystyle{apalike}
\bibliography{paper-ref}

\newpage
\onehalfspacing
\input{appendix.tex}

\end{document}

%% file: abstract.tex
Instrumental variables (IVs) are widely used to estimate causal effects from non-randomized data. A canonical example is a randomized trial with noncompliance, in which the randomized treatment assignment serves as an IV for the non-ignorable treatment received. Under a monotonicity assumption, a valid IV nonparametrically identifies the average treatment effect among a latent complier subgroup, whose generalizability is often under debate. In many studies, there exist multiple versions of an IV, for instance, different nudges to take the same treatment in different study sites in a multicenter clinical trial. These different versions of an IV may result in different compliance rates and offer a unique opportunity to study IV estimates' generalizability. In this article, we introduce a novel nested IV assumption and study identification of the average treatment effect among two latent subgroups: always-compliers and switchers, who are defined based on the joint potential treatment received under two versions of a binary IV. We derive the efficient influence function for the SWitcher Average Treatment Effect (SWATE) under a nonparametric model and propose efficient estimators. We then propose formal statistical tests of the generalizability of IV estimates under the nested IV framework. The proposed tests are flexible nonparametric generalizations of classical overidentification tests that allow estimating nuisance parameters using machine learning tools. We apply the proposed method to the Prostate, Lung, Colorectal and Ovarian (PLCO) Cancer Screening Trial and study the causal effect of colorectal cancer screening and its generalizability. 

%% file: 1_Introduction.tex
\section{Introduction}
\label{sec: intro}\vspace{-6pt}
\subsection{Instrumental variable (IV); complier average treatment effect; generalizability}
\label{subsec: intro IV and generalizability}\vspace{-6pt}
Randomized controlled trials (RCTs) are widely regarded as the gold standard for establishing causal effects. When RCTs are impractical or unethical, researchers often seek to draw causal conclusions from non-randomized studies. This task is challenging for many reasons, with unmeasured confounding being a primary concern. To mitigate bias due to unmeasured confounders, researchers frequently turn to quasi-experimental methods, among which instrumental variable (IV) approaches are some of the most widely used \citep{AIR1996, hernan2006instruments, baiocchi2014instrumental}. A variable is a valid IV if it is correlated with the treatment, has no direct effect on the outcome under study, and is independent of unmeasured treatment-outcome confounders (see, e.g., \citealp{baiocchi2014instrumental}). IV methods were popularized in econometrics through the framework of linear structural equations with omitted variables (see, e.g., \citealp{goldberger1972structural}). In a seminal paper, \citet{AIR1996} formalized an IV within the potential outcomes (PO) framework \citep{neyman1923application,rubin1974estimating}, defined the complier average treatment effect (CATE) as a central estimand, and studied its nonparametric identification under a monotonicity assumption, which states that the level of treatment received by an individual cannot decrease as the level of the IV increases. The concept of identifying and estimating causal effects within a latent subpopulation defined by potential outcomes under both treatment assignments---often referred to as ``cross-world” potential outcomes---was termed principal stratification and later generalized to other settings, including surrogate endpoints and truncation by death \citep{frangakis2002principal}.

A central critique of the CATE framework, and of principal stratification more broadly, is that the CATE pertains only to a latent subpopulation and generally cannot be generalized to the entire population without additional assumptions; see, for example, the extensive discussions in \citet{deaton2009instruments} and \citet{joffe2011principal}. One line of work seeks to bridge the CATE with the average treatment effect (ATE) or the average treatment effect on the treated (ATT) by imposing additional identification assumptions under which a valid IV point-identifies ATE or ATT. Such assumptions enforce homogeneity or no-interaction-type restrictions on the data-generating process \citep{swanson2018partial, levis2024nonparametric}.

For example, with a binary IV and a binary treatment, \citet{robins1994correcting} proposed the no-current-treatment-interaction assumption, which asserts the absence of an additive interaction between the IV and treatment in the model for the conditional average treatment effect. An alternative is the homogeneous selection bias assumption of \citeauthor{tchetgen2013alternative} \citeyearpar{tchetgen2013alternative}, which posits that the selection bias function is invariant to the value of the IV on an additive scale and thereby identifies the ATT using a valid IV. \citet{wang2018bounded} further showed that, in the absence of additive interactions between the unmeasured confounder and the IV or the treatment, the ATE is identifiable via an IV. These assumptions are typically not empirically verifiable and are scale-specific (e.g., additive or multiplicative). Under the principal stratification framework, an assumption known as principal ignorability, or one of its variants, is required for generalizability. This assumption posits that, conditional on baseline covariates, mean potential outcomes do not differ across principal strata \citep{jo2009use,ding2017principal}. Unfortunately, principal ignorability is likewise not empirically testable.

\vspace{-6pt}
\subsection{A clinical trial with two versions of an IV; nested IV}\label{subsec: intro PLCO and nested IV}\vspace{-6pt}
A canonical example of an instrumental variable (IV) arises in a randomized controlled trial (RCT) with noncompliance, in which randomized treatment assignment serves as an encouragement to receive treatment and thus functions as an IV for actual treatment uptake. The Prostate, Lung, Colorectal, and Ovarian (PLCO) Cancer Screening Trial was a multiphasic RCT conducted from 1993 to 2011 across 10 U.S. institutions. Participants randomized to the intervention arm were offered cancer screening examinations, whereas those to the control arm received usual medical care. {The clinical rationale for cancer screening is to identify and remove lesions before they progress to invasive cancer.}

Two consent procedures were implemented in this trial. Seven centers used a single-consent process, where enrollment and randomization were consented simultaneously. The remaining three centers initially used a dual-consent process, in which participants were not informed of the study design or intervention type at initial enrollment. Those randomized to the intervention arm were later informed of their assignment and asked for a second consent. Compliance under the single-consent process was 91.3\%, versus 71.2\% under dual consent. Due to high refusal rates, the three centers later adopted the single-consent process. In a post hoc analysis, \citet{marcus2014non} found that the consent process significantly influenced receipt of cancer screening, even after adjusting for baseline characteristics including age, race/ethnicity, gender, education, BMI, and comorbidity score.

The transition from the dual to the single consent process effectively created a natural experiment that strengthened the encouragement to comply with treatment assignment. In light of the findings of \citeauthor{marcus2014non} \citeyearpar{marcus2014non}, it is reasonable to assume that an individual who complied under the dual-consent process would also comply under the single-consent process. We refer to this assumption---that compliance under one binary IV implies compliance under a strengthened binary IV---as the nested IV assumption.

\vspace{-6pt}
\subsection{Implications of a nested IV assumption}\label{subsec: intro implication of nested IV}\vspace{-6pt}
While a conventional IV analysis defines latent subpopulations---compliers, always-takers, never-takers, and defiers---based on potential treatment received under two realizations of a single binary IV \citep{AIR1996}, the nested IV framework defines subpopulations based on potential treatment received under $2 \times 2 = 4$ realizations of two binary IVs and imposes structure on these latent subpopulations in a manner analogous to the monotonicity assumption in \citet{AIR1996}. Under the nested IV framework, the average treatment effects for two latent subpopulations, referred to as \emph{always-compliers} and \emph{switchers}, can be nonparametrically identified under suitable conditions. 

\emph{Always-compliers} are participants who would comply with treatment regardless of which IV is used; for example, individuals who would comply irrespective of the consent process employed. In contrast, a distinct latent subpopulation termed switchers consists of participants who would be noncompliers (i.e., always-takers or never-takers under the monotonicity assumption of \citeauthor{AIR1996} \citeyearpar{AIR1996}) under the weaker IV but become compliers under a strengthened IV. In the PLCO trial, these are participants who would not comply under the dual-consent process but would comply under the single-consent process.

Beyond identifying and estimating causal effects for always-compliers and switchers, comparing effects across these two disjoint subpopulations---possibly conditional on baseline covariates---provides insight into the generalizability of IV-based estimates and enables formal testing of no-interaction and principal-ignorability-type assumptions. The proposed tests are flexible, nonparametric generalizations of classical overidentification tests \citep{sargan1958estimation} and incorporate machine learning tools for the estimation of nuisance parameters. To the best of our knowledge, this progress towards formally assessing the generalizability of IV estimates represents a novel and valuable contribution to the literature. 

The remainder of the article is organized as follows. Section \ref{sec: nested IV design} describes the nested IV framework, defines the switcher average treatment effect, and presents the main identification results. Sections \ref{sec: connection and extension} and \ref{sec: estimation} develop extensions and propose estimators. Section \ref{sec: test homogeneity} studies hypothesis testing procedures for evaluating treatment effect homogeneity among latent subpopulations. Numerical experiments are reported in Section \ref{sec: simulation}. Section \ref{sec: case study} applies the proposed method to the PLCO study, and Section \ref{sec: discussion} concludes with a discussion.

%% file: 2.Estimand_and_identification.tex
\vspace{-6pt}\section{Nested instrumental variables framework}\label{sec: nested IV design}
\vspace{-6pt}

{\color{black}
\subsection{Potential outcomes}\label{subsec: design two version of Z}\vspace{-6pt}

We define potential outcomes and formulate causal estimands within the Neyman--Rubin potential outcomes framework \citep{neyman1923application,rubin1974estimating}. Suppose we observe an independent and identically distributed sample,
\[
\mathcal{D} = \{O_i=(Z_i,\boldsymbol{X}_i,D_i,Y_i),~i = 1,\ldots,n\},
\]
drawn from the true data-generating law $P_0 \in \mathcal{P}$, where $\mathcal{P}$ denotes a nonparametric model for the observed data distribution. Throughout the article, we write $\mathbb{E}_{P}[\cdot]$ for expectation under a distribution $P \in \mathcal{P}$ and $\mathbb{E}_{P_0}[\cdot]$ for expectation under the true data-generating law.

For participant $i$, let $\boldsymbol{X}_i$ denote a vector of baseline covariates, $Z_i$ a categorical instrumental variable taking values in $\{0_a,0_b,1_a,1_b\}$, $D_i \in \{0,1\}$ a binary indicator of the observed treatment received, and $Y_i$ the observed real-valued outcome of interest. In the PLCO trial, $0_a$ and $1_a$ correspond to randomization to standard of care or intervention under a dual-consent process, whereas $0_b$ and $1_b$ correspond to randomization under a single-consent process. By defining $Z_i$ as a four-level IV, rather than a binary IV, we explicitly allow for distinct versions of encouragement to intervention and control.

We further define a derived variable $G_i$ from $Z_i$ such that $G_i = a$ if $Z_i \in \{0_a,1_a\}$ and $G_i = b$ if $Z_i \in \{0_b,1_b\}$. For the Henry Ford Health System in the PLCO study, $G_i$ represents the type of consent process used (dual versus single consent), or equivalently, the trial stage (before versus after $1997$). 
An equivalent formulation conceptualizes treatment assignment as a two-dimensional vector IV, $(Z^\prime,G)$, where both $Z^\prime$ and $G$ are binary variables. One may then define
\[
Z = 1_g \iff (Z^\prime = 1, G = g), \qquad
Z = 0_g \iff (Z^\prime = 0, G = g),
\]
for $g \in \{a,b\}$, thereby mapping the vector-IV representation to the four-level categorical IV representation. In some applications, it may be more natural to conceptualize the intervention directly through the four-level variable $Z$, whereas in others the vector formulation $(Z^\prime,G)$ may provide greater clarity. For example, in a study of rationed fertility in China, \citet{guo2025rationed} used twinning at the second birth as an instrumental variable for realized fertility (having at least three children) before and after implementation of a birth-control policy. In that setting, one may define $Z^\prime$ as twinning status and $G$ as the policy regime, and then construct the induced categorical IV $Z$ from the pair $(Z^\prime,G)$. 

It is important to emphasize that the two representations of treatment assignment, $Z \in \{0_a, 0_b, 1_a, 1_b\}$ and $(Z^\prime, G)$, where $Z^\prime \in \{0,1\}$ and $G \in \{a,b\}$, are in one-to-one correspondence and simply represent different encodings of the same treatment assignment variable. In both representations, $G$ is an integral component of the treatment assignment.

We refer to the tuple $(0_g,1_g),~g \in \{a, b\}$, as an IV dose pair \citep{sun2025pairwise}. Participants randomized to intervention may decline the assigned screening, while those randomized to control may nevertheless choose to receive screening. Thus, randomized assignment functions only as an encouragement---or nudge---toward screening, with participants retaining full autonomy over treatment uptake. Let $D(Z=z)$ denote the potential treatment received under IV assignment $Z=z$, where $z \in \{0_a,1_a,0_b,1_b\}$. Following \citet{AIR1996}, for a given IV dose pair $\{0_g,1_g\}$, we classify a participant as a complier if $\{D(0_g),D(1_g)\}=(0,1)$, an always-taker if $\{D(0_g),D(1_g)\}=(1,1)$, a never-taker if $\{D(0_g),D(1_g)\}=(0,0)$, and a defier if $\{D(0_g),D(1_g)\}=(1,0)$, for $g\in\{a,b\}$. Finally, let $Y(D = d, Z = z)$ denote the potential outcome of $Y$ when a participant is assigned treatment $D = d$ and IV value $Z = z$. We use $Y(Z = z)$ as shorthand for $Y(D(Z = z), Z = z)$.   
}

\vspace{-6pt}
\subsection{Identification assumptions}\label{subsec: design notation and assumption}
Assumption \ref{ass: standard IV assumption} consists of standard IV assumptions and sometimes is referred to as \emph{core} IV assumptions \citep{AIR1996, baiocchi2014instrumental}.

\vspace{-6pt}
\begin{assumption}\label{ass: standard IV assumption}
    \textbf{(Standard IV assumptions).}\vspace{-6pt}
    \begin{enumerate}[label=(\emph{\roman*}).]
        \item \textbf{(Consistency).} $D = D(Z=z)$ if $Z=z$ and $Y = Y(D=d,Z=z)$ if $D=d$ and $Z=z$, for $d\in \{0,1\}$ and $z\in \{0_a, 1_a, 0_b, 1_b\}$.\vspace{-10pt}
        \item {\textbf{(Exclusion restriction)}. $Y(D=d,Z=z) = Y(D=d)$, for $d\in \{0,1\}$ and $z \in \{1_a,1_b,0_a,0_b\}$.} \vspace{-10pt}
    \item \textbf{(IV independence).} 
 $Z\indep \{D(Z=z),Y(D=d)\} \mid \boldsymbol{X} = \boldsymbol{x}, G=g$, for $d\in \{0,1\}$, $g \in \{a, b\}$, and $z \in \{1_a,1_b,0_a,0_b\}$.  \vspace{-10pt}
    \item \textbf{(IV relevance).} 
 $\mathbb{E}_{P_0}[D\mid Z = 1_g, \boldsymbol{X}] - \mathbb{E}_{P_0}[D \mid Z = 0_g, \boldsymbol{X}]\neq 0$ and $P_0(Z=z \mid \boldsymbol{X}) > 0$ for $g\in \{a,b\}$ and $z\in \{0_a,1_a,0_b,1_b\}$.  
 \end{enumerate}
\end{assumption} \vspace{-6pt}

In an RCT---particularly a double-blinded one---treatment assignment is generally expected to influence clinical outcomes only through its effect on the treatment actually received; thus, the exclusion restriction (Assumption \ref{ass: standard IV assumption}(\romannumeral2)) is typically plausible. { IV independence (Assumption \ref{ass: standard IV assumption}(\romannumeral3)) is ensured by design in a randomized trial, although chance covariate imbalance may still occur in any given trial.} Finally, IV relevance (Assumption \ref{ass: standard IV assumption}(\romannumeral4)) is empirically verifiable and holds in most clinical trials.

{
\begin{remark}[Partial exclusion restriction]\label{remark: partial ER}
  In some settings, researchers cannot rule out a direct effect of $Z \in \{1_a, 0_a\}$ versus $Z \in \{1_b, 0_b\}$ on the outcome. To accommodate this possibility, Assumption \ref{ass: standard IV assumption}(\romannumeral2) can be relaxed to the following partial exclusion restriction:
\[
Y(D = d,G = g) := Y(D = d,Z = 1_g) = Y(D = d,Z = 0_g),~\text{for}~g\in\{a, b\},
\]
while not requiring a priori that $Y(D = d, G = a) = Y(D = d, G = b)$, under which $Y(D = d, G = g)$ could be written simply as $Y(D = d)$. In the main article, we establish identification results under Assumption \ref{ass: standard IV assumption}(\romannumeral2). A parallel set of results under the weaker partial exclusion restriction is provided in Supplemental Material \ref{subsec: G has direct effect}.
\end{remark}
}
In addition to Assumption \ref{ass: standard IV assumption}, we consider the following \emph{nested IV assumptions}:
\vspace{-6pt}
\begin{assumption}\label{ass: nested IV assumptions}
    \textbf{(Nested IV assumptions).}\vspace{-6pt}
    \begin{enumerate}[label=(\emph{\roman*}).]
    \item { \textbf{(Mean effect exchangeability over $G$).} } $\mathbb{E}_{P_0}[Y(Z = 1_g) - Y(Z = 0_g)\mid G=a, \boldsymbol{X}] = \mathbb{E}_{P_0}[Y(Z = 1_g) - Y(Z = 0_g)\mid G = b, \boldsymbol{X}]$ and $\mathbb{E}_{P_0}[D(Z = 1_g) - D(Z = 0_g)\mid G=a, \boldsymbol{X}] = \mathbb{E}_{P_0}[D(Z = 1_g) - D(Z = 0_g)\mid G = b, \boldsymbol{X}]$, for $g \in \{a, b\}.$ \vspace{-6pt}
    
    \item \textbf{(Partial monotonicity).} 
 $D(Z=1_g)\geq D(Z=0_g)$, for $g \in \{a  , b\}$. \vspace{-6pt}
 \item \textbf{(Nonequal compliance).} 
      $\mathbb{E}_{P_0}[D\mid Z = 1_a, \boldsymbol{X} = \boldsymbol{x}]-\mathbb{E}_{P_0}[D\mid Z = 0_a, \boldsymbol{X} = \boldsymbol{x}] \\
      \neq  \mathbb{E}_{P_0}[D\mid Z = 1_b, \boldsymbol{X} = \boldsymbol{x}]-\mathbb{E}_{P_0}[D\mid Z = 0_b, \boldsymbol{X} = \boldsymbol{x}]$.
   \vspace{-6pt}
 \item {\small\textbf{(Nested IV).} 
    $(D(Z=0_a), D(Z=1_a)) = (0, 1)$ implies $(D(Z=0_b), D(Z=1_b)) = (0, 1)$.}
 \end{enumerate}
 \end{assumption}
\vspace{-6pt}

{ 
We highlight two scenarios in which Assumption 2(\romannumeral1) holds. First, the assumption holds when $G$ is exogenous, meaning that assignment to $Z \in \{0_a, 1_a\}$ versus $Z \in \{0_b, 1_b\}$ occurs effectively at random. For example, if the midtrial shift from dual to single consent occurred essentially by happenstance—particularly for participants who enrolled immediately before and after the transition—then Assumption 2(\romannumeral1) would be satisfied. Furthermore, for an ordinal IV $\widetilde{Z}$ with three levels $\{0, 1, 2\}$, where $0<1<2$ represents the level of IV strength, if we define $0_a = 0_b = 0$, $1_a = 1$, and $1_b = 2$, then $Z \in \{0, 1\}$ ($G = a$) as an IV is nested within the IV $Z \in \{0, 2\}$ ($G = b$) provided that $D(Z=2)\geq D(Z=1)\geq D(Z=0)$. In this case, $G$ is exogenous, provided that $\widetilde{Z}$ itself is a valid IV. 

Second, the assumption would also hold if $G$ is not an effect modifier. In the PLCO study, Assumption 2(\romannumeral1) would hold for $g = a$ if, conditional on $\boldsymbol{X}$, the effect of randomized treatment assignment under the dual-consent process (i.e., $Z = 1_a$ versus $Z = 0_a$) on both treatment received and the outcome---had the dual-consent process not been replaced by a single-consent process and instead continued after 1997---remained the same as its effect before 1997. In other words, Assumption 2(\romannumeral1) stipulates that the effect of the \emph{same} version of treatment assignment is stable across $G = a$ and $G = b$, conditional on $\boldsymbol{X}$. 

{\color{black}
It is helpful to compare Assumption \ref{ass: nested IV assumptions}(\romannumeral1) with the following  exchangeability assumption:
\begin{equation}\label{eq: mean exchangeability under usual SUTVA}
    \begin{split}
        \mathbb{E}_{P_0}[Y(Z = 1_a) - Y(Z = 0_a)\mid G=a, \boldsymbol{X}] = \mathbb{E}_{P_0}[Y(Z = 1_b) - Y(Z = 0_b)\mid G = b, \boldsymbol{X}],\\ \mathbb{E}_{P_0}[D(Z = 1_a) - D(Z = 0_a)\mid G=a, \boldsymbol{X}] = \mathbb{E}_{P_0}[D(Z = 1_b) - D(Z = 0_b)\mid G = b, \boldsymbol{X}],
    \end{split}
\end{equation}
which imposes equality on the effects induced by two different versions of treatment assignment, $Z \in \{1_a, 0_a\}$ versus $Z \in \{1_b, 0_b\}$, across $G$. Comparing to Equation \eqref{eq: mean exchangeability under usual SUTVA}, Assumption \ref{ass: nested IV assumptions}(\romannumeral1) only requires stability for the same version of the treatment assignment across $G$.
}


Lastly, we emphasize a subtle but important distinction between Assumption 2(\romannumeral1) and the exchangeability-type assumption in \citet{dahabreh2022generalizing}. The exchangeability-type assumption in \citet{dahabreh2022generalizing} pertains to generalizing causal effects from a clinical trial to a broader target population. Their setting includes data from both trial participants and nonparticipants. In contrast, our setup only includes observations with $G \in \{a, b\}$. Consequently, the estimands---later defined with respect to $\mathbb{E}_{P_0}[\cdot]$, either unconditionally or conditional on $\boldsymbol{X}$---are defined internally (i.e., implicitly conditional on $G \in \{a, b\}$) and should not be generalized to other contexts without additional assumptions. 
}

{\color{black}
Assumption \ref{ass: nested IV assumptions}(\romannumeral2) states that there are no defiers with respect to either version of the IV. Importantly, we do not impose any ordering between $D(0_a)$ and $D(0_b)$, nor between $D(0_a)$ and $D(1_b)$, $D(1_a)$ and $D(0_b)$, or $D(1_a)$ and $D(1_b)$. In contrast, the standard monotonicity assumption for an ordinal IV requires that, for any $z,z'$ in the support of $Z$, either $D(z)\ge D(z')$ or $D(z)\le D(z')$ holds. Therefore, Assumption \ref{ass: nested IV assumptions}(\romannumeral2) is weaker than the usual monotonicity assumption for an ordinal IV \citep{imbenslate}.
}

Assumption \ref{ass: nested IV assumptions}(\romannumeral3) states that the compliance rates under $\{0_a, 1_a\}$ and $\{0_b, 1_b\}$ are different; this would likely hold when the IV is of different versions in two strata. 

Finally, Assumption \ref{ass: nested IV assumptions}(\romannumeral4) states that if a participant is a complier under the IV pair $\{0_a, 1_a\}$, then the participant remains a complier under the IV pair $\{0_b, 1_b\}$. {  Equivalently, a noncomplier under $\{0_b, 1_b\}$ would remain a noncomplier under $\{0_a, 1_a\}$.} This assumption is central to our identification results and is most likely to hold when the IV pair  $\{0_b, 1_b\}$ is \emph{stronger}, or at least no weaker, than the IV pair $\{0_a, 1_a\}$. Empirically, the compliance rates can be calculated for both IV pairs, either marginally or conditional on a set of covariates, allowing the plausibility of Assumption \ref{ass: nested IV assumptions}(\romannumeral4) to be assessed. 

\vspace{-6pt}
\subsection{Principal strata; causal estimands; identification results}\label{subsec: identification results}
A total of $2^4 = 16$ possible subgroups can be defined based on the joint potential outcomes \[
\{D(Z=0_a), D(Z=1_a), D(Z=0_b), D(Z=1_b)\};
\]
see Table \ref{tb: principle_stratum}. Assumptions \ref{ass: nested IV assumptions}(\romannumeral2) (partial monotonicity) and \ref{ass: nested IV assumptions}(\romannumeral4) (nested IV) together eliminate $9$ out of $16$ principal strata. The population then contains the following $6$ subgroups:  
\begin{enumerate}
    \item \textbf{A}lways-\textbf{A}lways-\textbf{T}akers (\textbf{AAT}): Always-takers under both IV pairs $\{0_a, 1_a\}$ and $\{0_b, 1_b\}$;\vspace{-6pt}

    \item \textbf{SW}itchers (\textbf{SW}): Always-takers or never-takers under the IV pair $\{0_a, 1_a\}$ but compliers under the IV pair $\{0_b, 1_b\}$;\vspace{-6pt}

    \item \textbf{A}lways-\textbf{T}aker-\textbf{N}ever-\textbf{T}aker (\textbf{AT-NT}): Always-takers under the IV pair $\{0_a, 1_a\}$ but never-takers under the IV pair $\{0_b, 1_b\}$;\vspace{-6pt}

    \item \textbf{A}lways-\textbf{CO}mplier (\textbf{ACO}): Compliers under both IV pairs $\{0_a, 1_a\}$ and $\{0_b, 1_b\}$;\vspace{-6pt}

    \item \textbf{N}ever-\textbf{T}aker-\textbf{A}lways-\textbf{T}aker (\textbf{NT-AT}): Never-takers under the IV pair $\{0_a, 1_a\}$ but always-takers under the IV pair $\{0_b, 1_b\}$;\vspace{-6pt}

    \item \textbf{A}lways-\textbf{N}ever-\textbf{T}aker (\textbf{ANT}): Never-takers under both IV pairs $\{0_a,1_a\}$ and $\{0_b, 1_b\}$.\vspace{-6pt}
\end{enumerate}

\begin{table}[htbp]
  \centering
  \caption{Latent populations and their existence under Assumption \ref{ass: nested IV assumptions}(\romannumeral2) (partial monotonicity) and Assumption \ref{ass: nested IV assumptions}(\romannumeral4) (nested IV). NT: never-taker. AT: always-taker. CO: complier. DF: defier. ANT: always-never-taker. SW: switcher. AT-NT: always-taker-never-taker. ACO: always-complier. NT-AT: never-taker-always-taker. AAT: always-always-taker.}
   \resizebox{\textwidth}{!}{ \begin{tabular}{ccccccccc}
    \toprule
    $D(Z=0_a)$ & $D(Z=1_a)$ & $G=a$ & $D(Z=0_b)$ & $D(Z=1_b)$ & $G = b$ & Existence & \makecell{Which identification\\assumptions are violated?} & $S$ \\
    \midrule
    0     & 0  &NT   & 0     & 0 &NT    & Yes   &  None     & ANT \\
    1     & 0  & DF  & 0     & 0   & NT  & No    & Partial monotonicity &  \\
    0     & 1  & CO   & 0     & 0 &NT    & No    & Nested IV &  \\
    0     & 0  & NT   & 1     & 0  &DF    & No    & Partial monotonicity &  \\
    0     & 0   &NT   & 0     & 1 &CO    & Yes   &   None    & SW \\
    1     & 1   &AT   & 0     & 0   &NT  & Yes   &  None     & AT-NT \\
    1     & 0  &DF   & 1     & 0   &DF  & No    & Partial monotonicity &  \\
    1     & 0  &DF    & 0     & 1  &CO   & No    & Partial monotonicity &  \\
    0     & 1  &CO   & 1     & 0  &DF   & No    & Partial monotonicity \& Nested IV &  \\
    0     & 1  &CO   & 0     & 1  &CO   & Yes   &   None    & ACO \\
    0     & 0  &NT   & 1     & 1 &AT    & Yes   &  None     & NT-AT \\
    1     & 1   &AT  & 1     & 0  &DF   & No    & Partial monotonicity &  \\
    1     & 1  &AT   & 0     & 1 &CO    & Yes   &   None    & SW \\
    1     & 0  &DF   & 1     & 1  &AT   & No    & Partial monotonicity &  \\
    0     & 1  &CO   & 1     & 1  &AT   & No    & Nested IV &  \\
    1     & 1  &AT   & 1     & 1  &AT   & Yes   &   None    & AAT \\
    \bottomrule
    \end{tabular}}%
  \label{tb: principle_stratum}%
\end{table}%


\noindent Let $S$ denote principal stratum. Define the conditional switcher average treatment effect: 
\begin{align*}
\text{SWATE}_{P_0}(\boldsymbol{X}) &:=\mathbb{E}_{P_0}[Y(D=1)-Y(D=0)\mid S=\text{SW}, \boldsymbol{X}],
\end{align*}
and the conditional always-complier average treatment effect:
\begin{align*}
 \text{ACOATE}_{P_0}(\boldsymbol{X}) &:=\mathbb{E}_{P_0}[Y(D=1)-Y(D=0)\mid S=\text{ACO}, \boldsymbol{X}].
\end{align*}
\noindent Importantly, both $\text{SWATE}_{P_0}(\boldsymbol{X})$ and $\text{ACOATE}_{P_0}(\boldsymbol{X})$ are the average ``per-protocol" effect of $D$ on  $Y$, although on different subpopulations, within strata defined by observed covariates $\boldsymbol{X}$. Averaging $\text{SWATE}_{P_0}(\boldsymbol{X})$ and $\text{ACOATE}_{P_0}(\boldsymbol{X})$ then yields the SWitcher Average Treatment Effect (SWATE): 
\begin{align*}
    &\text{SWATE}_{P_0}:=\mathbb{E}_{P_0}[Y(D=1)-Y(D=0)\mid S=\text{SW}],
\end{align*}
and the Always-COmplier Average Treatment Effect (ACOATE):
\begin{align*}
    &\text{ACOATE}_{P_0}:=\mathbb{E}_{P_0}[Y(D=1)-Y(D=0)\mid S=\text{ACO}].
\end{align*}

{ 
\begin{remark}[SWATE and ACOATE under partial exclusion restriction]
   Under the weaker partial exclusion restriction discussed in Remark \ref{remark: partial ER}, an additional no-interaction assumption is required to ensure that $Y(D = 1)$ and $Y(D = 0)$---and hence the SWATE and ACOATE---are well defined and identified; see Supplemental Material \ref{subsec: G has direct effect} for details. 
\end{remark}
}
\noindent Theorem \ref{thm_identification} shows that the ACOATE and SWATE can be nonparametrically identified from observed data under Assumptions \ref{ass: standard IV assumption} and \ref{ass: nested IV assumptions}.

\begin{theorem}
    Under Assumptions \ref{ass: standard IV assumption}(\romannumeral1)-(\romannumeral4) and \ref{ass: nested IV assumptions}(\romannumeral1)-(\romannumeral4),   $\text{SWATE}_{P_0} (\boldsymbol{X})$, $\text{ACOATE}_{P_0}(\boldsymbol{X})$, $\text{SWATE}_{P_0}$, and $\text{ACOATE}_{P_0}$ can be nonparametrically identified as follows:
    \small\begin{align*}
    &\text{ACOATE}_{P_0}(\boldsymbol{X}) = \theta_{\text{ACO},P_0}(\boldsymbol{X}):=\frac{\delta_{a,P_0}(\boldsymbol{X})}{\eta_{a,P_0}(\boldsymbol{X})};
        ~~\text{SWATE}_{P_0}(\boldsymbol{X}) = \theta_{\text{SW},P_0}(\boldsymbol{X}):=\frac{\delta_{b,P_0}(\boldsymbol{X})-\delta_{a,P_0}(\boldsymbol{X})}{\eta_{b,P_0}(\boldsymbol{X})-\eta_{a,P_0}(\boldsymbol{X})};\\
        &\text{ACOATE}_{P_0} = \Psi_{\text{ACO},P_0}:= \frac{\mathbb{E}_{P_0}[\delta_{a,P_0}(\boldsymbol{X})]}{\mathbb{E}_{P_0}[\eta_{a,P_0}(\boldsymbol{X})]};
        ~~\text{SWATE}_{P_0} = \Psi_{\text{SW},P_0}:=\frac{\mathbb{E}_{P_0}[\delta_{b,P_0}(\boldsymbol{X})-\delta_{a,P_0}(\boldsymbol{X})]}{\mathbb{E}_{P_0}[\eta_{b,P_0}(\boldsymbol{X})-\eta_{a,P_0}(\boldsymbol{X})]},
    \end{align*}
    \normalsize where, for $P\in \mathcal{P}$ and $g \in \{a, b\}$, $\delta_{g, P}(\boldsymbol{X}) := \mathbb{E}_P[Y\mid Z=1_g, \boldsymbol{X}]-\mathbb{E}_P[Y\mid Z=0_g, \boldsymbol{X}]$ and
    $\eta_{g,P}(\boldsymbol{X}): = \mathbb{E}_P[D\mid Z=1_g, \boldsymbol{X}]-\mathbb{E}_P[D\mid Z=0_g, \boldsymbol{X}].$
    \label{thm_identification}
\end{theorem}
\noindent Unlike ACOATE or SWATE, the causal effect in other latent subgroups, including AAT, AT-NT, NT-AT, and ANT, cannot be nonparametrically identified.

\begin{remark}[Alternative identification]
 An alternative identification strategy---assuming (1) no effect modification by $G$ within principal strata and (2) stable principal strata composition across $G$ within baseline covariates---can also identify the estimands in Theorem \ref{thm_identification}; see Supplemental Material \ref{subsec: alternative nested IV assumption} for details.
\end{remark}

{\color{black}
\begin{remark}[Alternative frameworks]
Supplemental Material S1 compares our proposed framework to alternative frameworks that conceptualize the trial stage $G$ as a baseline covariate. We contrast Rubin's science table \citet{rubin2005causal} corresponding to each framework and discuss why the proposed framework is most suitable for the PLCO trial scenario as well as enabling identification and estimation of novel estimands like ACOATE and SWATE.
\end{remark}
}

\vspace{-10pt}
{\color{black}
\subsection{Interpretation of the conventional Wald estimand}
\label{subsec: convenctional Wald SUTVA violated}\vspace{-6pt}
A conventional IV analysis that ignores different versions of the IV would collapse them into a single binary IV $Z^*$, defined by $Z^*=1$ if $Z\in \{1_a, 1_b\}$, and $0$ if $Z\in \{0_a, 0_b\}$. Under this construction, one could consider the following conditional Wald estimator-style statistical functional\citep{FROLICH200735}:
\begin{equation}\label{eq: conditional wald SUTVA violate}
    \theta_{\text{CW},P_0}(\boldsymbol{X}) :=\frac{\mathbbm{E}_{P_0}[Y\mid Z^*=1, \boldsymbol{X} ]-\mathbbm{E}_{P_0}[Y\mid Z^*=0, \boldsymbol{X}]}{\mathbbm{E}_{P_0}[D\mid Z^*=1,\boldsymbol{X}]-\mathbbm{E}_{P_0}[D\mid Z^*=0,\boldsymbol{X}]}.
\end{equation}

It is of interest to understand what estimand the statistical functional in Equation \eqref{eq: conditional wald SUTVA violate} targets. Below, we provide two interpretations.

First, under standard IV assumptions (Assumption \ref{ass: standard IV assumption}) and an additional testable assumption $G \indep Z^*\mid\boldsymbol{X}$, Equation \eqref{eq: conditional wald SUTVA violate} can be understood as the average ``per-protocol" effect, $Y(D = 1) - Y(D = 0)$, among the subpopulation $\{D(Z \in \{1_a, 1_b\}, G) > D(Z \in \{0_a, 0_b\}, G)\}$, where we use the vector-IV representation of potential outcomes. It is important to note that this subpopulation is defined based on the observed value of $G$ and therefore is not fixed prior to treatment assignment. Consequently, it does not constitute a principal stratum in the traditional Frangakis--Rubin sense \citep{frangakis2002principal}.

The assumption $G \indep Z^* \mid \boldsymbol{X}$ requires that treatment assignment probabilities do not depend on $G$, conditional on $\boldsymbol{X}$: $P_0(Z=1_a \mid G=a, \boldsymbol{X}) = P_0(Z=1_b \mid G=b, \boldsymbol{X})$ and $P_0(Z=0_a \mid G=a, \boldsymbol{X}) = P_0(Z=0_b \mid G=b, \boldsymbol{X})$. This assumption is testable and holds in the PLCO trial, where $Z$ is randomized within each stage. 

Second, Proposition \ref{prop: violation of SUTVA} states that, if we are willing to further make the nested IV assumptions (Assumption \ref{ass: nested IV assumptions}), then this ``conventional" Wald estimand can be interpreted as a weighted average of the conditional ACOATE and the conditional SWATE:
\begin{proposition}\label{prop: violation of SUTVA}
    Under Assumption \ref{ass: standard IV assumption}, Assumption \ref{ass: nested IV assumptions}, and an additional assumption $G\indep Z^*\mid \boldsymbol{X}$, we have
   $
        \theta_{\text{CW},P_0}(\boldsymbol{X}) = \omega_1 \cdot \text{SWATE}_{P_0}(\boldsymbol{X})+\omega_2\cdot\text{ACOATE}_{P_0}(\boldsymbol{X}),
   $
    where
    \begin{align*}
        &\omega_1 = \frac{P_0(S=\text{SW}\mid \boldsymbol{X})P_0(G=b\mid \boldsymbol{X})}{P_0(S=\text{SW}\mid \boldsymbol{X})P_0(G=b\mid \boldsymbol{X})+P_0(S = \text{ACO}\mid \boldsymbol{X})},~~\text{and}\\
        &\omega_2= \frac{P_0(S = \text{ACO}\mid \boldsymbol{X})}{P_0(S = \text{SW}\mid \boldsymbol{X})P_0(G=b\mid \boldsymbol{X})+P_0(S=\text{ACO}\mid \boldsymbol{X})}.
    \end{align*}
\end{proposition}

Finally, we note that the marginalized Wald estimand,
\begin{align*}
    \frac{\mathbb{E}_{P_0}[\mathbb{E}_{P_0}[Y\mid Z^*=1,\boldsymbol{X}]-\mathbb{E}_{P_0}[Y\mid Z^*=0,\boldsymbol{X}]]}{\mathbb{E}_{P_0}[\mathbb{E}_{P_0}[D\mid Z^*=1,\boldsymbol{X}]-\mathbb{E}_{P_0}[D\mid Z^*=0, \boldsymbol{X}]]},
\end{align*}
does not, in general, equal a weighted average of SWATE and ACOATE.
}
\subsection{Practical relevance of ACOATE and SWATE}

\label{subsec: relevance of ACOATE and SWATE}

The causal estimands $\text{ACOATE}$ and $\text{SWATE}$ are examples of principal treatment effects \citep{frangakis2002principal}. As discussed in Section \ref{subsec: intro IV and generalizability}, the practical relevance of principal stratum effects has often been subject to scrutiny \citep{pearl2011principal,dawid2012imagine}. In response to the critiques of \citet{pearl2011principal}, we argue that SWATE and ACOATE provide valuable tools in a comprehensive IV analysis by facilitating the assessment of effect homogeneity and generalizability. For instance, researchers can obtain a crude but informative evaluation by directly comparing ACOATE and SWATE, potentially conditional on $\boldsymbol{X}$. This approach may offer greater power than directly comparing treatment effects among compliers under two versions of the IV. We will formalize this testing strategy in Section \ref{sec: test homogeneity}.

{ 
Moreover, ACOATE and SWATE are meaningful causal parameters in their own right. For one thing, SWATE is relevant for implementation science, whose goal is to bridge the gap between clinical trials that generate evidence and the real-world adoption of evidence-based interventions. If a substantial treatment effect is revealed among switchers, this would motivate more proactive efforts to implement the intervention in healthcare systems and communities. Conversely, if no meaningful effect is detected among switchers but a substantial effect is observed among always-compliers, this information is equally valuable and can help guide the more efficient allocation of often limited implementation resources.} 

{  ACOATE and SWATE also underpin microeconomic theory. For example, the always-compliers and switchers defined in our paper correspond to Type A and Type B mothers in \citet{guo2025rationed}. \citet{guo2025rationed} developed an economic theory of desired and undesired fertility, which predicts that $\text{SWATE} - \text{ACOATE} > 0$. This prediction is then empirically validated using structural equation models.} 

Although individual study participants cannot be directly classified into latent subgroups, the distributions of observed covariates among always-compliers and switchers can nevertheless be characterized. Proposition \ref{prop_covariates_dis} provides an identification formula for functions of observed covariates within these two subpopulations.

\begin{proposition}
   Let $g(\cdot)$ denote a generic function with finite first moment. The mean of $g(\boldsymbol{X})$ among switchers and always-compliers can be identified as follows:
   {\small
    \begin{align*}
&\mathbb{E}_{P_0}[g(\boldsymbol{X})\mid S=\text{ACO}] = \mathbb{E}_{P_0}\Bigg[g(\boldsymbol{X})\frac{\eta_{a,P_0}(\boldsymbol{X})}{\mathbb{E}_{P_0}[\eta_{a,P_0}(\boldsymbol{X})]}\Bigg],\\
&\mathbb{E}_{P_0}[g(\boldsymbol{X})\mid S=\text{SW}] = \mathbb{E}_{P_0}\Bigg[g(\boldsymbol{X})\frac{\eta_{b,P_0}(\boldsymbol{X})-\eta_{a,P_0}(\boldsymbol{X})}{\mathbb{E}_{P_0}[\eta_{b,P_0}(\boldsymbol{X})-\eta_{a,P_0}(\boldsymbol{X})]}\Bigg]
    \end{align*}
    }
\label{prop_covariates_dis}
\end{proposition} \vspace{-35pt}
\noindent Based on Proposition \ref{prop_covariates_dis}, the mean and variance of each observed covariate---and more generally, certain functions of the observed covariates (e.g., a risk score)---can be characterized within always-compliers and switchers and compared with those in the overall population. {\color{black} Similar results characterizing moments of observed covariates within latent principal strata can be found in, for example, \citet{ABADIE2003231} and \citet{goff2025doesividentificationrestrict}. }

\vspace{-6pt}
\section{Extensions}
\label{sec: connection and extension}\vspace{-6pt}
\subsection{Connection to the vector IV literature}\label{subsec: connection to vector IV}\vspace{-6pt}

{\color{black}
Under the vector-IV representation (see Section \ref{subsec: design two version of Z}), we compare our partial monotonicity assumption (PM) with (1) the partial monotonicity assumption of \citet{mogstad2021causal} and \citet{blackwell2023noncompliance} (PM2), (2) the limited monotonicity assumption of \citet{hoff2025limited} (LiM), and (3) the vector monotonicity assumption of \citet{goff2024vector} (VM), when both $Z^\prime$ and $G$ are binary. In Supplemental Material \ref{sec: literature on multiple IV and monotonicity, supp}, we review these identification assumptions and discuss their relationships. We show that PM2 and VM are equivalent, while PM, PM2 (VM), and LiM imply different collections of principal strata. Moreover, the set of seven principal strata implied by PM together with the nested IV assumption is neither a superset nor a subset of the principal strata implied by PM2 (VM) or LiM, highlighting the novelty of the identification assumptions developed in this article. Supplemental Material \ref{sec: literature on multiple IV and monotonicity, supp} also discusses the differences between the SWATE and the principal stratum effects proposed in \citet{goff2024vector}.}
\vspace{-6pt}
{ 
\subsection{Sensitivity to violation of the mean effect exchangeability assumption}\label{subsec: violate exchangeability}
We assess deviations from Assumption 2(\romannumeral1) using a sensitivity analysis. To this end, we define four sensitivity functions,
$\{h^{Y,g}(\boldsymbol{x}), h^{D,g}(\boldsymbol{x}), g \in \{a,b\}\}$:
\vspace{-10pt}
{\small\begin{align*}
h^{Y,g}(\boldsymbol{x}) &= \mathbb{E}_{P_0}[Y(Z=1_g) - Y(Z=0_g) \mid G=a, \boldsymbol{X} = \boldsymbol{x}] - \mathbb{E}_{P_0}[Y(Z=1_g) - Y(Z=0_g) \mid G=b,  \boldsymbol{X} = \boldsymbol{x}],\\
h^{D,g}(\boldsymbol{x}) &= \mathbb{E}_{P_0}[D(Z=1_g) - D(Z=0_g) \mid G=a,  \boldsymbol{X} = \boldsymbol{x}] - \mathbb{E}_{P_0}[D(Z=1_g) - D(Z=0_g) \mid G=b,  \boldsymbol{X} = \boldsymbol{x}],
\end{align*}}

\vspace{-10PT}
\noindent which quantify deviations from the mean effect exchangeability assumption.

Proposition \ref{prop: sensitivity analysis for mean exchangeability} establishes identification results for fixed sensitivity functions.

\begin{proposition}
\label{prop: sensitivity analysis for mean exchangeability}
   Suppose Assumption \ref{ass: standard IV assumption} and Assumption \ref{ass: nested IV assumptions}(\romannumeral2), (\romannumeral3), and (\romannumeral4) hold. For fixed sensitivity functions $\{h^{Y,g}(\boldsymbol{X}), h^{D,g}(\boldsymbol{X}), ~g \in \{a,b\}\}$, the estimands $\text{SWATE}_{P_0}(\boldsymbol{X})$, $\text{ACOATE}_{P_0}(\boldsymbol{X})$, $\text{SWATE}_{P_0}$, and $\text{ACOATE}_{P_0}$ can be identified as
\begin{align*}
    &\text{ACOATE}_{P_0}(\boldsymbol{X}) = \frac{\delta_{a,P_0}(\boldsymbol{X}) - h^{Y,a}(\boldsymbol{X}) P_0(G=b \mid \boldsymbol{X})}{\eta_{a,P_0}(\boldsymbol{X}) - h^{D,a}(\boldsymbol{X}) P_0(G=b \mid \boldsymbol{X})},\\
    &\text{SWATE}_{P_0}(\boldsymbol{X}) = \frac{\delta_{b,P_0}(\boldsymbol{X}) - \delta_{a,P_0}(\boldsymbol{X}) + \big(h^{Y,b}(\boldsymbol{X}) P_0(G=a \mid \boldsymbol{X}) + h^{Y,a}(\boldsymbol{X}) P_0(G=b \mid \boldsymbol{X})\big)}{\eta_{b,P_0}(\boldsymbol{X}) - \eta_{a,P_0}(\boldsymbol{X}) + \big(h^{D,b}(\boldsymbol{X}) P_0(G=a \mid \boldsymbol{X}) +h^{D,a}(\boldsymbol{X}) P_0(G=b \mid \boldsymbol{X})\big)},\\
    &\text{ACOATE}_{P_0} = \frac{\mathbb{E}_{P_0}[\delta_{a,P_0}(\boldsymbol{X}) - h^{Y,a}(\boldsymbol{X}) P_0(G=b \mid \boldsymbol{X})]}{\mathbb{E}_{P_0}[\eta_{a,P_0}(\boldsymbol{X}) - h^{D,a}(\boldsymbol{X}) P_0(G=b \mid \boldsymbol{X})]},\\
    &\text{SWATE}_{P_0} = \frac{\mathbb{E}_{P_0}[\delta_{b,P_0}(\boldsymbol{X}) - \delta_{a,P_0}(\boldsymbol{X}) + (h^{Y,b}(\boldsymbol{X}) P_0(G=a \mid \boldsymbol{X}) + h^{Y,a}(\boldsymbol{X}) P_0(G=b \mid \boldsymbol{X}))]}{\mathbb{E}_{P_0}[\eta_{b,P_0}(\boldsymbol{X}) - \eta_{a,P_0}(\boldsymbol{X}) + (h^{D,b}(\boldsymbol{X}) P_0(G=a \mid \boldsymbol{X}) +h^{D,a}(\boldsymbol{X}) P_0(G=b \mid \boldsymbol{X}))]}.
\end{align*}
\end{proposition}
}

\subsection{Sensitivity of the estimand to violation of the nested IV assumption}\label{subsec: violate nested IV}\vspace{-6pt}
We consider violations of the nested IV assumption. When the nested IV assumption fails, two additional principal strata, those with $\{D(Z=0_a)=0,D(Z=1_a)=1,D(Z=0_b)=0,D(Z=1_b)=0\}$ or $\{D(Z=0_a)=0,D(Z=1_a)=1,D(Z=0_b)=1,D(Z=1_b)=1\}$, come into existence. We refer to them as \textbf{D}e\textbf{F}iers to the \textbf{N}ested IV assumption (\textbf{DFN}).

\begin{proposition}
    Under Assumption \ref{ass: standard IV assumption} (\romannumeral1)-(\romannumeral4) and \ref{ass: nested IV assumptions}(\romannumeral1)-(\romannumeral3) but not \ref{ass: nested IV assumptions}(\romannumeral4), we have
    \begin{align}
      \theta_{\text{SW},P_0}(\boldsymbol{x}) =  \lambda_{P_0}(\boldsymbol{x})\cdot\text{SWATE}_{P_0}(\boldsymbol{x}) + \{1-\lambda_{P_0}(\boldsymbol{x})\}\cdot\text{DFNATE}_{P_0}(\boldsymbol{x}), \label{eq: violation of nested iv pair assumption}
    \end{align}
    where \vspace{-10pt}
    {\small
    \begin{align*}
        \lambda_{P_0}(\boldsymbol{x}) = \frac{P_0(S=\text{SW}\mid \boldsymbol{X}=\boldsymbol{x})}{P_0(S=\text{SW}\mid \boldsymbol{X}=\boldsymbol{x})+P_0(S=\text{DFN}\mid \boldsymbol{X}=\boldsymbol{x})}, \\
        \text{DFNATE}_{P_0}(\boldsymbol{x}) = \mathbbm{E}_{P_0}[Y(D=1)-Y(D=0)\mid S=\text{DFN}, \boldsymbol{X}=\boldsymbol{x}].
    \end{align*}}
    \label{prop: violation of nested IV pair assumption}
\end{proposition}
\vspace{-40pt}\noindent Proposition \ref{prop: violation of nested IV pair assumption} states that without Assumption \ref{ass: nested IV assumptions}(\romannumeral4), the statistical functional $\theta_{SW,P_0}$ equals a weighted average of the conditional average treatment effects for \textbf{SW} and \textbf{DFN}. If we further make Assumption \ref{ass: nested IV assumptions}(\romannumeral4), then $P_0(S=\text{DFN}\mid \boldsymbol{X}=\boldsymbol{x}) = 0$, $\lambda_{P_0}(\boldsymbol{x}) = 1$, and therefore the statistical functional $\theta_{\text{SW},P_0}(\boldsymbol{x})$ identifies $\text{SWATE}_{P_0}(\boldsymbol{\boldsymbol{x}})$. As $\lambda_{P_0}(\boldsymbol{x}) \in [0, 1]$, $\theta_{\text{SW},P_0}(\boldsymbol{x})$ lies in the convex hull of $\text{SWATE}_{P_0}(\boldsymbol{x})$ and $\text{DFNATE}_{P_0}(\boldsymbol{x})$. Proposition \ref{prop: violation of nested IV pair assumption} also implies that if we further assume $\text{SWATE}_{P_0}(\boldsymbol{x}) = \text{DFNATE}_{P_0}(\boldsymbol{x})$, then $\text{SWATE}_{P_0}(\boldsymbol{x})$ can still be identified by $\theta_{\text{SW},P_0}(\boldsymbol{x})$. Finally, rearranging \eqref{eq: violation of nested iv pair assumption} gives the following bias formula:
\begin{align}
    \text{Bias} := \theta_{\text{SW},P_0}(\boldsymbol{x})-\text{SWATE}_{P_0}(\boldsymbol{x})\nonumber
    = -\{1-\lambda_{P_0}(\boldsymbol{x})\}\left\{\text{SWATE}_{P_0}(\boldsymbol{x})-\text{DFNATE}_{P_0}(\boldsymbol{x})
    \right\},
    \label{eq: bias formula}
\end{align}
\noindent when the nested IV assumption is violated, and this expression can be used as a basis for a sensitivity analysis by specifying $\lambda_{P_0}(\boldsymbol{x})$ and $\text{DFNATE}_{P_0}(\boldsymbol{x})$ as sensitivity functions.

\vspace{-6pt}
{ 
\subsection{Partial identification region}
\label{subsec: partial identification}\vspace{-6pt}
In addition to identifying treatment effects for principal strata, valid IVs also allow for partial identification of treatment effects in the overall population; see, for example, \citet{swanson2018partial} for a review and \citet{duarte2024automated} and \citet{levis2025covariate} for recent developments. In Supplemental Material \ref{sec: partial identification, supp}, we discuss how to derive the partial identification region for ATE with binary outcome---namely, $P_0(Y(D=1)=1)-P_0(Y(D=0)=1)$---under Assumptions \ref{ass: standard IV assumption} and \ref{ass: nested IV assumptions}, by characterizing the set of $\{P_0(Y(D=1)=1), P_0(Y(D=0)=1)\}$ values that are consistent with the observed data distribution.

\vspace{-6pt}
\subsection{Multiple IV pairs}
\label{subsec: extension multiple IVs}\vspace{-6pt}
In Supplemental Material \ref{subsec: multiple IV pairs, supp}, we also discuss how the nested IV framework can be naturally extended to incorporate multiple IV pairs. We first formalize the nested IV assumption as a binary relation when $G$ takes more than two values. We then examine the case of three IV pairs, which gives rise to three possible structures: a fork, a collider, and a chain, in Supplemental Material \ref{subsec: three nested IV pairs, supp}. More general causal estimands, termed eager-switchers average treatment effect and reluctant-switchers average treatment effect, can be defined with three IV pairs; however, they cannot be identified without imposing additional assumptions unless they are reduced to the switcher average treatment effect studied in Section \ref{sec: nested IV design}; see Supplemental Material \ref{subsec: three nested IV pairs, supp} for details.
}

%% file: 3.Estimation.tex
\vspace{-15pt}
\section{Estimation and inference}
\label{sec: estimation}\vspace{-6pt}
Estimation of the ACOATE follows immediately by applying standard methods to the IV pair $(0_a, 1_a)$; see, e.g., \citet{FROLICH200735} and \citet{kennedy2022semiparametric}. {\color{black} In Supplemental Material \ref{subsec: wald estimator theorem, supp}, we also propose a simple Wald-type estimator for SWATE when the identification assumptions hold unconditionally on $\boldsymbol{X}$. In what follows, we focus on efficient estimation and inference of the SWATE under the assumptions in Section \ref{subsec: design notation and assumption}. }

\subsection{Nonparametric efficient estimators}
\label{subsec: EIF estimator}
\vspace{-6pt}
For a pathwise differentiable parameter, its nonparametric efficiency bound equals the variance of the canonical gradient at $P_0$ in a nonparametric model. Theorem \ref{thm_gradient} derives the canonical gradient and hence the nonparametric efficiency bound for $\Psi_{\text{SW}, P}$. 

\begin{theorem}\label{thm_gradient}
    The canonical gradient of $\Psi_{\text{SW}, P}$
    at $P_0$ equals $D_{P_0}$, where
    {\footnotesize\begin{equation*}
    \begin{split}
        D_{P}&=
        (z,\boldsymbol{x},d,y)\mapsto\frac{1}{\mathbb{E}_P[\eta_{b,P}(\boldsymbol{X})-\eta_{a,P}(\boldsymbol{X})]}\Bigg\{\frac{\mathbbm{1}\{z=1_b\}}{P(Z=1_b\mid \boldsymbol{X}=\boldsymbol{x})}\Big[y-\mathbb{E}_P[Y\mid Z=1_b,\boldsymbol{X}=\boldsymbol{x}]\Big]\\ 
        &-\frac{\mathbbm{1}\{z=0_b\}}{P(Z=0_b\mid \boldsymbol{X}=\boldsymbol{x})}\Big[y-\mathbb{E}_P[Y\mid Z=0_b,\boldsymbol{X}=\boldsymbol{x}]\Big]\\
        &-\frac{\mathbbm{1}\{z=1_a\}}{P(Z=1_a\mid \boldsymbol{X}=\boldsymbol{x})}\Big[y-\mathbb{E}_P[Y\mid Z=1_a,\boldsymbol{X}=\boldsymbol{x}]\Big] +\frac{\mathbbm{1}\{z=0_a\}}{P(Z=0_a\mid \boldsymbol{X}=\boldsymbol{x})}\Big[y-\mathbb{E}_P[Y\mid Z=0_a, \boldsymbol{X}=\boldsymbol{x}]\Big]\\
        &+\delta_{b,P}(\boldsymbol{x})-\delta_{a,P}(\boldsymbol{x})\Bigg\}- \frac{\Psi_{SW,P}}{\mathbb{E}_P[\eta_{b,P}(\boldsymbol{X})-\eta_{a,P}(\boldsymbol{X})]}\Bigg\{\frac{\mathbbm{1}\{z=1_b\}}{P(Z=1_b\mid \boldsymbol{X}=\boldsymbol{x})}\Big[d-\mathbb{E}_P(D\mid Z=1_b, \boldsymbol{X}=\boldsymbol{x})\Big] \\
        &-\frac{\mathbbm{1}\{z=0_b\}}{P(Z=0_b\mid \boldsymbol{X}=\boldsymbol{x})}\Big[d-\mathbb{E}_P(D\mid Z=0_b,\boldsymbol{X}=\boldsymbol{x})\Big]-\frac{\mathbbm{1}\{Z=1_a\}}{P(Z=1_a\mid \boldsymbol{X}=\boldsymbol{x})}\Big[d-\mathbb{E}_P(D\mid Z=1_a, \boldsymbol{X}=\boldsymbol{x})\Big] \\
        &+\frac{\mathbbm{1}\{Z=0_a\}}{P(Z=0_a \mid \boldsymbol{X}=\boldsymbol{x})}\Big[d-\mathbb{E}_P(D\mid Z=0_a,\boldsymbol{X}=\boldsymbol{x})\Big]+\eta_{b,P}(\boldsymbol{x})-\eta_{a,P}(\boldsymbol{x})\Bigg\}.
        \end{split}
    \end{equation*}}\normalsize
    Therefore, the nonparametric efficiency bound is $\mathbb{E}_{P_0}[D_{P_0}^2(Z,X,D,Y)]$.
\end{theorem}

{  The efficient influence function in Theorem \ref{thm_gradient} motivates both an asymptotically efficient one-step estimator and an estimating equation-based estimator based on cross-fitting.} The one-step estimator applies a first-order bias correction to a plug-in estimator \citep{Pfanzagl1990}, while the estimating equation-based estimator solves an estimating equation derived from the efficient influence function \citep{vanderLaan2003, chernozhukov2018double}. We focus on the estimating equation-based estimator and defer the one-step estimator to Supplemental Material \ref{subsec: additional information for estimation, supp}. Define
\[
\pi_{P} = (\pi_{0_a,P},\pi_{1_a,P},\pi_{0_b,P},\pi_{1_b,P}),
~\text{where}~\pi_{z,P} = P(Z=z\mid \boldsymbol{X} = \boldsymbol{x}),
\]
\[
\mu_{Y,P} = (\mu_{Y,0_a,P},\mu_{Y,1_a,P},\mu_{Y,0_b,P},
 \mu_{Y,1_b,P}),~\text{where}~ \mu_{Y,z,P} = \mathbb{E}_P[Y\mid Z = z, \boldsymbol{X} = \boldsymbol{x}], 
\]
\[
\mu_{D,P} = (\mu_{D,0_a,P},\mu_{D,1_a,P},\mu_{D,0_b,P},\mu_{D,1_b,P}),~\text{where}~\mu_{D,z,P} = \mathbb{E}_P[D\mid Z = z, \boldsymbol{X} = \boldsymbol{x}],
\]
for $z\in \{0_a,1_a,0_b,1_b\}$. The canonical gradient defines an estimating function of $\Psi_{P_0}$, denoted as 
$D'(o;\Psi_P,\pi_{P},\mu_{Y,P},\mu_{D,P})$, where
\small\begin{equation*}
    \begin{split}
&D'(o;\Psi_P,\pi_{P},\mu_{Y,P},\mu_{D,P})\\=&\Bigg\{\frac{\mathbbm{1}\{z=1_b\}}{\pi_{1_b,P}(\boldsymbol{x})}\Big[y-\mu_{Y,1_b,P}(\boldsymbol{x})\Big] 
        -\frac{\mathbbm{1}\{z=0_b\}}{\pi_{0_b,P}(\textbf{x})}\Big[y-\mu_{Y,0_b,P}(\boldsymbol{x})\Big]
        -\frac{\mathbbm{1}\{z=1_a\}}{\pi_{1_a,P}(\textbf{x})}\Big[y-\mu_{Y,1_a,P}(\boldsymbol{x})\Big] \\+&\frac{\mathbbm{1}\{z=0_a\}}{\pi_{0_a,P}(\textbf{x})}\Big[y-\mu_{Y,0_a,P}(\boldsymbol{x})\Big]
        +\delta_{b_P}(\boldsymbol{x})-\delta_{a,P}(\boldsymbol{x})\Bigg\}- \Psi_{SW,P}\Bigg\{\frac{\mathbbm{1}\{z=1_b\}}{\pi_{1_b,P}(\textbf{x})}\Big[d-\mu_{D,1_b,P}(\boldsymbol{x})\Big] \\
        -&\frac{\mathbbm{1}\{z=0_b\}}{\pi_{0_b,P}}\Big[d-\mu_{D,0_b,P}(\boldsymbol{x})\Big]-\frac{\mathbbm{1}\{Z=1_a\}}{\pi_{1_a,P}(\boldsymbol{x})}\Big[d-\mu_{D,1_a,P}(\boldsymbol{x})\Big] +\frac{\mathbbm{1}\{Z=0_a\}}{\pi_{0_a,P}(\boldsymbol{x})}\Big[d-\mu_{D,0_a,P}(\boldsymbol{x})\Big]\\
        &+\eta_{b,P}(\boldsymbol{x})-\eta_{a,P}(\boldsymbol{x})\Big\}.
        \end{split}
    \end{equation*}\normalsize
\noindent Consider a $K$-fold random partition $\{I_k\}_{k=1}^K$ of indices $[n]=\{1,...,n\}$, and define $I_k^c:=\{1,...,n\}- I_k$. For simplicity, we assume $I_k$'s have the same cardinality, denoted as $n_k$. For each $k \in [K]$, construct estimators $\widehat{\pi}_{n,k}$, $\widehat{\mu}_{D,n,k}$, and $\widehat{\mu}_{Y,n,k}$ of $\pi_{P_0}$, $\mu_{Y,P_0}$, and $\mu_{D,P_0}$, respectively, each based on sample $\{O_i\}_{i\in I_k^c}$. Let $\mathbb{P}_{n,k}$ be the empirical distribution of $\{O_i\}_{i\in I_k}$, and we use $\mathbb{P}_{n,k}\{\cdot\}$ to denote averaging with respect to the empirical measure $\mathbb{P}_{n,k}$. Let $\widehat{P}_{n,k}$ be any distribution such that $f(Z \mid \boldsymbol{X})$ is compatible with $\widehat{\pi}_{n,k}$, $f(Y \mid Z, \boldsymbol{X})$ is compatible with $\widehat{\mu}_{Y,n,k}$, $f(D \mid Z, \boldsymbol{X})$ is compatible with   $\widehat{\mu}_{D,n,k}$, and the marginal distribution of $\widehat{\mu}_{D,n,k}(z, \boldsymbol{X}_i)$ is given by its empirical distribution for $z\in\{0_a,1_a,0_b,1_b\}$ in the $k$-th fold. With this notation, $\widehat{\pi}_{n,k}$, $\widehat{\mu}_{D,n,k}$, and $\widehat{\mu}_{Y,n,k}$ can also be written as $\pi_{\hat{P}_{n,k}}$,  $\mu_{Y,\hat{P}_{n,k}}$ and $\mu_{D,\hat{P}_{n,k}}$, respectively. Observe that $\mathbb{E}_{P_0}[D'(O;\Psi_{P_0},\pi_{P_0},\mu_{Y,P_0},\mu_{D,P_0})]=0$. An estimating equation-based estimator $\widehat{\psi}_{ee}$ of the target parameter can be constructed as follows. 



\vspace{-6pt}
\begin{description}
    \item[Step I]~For each $k\in[K]$, solve the estimating equation 
    $\mathbbm{P}_{n,k}\{D'(\psi,\widehat{\pi}_{n,k},\widehat{\mu}_{Y,n,k},$ $\widehat{\mu}_{D,n,k})\} = 0$. Denote the solution as $\widehat{\psi}_{ee,k}$.\vspace{-6pt}

\item[Step II]~Construct the final cross-fitted estimating equation-based estimator as $\widehat{\psi}_{ee} = \frac{1}{K}\sum_{k=1}^K \widehat{\psi}_{ee,k}$.
\end{description}

\vspace{-6pt}
\noindent Theorem \ref{thm_one_step} establishes the property of $\widehat{\psi}_{ee}$. 

\begin{theorem}
    Under regularity conditions specified in Supplemental Material \ref{subsubsec: regularity conditions for ee and os, supp}, 
    we have
    \begin{equation}
        \sqrt{n}\left(\widehat{\psi}_{ee} - \Psi_{\text{SW},P_0}\right) = \sqrt{n}\mathbbm{P}_n D_{P_0} +o_p(1). \label{eq: ASL of psi_ee}    \end{equation}\label{thm_one_step}
\end{theorem}
\vspace{-20pt}
\noindent If we further assume $P_0D_{P_0}^2<\infty$, then we have:\vspace{-6pt}
\begin{equation*}
    \sqrt{n}\left(\widehat{\psi}_{ee} - \Psi_{\text{SW},P_0}\right) \rightsquigarrow \mathcal{N}\left(0,P_0D^2_{P_0}\right).
\end{equation*}\vspace{-6pt}
The asymptotic variance can be estimated by $\frac{1}{n}\sum_{k=1}^K n_k\mathbbm{P}_{n,k}D^2_{\widehat{P}_{n,k}}$. 

 \begin{remark}[Rate of convergence]
Supplemental Material \ref{subsec: additional information for estimation, supp} provides further details on constructing the one-step estimator $\widehat{\psi}_{os}$. Property \eqref{eq: ASL of psi_ee} also holds for $\widehat{\psi}{os}$ under a slightly different set of regularity conditions. Specifically, for the one-step estimator to be asymptotically linear, all nuisance function estimators must converge at a rate of $o_p(n^{-1/4})$. In contrast, the estimating equation-based estimator is rate doubly robust \citep{mixed_biased}, meaning that $\widehat{\psi}_{ee}$ is asymptotically linear with influence function $D_{P_0}$ as long as
     \begin{align*}
            \Big\Vert \pi_{z,P_0}(\cdot)-\pi_{z,\widehat{P}_{n,k}}(\cdot)\Big\Vert_{L^2(P_0)}\Big\Vert \mu_{B,z,P_0}(z,\cdot)-\mu_{B,z,\widehat{P}_{n,k}}(\cdot)\Big\Vert_{L^2(P_0)}=o_p(1/\sqrt{n}),
        \end{align*}
        for $z\in \{0_a,1_a,0_b,1_b\}$, $k \in \{1,...,K\}$, and $B=Y$ or $D$.
 \end{remark}

%% file: 4.Testing_homogeneity.tex
\section{Testable implications of homogeneity-type assumptions}
\label{sec: test homogeneity}

\subsection{Generalizing principal stratum effects}
\label{subsec: conditions for generalizability}
Let $\text{ATE}_{P_0}(\boldsymbol{x}) = \mathbb{E}_{P_0}[Y(D=1)-Y(D=0)\mid \boldsymbol{X} = \boldsymbol{x}]$ denote the conditional average treatment effect and $\text{ATE}_{P_0} = \mathbb{E}_{P_0}[\text{ATE}_{P_0}(\boldsymbol{X})]$ the average treatment effect. Proposition \ref{prop_homo} establishes two sufficient conditions that facilitate identifying $\text{ATE}_{P_0}(\boldsymbol{x})$ (and hence $\text{ATE}_{P_0}$) from principal stratum effect $\text{ACOATE}_{P_0}(\boldsymbol{x})$ or $\text{SWATE}_{P_0}(\boldsymbol{x})$.  

\begin{proposition}
\label{prop_homo}
Suppose one of the following assumptions holds:
 \begin{enumerate}[label=\emph{H(\roman*)}.]
 \item (Principal ignorability) {\small$\mathbb{E}_{P_0}[Y(D=d)\mid S=s, \boldsymbol{X}] = \mathbb{E}_{P_0}[Y(D=d) \mid \boldsymbol{X}]$ for $d\in \{0,1\}$ and all possible principal strata $s$}.
 \item (No unmeasured common effect modifier) \vspace{-6pt}
 \begin{align*}
     & \text{Cov}_{P_0}\Big[\Big(D(Z=1_a)-D(Z=0_a)\Big)\Big(Y(D=1)-Y(D=0)\Big) \mid \boldsymbol{X}\Big]=0,\\
     & \text{Cov}_{P_0}\Big[\Big(D(Z=1_b)-D(Z=0_b)\Big)\Big(Y(D=1)-Y(D=0)\Big)\mid  \boldsymbol{X}\Big]=0.
 \end{align*}
 \end{enumerate} \vspace{-6pt}
  Then we have 
    $
        \text{ATE}_{P_0}(\boldsymbol{x}) = \text{ACOATE}_{P_0}(\boldsymbol{x}) = \text{SWATE}_{P_0}(\boldsymbol{x})
    $.
\end{proposition}

\begin{remark}[Single IV setting]
    Assumption H(\romannumeral 1) and Assumption H(\romannumeral 2) have their counterparts in the setting with one binary IV; see, e.g., \citet{jo2009use} and \citet{wang2018bounded}. 
\end{remark}
Let
$
    \text{COATE}_{P_0}(\boldsymbol{X}) = \mathbb{E}_{P_0}[Y(D=1)-Y(D=0)\mid S\in\{\text{SW},\text{ACO}\}, \boldsymbol{X}]$ denote the conditional average treatment effect among compliers under IV pair $\{0_b, 1_b\}$. Then under Assumptions 1 and 2, we have $ \text{COATE}_{P_0}(\boldsymbol{X})=\theta_{\text{CO},P_0}(\boldsymbol{X}):=\frac{\delta_{b,P_0}(\boldsymbol{X})}{\eta_{b,P_0}(\boldsymbol{X})}
$.
 Observe that
\begin{equation*}
\begin{split}
    \theta_{\text{CO},P_0}(\boldsymbol{X}) &:=\frac{\delta_{b,P_0}(\boldsymbol{X})}{\eta_{b,P_0}(\boldsymbol{X})}\\ 
    &= \left\{\frac{\eta_{b,P_0}(\boldsymbol{X}) - \eta_{a,P_0}(\boldsymbol{X})}{\eta_{b,P_0}(\boldsymbol{X})}\right\} \times \theta_{\text{SW},P_0}(\boldsymbol{X}) + \left\{\frac{\eta_{a,P_0}(\boldsymbol{X})}{\eta_{b,P_0}(\boldsymbol{X})}\right\} \times \theta_{\text{ACO},P_0}(\boldsymbol{X}).
\end{split}
\end{equation*}
\noindent Under the nested IV setting, Proposition \ref{prop_homo} implies that Assumption H(\romannumeral 1) or H(\romannumeral 2) has the following three testable implications: $\theta^{(1)}_{P_0}\equiv \theta_{\text{ACO},P_0}(\boldsymbol{X}) - \theta_{\text{SW},P_0}(\boldsymbol{X}) =0 ~\text{a.s.}~P_0$, $\theta^{(2)}_{P_0} \equiv  \theta_{\text{ACO},P_0}(\boldsymbol{X}) - \theta_{\text{CO},P_0}(\boldsymbol{X})=0 ~\text{a.s.}~P_0$, and $\theta^{(3)}_{P_0} \equiv  \theta_{\text{SW},P_0}(\boldsymbol{X}) - \theta_{\text{CO},P_0}(\boldsymbol{X}) =0 ~\text{a.s.}~P_0$. In words, under either assumption, we would expect that, within strata of observed covariates $\boldsymbol{X}$, the average treatment effect among always-compliers equals that among the switchers and that among the compliers under the IV pair $\{0_b, 1_b\}$. 

Next, we propose formal tests of the null hypothesis: \vspace{-6pt}
\begin{equation}
\label{eq: H0 theta_0 = 0}\vspace{-6pt}
    H^{(j)}_{0}: \theta^{(j)}_{P_0}= 0~\text{a.s.}~ P_0~\text{against}~ H^{(j)}_1: \theta^{(j)}_{P_0} \neq  0,\qquad j \in \{1, 2, 3\}.
\end{equation} \vspace{-6pt}\noindent
If $\boldsymbol{X}$ is empty or low-dimensional, one may specify parametric models for $\theta_{\text{ACO},P_0}(\boldsymbol{X})$ and $\theta_{\text{SW},P_0}(\boldsymbol{X})$, along with additional conditional mean models for $Y$ and $D$. In this setting, the null hypothesis $H_0^{(j)}$ can be tested using standard overidentification tests, such as the Sargan-Hansen test \citep{sargan1958estimation}. However, when $\boldsymbol{X}$ is of moderate dimension, such tests can suffer from inflated Type I error (due to potential misspecification of nuisance models) and reduced power compared with procedures that accommodate more flexible alternatives. To address these limitations, we describe two flexible nonparametric testing procedures below.

{ 
\subsection{Testing the best least squares projection}
\label{subsec: test projection}
Let $\Gamma = \{\gamma(\boldsymbol{x};\beta),~\beta \in B \subset \mathbb{R}^{d_{\boldsymbol{x}}}\}$ denote a class of functions indexed by $\beta$, where $d_{\boldsymbol{x}}$ is the dimension of $\boldsymbol{X}$, and $\gamma(\boldsymbol{x};0) = 0$ for all $\beta$. Suppose that $\beta$ is well-defined in the sense that $\beta_1\neq \beta_2$ implies $\gamma(\cdot;\beta_1) \neq \gamma(\cdot;\beta_2)$. Furthermore, suppose $\Gamma$ is a closed subspace $L^2(P_0)$. The best least squares projection of $\theta^{(j)}_{P_0}$ onto $\Gamma$, denoted $\Pi(\theta^{(j)}_{P_0}\mid \Gamma)$, is defined as
\begin{equation*}
    \Pi(\theta^{(j)}_{P_0}\mid \Gamma) = \arg\min_{\gamma \in \Gamma} \Vert \gamma-\theta^{(j)}_{P_0} \Vert_{L^2(P_0)},
\end{equation*}
which admits the unique representation as $\gamma(\cdot,\beta^{(j)}_{P_0})$, with $\beta^{(j)}_{P_0}$ satisfying
\begin{equation}
\label{eq: projection}
    \beta^{(j)}_{P_0} = \arg\min_{\beta \in B}~\mathbb{E}_{P_0}\left[\theta^{(j)}_{P_0}(\boldsymbol{X})-\gamma(\boldsymbol{X};\beta_0)\right]^2.
\end{equation}
\noindent Under the null hypothesis $H^{(j)}_0$, we necessarily have $\beta^{(j)}_{P_0}=0$. 

Theorem \ref{thm: gradient for beta} derives the canonical gradient for $\beta^{(j)}_{P}$ based on which we will derive an asymptotically normal estimator for $\beta^{(j)}_{P_0}$ and then construct a Wald-type test for $H^{(j)}_0$.

\begin{theorem}
\label{thm: gradient for beta}
    Suppose $\partial \gamma(\boldsymbol{x};\beta)/\partial \beta$ exists and is continuous in $\beta$. Under a nonparametric model and for $j \in \{1, 2, 3\}$, the canonical gradient of $\beta^{(j)}_{P}$ at $P_0$ is $\phi^{(j)}_{P_0}$, defined as follows:
    \small\begin{align*}
        \phi^{(1)}_{P}:(z,\boldsymbol{x},d,y)\mapsto  C^{(1)}_{P}\frac{\partial\gamma(\boldsymbol{x};\beta)}{\partial \beta}\Big|_{\beta=\beta^{(1)}_P}\Bigg[D^{(1)}_{P}(z,\boldsymbol{x},d,y)-D^{(2)}_{P}(z,\boldsymbol{x},d,y)+\theta^{(1)}_{P}(\boldsymbol{x})-\gamma(\boldsymbol{x};\beta^{(1)}_P)\Bigg],\\
        \phi^{(2)}_{P}: (z,\boldsymbol{x},d,y)\mapsto C^{(2)}_{P}\frac{\partial\gamma(\boldsymbol{x};\beta)}{\partial \beta}\Big|_{\beta=\beta^{(2)}_P}\Bigg[D^{(1)}_{P}(z,\boldsymbol{x},d,y)-D^{(3)}_{P}(z,\boldsymbol{x},d,y)+\theta^{(2)}_{P}(\boldsymbol{x})-\gamma(\boldsymbol{x};\beta^{(2)}_P)\Bigg], \\
        \phi^{(3)}_{P}:(z,\boldsymbol{x},d,y)\mapsto  C^{(3)}_{P}\frac{\partial\gamma(\boldsymbol{x};\beta)}{\partial \beta}\Big|_{\beta=\beta^{(3)}_P}\Bigg[D^{(2)}_{P}(z,\boldsymbol{x},d,y)-D^{(3)}_{P}(z,\boldsymbol{x},d,y)+\theta^{(3)}_{P}(\boldsymbol{x})-\gamma(\boldsymbol{x};\beta^{(3)}_P)\Bigg],
    \end{align*} \normalsize
    where $C^{(j)}_{P}$ are constants, and the expressions of $D^{(1)}_{P}$, $D^{(2)}_{P}$, and $D^{(3)}_{P}$ can be found in Supplemental Material \ref{subsec: additional information, projection test, supp}.
\end{theorem}
\noindent Below we focus on the case when $\gamma(\boldsymbol{X};\beta) = \boldsymbol{X}^T\beta $. Analogous to the development in Section \ref{subsec: EIF estimator},  $\beta^{(j)}_{P_0}$ can be estimated based on estimating equations with  flexible nuisance estimator and cross-fitting.  Our proposed test is constructed based on the following steps: 

\begin{description}
    \item[Step I:]~For each $k\in[K]$, construct the estimating equation-based estimator $\widehat{\beta}^{(j)}_{ee,k}$, where the estimating equation is defined based on the canonical gradient $\phi^{(j)}_{P_0}$;

    

\item[Step II:]~Calculate $ \widehat{\beta}^{(j)}_{ee} = \frac{1}{K}\sum_{k=1}^n \widehat{\beta}^{(j)}_{ee,k}$ and estimate the asymptotic variance $\widehat{\Sigma}^{(j)} = \frac{1}{n}\sum_{k=1}^K n_k \mathbb{P}_{n,k}\left\{ {\phi^{(j)}}^\top_{\widehat{P}_{n,k}}\phi^{(j)}_{\widehat{P}_{n,k}}\right\}$;

\item[Step \uppercase\expandafter{\romannumeral3}:] ~Construct the test statistic $W^{(j)} =n{\widehat{\beta}_{ee}^{(j), \top}}\widehat{\Sigma}^{(j), -1}\widehat{\beta}_{ee}^{(j)}$
and define the test: 
\[
T^{(j)}_{\text{Wald,$\alpha$}}=\mathbbm{1}\left\{W^{(j)} > \chi_{1-\alpha}^2(d_{\boldsymbol{x}})\right\},
\] 
where $\chi_{1-\alpha}^2(d_{\boldsymbol{x}})$ is the $1-\alpha$ quantile of $\chi^2$-distribution with $d_{\boldsymbol{x}}$ degrees of freedom.
\end{description}

Theorem \ref{thm_one_step_test} establishes the properties of $ \widehat{\beta}^{(j)}_{ee}$ and the test $T^{(j)}_{\text{Wald},\alpha}$. 

\begin{theorem}
    Under regularity conditions specified in Supplemental Material \ref{subsec: additional information, np test, supp},
    \begin{enumerate}
        \item For each $j \in \{1, 2, 3\},$ the estimator $\widehat{\beta}_{ee}^{(j)}$ is consistent and asymptotically normal, with its asymptotic variance achieving the efficiency bound:
        \begin{equation*}
\sqrt{n}\left(\widehat{\beta}_{ee}^{(j)} - \beta^{(j)}_{P_0}\right) \rightsquigarrow \mathcal{N}\left(0,~ \mathbb{E}_{P_0}\left[{\phi^{(j)}}^\top_{P_0}\phi^{(j)}_{P_0}\right]\right).
        \end{equation*}
        \item For each $j \in \{1, 2, 3\},$ the test $T^{(j)}_{\text{Wald},\alpha}$ is asymptotically of size-$\alpha$ for testing $H_0^{(j)}$.
    \end{enumerate}
    \label{thm_one_step_test}
\end{theorem}

\subsection{Nonparametric $L^p$ norm-based test}
\label{subsec: test KS}

Next, we introduce a class of nonparametric $L^p$ norm-based tests, following the framework of \citet{westling2022nonparametric}, for testing $H_0^{(j)},~j \in \{1,2,3\}$. Define $\Omega^{(j)}_{P_0}(c):=\int \theta^{(j)}_{P_0}(x)\mathbbm{1}_c(x)dP_0(x)=0$, where $\mathbbm{1}_c(x) = \mathbbm{1}\left\{x\in (-\infty,c_1]\times ... \times (-\infty,c_{d_{\boldsymbol{x}}}]\right\}$. In Supplemental Material \ref{subsec: the lemma, supp}, we show that testing $H_0^{(j)}: \theta^{(j)}_{P_0} = 0$ a.s. $P_0$ is equivalent to testing $\Omega^{(j)}_{P_0}(c) = 0$ for any $c\in \mathcal{X}$, where $\mathcal{X}$ denotes the support of $\boldsymbol{X}$. 

An advantage of basing the test on $\Omega^{(j)}_{P_0}(c)$ is that $\Omega^{(j)}_{P_0}(c)$ is $\sqrt{n}$-estimable for any fixed $c$ without requiring additional parametric assumptions. In contrast, even when all the nuisance parameters and potential outcomes are known, $\theta^{(j)}_{P_0}$ is in general not $\sqrt{n}$-estimable.


Proposition \ref{prop: gradient Omega(c)} derives the canonical gradient for $\Omega^{(j)}_P(c)$, based on which we will construct an efficient, estimating equation-based estimator of $\Omega^{(j)}_{P_0}(c)$ for any fixed $c$.


\begin{proposition}
\label{prop: gradient Omega(c)}
    The canonical gradient for $\Omega^{(j)}_P(c)$ at $P_0$ is $D^{(j)*}_{P_0}(c)$, where
    \begin{align*}
        &D^{(1)*}_P(c):o\mapsto D^{(1)}_{P}(z,x,d,y)\mathbbm{1}_{c}(x)-D^{(2)}_{P}(g,z,x,d,y)\mathbbm{1}_{c}(x)+\theta^{(1)}_{P}(x)\mathbbm{1}_{c}(x)-\Omega^{(1)}_P(c),\\
        &D^{(2)*}_P(c):o\mapsto D^{(1)}_{P}(z,x,d,y)\mathbbm{1}_{c}(x)-D^{(3)}_{P}(g,z,x,d,y)\mathbbm{1}_{c}(x)+\theta^{(2)}_{P}(x)\mathbbm{1}_{c}(x)-\Omega_P^{(2)}(c), \\
        &D^{(3)*}_P(c):o\mapsto D^{(2)}_{P}(z,x,d,y)\mathbbm{1}_{c}(x)-D^{(3)}_{P}(g,z,x,d,y)\mathbbm{1}_{c}(x)+\theta^{(3)}_{P}(x)\mathbbm{1}_{c}(x)-\Omega^{(3)}_P(c).
    \end{align*}
\end{proposition}

\noindent For each $j \in \{1,2,3\}$, let $\widehat{\Omega}^{(j)}_{ee}(c)$ denote an estimating equation-based estimator of $\Omega^{(j)}_{P_0}(c)$ by treating the canonical gradients as the estimating functions. The estimator $\widehat{\Omega}^{(j)}(c)$ admits the following asymptotic expansion:



\begin{equation*}
    \widehat{\Omega}^{(j)}(c) - {\Omega}^{(j)}_{P_0}(c) = \mathbb{P}_nD^{(j)*}_{P_0}(c) + r_n(c),
\end{equation*}

\noindent where $D^{(j)*}_{P_0}(c)$ is the canonical gradient of ${\Omega}^{(j)}_{P}(c)$ at $P_0$. If we further have $\sup_{c\in \mathcal{X}}|r_n(c)| = o_{p}(n^{-1/2})$ and $\{D^{(j)*}_{P_0}(c), c\in \mathcal{X}\}$ is $P_0$-Donsker, then the stochastic process 
\[
\mathbb{H}^{(j)}_n:=\left\{\sqrt{n}\{\widehat{\Omega}^{(j)}_{}(c) - {\Omega}^{(j)}_{P_0}(c)\}:c\in \mathcal{X}\right\}
\]
converges weakly in $l^\infty(\mathcal{X})$ to a tight, mean-zero Gaussian process $\mathbb{H}$, with covariance function $\Sigma^{(j)}(s,t) = \mathbb{E}_{P_0}[D_{P_0}^{(j)*}(s)D_{P_0}^{(j)*}(t)]$, where $(s,t)\in \mathcal{X}\times \mathcal{X}$. Finally, by the continuous mapping theorem, $\Vert\sqrt{n}\{\widehat{\Omega}^{}_{}(c) - {\Omega}^{}_{P_0}(c)\}\Vert_{P_0,p}\rightsquigarrow \Vert \mathbb{H} \Vert_{P_0,p}$ for $p\geq 1$ or $p=\infty$, which will be the basis of our proposed $L^p$ norm-based test. Below, we summarize the testing procedure:

\begin{description}
    \item[Step I:]~For each $k\in[K]$ and a fixed $c$, construct $\widehat{\Omega}^{(j)}_{ee}(c)$.
    \item [Step \uppercase\expandafter{\romannumeral2}:]~Define $\mathcal{X}_n = \{\boldsymbol{X}_1,\dots,\boldsymbol{X}_n\}$. Let $Q^{(j)}_{n,\alpha}$ be the $1-\alpha$ quantile of 
        $\Vert\widehat{\mathbb{H}}^{(j)}_n\Vert_{P_n,p}$, where $(\widehat{\mathbb{H}}^{(j)}_n(\boldsymbol{X}_1),\dots,\widehat{\mathbb{H}}^{(j)}_n(\boldsymbol{X}_n))$ is a mean-zero multivariate normal distribution with $(s,t)$-th element of  covariance matrix given by $\Sigma^{(j)}_{n}(X_s,X_j)$, where $\Sigma^{(j)}_{n}(s,t) =$ $\frac{1}{K}\sum_{k=1}^K \mathbbm{P}_{n,k}$ $\left\{D^{(j)*}_{\widehat{P}_{n,k}}(s)D^{(j)*}_{\widehat{P}_{n,k}}(t)\right\}$, $1\leq s,t\leq n$, conditional on data $(O_1,...,O_n)$. 
\item [Step \uppercase\expandafter{\romannumeral3}:]~Define the test: 
\[T^{(j)}_{np,\alpha} = \mathbbm{1}\left\{\sup_{i}\left\Vert \widehat{\Omega}^{(j)}(\boldsymbol{X}_i)\right\Vert_{P_n,p}>Q^{(j)}_{n,\alpha}\right\}.
\]
\end{description}

Theorem \ref{thm: T_np properties} summarizes the key properties of $T^{(j)}_{{np},\alpha}$ for $j \in \{1,2,3\}$.

\begin{theorem}
\label{thm: T_np properties}
    Under regularity conditions specified in Supplemental Material \ref{subsec: NP test proof, supp}, (\romannumeral 1). $T^{(j)}_{np,\alpha}$ is asymptotically of size $\alpha$. (\romannumeral 2). The asymptotic power of $T^{(j)}_{np,\alpha}$ is 1. (\romannumeral 3).$T^{(j)}_{np,\alpha}$ has non-trivial power under local alternative distributions $P_n$:
        \begin{equation*}
            \lim_{n\rightarrow \infty} \int \Bigg[n^{1/2}\Big(dP_n^{1/2}-dP_0^{1/2}\Big)-\frac{1}{2}hdP_0^{1/2}\Bigg]=0,
        \end{equation*}
        where $h$ is a score function with $P_0h=0$ and $P_0h^2<\infty$.
\end{theorem}

\begin{remark}[Choice of $p$]
Two choices of $p$ are often adopted in the literature: $p=2$, corresponding to the Cramer-von Mises-type test, and $p=\infty$, corresponding to the Kolmogorov-Smirnov-type test.
\end{remark}

\begin{remark}[Computation cost]
The proposed non-parametric $L^p$ norm-based tests are computationally challenging when the dimension of $\boldsymbol{X}$ is moderate, because the procedure simulates a Gaussian process indexed by a subset of $\mathbb{R}^{d_x}$. The computation cost of the projection-based test, on the other hand, is essentially agnostic to the dimension of $\boldsymbol{X}$.
\end{remark}

{\color{black}
\begin{remark}[Combining different tests]
The three null hypotheses in \eqref{eq: H0 theta_0 = 0} are all testable implications of the same homogeneity-type assumptions in Proposition~\ref{prop_homo}. Hence, in applications, it may be useful to combine evidence across the tests $\{T^{(j)}_{np,\alpha}: j=1,2,3\}$  {and/or} $
\{T^{(j)}_{\mathrm{Wald},\alpha}: j=1,2,3\}$. A simple approach is to use a multiple-testing correction, such as the Bonferroni correction. A potentially more powerful approach is to exploit the joint asymptotic distribution of the test statistics. For example, for the projection-based tests, we can stack the estimated coefficients $\widehat{\beta}_{\text{ee}} = ({\widehat{\beta}^{(1),\top}}_{\text{ee}},{\widehat{\beta}^{(2),\top}}_{\text{ee}},{\widehat{\beta}^{(3),\top}}_{\text{ee}})^\top$, with influence function ${\phi}_{P_0} = (\phi^{(1),\top}_{P_0},\phi^{(2),\top}_{{P}_{0}},\phi^{(3),\top}_{{P}_{0}})^\top$. Under the null, we know  $\widehat{\beta}_{\text{ee}}$ converges to a zero mean normal distribution with asymptotic covariance matrix $P_0[{\phi}_{P_0}{\phi}^\top_{P_0}]$. Based on this zero mean normal asymptotic distribution, we can construct Wald type test to jointly test if all the projection coefficients are zero.
\end{remark}
}
}

%% file: 5.Simulation.tex
\vspace{-10pt}
\section{Simulation study}
\label{sec: simulation}
This simulation study has three objectives. First, we evaluate the finite-sample performance of the proposed estimators for SWATE. Second, we assess the Type I error rates of each projection-based test for the null hypotheses $H_0^{(j)}$. Since the test ${T}_{\text{Wald},\alpha}^{(j)}$, developed for $H_0^{(j)}$ ($j = 1,2,3$), can each be used to test the same null hypotheses (Assumption H(i) or H(ii)), we additionally compare their power. Third, we compare both the Type I error rates and power of the projection-based tests with those of two nonparametric $L^p$ norm-based tests.

\subsection{Estimation of SWATE}
\label{subsec: simulation 1}

{  Our simulation study used a factorial design that varied sample size, the proportion of switchers, the SWATE effect size, and outcome type. The data-generating mechanisms are provided in Supplemental Material \ref{subsec: main simu details, supp}.} For each simulated dataset, we constructed the proposed estimator $\widehat{\psi}_{ee}$ with all nuisance functions estimated using the \texttt{R} \citep{r2013r} package \texttt{SuperLearner} \citep{van2007super}. Specifically, we included the random forest and generalized linear models in the SuperLearner library. For each setting, we repeated the simulation $1000$ times. \textcolor{black}{The standard error of $\widehat{\psi}_{ee}$ was estimated based on the efficient influence function, as described in Section \ref{subsec: EIF estimator}.}

Figure S1 displays the sampling distributions of $\widehat{\psi}_{ee}$ for various combinations of sample size and switcher proportion, assuming a continuous outcome and $\beta = (2, 2, 2)$. The true SWATE values are indicated by red dashed lines. Overall, the sampling distributions are approximately normal, except in scenarios with both a small sample size and a low switcher proportion (e.g., $n = 1000$ with 11\% switchers). For a fixed switcher proportion, the distributions become more concentrated around the true values as the sample size increases. Similarly, for a fixed sample size, higher switcher proportions result in greater concentration around the true values.

Table \ref{tab:simu_SWATE_ee_con} reports the bias, relative bias, and coverage probability across different settings, showing consistent trends. When the switcher proportion is small (e.g., 11\%), a large sample size (e.g., $n = 10000$) is required for the estimator to achieve low relative bias and nominal coverage, reflecting weak IV bias \citep{bound1995problems,andrews2019weakiv}. Conversely, when the switcher proportion is high (e.g., 66\%), the relative bias is small and the 95\% CI coverage is close to nominal even for moderately sized samples; for example, with $\beta = (2, 2, 2)$ and $n = 1000$, the relative bias is 2.1\% and the 95\% CI coverage is 94.1\%.

Table \ref{tab:simu_SWATE_ee_con} also reports the frequency of estimator instability, which occurs only when both the sample size and switcher proportion are small. For instance, when $n = 2000$ and the switcher proportion is 32\%, none of the 1000 simulated datasets produced an unstable estimator. The one-step estimator, by comparison, was slightly more prone to instability than the estimating equation-based estimator (see Table \ref{tab:simu_SWATE_os_con} in Supplemental Material \ref{subsec: simulation, estimators, supp}).
 
Overall, the simulation results indicate that when the product of the sample size and the switcher proportion exceeds 500, the proposed estimator performs well, exhibiting low relative bias and approximately nominal coverage. Similar qualitative patterns were observed for a binary outcome; see Supplemental Material \ref{subsec: simulation, estimators, supp} for details.

\begin{table}[htbp]
  \centering
  \footnotesize
  \caption{Simulation results  for $\widehat{\psi}_{ee}$ when the outcome is continuous. Column \textbf{SWATE} reports the true values of SWATE under each setting. Column \textbf{Estimate} reports the average point estimates (excluded if the absolute value exceeded $500$). Column \textbf{SE Est} reports the $90\%$ Winsorized mean of standard error estimates. Column \textbf{Acc. Rate} reports the proportion of point estimates not truncated. }
  \resizebox{0.9\textwidth}{!}{\begin{tabular}{cccccccccc}
    \toprule
    $\beta$  & SW \% & SWATE & Sample size & Estimate & Rel. Bias & Bias  & SE Est & Coverage & Acc. Rate \\
    \midrule
    \multirow{16}[8]{*}{(2,2,2)} & \multirow{4}[2]{*}{11\%} & \multirow{4}[2]{*}{0.92} & 1000  & -1.202 & -0.610 & -0.559 & 43.614 & 0.906 & $99.8\%$ \\
          &       &       & 2000  & 0.765 & -2.244 & -2.06 & 10.774 & 0.923 & $99.9\%$ \\
          &       &       & 5000  & 1.052 & 0.147 & 0.135 & 1.018 & 0.941 & $100\%$ \\
          &       &       & 10000 & 1.019 & 0.110  & 0.101 & 0.546 & 0.938 & $100\%$ \\
\cmidrule{2-10}          & \multirow{4}[2]{*}{22\%} & \multirow{4}[2]{*}{1.02} & 1000  & 0.799 & -0.216 & -0.220 & 4.852 & 0.909 & $100\%$\\
          &       &       & 2000  & 1.130  & 0.109 & 0.111 & 0.966 & 0.949 & $100\%$ \\
          &       &       & 5000  & 1.081 & 0.061 & 0.062 & 0.43  & 0.961 & $100\%$ \\
          &       &       & 10000 & 1.053 & 0.033 & 0.034 & 0.273 & 0.954 & $100\%$ \\
\cmidrule{2-10}          & \multirow{4}[2]{*}{32\%} & \multirow{4}[2]{*}{1.34} & 1000  & 1.838 & 0.334 & 0.461 & 1.259 & 0.942 & $100\%$ \\
          &       &       & 2000  & 1.486 & 0.079 & 0.108 & 0.606 & 0.962 & $100\%$ \\
          &       &       & 5000  & 1.383 & 0.004 & 0.005 & 0.304 & 0.959 & $100\%$ \\
          &       &       & 10000 & 1.378 & $<0.001$     & $<0.001$     & 0.202 & 0.961 & $100\%$ \\
\cmidrule{2-10}          & \multirow{4}[2]{*}{66\%} & \multirow{4}[2]{*}{1.56} & 1000  & 1.590  & 0.021 & 0.033 & 0.341 & 0.941 & $100\%$ \\
          &       &       & 2000  & 1.553 & -0.003 & -0.004 & 0.210  & 0.949 & $100\%$ \\
          &       &       & 5000  & 1.560  & 0.002 & 0.003 & 0.122 & 0.946 & $100\%$ \\
          &       &       & 10000 & 1.560  & 0.002 & 0.003 & 0.083 & 0.949 & $100\%$ \\
    \midrule
    \multirow{16}[8]{*}{(4,4,4)} & \multirow{4}[2]{*}{11\%} & \multirow{4}[2]{*}{0.91} & 1000  & 1.518 & 0.674 & 0.612 & 51.315 & 0.924 & $100\%$ \\
          &       &       & 2000  & 1.273 & 3.698 & 3.354 & 9.969 & 0.937 & $99.9\%$ \\
          &       &       & 5000  & 1.168 & 0.288 & 0.261 & 1.216 & 0.953 & $100\%$ \\
          &       &       & 10000 & 1.111 & 0.225 & 0.204 & 0.607 & 0.931 & $100\%$ \\
\cmidrule{2-10}          & \multirow{4}[2]{*}{22\%} & \multirow{4}[2]{*}{1.15} & 1000  & 1.258 & 0.097 & 0.111 & 5.562 & 0.921 & $100\%$ \\
          &       &       & 2000  & 1.328 & 0.158 & 0.181 & 1.298 & 0.964 & $100\%$ \\
          &       &       & 5000  & 1.213 & 0.058 & 0.066 & 0.546 & 0.968 & $100\%$ \\
          &       &       & 10000 & 1.152 & 0.004 & 0.005 & 0.340  & 0.953 & $100\%$ \\
\cmidrule{2-10}          & \multirow{4}[2]{*}{32\%} & \multirow{4}[2]{*}{1.70} & 1000  & 1.814 & 0.068 & 0.115 & 1.721 & 0.944 & $100\%$ \\
          &       &       & 2000  & 1.966 & 0.157 & 0.267 & 0.853 & 0.962 & $100\%$ \\
          &       &       & 5000  & 1.741 & 0.024 & 0.041 & 0.425 & 0.969 & $100\%$ \\
          &       &       & 10000 & 1.735 & 0.021 & 0.036 & 0.278 & 0.961 & $100\%$ \\
\cmidrule{2-10}          & \multirow{4}[2]{*}{66\%} & \multirow{4}[2]{*}{2.32} & 1000  & 2.336 & 0.009 & 0.020  & 0.574 & 0.953 & $100\%$ \\
          &       &       & 2000  & 2.336 & 0.009 & 0.021 & 0.350  & 0.964 & $100\%$ \\
          &       &       & 5000  & 2.302 & -0.006 & -0.014 & 0.202 & 0.960  & $100\%$ \\
          &       &       & 10000 & 2.306 & -0.004 & -0.010 & 0.138 & 0.973 & $100\%$ \\
    \bottomrule
    \label{tab:simu_SWATE_ee_con}
    \end{tabular}}
\end{table}%

\subsection{Comparing tests of the homogeneity-type assumptions}
\label{subsec: simu test homogeneity}
We evaluated and compared the level and power of the best linear projection-based test, $L^2$ norm-based test (Cramer-von Mises-type), and $L^\infty$ norm-based tests (Kolmogorov-Smirnov-type) under various data generating processes. Across all settings, we did not observe Type I error rate inflation at the usual $0.05$ level for the proposed tests. Projection-based tests were computationally more efficient and generally achieved comparable or superior power relative to the $L^2$ and $L^\infty$ norm-based tests. 

We also found that when the proportion of always-compliers is high, ${T}^{(2)}_{\text{Wald},0.05}$ is generally the most powerful among $\{T^{(j)}_{\text{Wald},0.05},~j = 1, 2, 3\}$. Conversely, when the proportion of switchers is high, ${T}^{(3)}_{\text{Wald},0.05}$ exhibits the greatest power. Additional simulation results are provided in Supplemental Material \ref{subsec: simu test homogeneity, np test, supp}.

%% file: 6.Case_study.tex
\section{Revisiting the PLCO study}
\label{sec: case study}
We applied our nested IV framework to re-analyze the Prostate, Lung, Colorectal, and Ovarian (PLCO) Cancer Screening Trial, focusing on the Henry Ford Health System, which transitioned from a dual- to a single-consent process in 1997. We included participants who enrolled before 1997 ($G = a$) and after 1997 ($G = b$), excluding those enrolled during 1997. Table \ref{tab:descriptive statistics, supp} in Supplemental Material \ref{sec: additional real data, supp} summarizes key baseline covariates and treatment uptake (attendance at the first scheduled cancer screening) by stage and treatment assignment. Participants enrolled before and after 1997 differed moderately in baseline covariates; within each stage, the two arms were comparable due to randomization. \textcolor{black}{The endpoint is primary invasive colorectal cancer, the key secondary endpoint used in \citet{schoen2012colorectal}.} 

Following the switch to a single-consent process, compliance increased substantially. The unadjusted compliance rate rose from 50.9\% before 1997 to 80.1\% after 1997. After adjusting for covariates, the estimated compliance rates were 52.6\% before 1997 and 84.9\% after 1997. These results indicate that (1) the moderate differences in baseline covariates between the two stages did not fully account for the marked improvement in compliance, and (2) under the nested IV framework, approximately 32.3\% of participants ($84.9\% - 52.6\%$) are classified as ``switchers.”

Who were the switchers---participants who became compliers after the change in the consent process? Table \ref{tab:characterize SW and AC} in the Supplemental Material summarizes the mean of each baseline covariate for the switcher and always-complier groups, using Proposition \ref{prop_covariates_dis}. Interestingly, although always-compliers and switchers were similar in age and smoking status, switchers were more likely to be female, white, and have a BMI below 25.

We then estimated the SWATE using both the proposed one-step estimator and the estimating equation-based estimator, with 15-year colorectal cancer incidence as the outcome of interest. To implement the estimators, we used \textsf{SuperLearner} with a library comprising \texttt{glm} and \texttt{randomforest} to estimate the nuisance functions. 

Using the one-step estimator, the estimated 15-year risk difference was $0.9\%$ ($-1.8\%$, $3.6\%$) for switchers and $-1.1\%$ ($-2.3\%$, $0.1\%$) for always-compliers. Results from the estimating equation-based estimator were similar: $1.1\%$ ($-1.6\%$, $3.8\%$) for switchers and $-1.1\%$ ($-2.3\%$, $0.1\%$) for always-compliers. By comparison, the Wald estimator with bootstrapped confidence intervals was less precise: $1.1\%$ ($-1.8\%$, $4.0\%$) for switchers and $-0.7\%$ ($-1.8\%$, $0.4\%$) for always-compliers. 

Finally, we tested the null hypothesis that the effect of cancer screening on colorectal cancer incidence was the same across all principal strata, within strata defined by six observed covariates: age, sex, race, education, smoking status, and BMI. The test statistics were $7.31$, $9.29$, and $7.75$ for three best-projection-based tests ${T}_{\text{Wald},0.05}^{(1)}$, ${T}_{\text{Wald},0.05}^{(2)}$, and ${T}_{\text{Wald},0.05}^{(3)}$, respectively. The $95\%$ quantile of a  $\chi^2(9)$ distribution is $16.92$; therefore, none of these tests provided evidence of additional effect heterogeneity among latent principal strata within the covariate-defined strata.

In Supplemental Material \ref{subsec: case study analysis of multiple sites, supp}, we present an integrated analysis of data from three clinical sites participating in the trial: Henry Ford during the dual consent stage ($G = a$), University of Colorado ($G = b$), and University of Alabama at Birmingham ($G = c$). We adopt the nested IV assumption that the IV pair $\mathcal{Z}_a = \{1_a,0_a\}$ is nested within both $\mathcal{Z}_b = \{1_b,0_b\}$ and $\mathcal{Z}_c = \{1_c,0_c\}$. This defines two separate switcher populations: one associated with $\mathcal{Z}_a$ and $\mathcal{Z}_b$, and the other with $\mathcal{Z}_a$ and $\mathcal{Z}_c$. Using our proposed method, we characterized the switcher populations and found them to be similar to each other and to the switcher population reported in Table \ref{tab:characterize SW and AC}. Furthermore, we found little evidence of additional effect heterogeneity beyond that captured by observed covariates.

%% file: 7.Discussion.tex
\vspace{-6pt}
\section{Discussion}
\label{sec: discussion}
In this work, we show that the switcher average treatment effect can be nonparametrically identified under a novel nested IV framework, and we develop efficient estimators for it. We also propose a projection-based test and a nonparametric test to formally assess effect-homogeneity-type assumptions within the nested IV framework.

{\color{black} If we further assume that $D(0_a) = D(0_b)$ with probability 1, together with unconditional randomization of $G$, the proposed framework reduces to a three-level ordinal IV setting. However, in many clinical trials where the control arm represents the standard of care, the definition of ``standard of care" may vary across study periods or clinical settings. For example, in the PLCO trial, participants assigned to the standard-of-care arm may have been more likely to undergo cancer screening after 1997, as public awareness and acceptance of screening increased over time. Consequently, the assumption $D(0_a) = D(0_b)$ with probability 1 may be overly restrictive in these settings. Furthermore, the mean effect changeability assumption considered in the article applies more broadly than unconditional randomization of (G), because it (1) permits conditioning on baseline covariates and (2) operates on the mean difference scale. The proposed nested IV framework therefore meaningfully extends the three-level ordinal IV literature. In particular, it demonstrates that, even after dropping the assumption $D(0_a) = D(0_b)$ and relaxing the requirement of unconditional randomization of $G$, novel and practically relevant estimands such as ACOATE and SWATE remain nonparametrically identifiable, while treatment effect heterogeneity can still be meaningfully evaluated.
}

Interestingly, the resulting conditional switcher average treatment effect (SWATE) takes a form similar to the conditional ATE in the instrumented difference-in-differences framework \citep{ye_iv_did}. A key distinction, however, is that our SWATE integrates over the covariate distribution among switchers, whereas the marginalized estimand in \citeauthor{ye_iv_did} \citeyearpar{ye_iv_did} integrates over the covariate distribution of the full study population. Moreover, the richness of the principal strata in our setting substantially complicates modeling of both treatment uptake and potential outcomes. For this reason, we focus on nonparametric estimators that enable flexible estimation of the required nuisance functions.

{  We note several limitations of the proposed framework. First, the nested IV assumption is not appropriate in all settings. In a seminal study of the effect of attending a community college versus a four-year college on educational attainment (specifically, whether the difference exceeds two years), \citet{rouse1995democratization} considered two IVs: differential tuition and differential distance. In this example, it is not reasonable to assume that compliers under the differential tuition IV would also comply under the differential distance IV, or vice versa; hence, the nested IV assumption is inappropriate. Similarly, when using genetic variants (SNPs) as IVs for certain phenotypes, researchers generally lack substantive justification to assume that one SNP is nested within another. In such settings, one possible way to leverage the proposed framework is to first define a strengthened IV \citep{baiocchi2010building, heng2023instrumental}---for example, by classifying individuals as ``doubly encouraged” or ``doubly discouraged” if both tuition-based and distance-based IVs equal 1 or 0, respectively---and then nesting the tuition-based IV (or the distance-based IV) within this strengthened IV.

Mean effect exchangeability over the stratification variable $G$ is another key assumption of the proposed framework. This assumption asserts stability of the intention-to-treat effect for \emph{the same IV} among participants with \emph{the same observed covariates} across different levels of $G$. The assumption is automatically satisfied when $G$ is randomized, for example, when the nested IV framework is applied to a valid ordinal IV. When $G$ is not randomized, it is essential to include as many potential effect modifiers as possible in $\boldsymbol{X}$ to ensure that the assumption holds approximately. Violations may arise when $G$ represents factors that modify the treatment effect. For example, in a multicenter trial, differences in the conditional $ITT$ between a high-compliance center and a low-compliance center may reflect center-specific characteristics (e.g., patient volume) rather than differences in compliance itself. If such factors are identified, the nested IV framework would no longer be appropriate without further conditioning on those center-level characteristics. In particular, the mean effect exchangeability assumption would be invalid if treatment protocols (e.g., dosing or procedural details) differ across studies, in which case an integrated analysis of multiple studies under a nested IV framework would not be appropriate.

In this article, ``generalizability” refers to generalizing the causal effect from one subpopulation to another, rather than from one context (e.g., a trial setting) to another (e.g., a real-world setting). For instance, the ACOATE and SWATE estimated for the PLCO trial are both defined under a trial setting, and the ACOATE and SWATE estimated in \citet{guo2025rationed} correspond to a specific historical period in China. We cannot generalize these effects to other contexts without additional data or assumptions. Nevertheless, the effect heterogeneity detected under the nested IV framework remains meaningful and may suggest the presence of similar heterogeneity in other contexts.

Statistical tests typically assess a particular feature of the data-generating process while holding other features fixed through assumptions. The proposed homogeneity test is no exception. When both the core IV assumptions and the nested IV assumption hold, rejection of the proposed test provides evidence of effect heterogeneity. However, the test may also be rejected if the core IV assumptions and/or the nested IV assumption are violated. In Supplemental Material \ref{subsec: simulation when IV assumption fail, supp}, we present additional simulations in which the exclusion restriction fails for both IVs. In these scenarios, the empirical size of the proposed test exceeds 0.05 at the nominal 0.05 significance level, even when the causal effect is homogeneous across always-compliers and switchers.
}

\vspace{-0.5cm}
\section*{Data availability}
\vspace{-0.5cm}
The data that support the findings of this study are available from NCI Cancer Data Access System \url{https://cdas.cancer.gov/datasets/plco/22/}. Data are available from the authors with the permission of NCI.

%% file: 10_Acknowledgement.tex




%% file: appendix.tex
 \begin{center}
     {\Large Supporting Information to ``Nested Instrumental Variables Analysis: Switcher Average Treatment
Effect, Identification, Efficient Estimation and Generalizability"} 
 \end{center}
\renewcommand{\thesection}{S\arabic{section}}
\setcounter{page}{1}
\setcounter{section}{0}

\renewcommand{\thefigure}{S.\arabic{figure}}

\setcounter{figure}{0}

\renewcommand{\thedefinition}{S\arabic{definition}}

\setcounter{definition}{0}

\renewcommand{\theproposition}{S\arabic{proposition}}

\setcounter{proposition}{0}

\renewcommand{\theassumption}{S\arabic{assumption}}

\setcounter{assumption}{0}

\renewcommand{\thetheorem}{S\arabic{theorem}}

\setcounter{theorem}{0}

\renewcommand{\theexample}{S\arabic{example}}

\setcounter{example}{0}

\renewcommand{\theremark}{S\arabic{remark}}

\setcounter{remark}{0}

\renewcommand{\thelemma}{S\arabic{lemma}}

\setcounter{lemma}{0}

\renewcommand{\thetable}{S\arabic{table}}
\setcounter{table}{0}

{\color{black}
\section{Alternative formulation of the problem; Rubin's science tables}
\label{sec: three frameworks, supp}

\subsection{The proposed framework and role of $G$}
\label{subsec: framework I, supp}

In the PLCO trial, we conceptualize treatment assignment as a categorical IV taking values in $\{0_a, 0_b, 1_a, 1_b\}$. This four-level, categorical IV can be equivalently expressed as $(Z^\prime, G)$, where $Z^\prime$ and $G$ are derived, binary variables, so that $Z = 1_g$ can be equivalently written as $(Z^\prime = 1, G = g),~g \in \{a, b\}$, and $Z = 0_g$ can be equivalently written as $(Z^\prime = 0, G = g),~g \in \{a, b\}$.

Table~\ref{tab:science_table framework I} summarizes Rubin's science table \citep{rubin2005causal} under this framework. For ease of presentation, we omit the potential treatment outcomes $D_i(z)$ from the science table. We refer to this proposed framework as \textsf{Framework I}.

Alternatively, one may first conceptualize treatment assignment as a two-dimensional vector IV, $(Z^\prime, G)$, where both $Z^\prime$ and $G$ are binary variables. Then, one may define
\[
Z = 1_g = 1\{Z^\prime = 1, G = g\}, \qquad\text{and} \qquad 
Z = 0_g = 1\{Z^\prime = 0, G = g\},
\]
for $g \in \{a,b\}$, to map the vector-IV representation to a four-level categorical IV representation.

The mapping between $Z \in \{0_a, 0_b, 1_a, 1_b\}$ and $(Z^\prime \in \{1, 0\}, G \in \{a, b\})$ is one-to-one: the two formulations are simply alternative encodings of the same four-level IV. \emph{Either way, $G$ is part of treatment assignment.} Consequently, they lead to the same potential outcomes framework (\textsf{Framework I}).
    
While $Z \in \{0_a, 0_b, 1_a, 1_b\}$ and $(Z^\prime \in \{1, 0\}, G \in \{a, b\})$ are two equivalent, one-to-one encodings of the same IV structure, one formulation may be more natural than the other depending on the context. In some settings, it is easier to conceptualize the intervention directly in terms of a four-level variable $Z$, whereas in others it is more natural to begin with the vector formulation $(Z^\prime, G)$. In the PLCO trial, for example, we first define $Z \in \{0_a, 0_b, 1_a, 1_b\}$, where $1_a$ and $1_b$ (resp. $0_a$ and $0_b$) correspond to randomization to treatment (resp. standard of care) under different consent types, and then view $(Z^\prime, G)$ as derived from $Z$. As a second example, when $Z$ is an ordinal IV with many levels, then it is more convenient to first define $Z$ and regard $(Z^\prime, G)$ as a derived representation. In contrast, in the rationed fertility example of \citet{guo2025rationed}, it is conceptually clearer to first define $Z^\prime$ as twinning status and $G$ as the implementation status of the birth control policy, and then define $Z$ as the induced categorical IV derived from $(Z^\prime, G)$.

\begin{table}[htbp]
\centering
\caption{Rubin causal model science table under \textsf{Framework I}, the proposed framework. \textsf{Framework I} views treatment assignment $Z$ as a categorical IV taking values in $\{0_a, 1_a, 0_b, 1_b\}$, and $G$ (not shown here) is a derived variable such that $G_i = a$ if $Z_i \in \{0_a, 1_a\}$ and $G_i = b$ if $Z_i \in \{0_b, 1_b\}$. 
}
\begin{tabular}{c c c c c ccc}
\hline
Unit  & Baseline covariates  & \multicolumn{4}{c}{Potential outcomes} & Treatment &Observed\\ \cline{3-6}
$i$ & $X_i$ &  $Y_i(0_a)$ &  $Y_i(1_a)$ &  $Y_i(0_b)$ &  $Y_i(1_b)$ &received &outcome\\
\hline
1 & $x_1$  & $Y_1(0_a)$ & $Y_1(1_a)$ & $Y_1(0_b)$ & $Y_1(1_b)$ & $1_a$ & $Y_1(1_a)$\\
2 & $x_2$  & $Y_2(0_a)$ & $Y_2(1_a)$ & $Y_2(0_b)$ & $Y_2(1_b)$ & $0_b$ & $Y_2(0_b)$\\
3 & $x_3$  & $Y_3(0_a)$ & $Y_3(1_a)$ & $Y_3(0_b)$ & $Y_3(1_b)$ & $0_a$ &$Y_3(0_a)$\\
\vdots & \vdots & \vdots & \vdots &\vdots & \vdots & \vdots & \vdots \\
\hline
\end{tabular}
\label{tab:science_table framework I}
\end{table}

\subsection{Two alternative frameworks}
\label{subsec: frameworks II and III, supp}
An alternative formulation treats $G_i \in \{a,b\}$ as a non-manipulable baseline covariate, analogous to other baseline covariates $\boldsymbol{X}$, while still viewing treatment assignment as a four-level categorical IV. Under this formulation, treatment assignments $\{0_b,1_b\}$ are not feasible for units with $G_i=a$ (for example, in the PLCO trial, a participant enrolled before $1997$ ($G_i=a$) could never be randomized under the single-consent process corresponding to $Z \in \{0_b,1_b\}$), and similarly, assignments $\{0_a,1_a\}$ are not feasible for units with $G_i=b$. Consequently, for units with $G_i=a$ (resp. $G_i=b$), the potential outcomes $Y_i(0_b)$ and $Y_i(1_b)$ (resp. $Y_i(0_a)$ and $Y_i(1_a)$) may be viewed either (1) as undefined (NA) or (2) as conceptually well defined counterfactuals but never observable. In either case, quantities involving $(D_i(0_b), D_i(1_b), Y_i(0_b), Y_i(1_b))$ (resp. $(D_i(0_a), D_i(1_a), Y_i(0_a), Y_i(1_a))$) cannot be estimated for units with $G_i = a$ ($G_i = b$). 

In this scenario, one may still define a binary instrument $Z_i^\ast \in \{0,1\}$, where ``1'' corresponds to $\{1_a,1_b\}$, and ``0'' corresponds to $\{0_a,0_b\}$. Assigning $Z_i^\ast = 1$ to all units would amount to assigning $1_a$ to units with $G_i = a$ and $1_b$ to those with $G_i = b$. We refer to this setup as \textsf{Framework II}.

Table~\ref{tab:science_table framework II} below summarizes Rubin's science table corresponding to \textsf{Framework II}, where, each unit $i$ has either $(Y_i(0_a), Y_i(1_a))$ or $(Y_i(0_b), Y_i(1_b))$, but not both, defined. Similarly, each unit $i$ has $(D_i(0_a), D_i(1_a))$ or $(D_i(0_b), D_i(1_b))$, but not both, defined (omitted in the science table). 

\begin{table}[H]
\centering
\caption{Rubin causal model science table under \textsf{Framework II}. \textsf{Framework II} views treatment assignment $Z$ as a categorical IV taking values in $\{0_a, 1_a, 0_b, 1_b\}$, and $G$ as a baseline covariate.}
\begin{tabular}{c c c c c ccc}
\hline
Unit & Baseline covariates & \multicolumn{4}{c}{Potential Outcomes} & Treatment &Observed\\ \cline{3-6}
$i$ & $(X_i, G_i)$ &  $Y_i(0_a)$ &  $Y_i(1_a)$ &  $Y_i(0_b)$ &  $Y_i(1_b)$ &received &outcome\\
\hline
1 & $(x_1, a)$ & $Y_1(0_a)$ & $Y_1(1_a)$ & NA & NA & $1_a$ &$Y_1(1_a)$ \\
2 & $(x_2, b)$ & NA & NA & $Y_2(0_b)$ & $Y_2(1_b)$ & $0_b$ &$Y_2(0_b)$ \\
3 & $(x_3, a)$ & $Y_3(0_a)$ & $Y_3(1_a)$ & NA & NA & $0_a$ &$Y_3(0_a)$ \\
\vdots & \vdots & \vdots & \vdots \\
\hline
\end{tabular}
\label{tab:science_table framework II}
\end{table}

A third framework, which is the most simplified formulation, treats $G$ as a baseline covariate and does not distinguish between different versions of treatment assignment (i.e., \emph{a priori} assuming $1_a = 1_b = 1$ and $0_a = 0_b = 0$). Table~\ref{tab:science_table framework III} summarizes the science table corresponding to \textsf{Framework III}.

\begin{table}[htbp]
\centering
\caption{Rubin causal model science table under \textsf{Framework III}. \textsf{Framework III} views treatment assignment $Z$ as a binary IV taking values in $\{0, 1\}$, and $G$ as a baseline covariate.}
\begin{tabular}{c c c c c c}
\hline
Unit & Baseline covariates & \multicolumn{2}{c}{Potential Outcomes} & Treatment &Observed\\ \cline{3-4}
$i$ & $(X_i, G_i)$ &  $Y_i(0)$ &  $Y_i(1)$ &received &outcome\\
\hline
1 & $(x_1, a)$ & $Y_1(0)$ & $Y_1(1)$ & $1$ &$Y_1(1)$ \\
2 & $(x_2, b)$ & $Y_2(0)$ & $Y_2(1)$ & $0$ &$Y_2(0)$ \\
3 & $(x_3, a)$ & $Y_3(0)$ & $Y_3(1)$ & $0$ &$Y_3(0)$ \\
\vdots & \vdots & \vdots & \vdots \\
\hline
\end{tabular}
\label{tab:science_table framework III}
\end{table}

\subsection{A comparison of the three frameworks}
\label{subsec: comparison, supp}
In the PLCO trial, one can substantively distinguish the nature of encouragement to receive treatment (through the consent process) before and after $1997$, which motivates defining $\{1_a, 1_b\}$ rather than a single level $1$. The same is also true, for instance, in an integrated analysis of data from multiple centers or even multiple studies, where ``encouragement to take treatment" often encodes different elements (follow-up calls, distributions of brochures, etc) and better not be collapsed into one. Empirically, after adjusting for observed covariates, the rate of treatment uptake in the encouraged arm was dramatically different before and after implementing the single consent process \citep{marcus2014non}, further supporting defining $\{1_a, 1_b\}$ as opposed to $1$. 

Likewise, randomization to ``standard of care,'' represented by $\{0_a, 0_b\}$, may also carry different interpretations across the two periods. For example, the standard of care after $1997$ may have involved a higher baseline rate of cancer screening than before $1997$, reflecting the possibility that cancer screening had become more routine or more widely accepted by the general public. Hence, retaining $\{0_a,0_b\}$ as distinct levels, rather than collapsing them into a single level $0$, is a safer choice.

For these reasons, we rule out \textsf{Framework III} in what follows.

Next, we discuss \textsf{Framework II} and \textsf{Framework I}, both of which consider a categorical IV $Z \in \{1_a, 1_b, 0_a, 0_b\}$, although they conceptualize $G$ in a different way. Comparing \textsf{Framework II} with \textsf{Framework I} raises two related questions: first, whether \textsf{Framework II} is well defined; and second, how the scientific questions that can be investigated under \textsf{Framework II} differ from those that can be investigated under \textsf{Framework I}.

In the main article, we have discussed what estimands can be defined and identified under \textsf{Framework I} (ACOATE, SWATE, and COATE, and their conditional on $\boldsymbol{X}$ version), so below we will focus on \textsf{Framework II}. 

Under \textsf{Framework II}, we can define causal contrasts $E_{P_a}[Y(1_a) - Y(0_a)]$ and $E_{P_b}[Y(1_b) - Y(0_b)]$ (and similarly the unconditional and conditional complier/local average treatment effect under $(1_a, 0_a)$ among $P_a$, and under $(1_b, 0_b)$ among $P_b$), where $P_a$ and $P_b$ denote the marginal distribution of $(X, G = a)$ and  $(X, G = b)$, respectively. 

If we regard the potential outcomes $(D_i(0_g), D_i(1_g), Y_i(0_g), Y_i(1_g))$ as undefined for units with $G_i = g'$, where $g' \neq g$, then principal strata such as switchers or always-compliers are themselves not well defined, since their definitions simultaneously involve potential outcomes under both $g$ and $g'$, namely $D_i(0_g)$, $D_i(0_{g'})$, $D_i(1_g)$, and $D_i(1_{g'})$. Alternatively, if these potential outcomes are viewed as conceptually well defined but never observable for units with $G_i = g'$, where $g' \neq g$, then principal strata and related estimands, including switchers, always-compliers, SWATE, and ACOATE, remain well defined. However, the usual ``positivity assumption" is violated (e.g., $P(Z = 1_g \mid G = g', \boldsymbol{X}) = 0$, when $g \neq g'$). Note that the positivity assumption is different (conditional only on $\boldsymbol{X}$, not $(\boldsymbol{X}, G)$), and it is not violated under Framework I (see Assumption 1(iv) in the main article).

In order to still identify these estimands, one needs to assume the baseline covariate $G$ is not an effect modifier, e.g.,
\[
\mathbb{E}[Y(1_a) - Y(0_a) \mid G = a, \mathbf{X}] = \mathbb{E}[Y(1_a) - Y(0_a) \mid G = b, \boldsymbol{X}].
\]
We note that the mathematical form of this assumption is the same as the mean effect exchangeability assumption; however, here, $G$ is a baseline covariate, while previously, $G$ is a treatment component.

To summarize, \textsf{Framework II} is well defined and can address assorted questions, including estimating the intention-to-treat effect (or total effect) of $1_a$ versus $0_a$ among the subpopulation with $G = a$ and of $1_b$ versus $0_b$ among the subpopulation with $G = b$, as well as the complier (or local) average treatment effect among the subpopulation with $G = a$ ($LATE_a$) and among the subpopulation with $G = b$ ($LATE_b$). However, novel principal strata like switchers and always-compliers and novel causal estimands like SWATE and ACOATE maynot be defined or identified under \textsf{Framework II}. 

Table~\ref{tab: comparison three frameworks} summarizes key aspects of three frameworks.

\begin{table}[ht]
    \centering
    \begin{tabular}{cccc}\hline
        &\textsf{Framework I} &\textsf{Framework II} &\textsf{Framework III}  \\
    \makecell{Treatment \\assignment $Z$} & $\{1_a, 0_a, 1_b, 0_b\}$ & $\{1_a, 0_a, 1_b, 0_b\}$ & $\{1, 0\}$ \\[10pt]
    
    Role of $G$ & \makecell{Derived variable\\ from $Z$}& Baseline covariate & Baseline covariate \\[10pt]
    \makecell{Cohort or \\Population} & \makecell{Characterized\\ by $\boldsymbol{X}$} & \makecell{Characterized \\by $(\boldsymbol{X}, G)$} & \makecell{Characterized \\by $(\boldsymbol{X}, G)$}\\[10pt]
    \makecell{Potential \\outcomes} & \makecell{$D_i(z), Y_i(z),$ \\ $~z \in \{1_a, 0_a, 1_b, 0_b\}$}
    & \makecell{$D_i(z), Y_i(z),~z \in \{1_a, 0_a\}$ \\ or $D_i(z), Y_i(z),~z \in \{1_b, 0_b\}$,\\ but not both} & \makecell{$D_i(z), Y_i(z),$ \\ $~z \in \{1, 0\}$} \\[10pt]
    \makecell{Principal \\strata\\ estimands} & \makecell{ACOATE, SWATE\\ COATE all defined on\\ the entire population} & \makecell{LATE of $1_a$ vs $0_a$ among $G=a$\\ LATE of $1_b$ vs $0_b$ among $G=b$} & \makecell{LATE of $1$ vs $0$ \\ among the entire \\population} \\ \hline
    \end{tabular}
    \caption{A summary of \textsf{Framework I}, \textsf{Framework II}, and \textsf{Framework III}.}
    \label{tab: comparison three frameworks}
\end{table}

\subsection{Additional discussion on SUTVA}

One of the reviewers pointed out that one can define
$Y_i(Z \in \{1_a, 1_b\}) = Y_i(Z \in \{1_a, 1_b\}, G_i)$. When we view $G$ as part of the baseline covariate as in \textsf{Framework II}, then reviewer's notion of 
\[
Y_i(Z \in \{1_a, 1_b\}) = Y_i(Z \in \{1_a, 1_b\}, G_i)
\]
is well defined, and $Y_i(Z \in \{1_a, 1_b\}, G_i)$ simplifies to $Y_i(1_a)$ for units with baseline covariate $G_i = a$ and $Y_i(1_b)$ for $G_i = b$, as we formalized in Table \ref{tab:science_table framework II}. 

Under our proposed \textsf{Framework I}, we would like to differentiate between two types of potential outcomes: 
\begin{description}
    \item[Primitive potential outcomes.] Primitive or basic potential outcomes do not depend on the treatment assignment mechanism. They preexist before intervention, just like baseline covariates, and  are what Rubin refers to as ``science" in Rubin's science table. Some, but not all, primitive potential outcomes are revealed by the treatment assignment mechanism. In our proposed framework, the primitive potential outcomes consist of $Y_i(Z = z),~ z\in\{0_a, 1_a, 0_b, 1_b\}$ or $Y_i(Z^\prime = z, G = g),~z \in \{0, 1\}, g \in \{a, b\}$. 

    When referring to SUTVA,  we refer to the ``no hidden version of treatment" part of SUTVA, and the SUTVA is a property of ``primitive" potential outcomes. 
    
    For these primitive potential outcomes, defining $Y_i(Z \in \{1_a, 1_b\})$ would violate SUTVA, as $1_a$ and $1_b$ constitute two versions of treatment assignments; similarly,
    writing $Y_i(Z^\prime = z, G \in \{a, b\})$ would violate SUTVA if   $Y_i(Z^\prime = z, G = a) \neq Y_i(Z^\prime = z, G = b)$.

    \item[Derived potential outcomes.] Derived potential outcomes are potential outcomes that depend on both primitive potential outcomes and the treatment assignment mechanism. As the reviewer correctly pointed out, potential outcomes like $Y_i(Z' = z, G = G_i)$ are well-defined: $Y_i(Z' = z, G = G_i)$ is a function of both the primitive potential outcomes and treatment assignment mechanism.   
\end{description}

\subsection{Additional causal parameters}
\label{subsec: latex_a and late_b, supp}
We discuss two additional causal parameters in this section. Under \textsf{Framework I}, we can define the estimand
\begin{align*}
    \text{LATE}_a := &\mathbb{E}\{Y(D = 1)-Y(D = 0) \mid D(Z=1_a)>D(Z=0_a), G=a\} \\
    =&\mathbb{E}\{Y(D = 1)-Y(D = 0) \mid D(Z=1_a)>D(Z=0_a), Z \in \{0_a, 1_a\}\},
\end{align*}
which is analogous to the ``average treatment effect on the treated," except that it is the average ``per-protocol effect" among those with observed assignment component treatment $G = a$ (or equivalently assigned $Z \in \{0_a, 1_a\}$) in the factual world and are compliers under $(0_a, 1_a)$. We can similarly define $\text{LATE}_b$. 

Identifying these two estimands, under \textsf{Framework I}, does not require the assumption of mean effect exchangeability over $G$, as shown in the derivation below (omitting baseline covariates $\boldsymbol{X}$):
\begin{align*}
    &\mathbbm{E}_{P_0}[Y\mid Z=1_a,G=a]- \mathbbm{E}_{P_0}[Y\mid Z=0_a,G=a]\\
    =&\mathbbm{E}_{P_0}[Y(Z=1_a)\mid Z=1_a,G=a]- \mathbbm{E}_{P_0}[Y(Z=0_a)\mid Z=0_a,G=a]\\
    =&\mathbbm{E}_{P_0}[Y(Z=1_a)-Y(Z=0_a)\mid G=a]\\
    =&\mathbbm{E}_{P_0}[Y(Z=1_a)-Y(Z=0_a)\mid D(Z=1_a)=1, D(Z=0_a)=0, G=a]\\
    &\times P_0(D(Z=1_a)=1, D(Z=0_a)=0\mid G=a)\\
    =&\mathbbm{E}_{P_0}[Y(D=1)-Y(D=0)\mid D(Z=1_a)=1, D(Z=0_a)=0, G=a]\\
    &\times P_0(D(Z=1_a)=1, D(Z=0_a)=0\mid G=a),
\end{align*}
where the first equality is due to consistency, the second equality is due to IV independence, the third equality is due to partial monotonicity, and the last equality is due to exclusion restriction. Similarly, we have
\begin{align*}
    &\mathbbm{E}_{P_0}[D\mid Z=1_a,G=a]- \mathbbm{E}_{P_0}[D\mid Z=0_a,G=a]\\
    =&\mathbbm{E}_{P_0}[D(Z=1_a)\mid Z=1_a,G=a]- \mathbbm{E}_{P_0}[D(Z=0_a)\mid Z=0_a,G=a]\\
    =&\mathbbm{E}_{P_0}[D(Z=1_a)-D(Z=0_a)\mid G=a]\\
    =&P_0(D(Z=1_a)=1, D(Z=0_a)=0\mid G=a).
\end{align*}
Put together, we have the following identification result:
\begin{align*}
    \text{LATE}_a = \frac{\mathbbm{E}_{P_0}[Y\mid Z=1_a,G=a]- \mathbbm{E}_{P_0}[Y\mid Z=0_a,G=a]}{\mathbbm{E}_{P_0}[D\mid Z=1_a,G=a]- \mathbbm{E}_{P_0}[D\mid Z=0_a,G=a]}.
\end{align*}

Under \textsf{Framework I}, the estimands $\text{LATE}_a$ and $\text{LATE}_b$ are defined by conditioning on both a principal stratum and the observed treatment assignment ($G=a$ or $b$, equivalently, $Z \in {0_a,1_a}$ or $Z \in {0_b,1_b}$). We view these estimands as less conceptually clean than estimands defined solely in terms of potential outcomes, such as the SWATE and ACOATE. Because the interpretations of $\text{LATE}_a$ and $\text{LATE}_b$ are themselves tied to observed treatment assignment and are therefore less interpretable, comparisons between them, while formally assessing treatment effect heterogeneity across subpopulations, inherit these interpretational challenges.

\subsection{Interpretation of the traditional Wald estimand}
Under standard IV assumptions, equation \eqref{eq: conditional wald SUTVA violate} can be understood as the average ``per-protocol" effect, $Y(D = 1) - Y(D = 0)$, among the subpopulation $\{D(Z^* \in \{1_a, 1_b\}, G) > D(Z^* \in \{0_a, 0_b\}, G)\}$, where we use the vector-IV representation of potential outcomes. To see this,
we have
\begin{align*}
    &\mathbbm{E}_{P_0}[Y(Z^* =1, G)-Y(Z^* =0, G)|\boldsymbol{X}]\\
    =&\mathbbm{E}_{P_0}[Y(Z^* =1, G)-Y(Z^* =0, G)\vert D(Z^* =1, G) > D(Z^*=0, G),\boldsymbol{X}]\\
    &\times P_0(D(Z^*=1, G) > D(Z^*=0, G)|\boldsymbol{X})+\\
    &\mathbbm{E}_{P_0}[Y(Z^*=1, G)-Y(Z^*=0, G)\vert D(Z^*=1, G) = D(Z^*=0, G),\boldsymbol{X}]\\
    &\times P_0(D(Z^*=1, G) = D(Z^*=0, G)|\boldsymbol{X})+\\
    &\mathbbm{E}_{P_0}[Y(Z^*=1, G)-Y(Z^*=0, G)\vert D(Z^*=1, G) < D(Z^*=0, G),\boldsymbol{X}]\\
    &\times P_0(D(Z^*=1, G) < D(Z^*=0, G)|\boldsymbol{X})\\
    =&\mathbbm{E}_{P_0}[Y(D=1)-Y(D=0)\vert D(Z^* =1, G) > D(Z^*=0, G),\boldsymbol{X}]\\
    &\times P_0(D(Z^*=1, G) > D(Z^*=0, G)|\boldsymbol{X}),
\end{align*}
where the last equality is because
\begin{align*}
    &P_0(D(Z^*=1, G) < D(Z^*=0, G)|\boldsymbol{X})\\
    =&P_0(D(Z^*=1, G=a) <D(Z^*=0, G=a)|G=a,\boldsymbol{X})P(G=a|\boldsymbol{X})+\\
    &P_0(D(Z^*=1, G=b) < D(Z^*=0, G=b)|G=b,\boldsymbol{X})P(G=b|\boldsymbol{X})\\
    =&0.
\end{align*}
and 
\begin{align*}
    \mathbbm{E}_{P_0}[Y(Z^*=1, G)-Y(Z^*=0, G)\vert D(Z^*=1, G) = D(Z^*=0, G),\boldsymbol{X}]=0.
\end{align*}

Furthermore, we have
\begin{align*}
    &P_0(D(Z^*=1, G) > D(Z^*=0, G)|\boldsymbol{X}) \\
    =&\mathbbm{E}_{P_0}[D(Z^*=1, G=a)|\boldsymbol{X}]-\mathbbm{E}_{P_0}[D(Z^*=0, G=a)|\boldsymbol{X}]\\
    =&\mathbbm{E}_{P_0}[D(Z^*=1, G=a)|G=a,\boldsymbol{X}]P(G=a|\boldsymbol{X})+\mathbbm{E}_{P_0}[D(Z^*=1, G=b)|G=b,\boldsymbol{X}]P(G=b|\boldsymbol{X})\\
    &-[\mathbbm{E}_{P_0}[D(Z^*=0, G=a)|G=a,\boldsymbol{X}]P(G=a|\boldsymbol{X})+\mathbbm{E}_{P_0}[D(Z^*=0, G=b)|G=b,\boldsymbol{X}]P(G=b|\boldsymbol{X})]\\
    =&\mathbbm{E}_{P_0}[D(Z^*=1, G=a)|Z^*=1,G=a,\boldsymbol{X}]P(G=a|\boldsymbol{X})\\
    &+\mathbbm{E}_{P_0}[D(Z^*=1, G=b)|Z^*=1,G=b,\boldsymbol{X}]P(G=b|\boldsymbol{X})\\
    &-[\mathbbm{E}_{P_0}[D(Z^*=0, G=a)|Z^*=0,G=a,\boldsymbol{X}]P(G=a|\boldsymbol{X})\\
    &+\mathbbm{E}_{P_0}[D(Z^*=0, G=b)|Z^*=0,G=b,\boldsymbol{X}]P(G=b|\boldsymbol{X})]\\
    =&\mathbbm{E}_{P_0}[D|Z^*=1,G=a,\boldsymbol{X}]P(G=a|\boldsymbol{X})+\mathbbm{E}_{P_0}[D|Z^*=1,G=b,\boldsymbol{X}]P(G=b|\boldsymbol{X})\\
    &-[\mathbbm{E}_{P_0}[D|Z^*=0,G=a,\boldsymbol{X}]P(G=a|\boldsymbol{X})+\mathbbm{E}_{P_0}[D|Z^*=0,G=b,\boldsymbol{X}]P(G=b|\boldsymbol{X})]\\
    =&\mathbbm{E}_{P_0}[D|Z^*=1,\boldsymbol{X}]-\mathbbm{E}_{P_0}[D|Z^*=0,\boldsymbol{X}].
\end{align*}

For the third equality, we used IV independence, for the fourth equality, we used consistency, for the last equality, we used the additional assumption $Z^*\indep G|\boldsymbol{X}$.

Similarly, we have
\begin{align*}
    &\mathbbm{E}_{P_0}[Y(Z^* =1, G)-Y(Z^* =0, G)|\boldsymbol{X}]\\
    =&\mathbbm{E}_{P_0}[Y|Z^*=1,\boldsymbol{X}]-\mathbbm{E}_{P_0}[Y|Z^*=0,\boldsymbol{X}].
\end{align*}

}

\section{Connections to the vector IV literature}
\label{sec: literature on multiple IV and monotonicity, supp}
\subsection{Monotonicity assumptions}

An alternative approach for defining the potential outcomes is to use a 
vector-valued instrumental variable $(Z', G)$, where $Z' \in \{0,1\}$ 
indicates whether an individual is assigned to the treatment group or the 
control group, and $G$ is the variable taking values in 
$\{a, b\}$. For $g \in \{a, b\}$, one may alternatively write 
$D(Z = 1_g)$ as $D(Z' = 1, G = g)$ and $D(Z = 0_g)$ as 
$D(Z' = 0, G = g)$. Under this vector-valued IV formulation, we may compare 
our partial monotonicity assumption (denoted PM) to 
(1) the partial monotonicity assumption in \citet{mogstad2021causal} and 
\citet{blackwell2023noncompliance,blackwell2025bounds} (denoted PM2); 
(2) the limited monotonicity assumption in \citet{hoff2025limited} 
(denoted LiM); and (3) the vector monotonicity assumption in 
\citet{goff2024vector} (denoted VM), when both $Z'$ and $G$ are binary.

Definition \ref{def: mono assumptions} states each monotonicity assumption using the notation adopted in the current manuscript.
\begin{definition}\label{def: mono assumptions}
Monotonicity assumptions PM, PM2, LiM, and VM are defined as follows:
    \begin{enumerate}
        \item \textbf{(PM).} $D(Z'=1,G=g)\geq D(Z'=0,G=g)$ for $g\in\{a,b\}$.
        \item \textbf{(PM2).} $D(Z'=1,G=g)\geq D(Z'=0,G=g)$ for $g\in\{a,b\}$  and $D(Z'=z,G=b)\geq D(Z'=z,G=a)$ for $z\in\{0,1\}$. 
        \item \textbf{(LiM).} $D(Z'=1,G=b)\geq D(Z'=0,G=a)$.
        \item \textbf{(VM).} $D(Z'=1,G=g)\geq D(Z'=0,G=g)$ for $g\in\{a,b\}$ and $D(Z'=z,G=b)\geq D(Z'=z,G=a)$ for $z\in\{0,1\}$.
    \end{enumerate}
\end{definition}
\textcolor{black}{When defining \textbf{PM2}, we imposed similar direction constraint as \citet[Equation 7]{mogstad2021causal}. When defining \textbf{VM}, we imposed the order $b\geq a$.} 

Under the setting considered in this article, the relationships among the four monotonicity assumptions are as follows:
\begin{enumerate}
\item \textbf{PM2} and \textbf{VM} are equivalent.
\item \textbf{PM2} implies both \textbf{PM} and \textbf{LiM}.
\item \textbf{PM} and \textbf{LiM} do not imply each other.
\end{enumerate}

\noindent Importantly, \textbf{PM}, \textbf{PM2} (\textbf{VM}), and \textbf{LiM} imply the existence of different principal strata, which is summarized in Table~\ref{tab: Monotonicity assumption}. \textbf{LiM} is the weakest assumption and implies $12$ out of $16$ possible principal strata. \textbf{PM2} (\textbf{VM}) implies $6$ principal strata, which is also a proper subset of the $9$ principal strata implied by \textbf{PM}. \textbf{PM}, together with the key nested IV assumption considered in the current article, implies $7$ principal strata, including Always-Taker-Never-Taker (AT-NT) and Always-Taker-Complier (a particular type of Switcher), that do not exist under \textbf{PM2} (\textbf{VM}). Complier-Always-Taker, which exists under \textbf{PM2} (\textbf{VM}), no longer exists under \textbf{PM} + Nested IV.

\begin{table}[htbp]
  \centering
  \caption{Principal strata under different monotonicity assumptions. $D(z,g)$ is a shorthand of $D(Z'=z,G=g)$.}
    \begin{tabular}{cccccccc}
    \toprule
    $D(0,a)$ & $D(1,a)$ & $D(0,b)$ & $D(1,b)$ & PM    & PM+Nested IV & PM2 (VM)   & LiM \\
    \midrule
    0     & 0     & 0     & 0     & Yes   & Yes   & Yes   & Yes \\
    1     & 0     & 0     & 0     &       &       &       &  \\
    0     & 1     & 0     & 0     & Yes   &       &       & Yes \\
    0     & 0     & 1     & 0     &       &       &       & Yes \\
    0     & 0     & 0     & 1     & Yes   & Yes   & Yes   & Yes \\
    1     & 1     & 0     & 0     & Yes   & Yes   &       &  \\
    1     & 0     & 1     & 0     &       &       &       &  \\
    1     & 0     & 0     & 1     &       &       &       & Yes \\
    0     & 1     & 1     & 0     &       &       &       & Yes \\
    0     & 1     & 0     & 1     & Yes   & Yes   & Yes   & Yes \\
    0     & 0     & 1     & 1     & Yes   & Yes   & Yes   & Yes \\
    1     & 1     & 1     & 0     &       &       &       &  \\
    1     & 1     & 0     & 1     & Yes   & Yes   &       & Yes \\
    1     & 0     & 1     & 1     &       &       &       & Yes \\
    0     & 1     & 1     & 1     & Yes   &       & Yes   & Yes \\
    1     & 1     & 1     & 1     & Yes   & Yes   & Yes   & Yes \\
    \bottomrule
    \end{tabular}%
  \label{tab: Monotonicity assumption}%
\end{table}%

\subsection{Related estimands}
In this section, we review some estimands proposed in vector IV literature. \citet{mogstad2021causal} introduced several new principal strata, including eager complier, reluctant complier, $Z'$ complier, $G$ complier (They used the notations $Z_1$ complier and $Z_2$ complier.) Those principal strata are defined as follows:
\begin{enumerate}
    \item $Z'$ compliers: participants with $(D(Z'=0,G=a)=0,D(Z'=0,G=b)=0,D(Z'=1,G=a)=1,D(Z'=1,G=b)=1)$, denoted $S=\text{$Z'$-complier}$;
    \item $G$-complier: participants with $(D(Z'=0,G=a)=0,D(Z'=0,G=b)=1,D(Z'=1,G=a)=0,D(Z'=1,G=b)=1)$, denoted $S=\text{$G$-complier}$;
    \item Reluctant compliers: participants with $(D(Z'=0,G=a)=0,D(Z'=0,G=b)=0,D(Z'=1,G=a)=0,D(Z'=1,G=b)=1)$, denoted $S=\text{RC}$;
    \item Eager compliers: participants with $(D(Z'=0,G=a)=0,D(Z'=0,G=b)=1,D(Z'=1,G=a)=1,D(Z'=1,G=b)=1)$, denoted $S=\text{EC}$. 
\end{enumerate}

\citet{hoff2025limited} introduced a new principal treatment effect, termed the combined compliers local average treatment effect (CC-LATE). The combined compliers (denoted $S=CC$) are defined as indivisuals with $D(Z'=0,G=a)=0$ and $D(Z'=1,G=b)=1$. Under standard conditions , the CC-LATE, 
\begin{align*}
    \mathbbm{E}_{P_0}[Y(D=1)-Y(D=0)\mid S=\text{CC}],
\end{align*}
can be identified via the following identification formula \citep{hoff2025limited}:
\begin{align*}
     \text{CC-LATE} = \frac{\mathbbm{E}_{P_0}[Y\mid Z'=1,G=b]-\mathbbm{E}_{P_0}[Y\mid Z'=0,G=a]}{\mathbbm{E}_{P_0}[D\mid Z'=1,G=b]-\mathbbm{E}_{P_0}[D\mid Z'=0,G=a]}.
 \end{align*}

The CC-LATE defined by \citet{hoff2025limited} is equivalent to ACLATE defined by \citet{goff2024vector} with two IVs.  Furthermore, according to \citet{goff2024vector}, the average treatment effect on $Z'$ compliers, $G$ compliers, RC and EC are not identifiable in general. However, certain combinations of the principal strata are identifiable. For example, the average treatment effect on $\{D(1,a)=1,D(0,a)=0\}$, which is a union of $Z'$ compliers and EC, can be identified. The average treatment effect on $\{D(1,a)=1,D(0,a)=0\}$ is denoted as $PTE_{j}(z_{-j})$-type parameter in \citet{goff2024vector}'s notation.

\section{Extension to multiple IV pairs}
\subsection{Nested IV relation}
\label{subsec: multiple IV pairs, supp}
The nested IV framework can be extended to multiple IV pairs. Let $Z\in \mathcal{Z}$ denote an IV, where $\mathcal{Z}$ is the support of $Z$. Throughout this subsection, we assume consistency, exclusion restriction and IV independence assumptions hold for any $z\in\mathcal{Z}$, so $Z$ is a valid IV. Let $\mathcal{G}$ denote an index set, and $0_g$ and $1_g$ denote two distinct elements of $\mathcal{Z}$, where $g\in\mathcal{G}$ is the index. The ordered pair $(0_g, 1_g)$ is referred to as an IV pair, denoted as $\mathcal{Z}_g$, and we have $D(1_g)\geq D(0_g)$ by the monotonicity assumption. 

Definition \ref{def: IV and nested IV} formally defines a nested IV relation.

\begin{definition}
 \textbf{(Nested IV relation).} Let $g,g'\in\mathcal{G}$. IV pair $\mathcal{Z}_{g}$ is said to be nested within IV pair $\mathcal{Z}_{g'}$, denoted as $\mathcal{Z}_g \preceq \mathcal{Z}_{g'}$, if $\{(D(0_g),D(1_g))=(0,1)\}\subset \{(D(0_{g'}),D(1_{g'}))=(0,1)\}$. We also define $\mathcal{Z}_g  \succeq \mathcal{Z}_{g'}$ if $\mathcal{Z}_{g'}  \preceq \mathcal{Z}_{g}$, and $\mathcal{Z}_g  \sim \mathcal{Z}_{g'}$  if both $\mathcal{Z}_g  \succeq \mathcal{Z}_{g'}$ and  $\mathcal{Z}_g  \preceq \mathcal{Z}_{g'}$ hold.
 \label{def: IV and nested IV}
\end{definition}

The nested IV relation is a binary relation defined on the set of all IV pairs $\mathcal{F}:=\{\mathcal{Z}_g, g\in\mathcal{G}\}$. Proposition \ref{prop: nested IV relation} summarizes the key properties of this relation, which follow directly from the Definition \ref{def: IV and nested IV}.

\begin{proposition}\label{prop: nested IV relation}
     The nested IV relation $\preceq$, defined in Definition \ref{def: IV and nested IV}, is reflexive, transitive, but not necessarily complete. The relation ``$\sim$" defines an equivalence relation on $\mathcal{F}$.
\end{proposition}

Suppose the IV pair $\mathcal{Z}_g$ is nested within the IV pair $\mathcal{Z}_{g'}$, the switcher subpopulation associated with IV pairs $\mathcal{Z}_g,\mathcal{Z}_{g'}$ is defined as
\begin{align*}
        \{D(1_{g'})=1,D(0_{g'})=0\}\backslash\{D(1_{g})=1,D(0_{g})=0\}.
\end{align*}


Formalizing the nested IV assumption as a binary relation is  conceptually useful when $G$ takes multiple values. Below, we present three illustrative examples.

\begin{example}[Multicenter clinical trials]
When the objective is to pool data across clinical sites with varying degrees of compliance to estimate the switcher treatment effect, $G$ represents different clinical sites. For instance, in the PLCO study, when the Henry Ford Health System adopted a dual consent process, two other sites---the University of Colorado (UC) and the University of Alabama at Birmingham (UAB)---both adopted a single consent process and had high compliance rates. As we demonstrate in the case study (Section~\ref{subsec: case study analysis of multiple sites, supp}), one may conduct an integrated data analysis by defining $G\in\{a,b,c\}$, where $G=a$ corresponds to Henry Ford, $G=b$ to UC, and $G=c$ to UAB. In this analysis, $\mathcal{Z}=\{0_a,1_a,0_b,1_b,0_c,1_c\}$, and it is reasonable to assume that the IV pair $(0_a,1_a)$ is nested within both $(0_b,1_b)$, denoted $(0_a,1_a)\preceq(0_b,1_b)$, and $(0_c,1_c)$, denoted $(0_a,1_a)\preceq(0_c,1_c)$, without imposing any relation between $(0_b,1_b)$ and $(0_c,1_c)$, because the nested IV assumption may not hold for UC and UAB participants. When viewed as a partial order, both $(0_b,1_b)$ and $(0_c,1_c)$ are maximal elements.
\end{example}

\begin{example}[Ordinal IV]
An ordinal IV can be understood as a collection of IV pairs equipped with a nested IV relation. For instance, let $Z \in \{0,1,2\}$ be an ordinal IV with three levels. Suppose $D(Z)$ is binary, $D(z) \ge D(z')$ whenever $z \ge z'$, and the population consists of a mixture of units with 
\[
(D(0),D(1),D(2)) \in \{(0,0,0),\, (0,0,1),\, (0,1,1),\, (1,1,1)\}.
\]
This three-level ordinal IV can be decomposed into the following three IV pairs: $(0,1)$, $(1,2)$, and $(0,2)$. Among these three IV pairs, we have the following nested IV relations:
\[
(0,1) \preceq (0,2), \qquad (1,2) \preceq (0,2),
\]
while neither $(0,1) \succeq (1,2)$ nor $(0,1) \preceq (1,2)$ holds.
\end{example}


\begin{example}[Strengthening an IV]
A nested IV relation helps understand how an ordinal or continuous IV could be strengthened in a matched cohort study \citep{baiocchi2010building,chen2024manipulatingcontinuousinstrumentalvariable}. Let $Z\in[0,1]$ denote a continuous IV. Suppose the treatment uptake $D(z)$ is determined by some latent continuous variable $D^*(z)$ via $D(z)=\mathbbm{1}\{D^*(z)\geq \tfrac{3}{4}\}$. Suppose the population consists of (1) those with $D^*(z)=z$; and (2) those with $D^*(z)=\tfrac{1}{2}z+\tfrac{1}{2}$. The set of IV pairs can then be defined as $\{(a,b): 0\leq a\leq b\leq 1\}$. One may easily verify, for instance, that $(0,\tfrac{1}{3})$ is nested in $(0,\tfrac{2}{3})$, and $(\tfrac{2}{3},1)$ is nested in $(\tfrac{1}{3},1)$. Consider four participants with the same observed covariates to be matched, with $Z\in\{0,\tfrac{1}{3},\tfrac{2}{3},1\}$. By pairing the participant with $Z=0$ with the one having $Z=\tfrac{2}{3}$, and pairing $Z=\tfrac{1}{3}$ with $Z=1$, one can construct a matched cohort with a higher compliance rate, compared to pairing $Z=0$ with $Z=\tfrac{1}{3}$ and $Z=\tfrac{2}{3}$ with $Z=1$, as advocated in \citet{baiocchi2010building}.
\end{example}

\subsection{Generalized SWATE estimands with three IV pairs}
\label{subsec: three nested IV pairs, supp}
{ 
In this subsection, we generalize the SWATE to settings with three IV pairs. Suppose we have three IV pairs, denoted by $\{a,b,c\}$. There are three possible hierarchical structures among these three pairs, which we refer to as the \emph{fork}, \emph{collider}, and \emph{chain} structures (see also Figure~\ref{fig: structures for three nested IV pairs}):
\begin{enumerate}
    \item Fork: $\{1_a,0_a\}$ is nested within both $\{1_b,0_b\}$ and $\{1_c,0_c\}$.
    \item Collider: $\{1_a,0_a\}$ and $\{1_b,0_b\}$ are each nested $\{1_c,0_c\}$.
    \item Chain: $\{1_a,0_a\}$ is nested within $\{1_b,0_b\}$, which is in turn nested within $\{1_c,0_c\}$.
\end{enumerate}

\begin{figure}[ht]
    \centering
    \includegraphics[width=1\linewidth]{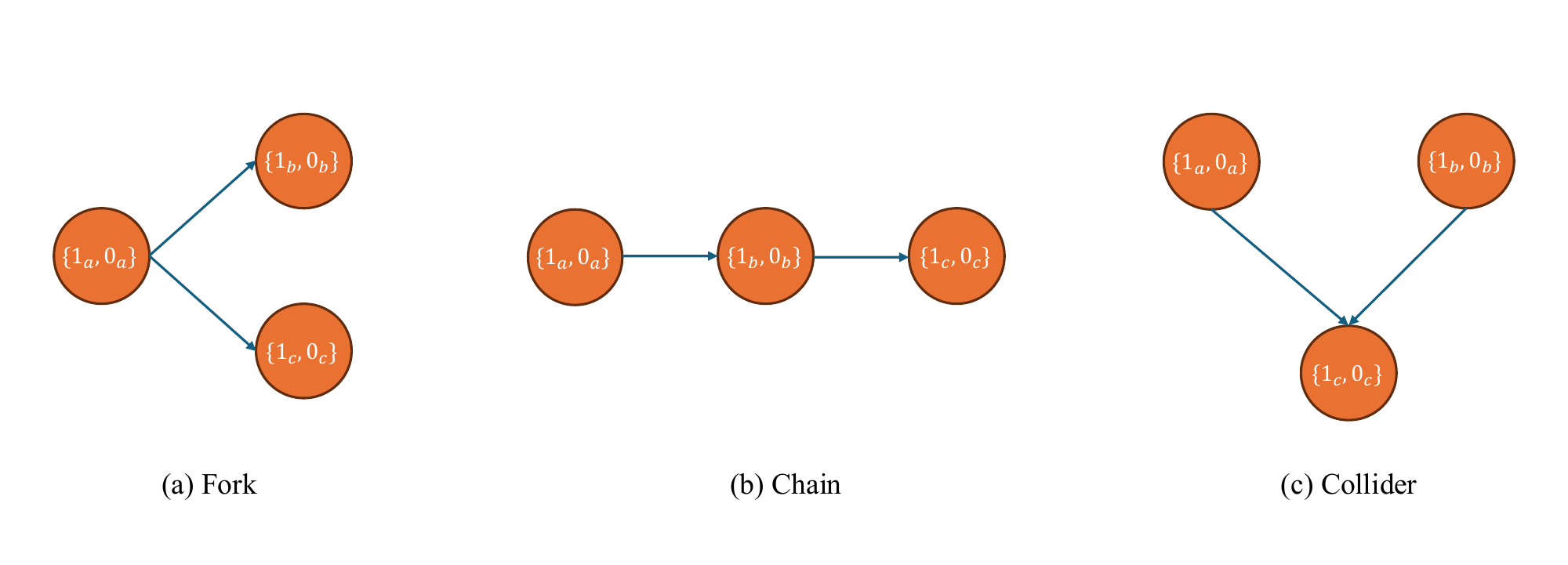}
    \caption{Possible structures for three nested IV pairs.}
    \label{fig: structures for three nested IV pairs}
\end{figure}

\noindent Under three nested IV pairs, we can define more general switcher subgroups. However, these more complex switcher subgroups are not identifiable in general. The switcher-related principal strata are listed in Table~\ref{tab: Switcher-related principal strata under three nested IV pairs}. For simplicity, we use the notation $\text{CO\text{-}CO\text{-}CO}$ to denote the subgroup that is a complier under all three IV pairs $\{1_a,0_a\}$, $\{1_b,0_b\}$, and $\{1_c,0_c\}$. Other notations are defined analogously. We use the notation $\text{SW}(g,g')$ to denote the subgroup that is a non-complier under IV pair $\{1_g,0_g\}$ and a complier under IV pair $\{1_{g'},0_{g'}\}$.

Under the fork structure, we can identify the subgroup $A1$, since it is the complier subgroup under the IV pair $\{1_a,0_a\}$. Similarly, we can identify $A2\cup A4$ and $A2\cup A3$, as they are equivalent to $\text{SW}(a,b)$ and $\text{SW}(a,c)$, respectively. The subgroup $A2$ can be thought of as a generalized switcher subgroup, which includes individuals who are not compliers under the weakest IV pair $\{1_a,0_a\}$, but are compliers under both $\{1_b,0_b\}$ and $\{1_c,0_c\}$. We call this subgroup \emph{Eager-SWitchers (ESW)}. Similarly, the union of $A3$ and $A4$ can also be thought of as a generalized switcher subgroup, which includes individuals who are not compliers under $\{1_a,0_a\}$ but are compliers under either $\{1_b,0_b\}$ or $\{1_c,0_c\}$. We call this subgroup \emph{Reluctant-SWitchers (RSW)}.

However, the average treatment effects among ESW and RSW are not identifiable under the fork structure. To see this, consider the following equations (omitting covariates) for the outcome and treatment:

\begin{align*}
    &\mathbb{E}_{P_0}[Y\mid Z=1_a]-\mathbb{E}_{P_0}[Y\mid Z=0_a] = P(A1)\,\mathbb{E}_{P_0}[Y(1)-Y(0)\mid Z=0, A1],\\
    &\mathbb{E}_{P_0}[Y\mid Z=1_b]-\mathbb{E}_{P_0}[Y\mid Z=0_b] = P(A1)\,\mathbb{E}_{P_0}[Y(1)-Y(0)\mid Z=0, A1]\\
    &\quad +P(A2)\,\mathbb{E}_{P_0}[Y(1)-Y(0)\mid Z=0, A2]+P(A4)\,\mathbb{E}_{P_0}[Y(1)-Y(0)\mid Z=0, A4],\\
    &\mathbb{E}_{P_0}[Y\mid Z=1_c]-\mathbb{E}_{P_0}[Y\mid Z=0_c] = P(A1)\,\mathbb{E}_{P_0}[Y(1)-Y(0)\mid Z=0, A1]\\
    &\quad +P(A2)\,\mathbb{E}_{P_0}[Y(1)-Y(0)\mid Z=0, A2]+P(A3)\,\mathbb{E}_{P_0}[Y(1)-Y(0)\mid Z=0, A3],\\
    &\mathbb{E}_{P_0}[D\mid Z=1_a]-\mathbb{E}_{P_0}[D\mid Z=0_a] = P(A1),\\
    &\mathbb{E}_{P_0}[D\mid Z=1_b]-\mathbb{E}_{P_0}[D\mid Z=0_b] = P(A1)+P(A2)+P(A4),\\
    &\mathbb{E}_{P_0}[D\mid Z=1_c]-\mathbb{E}_{P_0}[D\mid Z=0_c] = P(A1)+P(A2)+P(A3).
\end{align*}

From the above system, we can identify $P(A1)$ and $\mathbb{E}_{P_0}[Y(1)-Y(0)\mid Z=0, A1]$. However, since there are only four remaining equations but six unknowns (the proportions and treatment effects of $A2$, $A3$, and $A4$), the remaining quantities cannot be identified without imposing additional assumptions.

Under the chain structure, we can similarly define \emph{Eager-SWitchers (ESW)} to be $B1$ and \emph{Reluctant-SWitchers (RSW)} to be $B3$. In this case, ESW and RSW can be reduced to $\text{SW}(a,b)$ and $\text{SW}(a,c)$, respectively; therefore, their average treatment effects can be identified.

Under the collider structure, we can similarly define \emph{Eager-SWitchers (ESW)} to be $C3$ and \emph{Reluctant-SWitchers (RSW)} to be $C4$. In this case, ESW can be identified since it is equivalent to $\text{SW}(a,b)$, whereas RSW remains unidentifiable.

To summarize, these generalized switcher effects can be identified only when they can be reduced to the SWATE as previously defined.

\begin{table}[htbp]
  \centering
  \caption{Switcher-related principal strata under three nested IV pairs. $\text{AT/NT}$ represents the union of always-takers and never-takers. }
    \begin{tabular}{ccccc}
    \toprule
          & $\{1_a,0_a\}$ &$ \{1_b,0_b\}$ & $\{1_c,0_c\}$ & Notations \\
    \midrule
    \multirow{5}[2]{*}{Fork} & CO    & CO    & CO    & A1 \\
          & AT/NT & CO    & CO    & A2 \\
          & AT/NT & AT/NT & CO    & A3 \\
          & AT/NT & CO    & AT/NT & A4 \\
          & AT/NT & AT/NT & AT/NT & A5 \\
    \midrule
    \multirow{4}[2]{*}{Chain} & CO    & CO    & CO    & B1 \\
          & AT/NT & CO    & CO    & B2 \\
          & AT/NT & AT/NT & CO    & B3 \\
          & AT/NT & AT/NT & AT/NT & B4 \\
    \midrule
    \multirow{5}[2]{*}{Collider} & CO    & CO    & CO    & C1 \\
          & CO    & AT/NT & CO    & C2 \\
          & AT/NT & CO    & CO    & C3 \\
          & AT/NT & AT/NT & CO    & C4 \\
          & AT/NT & AT/NT & AT/NT & C5 \\
    \bottomrule
    \end{tabular}%

  \label{tab: Switcher-related principal strata under three nested IV pairs}%
\end{table}%

\section{Additional details on identification}
\label{sec: additional identification details, supp}
\subsection{Extension to the case where $G$ has a direct effect on $Y$}
\label{subsec: G has direct effect}
{ 
In this subsection, we consider the case in which $G$ has a direct effect on Y. Under this setting, the core IV assumptions can be expressed as follows:
\begin{assumption}\label{ass: standard IV assumption, supp}
    \textbf{(Standard IV assumptions).}\vspace{-6pt}
    \begin{enumerate}[label=(\emph{\roman*}).]
        \item \textbf{(Consistency).} $D = D(Z=z)$ if $Z=z$, $Y = Y(D=d,Z=z)$ if $D=d$ and $Z=z$, for $d\in \{0,1\}$ and $z\in \{0_a, 1_a, 0_b, 1_b\}$.\vspace{-6pt}
        \item \textbf{(Partial exclusion restriction)}. $Y(D=d,Z=1_g) = Y(D=d,Z=0_g)$, for $d\in \{0,1\}$ and $g\in \{a,b\}$.\vspace{-6pt}
  
    \item \textbf{(IV independence).} 
 $Z\indep \{D(Z=z),Y(D=d)\} \mid \boldsymbol{X} = \boldsymbol{X}, G=g$, $\forall~d$, $z$ and $g$.\vspace{-6pt}
    \item \textbf{(IV relevance).} 
 $\mathbb{E}_{P_0}[D\mid Z = 1_g, \boldsymbol{X}] - \mathbb{E}_{P_0}[D \mid Z = 0_g, \boldsymbol{X}]\neq 0$ and $P_0(Z=z \mid \boldsymbol{X}) > 0$ for $g\in \{a,b\}$, $z\in \{0_a,1_a,0_b,1_b\}$.  
 \end{enumerate}
\end{assumption} 

\noindent Compared with Assumption~1 in the main article, only the exclusion restriction assumption is relaxed to a partial exclusion restriction assumption in Assumption~\ref{ass: standard IV assumption, supp}. Under this partial exclusion restriction, the potential outcomes can no longer be written simply as $Y(D=d)$, because $Y(D=d, Z=1_a)$ may differ from $Y(D=d, Z=1_b)$. However, we can express the potential outcome as $Y(D=d, G=g)$, which we define to be equal to both $Y(D=d, Z=1_g)$ and $Y(D=d, Z=0_g)$ under partial exclusion restriction.

The nested IV assumptions are not modified under this setting and are restated below for completeness.

\begin{assumption}\label{ass: nested IV assumptions, supp}
    \textbf{(Nested IV assumptions).}\vspace{-6pt}
    \begin{enumerate}[label=(\emph{\roman*}).]
    \item {\textbf{(Mean effect exchangeability over $G$).} } $\mathbb{E}_{P_0}[Y(Z = 1_g) - Y(Z = 0_g)\mid G=a, \boldsymbol{X}] = \mathbb{E}_{P_0}[Y(Z = 1_g) - Y(Z = 0_g)\mid G = b, \boldsymbol{X}]$ and $\mathbb{E}_{P_0}[D(Z = 1_g) - D(Z = 0_g)\mid G=a, \boldsymbol{X}] = \mathbb{E}_{P_0}[D(Z = 1_g) - D(Z = 0_g)\mid G = b, \boldsymbol{X}]$, $g \in \{a, b\}.$
    
    \item \textbf{(Partial monotonicity).} 
 $D(Z=1_g)\geq D(Z=0_g)$, $g \in \{a  , b\}$. \vspace{-6pt}
 \item \textbf{(Non-equal compliance).} 
      $\mathbb{E}_{P_0}[D\mid Z = 1_a, \boldsymbol{X} = \boldsymbol{x}]-\mathbb{E}_{P_0}[D\mid Z = 0_a, \boldsymbol{X} = \boldsymbol{x}] \\
      \neq  \mathbb{E}_{P_0}[D\mid Z = 1_b, \boldsymbol{X} = \boldsymbol{x}]-\mathbb{E}_{P_0}[D\mid Z = 0_b, \boldsymbol{X} = \boldsymbol{x}]$.
   \vspace{-6pt}
 \item {\small\textbf{(Nested IV).} 
    $\{D(Z=0_a), D(Z=1_a)\} = (0, 1)$ implies $\{D(Z=0_b), D(Z=1_b)\} = (0, 1)$.}
 \end{enumerate}
 \end{assumption}

To ensure that SWATE and ACOATE are well defined and identifiable, an additional assumption of no additive interaction between $D$ and $G$ is required.
 
\begin{assumption}\label{ass: no interaction of G}
\textbf{(No interaction between $D$ and $G$).} 
For $s \in \{\text{SW}, \text{ACO}\}$,
\begin{align*}
    &\mathbb{E}_{P_0}\big[Y(D=1,G=a) - Y(D=0,G=a) \mid S=s, \boldsymbol{X}\big] \\
=~&\mathbb{E}_{P_0}\big[Y(D=1,G=b) - Y(D=0,G=b) \mid S=s, \boldsymbol{X}\big].
\end{align*}
\end{assumption}

\noindent Assumption~\ref{ass: no interaction of G} is analogous to the ``no-interaction-type'' assumption in \citet{wang2018bounded} in a single-IV setting. This assumption essentially states that $G$ does not modify the per-protocol effect of $D$ on $Y$. For example, in the PLCO trial studied in that paper, Assumption~\ref{ass: no interaction of G} would hold if the effect of receiving the first scheduled colorectal cancer screening on colorectal cancer risk remained the same before 1997 and after 1997 for participants with the same observed covariates \emph{and} within the same principal stratum. However, by introducing notation of the form $Y(D=d, G=g)$, we still allow the cancer incidence to vary between the periods before 1997 and after 1997 (i.e., $Y(D = d, G = a)$ may not equal $Y(D = d, G = b)$,~ for $d \in \{0, 1\}$).

Under Assumption~\ref{ass: no interaction of G}, the notation 
\[
\mathbb{E}_{P_0}[Y(D=1) - Y(D=0) \mid S=s, \boldsymbol{X}]
\] 
can be used to represent either 
$\mathbb{E}_{P_0}\big[Y(D=1,G=a) - Y(D=0,G=a) \mid S=s, \boldsymbol{X}\big]$ or
$\mathbb{E}_{P_0}\big[Y(D=1,G=b) - Y(D=0,G=b) \mid S=s, \boldsymbol{X}\big]$.

Therefore, the SWATE and ACOATE remain well defined. Moreover, under this assumption, SWATE and ACOATE can be identified using the same formulas as those provided in Theorem~1 of the main article. We formalize this as follows:

\begin{theorem}
   Suppose Assumptions~\ref{ass: standard IV assumption, supp}, 
\ref{ass: nested IV assumptions, supp}, and \ref{ass: no interaction of G} hold. 
Then, under these assumptions, the SWATE and ACOATE can be identified using the same formulas as those provided in Theorem~\ref{thm_identification}. \label{thm: G direct effect,1,supp}
\end{theorem}

\subsection{An alternative identification result}
\label{subsec: alternative nested IV assumption}
In the main article, a key identification assumption is the mean effect exchangeability assumption. 
Below, we present an alternative to the mean effect exchangeability assumption that still allows for the identification of the SWATE and ACOATE.

\begin{assumption}\label{ass: nested IV assumptions, v2, supp}
    \textbf{(Nested IV assumptions).}\vspace{-6pt}
    \begin{enumerate}[label=(\emph{\roman*}).]
    \item {\textbf{(No effect modification by $G$).} } For $s\in \{\text{ACO,SW}\}$, $\mathbb{E}_{P_0}[Y(D=1) - Y(D=0)\mid G=a, S=s, \boldsymbol{X}] = \mathbb{E}_{P_0}[Y(D=1) - Y(D=0)\mid G = b, S=s, \boldsymbol{X}]$ and $P_0(S=s\mid G=a,\boldsymbol{X}) = P_0(S=s\mid G=b,\boldsymbol{X})$.
    
    \item \textbf{(Partial monotonicity).} 
 $D(Z=1_g)\geq D(Z=0_g)$, $g \in \{a  , b\}$. \vspace{-6pt}
 \item \textbf{(Non-equal compliance).} 
      $\mathbb{E}_{P_0}[D\mid Z = 1_a, \boldsymbol{X} = \boldsymbol{x}]-\mathbb{E}_{P_0}[D\mid Z = 0_a, \boldsymbol{X} = \boldsymbol{x}] \\
      \neq  \mathbb{E}_{P_0}[D\mid Z = 1_b, \boldsymbol{X} = \boldsymbol{x}]-\mathbb{E}_{P_0}[D\mid Z = 0_b, \boldsymbol{X} = \boldsymbol{x}]$.
   \vspace{-6pt}
 \item {\small\textbf{(Nested IV).} 
    $\{D(Z=0_a), D(Z=1_a)\} = (0, 1)$ implies $\{D(Z=0_b), D(Z=1_b)\} = (0, 1)$.}
 \end{enumerate}
 \end{assumption}

\noindent Assumption \ref{ass: nested IV assumptions, v2, supp}~(\romannumeral1) has two parts. The first part is saying that the ``per-protocol" effect of $D$ on $Y$ is not modified by $G$. For instance, in our case study, it would hold if the effect of receiving the first scheduled colorectal cancer screening on colorectal cancer risk remained the same before 1997 and after 1997 for participants with the same observed covariates \emph{and} within the same principal stratum. The second part of Assumption \ref{ass: nested IV assumptions, v2, supp}~(\romannumeral1) is saying that the proportion of always-compliers and switchers, within the stratum defined by observed covariates, does not change in two stages: $G = a$ vs $G = b$. 

\begin{theorem}
    Suppose Assumptions~\ref{ass: standard IV assumption, supp} and \ref{ass: nested IV assumptions, v2, supp} hold. 
Then, the SWATE and ACOATE can be identified using the same formulas as those provided in Theorem~\ref{thm_identification}. \label{thm: G direct effect,2,supp}

\end{theorem}

}

\section{Partial identification of the ATE when the outcome is binary}
\label{sec: partial identification, supp}
We consider how to derive the partial identification region of $P(Y(D=1)=1)- P(Y(D=0)=1)$ under different sets of assumptions, assuming $Y$ is a binary outcome and no additional covariates are available. This discrete variable setting is canonical in the partial identification literature for instrumental variables \citep{richardson2024assumptions, song2024instrumental}. We will consider the following assumptions:

\begin{assumption}\label{ass: standard IV assumption, partial, supp}
    \begin{enumerate}[label=(\emph{\roman*}).]
    \item \textbf{(IV independence).} 
 $Z\indep \{D(Z=z),Y(D=d)\} $, $\forall~d$ and $z$.\vspace{-6pt}
 \item \textbf{(Partial monotonicity).} 
 $D(Z=1_g)\geq D(Z=0_g)$, $g \in \{a  , b\}$. 
  \item {\small\textbf{(Nested IV).} 
    $\{D(Z=0_a), D(Z=1_a)\} = (0, 1)$ implies $\{D(Z=0_b), D(Z=1_b)\} = (0, 1)$.}
 \end{enumerate}
 \label{ass: partial identification, supp}
\end{assumption} 

Our goal is to derive an identification region for the ATE 
\[
P(Y(D=1)=1) - P(Y(D=0)=1),
\] 
which is compatible with the observed data distribution 
\[
P(Y=y, D=d \mid Z=z), \quad y \in \{0,1\},\ d \in \{0,1\},\ z \in \{0_a,0_b,1_a,1_b\},
\] 
(1) under Assumption (\romannumeral1) alone; and (2) under Assumptions (\romannumeral1), (\romannumeral2), and (\romannumeral3).

We first consider imposing Assumption (\romannumeral1) alone. To simplify notation, we write $NA$ as a shorthand for the principal stratum that is a Never-taker under the IV pair $\{0_a,1_a\}$ and an Always-taker under the IV pair $\{0_b,1_b\}$. Similarly, we can define shorthands for other principal strata. Furthermore, we use the notation 
\[
N\cdot = \{NA, NN, NC, ND\}, \quad \cdot N = \{AN, NN, CN, DN\},
\] 
and other analogous notations for other combinations. 

For the observed conditional probabilities, for $y \in \{0,1\}$, we have:
\begin{equation}
\begin{aligned}
    &P_0(Y=y,D=1 \mid Z=1_a) = \sum_{s \in C\cdot \cup A\cdot} P_0(Y(D=1)=y, S=s \mid Z=1_a),\\
    &P_0(Y=y,D=0 \mid Z=1_a) = \sum_{s \in N\cdot \cup D\cdot} P_0(Y(D=0)=y, S=s \mid Z=1_a),\\
    &P_0(Y=y,D=1 \mid Z=1_b) = \sum_{s \in \cdot C \cup \cdot A} P_0(Y(D=1)=y, S=s \mid Z=1_b),\\
    &P_0(Y=y,D=0 \mid Z=1_b) = \sum_{s \in \cdot N \cup \cdot D} P_0(Y(D=0)=y, S=s \mid Z=1_b),\\
    &P_0(Y=y,D=1 \mid Z=0_a) = \sum_{s \in A\cdot \cup D\cdot} P_0(Y(D=1)=y, S=s \mid Z=0_a),\\
    &P_0(Y=y,D=0 \mid Z=0_a) = \sum_{s \in N\cdot \cup C\cdot} P_0(Y(D=0)=y, S=s \mid Z=0_a),\\
    &P_0(Y=y,D=1 \mid Z=0_b) = \sum_{s \in \cdot A \cup \cdot D} P_0(Y(D=1)=y, S=s \mid Z=0_b),\\
    &P_0(Y=y,D=0 \mid Z=0_b) = \sum_{s \in \cdot N \cup \cdot C} P_0(Y(D=0)=y, S=s \mid Z=0_b).
\end{aligned}
\label{eq: constraints 1}
\end{equation}

Therefore, there are a total of 16 constraints for the observed data.

Furthermore, the law of total expectation implies that 
\[
P(Y(D=d)=y \mid Z=z) = \sum_{s \in N\cdot \cup A\cdot \cup C\cdot \cup D\cdot} P(Y(D=d)=y, S=s \mid Z=z),
\] 
so that we have the following equalities: 

\scriptsize
\begin{equation}
\begin{aligned}
    &P_0(Y(D=1)=y \mid Z=1_a) - P_0(Y=y,D=1 \mid Z=1_a) = \sum_{s \in N\cdot \cup D\cdot} P_0(Y(D=1)=y, S=s \mid Z=1_a),\\
    &P_0(Y(D=0)=y \mid Z=1_a) - P_0(Y=y,D=0 \mid Z=1_a) = \sum_{s \in C\cdot \cup A\cdot} P_0(Y(D=0)=y, S=s \mid Z=1_a),\\
    &P_0(Y(D=1)=y \mid Z=1_b) - P_0(Y=y,D=1 \mid Z=1_b) = \sum_{s \in \cdot N \cup \cdot D} P_0(Y(D=1)=y, S=s \mid Z=1_b),\\
    &P_0(Y(D=0)=y \mid Z=1_b) - P_0(Y=y,D=0 \mid Z=1_b) = \sum_{s \in \cdot C \cup \cdot A} P_0(Y(D=0)=y, S=s \mid Z=1_b),\\
    &P_0(Y(D=1)=y \mid Z=0_a) - P_0(Y=y,D=1 \mid Z=0_a) = \sum_{s \in N\cdot \cup C\cdot} P_0(Y(D=1)=y, S=s \mid Z=0_a),\\
    &P_0(Y(D=0)=y \mid Z=0_a) - P_0(Y=y,D=0 \mid Z=0_a) = \sum_{s \in A\cdot \cup D\cdot} P_0(Y(D=0)=y, S=s \mid Z=0_a),\\
    &P_0(Y(D=1)=y \mid Z=0_b) - P_0(Y=y,D=1 \mid Z=0_b) = \sum_{s \in \cdot N \cup \cdot C} P_0(Y(D=1)=y, S=s \mid Z=0_b),\\
    &P_0(Y(D=0)=y \mid Z=0_b) - P_0(Y=y,D=0 \mid Z=0_b) = \sum_{s \in \cdot A \cup \cdot D} P_0(Y(D=0)=y, S=s \mid Z=0_b).
\end{aligned}
\label{eq: constraints for PO}
\end{equation}
\normalsize

Under Assumption (\romannumeral1), the constraints \eqref{eq: constraints 1} become

\begin{equation}
\begin{aligned}
    &P_0(Y=y,D=1 \mid Z=1_a) = \sum_{s \in C\cdot \cup A\cdot} P_0(Y(D=1)=y, S=s),\\
    &P_0(Y=y,D=0 \mid Z=1_a) = \sum_{s \in N\cdot \cup D\cdot} P_0(Y(D=0)=y, S=s),\\
    &P_0(Y=y,D=1 \mid Z=1_b) = \sum_{s \in \cdot C \cup \cdot A} P_0(Y(D=1)=y, S=s),\\
    &P_0(Y=y,D=0 \mid Z=1_b) = \sum_{s \in \cdot N \cup \cdot D} P_0(Y(D=0)=y, S=s),\\
    &P_0(Y=y,D=1 \mid Z=0_a) = \sum_{s \in A\cdot \cup D\cdot} P_0(Y(D=1)=y, S=s),\\
    &P_0(Y=y,D=0 \mid Z=0_a) = \sum_{s \in N\cdot \cup C\cdot} P_0(Y(D=0)=y, S=s),\\
    &P_0(Y=y,D=1 \mid Z=0_b) = \sum_{s \in \cdot A \cup \cdot D} P_0(Y(D=1)=y, S=s),\\
    &P_0(Y=y,D=0 \mid Z=0_b) = \sum_{s \in \cdot N \cup \cdot C} P_0(Y(D=0)=y, S=s).
\end{aligned}
\label{eq: constraints 2}
\end{equation}

Furthermore, we have the following two sets of natural constraints: for $d,d' \in \{0,1\}$, 
\begin{equation}
\begin{aligned}
    &P_0(Y(D=d)=1, S=s) + P_0(Y(D=d)=0, S=s) \\
    =\ &P_0(Y(D=d')=1, S=s) + P_0(Y(D=d')=0, S=s).
\end{aligned}
\label{eq: natural constraints}
\end{equation}

The constraints \eqref{eq: constraints for PO} are simplified to
\small
\begin{equation}
    \begin{aligned}
        &P_0(Y(D=1)=y) -P_0(Y=y,D=1|Z=1_a) = \sum_{s\in N\cdot \cup D\cdot} P_0(Y(D=1)=y,S=s)\\
&P_0(Y(D=0)=y) -P_0(Y=y,D=0|Z=1_a) = \sum_{s\in C\cdot \cup A\cdot} P_0(Y(D=1)=y,S=s)\\
&P_0(Y(D=1)=y) -P_0(Y=y,D=1|Z=1_b) = \sum_{s\in \cdot N \cup \cdot D} P_0(Y(D=0)=y,S=s)\\
&P_0(Y(D=0)=y) -P_0(Y=y,D=1|Z=1_b) = \sum_{s\in \cdot C \cup \cdot A} P_0(Y(D=1)=y,S=s)\\
&P_0(Y(D=1)=y) -P_0(Y=y,D=1|Z=0_a) = \sum_{s\in N\cdot \cup C\cdot} P_0(Y(D=1)=y,S=s)\\
&P_0(Y(D=0)=y) -P_0(Y=y,D=0|Z=0_a) = \sum_{s\in A\cdot \cup D\cdot} P_0(Y(D=0)=y,S=s)\\
&P_0(Y(D=1)=y) -P_0(Y=y,D=1|Z=0_b) = \sum_{s\in \cdot N \cup \cdot C} P_0(Y(D=1)=y,S=s)\\
&P_0(Y(D=0)=y) -P_0(Y=y,D=0|Z=0_b) = \sum_{s\in \cdot A \cup \cdot D} P_0(Y(D=0)=y,S=s)
    \end{aligned}
    \label{eq: constraints for PO,2}
\end{equation}
\normalsize

Further imposing the partial monotonicity assumption and the nested IV assumption may improve the partial identification bounds. According to the partial monotonicity assumption, we have the following additional constraints:
\begin{align*}
    P_0(Y(D=d), S=s) = 0, \quad \text{for } s \in D\cdot \cup \cdot D.
\end{align*}
These constraints hold because the partial monotonicity assumption implies that there are no defiers for either IV pair. 

If the nested IV assumption also holds, then we additionally have the following constraints:
\begin{align*}
    P_0(Y(D=d), S=s) = 0, \quad \text{for } s \in \{CA, CN\}.
\end{align*}

We now state formal results.

\begin{definition}
    We say a pair of marginal probabilities $(p_0,p_1)$ is compatible with a joint distribution $P^*$ on $(Y(1),Y(0),D,Z)$ if $p_0 = P^*(Y(0)=1)$, $p_0 = P^*(Y(1)=1)$.
\end{definition}

\begin{definition}
    We say a pair of marginal probabilities $(p_0,p_1)$ is compatible with a joint distribution $P$ on $(Y,D,Z)$ and additional constraints if there exists a $P^*$ such that $P^*$ implies $P$, $(p_0,p_1)$ is compatible with $P^*$ and  $P^*$ satisfies the given constraints.
\end{definition}

\begin{proposition}
    Let $P_0$ be a fixed observed data distribution on $(Y,D,Z)$. Then a pair of marginal probabilities $(p_0,p_1)$ for the potential outcomes $(p_0,p_1)$ is compatible with $P_0$ under the IV independence assumption if and only if there exists non-negative functions $\{(h_{1,s},h_{0,s}), t \in \cdot\cdot\}$, such that the following constraints are satisfied for $y\in \{0,1\}$:
    \footnotesize
    \begin{equation}
    \label{eq: charaterization, supp}
        \begin{aligned}
        &P_0(Y=y,D=1|Z=1_a) = \sum_{s\in C\cdot\cup A\cdot}h_{1,s}(y),
    P_0(Y=y,D=0|Z=1_a) = \sum_{s\in N\cdot\cup D\cdot}h_{1,s}(y),\\
    &P_0(Y=y,D=1|Z=1_b) = \sum_{s\in \cdot C\cup \cdot A}h_{1,s}(y),
    P_0(Y=y,D=0|Z=1_b) = \sum_{s\in \cdot N\cup \cdot D}h_{1,s}(y),\\ 
    &P_0(Y=y,D=1|Z=0_a) = \sum_{s\in A\cdot\cup D\cdot}h_{0,s}(y),
    P_0(Y=y,D=0|Z=0_a) = \sum_{s\in N\cdot\cup C\cdot}h_{1,s}(y)),\\
    &P_0(Y=y,D=1|Z=0_b) = \sum_{s\in \cdot A\cup \cdot D}h_{0,s}(y),
    P_0(Y=y,D=0|Z=0_b) = \sum_{s\in \cdot N\cup \cdot C}h_{0,s}(y),\\
    &yp_1+(1-y)(1-p_1)-P_0(Y=y,D=1|Z=1_a) = \sum_{s\in N\cdot \cup D\cdot} h_{1,s}(y)\\
&yp_0+(1-y)(1-p_0) -P_0(Y=y,D=0|Z=1_a) = \sum_{s\in C\cdot \cup A\cdot} h_{1,s}(y),\\
&yp_1+(1-y)(1-p_1) -P_0(Y=y,D=1|Z=1_b) = \sum_{s\in \cdot N \cup \cdot D} h_{0,s}(y),\\
&yp_0+(1-y)(1-p_0) -P_0(Y=y,D=1|Z=1_b) = \sum_{s\in \cdot C \cup \cdot A} h_{1,s}(y)\\
&yp_1+(1-y)(1-p_1) -P_0(Y=y,D=1|Z=0_a) = \sum_{s\in N\cdot \cup C\cdot} h_{1,s}(y),\\
&yp_0+(1-y)(1-p_0) -P_0(Y=y,D=0|Z=0_a) = \sum_{s\in A\cdot \cup D\cdot} h_{0,s}(y)\\
&yp_1+(1-y)(1-p_1) -P_0(Y=y,D=1|Z=0_b) = \sum_{s\in \cdot N \cup \cdot C} h_{1,s}(y),\\
&yp_0+(1-y)(1-p_0) -P_0(Y=y,D=0|Z=0_b) = \sum_{s\in \cdot A \cup \cdot D} h_{0,s}(y),\\
&h_{1,s}(1)+ h_{1,s}(0) =  h_{0,s}(1)+ h_{0,s}(0), s\in \{\cdot\cdot\}.
    \end{aligned}
    \end{equation}
    
    \normalsize
\end{proposition}

\begin{proof}
We have already shown the ``only if" part. It remains to show the ``if" part. Let 
\[
\pi_s = h_{1,s}(1) + h_{1,s}(0) = h_{0,s}(1) + h_{0,s}(0).
\] 
Consider the conditional probability $(Y_1, Y_0, S)$ given $Z$ constructed as
\begin{align*}
    P^*(Y_1=y, Y_0=y', S=s \mid Z=z) = \frac{h_{1,s}(y) \, h_{0,s}(y')}{\pi_s}, \quad \text{if } \pi_s > 0.
\end{align*}
Under this construction, 
\[
P^*(Y_1=y, Y_0=y', S=s \mid Z=z) = P^*(Y_1=y, Y_0=y', S=s \mid Z=z'),
\] 
so the IV independence assumption holds. 

Furthermore, we have
\begin{align*}
    p_1 
    &= P_0(Y=1,D=1 \mid Z=1_a) + \sum_{s \in N\cdot \cup D\cdot} h_{1,s}(1) \\
    &= \sum_{s \in C\cdot \cup A\cdot} h_{1,s}(1) + \sum_{s \in N\cdot \cup D\cdot} h_{1,s}(1) \\
    &= \sum_s h_{1,s}(1) \\
    &= \sum_s \sum_{y'} \frac{h_{1,s}(1) h_{0,s}(y')}{\pi_s} \\
    &= \sum_s \sum_{y'} P^*(Y_1=1, Y_0=y', S=s \mid Z=z) \\
    &= P^*(Y_1=1 \mid Z=z) \\
    &= P^*(Y_1=1).
\end{align*}
Similarly, we can show that $p_0 = P^*(Y_0=1)$.

Next, we verify that $P^*$ is compatible with the observed data distribution. We will show that $P_0(Y=y,D=1 \mid Z=1_a) = P^*(Y=y,D=1 \mid Z=1_a)$; verification of other conditions is similar.  
\begin{align*}
    P^*(Y=y,D=1 \mid Z=1_a) 
    &= \sum_{s \in A\cdot \cup C\cdot} P^*(Y(1)=y, S=s \mid Z=1_a) \\
    &= \sum_{s \in A\cdot \cup C\cdot} \frac{h_{1,s}(y) \, (h_{0,s}(1)+h_{0,s}(0))}{\pi_s} \\
    &= \sum_{s \in A\cdot \cup C\cdot} h_{1,s}(y) \\
    &= P_0(Y=y, D=1 \mid Z=1_a).
\end{align*}

Therefore, the proposition is proved.
\end{proof}

In practice, the upper bound for $\text{ATE} = p_1 - p_0$ can be obtained by solving the following linear programming problem:
\begin{align*}
    &\max_{p_0, p_1, h_{s,d}(y), \, s \in \{\cdot\cdot\}, \, d \in \{0,1\}, \, y \in \{0,1\}} \; p_1 - p_0\\
    &\text{subject to the constraints in \eqref{eq: charaterization, supp}}.
\end{align*}
\noindent Similarly, the lower bound can be obtained by replacing $p_1 - p_0$ in the objective function by $p_0 - p_1$. The bounds for $\mathbbm{E}_{P_0}[Y(D=d)]$ can also be obtained by modifying the objective function of this optimization problem. Note that the IV assumptions are falsified if no value of the decision variables satisfies the imposed constraints.


\begin{proposition}
    Let $P_0$ be a fixed observed data distribution on $(Y,D,Z)$. Then a pair of marginal probabilities $(p_0,p_1)$ for the potential outcomes $(p_0,p_1)$ is compatible with $P_0$ under the IV independence assumption, the partial monotonicity assumption, and the nested IV assumption if and only if there exists non-negative functions $\{(h_{1,s},h_{0,s}), t \in \cdot\cdot\}$, such that the following constraints are satisfied for $y\in \{0,1\}$:
    \footnotesize
    \begin{equation}
    \label{eq: eq: charaterization2, supp}
         \begin{aligned}
        &P_0(Y=y,D=1|Z=1_a) = \sum_{s\in C\cdot\cup A\cdot}h_{1,s}(y),
    P_0(Y=y,D=0|Z=1_a) = \sum_{s\in N\cdot\cup D\cdot}h_{1,s}(y),\\
    &P_0(Y=y,D=1|Z=1_b) = \sum_{s\in \cdot C\cup \cdot A}h_{1,s}(y),
    P_0(Y=y,D=0|Z=1_b) = \sum_{s\in \cdot N\cup \cdot D}h_{1,s}(y),\\ 
    &P_0(Y=y,D=1|Z=0_a) = \sum_{s\in A\cdot\cup D\cdot}h_{0,s}(y),
    P_0(Y=y,D=0|Z=0_a) = \sum_{s\in N\cdot\cup C\cdot}h_{1,s}(y)),\\
    &P_0(Y=y,D=1|Z=0_b) = \sum_{s\in \cdot A\cup \cdot D}h_{0,s}(y),
    P_0(Y=y,D=0|Z=0_b) = \sum_{s\in \cdot N\cup \cdot C}h_{0,s}(y),\\
    &P_0(Y(D=1)=y) -P_0(Y=y,D=1|Z=1_a) = \sum_{s\in N\cdot \cup D\cdot} h_{1,s}(y)\\
&P^*(Y(D=0)=y) -P_0(Y=y,D=0|Z=1_a) = \sum_{s\in C\cdot \cup A\cdot} h_{1,s}(y),\\
&P_0(Y(D=1)=y) -P_0(Y=y,D=1|Z=1_b) = \sum_{s\in \cdot N \cup \cdot D} h_{0,s}(y),\\
&P_0(Y(D=0)=y) -P_0(Y=y,D=1|Z=1_b) = \sum_{s\in \cdot C \cup \cdot A} h_{1,s}(y)\\
&P_0(Y(D=1)=y) -P_0(Y=y,D=1|Z=0_a) = \sum_{s\in N\cdot \cup C\cdot} h_{1,s}(y),\\
&P_0(Y(D=0)=y) -P_0(Y=y,D=0|Z=0_a) = \sum_{s\in A\cdot \cup D\cdot} h_{0,s}(y)\\
&P_0(Y(D=1)=y) -P_0(Y=y,D=1|Z=0_b) = \sum_{s\in \cdot N \cup \cdot C} h_{1,s}(y),\\
&P_0(Y(D=0)=y) -P_0(Y=y,D=0|Z=0_b) = \sum_{s\in \cdot A \cup \cdot D} h_{0,s}(y),\\
&h_{1,s}(1)+ h_{1,s}(0) =  h_{0,s}(1)+ h_{0,s}(0), \text{ for }s\in \{\cdot\cdot\},\\
&h_{1,s}(y) = h_{0,s}(y) = 0, \text{ for }s\in D\cdot \cup \cdot D \cup \{CA,CN\}.
    \end{aligned}
    \end{equation}
   
    \normalsize
\end{proposition}

The proof is similar to the previous proof and thus omitted. 

The upper bound for $\text{ATE} = p_1-p_0$ can be obtained by solving the following linear programming problem:
\begin{align*}
    &\max_{p_0,p_1,h_{s,d}(y),s\in \{\cdot\cdot\}, d\in \{0,1\}, y \in \{0,1\}} p_0-p_1\\
    &\text{subject to constraints \ref{eq: eq: charaterization2, supp}}.
\end{align*}

The lower bound can be obtained by replacing $p_1 - p_0$ in the objective function by $p_0 - p_1$.

Furthermore, we briefly discuss how to obtain partial identification bounds for principal causal effects. Take always-always-taker average treatment effect as an example, it can be written as
\begin{align*}
    &\mathbbm{E}[Y(D=1)-Y(D=0)|S=AAT]\\
    =&P[Y(D=1)=1|S=AAT]-P[Y(D=0)=1|S=AAT]\\
    =&\frac{P(Y(D=1)=1,S=AAT)-P(Y(D=0)=1,S=AAT)}{P(S=AAT)}
\end{align*}

Therefore, in addition  to the constraints for $P(Y(D=1)=1,S=AAT)$ and $P(Y(D=0)=1,S=AAT)$, which we have derived before, we also need to derive constraints for $P(S=AAT)$, which are given below:
\begin{align*}
&P(S=\text{AAT})+P(S=\text{AT-NT})+P(S=\text{AT-CO}) = P(D=1|Z=0_a),\\
&P(S=\text{ANT})+P(S=\text{NT-AT})+P(S=\text{NT-CO}) = P(D=0|Z=1_a),\\
&P(S=\text{AAT})+P(S=\text{NT-AT}) = P(D=1|Z=0_b),\\
&P(S=\text{ANT})+P(S=\text{AT-NT}) = P(D=0|Z=1_b).
\end{align*}

Therefore, we can similarly by treating the joint probabilities of potential outcomes and principal strata and the marginal probabilities of principal strata as decision variables to define the corresponding optimization problem to derive the partial identification bound for principal causal effects.

The partial identification problem under Assumptions \ref{ass: partial identification, supp}  can also be fitted into the framework proposed by \citet{duarte2024automated}, where they proposed a general algorithm for calculating the partial identification bound for a given estimand and additional assumptions. Their framework works when the estimand can be written as a polynomial function and the model constraints can be written as a polynomial constraints.  In our case, since the estimand can be written as a linear function of the principal strata, the partial monotonicity assumption and nested IV assumption in our paper can be converted into linear constraints, their  approach works. \citet{richardson2024assumptions} give the explicit formula for the partial identification bounds of ATE when $Z$ is categorical and under the ignorability assumption. \citet{song2024instrumental} generalize the results in \citet{richardson2024assumptions}, and gives the partial identification results when $Y,D,Z$ are categorical variable. However,  both \citet{song2024instrumental}  and \citet{richardson2024assumptions} don't consider the case when additional constraints on the principal strata are available. The derivation of explicit formulas for partial identification bounds in the presence of constraints on principal strata is left for future work.

}





\section{Proofs}
\subsection{Proof of Theorem \ref{thm_identification}}

\begin{proof}
    For $SWATE$, Since we have
    \begin{align*}
    &\mathbb{E}_{P_0}[Y\mid Z=1_b,\boldsymbol{X}]-\mathbb{E}_{P_0}[Y\mid Z=0_b,\boldsymbol{X}]\\
     =&\mathbb{E}_{P_0}[Y(Z=1_b)\mid Z=1_b,G=b,\boldsymbol{X}]-\mathbb{E}_{P_0}[Y(Z=0_b)|Z=0_b,G=b, X]\\
        =&\mathbb{E}_{P_0}[Y(Z=1_b)-Y(Z=0_b)|G=b,\boldsymbol{X}]\\
        =&\mathbb{E}_{P_0}[Y(Z=1_b)-Y(Z=0_b)|\boldsymbol{X}]\\
        =&\mathbb{E}_{P_0}[Y(D=1)-Y(D=0)|S=\text{SW},\boldsymbol{X}]P_0(S=\text{SW}|\boldsymbol{X}) \\
        &+\mathbb{E}_{P_0}[Y(D=1)-Y(D=0)|S=\text{ACO},\boldsymbol{X}]P_0(S=\text{ACO}|\boldsymbol{X})
    \end{align*}
    and
    \begin{align*}
        &\mathbb{E}_{P_0}[Y|Z=1_a,\boldsymbol{X}]-\mathbb{E}_{P_0}[Y|Z=0_a,\boldsymbol{X}]\\
        =&\mathbb{E}_{P_0}[Y(Z=1_a)|Z=1_a,G=a,\boldsymbol{X}]-\mathbb{E}_{P_0}[Y(Z=0_a)|Z=0_a,G=a,\boldsymbol{X}]\\
        =&\mathbb{E}_{P_0}[Y(Z=1_a)-Y(Z=0_a)|G=a,\boldsymbol{X}]\\
        =&\mathbb{E}_{P_0}[Y(Z=1_a)-Y(Z=0_a)|\boldsymbol{X}]\\
        =&\mathbb{E}_{P_0}[Y(D=1)-Y(D=0)|S=\text{ACO},\boldsymbol{X}]P_0(S=\text{ACO}|\boldsymbol{X})
    \end{align*}
    therefore
    \begin{align*}
    &\mathbb{E}_{P_0}[Y(D=1)-Y(D=0)|S=\text{SW},\boldsymbol{X}]P_0(S=\text{SW}|\boldsymbol{X})\\
        =&(\mathbb{E}_{P_0}[Y|Z=1_b,\boldsymbol{X}]-\mathbb{E}_{P_0}[Y|Z=0_b,\boldsymbol{X}])-(\mathbb{E}_{P_0}[Y|Z=1_a,\boldsymbol{X}]-\mathbb{E}_{P_0}[Y|Z=0_a,\boldsymbol{X}])
    \end{align*}
Similarly, we have
\begin{align*}
    &\mathbb{E}_{P_0}[D|Z=1_b,\boldsymbol{X}]-\mathbb{E}_{P_0}[D|Z=0_b,\boldsymbol{X}]\\
    =&\mathbb{E}_{P_0}[D|Z=1_b,G=b,\boldsymbol{X}]-\mathbb{E}_{P_0}[D|Z=0_b,G=b,\boldsymbol{X}]\\
    =&\mathbb{E}_{P_0}[D(Z=1_b)-D(Z=0_b)|G=b,\boldsymbol{X}]\\
    =&\mathbb{E}_{P_0}[D(Z=1_b)-D(Z=0_b)|\boldsymbol{X}]\\
    =&P_0(S=\text{ACO}|\boldsymbol{X})+P_0(S=\text{SW}|\boldsymbol{X})
\end{align*}
and
\begin{align*}
&\mathbb{E}_{P_0}[D|Z=1_a,\boldsymbol{X}]-E[D|Z=0_a,\boldsymbol{X}]\\
    =&\mathbb{E}_{P_0}[D|Z=1_a,G=a,\boldsymbol{X}]-\mathbb{E}_{P_0}[D|Z=0_a,G=a, \boldsymbol{X}]\\
    =&\mathbb{E}_{P_0}[D(Z=1_a)-D(Z=0_a)|G=a,\boldsymbol{X}]\\
    =&\mathbb{E}_{P_0}[D(Z=1_a)-D(Z=0_a)|\boldsymbol{X}]\\
    =&P_0(S=\text{ACO}|\boldsymbol{X})
\end{align*}
    Therefore we have
    \begin{align*}
       (\mathbb{E}_{P_0}[D|Z=1_b,\boldsymbol{X}]-\mathbb{E}_{P_0}[D|Z=0_b,\boldsymbol{X}])- (\mathbb{E}_{P_0}[D|Z=1_a,\boldsymbol{X}]-\mathbb{E}_{P_0}[D|Z=0_a,\boldsymbol{X}]) = P_0(S=\text{SW}|\boldsymbol{X})
    \end{align*}
    Therefore,
    \begin{align*}
        SWATE_{P_0}(\boldsymbol{X})&:=\mathbb{E}_{P_0}[Y(D=1)-Y(D=0)|S=\text{SW},\boldsymbol{X}]\\
        & = \frac{(\mathbb{E}_{P_0}[Y|Z=1_b,\boldsymbol{X}]-\mathbb{E}_{P_0}[Y|Z=0_b,\boldsymbol{X}])-(\mathbb{E}_{P_0}[Y|Z=1_a,\boldsymbol{X}]-\mathbb{E}_{P_0}[Y|Z=0_a,\boldsymbol{X}])}{(\mathbb{E}_{P_0}[D|Z=1_b,\boldsymbol{X}]-\mathbb{E}_{P_0}[D|Z=0_b,\boldsymbol{X}])- (\mathbb{E}_{P_0}[D|Z=1_a,\boldsymbol{X}]-\mathbb{E}_{P_0}[D|Z=0_a,\boldsymbol{X}])}
    \end{align*}
    and
    \begin{align*}
        \text{SWATE}_{P_0}&= \int  \text{SWATE}_{P_0}(\boldsymbol{x})f_{P_0}(\boldsymbol{x}|S=SW)d \mu(\boldsymbol{x})\\
        &= \int \frac{\delta_{b,P_0}(\boldsymbol{x})-\delta_{a,P_0}(\boldsymbol{x})}{\eta_{b,P_0}(\boldsymbol{x})-\eta_{a,P_0}(\boldsymbol{x})}\frac{\eta_{b,P_0}(\boldsymbol{X})-\eta_{a,P_0}(\boldsymbol{X})}{P_0(S=\text{SW})}f_{P_0}(\boldsymbol{x})d \mu(\boldsymbol{X})\\
        & = \frac{\mathbb{E}_{P_0}[\delta_{b,P_0}(\boldsymbol{X})-\delta_{a,P_0}(\boldsymbol{X})]}{P_0(S=\text{SW})}\\
        & = \frac{\mathbb{E}_{P_0}[\delta_{b,P_0}(\boldsymbol{X})-\delta_{a,P_0}(\boldsymbol{X})]}{\mathbb{E}_{P_0}[\eta_{b,P_0}(\boldsymbol{X})-\eta_{a,P_0}(\boldsymbol{X})]}
    \end{align*}
where $f_{P_0}(\boldsymbol{x}|S=SW)$ and $f_{P_0}(\boldsymbol{X})$ are conditional density of and density functions of $X$ implied by $P_0$.
    Similarly,  for $\text{ACOATE}$, we have
    \begin{align*}
        \text{ACOATE}_{P_0}(\boldsymbol{X})&:=\mathbb{E}_{P_0}[Y(D=1)-Y(D=0)|S=\text{ACO},\boldsymbol{X}]\\
        & = \frac{\mathbb{E}_{P_0}[Y|Z=1_a,\boldsymbol{X}]-\mathbb{E}_{P_0}[Y|Z=0_a,\boldsymbol{X}]}{\mathbb{E}_{P_0}[D|Z=1_a,\boldsymbol{X}]-\mathbb{E}_{P_0}[D|Z=0_a,\boldsymbol{X}]}
    \end{align*}
    and
    \begin{equation*}
        \text{ACOATE}_{P_0} = \frac{\mathbb{E}_{P_0}[\delta_{a,P_0}(\boldsymbol{X})]}{\mathbb{E}_{P_0}[\eta_{a,P_0}(\boldsymbol{X})]}
    \end{equation*}
\end{proof}

\subsection{Proof of Proposition \ref{prop: violation of SUTVA}}
\begin{proof}
Direct calculation yields
\begin{align*}
    &\mathbbm{E}_{P_0}[Y|Z^*=1,\boldsymbol{X}]\\
    =&\mathbbm{E}_{P_0}[Y|Z=1_a,\boldsymbol{X}]P_0(Z=1_a|Z^*=1,\boldsymbol{X})+\mathbbm{E}_{P_0}[Y|Z=1_b,\boldsymbol{X}]P_0(Z=1_b|Z^*=1,\boldsymbol{X})\\
    =&\mathbbm{E}_{P_0}[Y|Z=1_a,\boldsymbol{X}]P_0(G=a|Z^*=1,\boldsymbol{X})+\mathbbm{E}_{P_0}[Y|Z=1_b,\boldsymbol{X}]P_0(G=b|Z^*=1,\boldsymbol{X})\\
    =&\mathbbm{E}_{P_0}[Y|Z=1_a,\boldsymbol{X}]P(G=a|\boldsymbol{X})+\mathbbm{E}_{P_0}[Y|Z=1_b,\boldsymbol{X}]P_0(G=b|\boldsymbol{X})\\
    =&\mathbbm{E}_{P_0}[Y(Z=1_a)|G=a,\boldsymbol{X}]P(G=a|\boldsymbol{X})+\mathbbm{E}_{P_0}[Y(Z=1_b)|G=b,\boldsymbol{X}]P_0(G=b|\boldsymbol{X}),
\end{align*}
where the second last inequality is due to the assumption that $G\indep Z^*|\boldsymbol{X}$. Similarly, we have
\begin{align*}
    &\mathbbm{E}_{P_0}[Y|Z^*=0,\boldsymbol{X}]\\
    =&\mathbbm{E}_{P_0}[Y(Z=0_a)|G=a,\boldsymbol{X}]P(G=a|\boldsymbol{X})+\mathbbm{E}_{P_0}[Y(Z=0_b)|G=b,\boldsymbol{X}]P_0(G=b|\boldsymbol{X}).
\end{align*}
Therefore, 
\begin{align*}
    &\mathbbm{E}_{P_0}[Y|Z^*=1,\boldsymbol{X}]-\mathbbm{E}_{P_0}[Y|Z^*=0,\boldsymbol{X}]\\
    =&\mathbbm{E}_{P_0}[Y(Z=1_a)-Y(Z=0_a)|G=a,\boldsymbol{X}]P(G=a|\boldsymbol{X})\\
    &+\mathbbm{E}_{P_0}[Y(Z=1_b)-Y(Z=0_b)|G=b,\boldsymbol{X}]P_0(G=b|\boldsymbol{X})\\
    =&\mathbbm{E}_{P_0}[Y(Z=1_a)-Y(Z=0_a)|\boldsymbol{X}]P_0(G=a|\boldsymbol{X})+\mathbbm{E}_{P_0}[Y(Z=1_b)-Y(Z=0_b)|\boldsymbol{X}]P(G=b|\boldsymbol{X})\\
    =&\mathbbm{E}_{P_0}[Y(D=1)-Y(D=0)|S=\text{ACO},\boldsymbol{X}]P_0(S=\text{ACO}|\boldsymbol{X})P_0(G=a|\boldsymbol{X})\\
    &+\mathbbm{E}_{P_0}[Y(D=1)-Y(D=0)|S=\text{SW},\boldsymbol{X}]P_0(S=\text{SW}|\boldsymbol{X})P_0(G=b|\boldsymbol{X})\\
    &+\mathbbm{E}_{P_0}[Y(D=1)-Y(D=0)|S=\text{ACO},\boldsymbol{X}]P_0(S=\text{ACO}|\boldsymbol{X})P_0(G=b|\boldsymbol{X})
\end{align*}
Similarly, we have
\begin{align*}
    &\mathbbm{E}_{P_0}[D|Z^*=1,\boldsymbol{X}]-\mathbbm{E}_{P_0}[D|Z^*=0,\boldsymbol{X}]\\
=&P_0(S=\text{ACO}|\boldsymbol{X})P_0(G=a|\boldsymbol{X})+P_0(S=\text{SW}|\boldsymbol{X})P_0(G=b|\boldsymbol{X})+P_0(S=\text{ACO}|\boldsymbol{X})P_0(G=b|\boldsymbol{X}).
\end{align*}
Therefore, 
\begin{align*}
        \theta_{\text{CW},P_0}(\boldsymbol{X}) = \omega_1 \cdot \text{SWATE}_{P_0}(\boldsymbol{X})+\omega_2\cdot\text{ACOATE}_{P_0}(\boldsymbol{X}),
    \end{align*}
    with 
    \begin{align*}
        &\omega_1 = \frac{P_0(S=\text{SW}\mid \boldsymbol{X})P_0(G=b\mid \boldsymbol{X})}{P_0(S=\text{SW}\mid \boldsymbol{X})P_0(G=b\mid \boldsymbol{X})+P_0(S = \text{ACO}\mid \boldsymbol{X})},~~\text{and}\\
        &\omega_2= \frac{P_0(S = \text{ACO}\mid \boldsymbol{X})}{P_0(S = \text{SW}\mid \boldsymbol{X})P_0(G=b\mid \boldsymbol{X})+P_0(S=\text{ACO}\mid \boldsymbol{X})}.
    \end{align*}
\end{proof}

\subsection{Proof of Proposition \ref{prop_covariates_dis}}
\begin{proof}
Direct calculation yields
    \begin{align*}
        &\mathbbm{E}_{P_0}[g(\boldsymbol{X})|S=\text{SW}]\\
        =& \int g(\boldsymbol{x})f_{P_0}(\boldsymbol{x}|S=\text{SW})d\mu(\boldsymbol{x})\\
        =&\int g(\boldsymbol{x})\frac{P_0(S=\text{SW}|\boldsymbol{X}=\boldsymbol{x})}{P_0(S=\text{SW})}f_{P_0}(\boldsymbol{x})d\mu(\boldsymbol{x})\\
        =&\mathbb{E}_{P_0}[g(\boldsymbol{X})\mid S=\text{SW}] = \mathbb{E}_{P_0}\Bigg[g(\boldsymbol{X})\frac{\eta_{b,P_0}(\boldsymbol{X})-\eta_{a,P_0}(\boldsymbol{X})}{\mathbb{E}_{P_0}[\eta_{b,P}(\boldsymbol{X})-\eta_{a,P_0}(\boldsymbol{X})]}\Bigg].
    \end{align*}
\end{proof}

\subsection{Proof of Proposition \ref{prop: sensitivity analysis for mean exchangeability}}
\begin{proof}
Direct calculation yields
\begin{align*}
    &\mathbb{E}_{P_0}[Y(Z=1_b)-Y(Z=0_b)|\boldsymbol{X}]\\
    =&\mathbb{E}_{P_0}[Y(Z=1_b)-Y(Z=0_b)|G=a,\boldsymbol{X}]P_0(G=a|\boldsymbol{X})\\
    &+\mathbb{E}_{P_0}[Y(Z=1_b)-Y(Z=0_b)|G=b,\boldsymbol{X}]P_0(G=b|\boldsymbol{X})\\
    =&[\mathbb{E}_{P_0}[Y(Z=1_b)-Y(Z=0_b)|G=b,\boldsymbol{X}]+h^{Y,b}(\boldsymbol{X})]P_0(G=a|\boldsymbol{X})\\
    &+\mathbb{E}_{P_0}[Y(Z=1_b)-Y(Z=0_b)|G=b,\boldsymbol{X}]P_0(G=b|\boldsymbol{X})\\
    =&\mathbb{E}_{P_0}[Y(Z=1_b)-Y(Z=0_b)|G=b,\boldsymbol{X}]+h^{Y,b}(\boldsymbol{X})P_0(G=a|\boldsymbol{X})\\
    =&\mathbb{E}_{P_0}[Y|Z=1_b,G=b,\boldsymbol{X}]-\mathbb{E}_{P_0}[Y|Z=0_b,G=b,\boldsymbol{X}]+h^{Y,b}(\boldsymbol{X})P_0(G=a|\boldsymbol{X})\\
    =&\mathbb{E}_{P_0}[Y|Z=1_b,\boldsymbol{X}]-\mathbb{E}_{P_0}[Y|Z=0_b,\boldsymbol{X}]+h^{Y,b}(\boldsymbol{X})P_0(G=a|\boldsymbol{X}).
\end{align*}
Similarly, we have
\begin{align*}
    &\mathbb{E}_{P_0}[Y(Z=1_a)-Y(Z=0_a)|\boldsymbol{X}]\\
    =&\mathbb{E}_{P_0}[Y(Z=1_a)-Y(Z=0_a)|G=a,\boldsymbol{X}]P_0(G=a|\boldsymbol{X})\\
    &+\mathbb{E}_{P_0}[Y(Z=1_a)-Y(Z=0_a)|G=b,\boldsymbol{X}]P_0(G=b|\boldsymbol{X})\\
    =&[\mathbb{E}_{P_0}[Y(Z=1_a)-Y(Z=0_a)|G=a,\boldsymbol{X}](\boldsymbol{X})]P_0(G=a|\boldsymbol{X})\\
    &+[\mathbb{E}_{P_0}[Y(Z=1_a)-Y(Z=0_a)|G=a,\boldsymbol{X}]-h^{Y,a}(\boldsymbol{X})]P_0(G=b|\boldsymbol{X})\\
    =&\mathbb{E}_{P_0}[Y(Z=1_a)-Y(Z=0_a)|G=a,\boldsymbol{X}]-h^{Y,a}(\boldsymbol{X})P_0(G=b|\boldsymbol{X})\\
    =&\mathbb{E}_{P_0}[Y|G=a,Z=1_a,\boldsymbol{X}]-\mathbb{E}_{P_0}[Y|G=a,Z=0_a,\boldsymbol{X}]-h^{Y,a}(\boldsymbol{X})P_0(G=b|\boldsymbol{X})\\
    =&\mathbb{E}_{P_0}[Y|Z=1_a,\boldsymbol{X}]-\mathbb{E}_{P_0}[Y|Z=0_a,\boldsymbol{X}]-h^{Y,a}(\boldsymbol{X})P_0(G=b|\boldsymbol{X}).
\end{align*}
Therefore,
    \begin{align*}
&\mathbb{E}_{P_0}[Y|Z=1_b,\boldsymbol{X}]-\mathbb{E}_{P_0}[Y|Z=0_b,\boldsymbol{X}]+h^{Y,b}(\boldsymbol{X})P_0(G=a|\boldsymbol{X})\\
=&\mathbb{E}_{P_0}[Y(Z=1_b)-Y(Z=0_b)|\boldsymbol{X}]\\
        =&\mathbb{E}_{P_0}[Y(D=1)-Y(D=0)|S=\text{SW},\boldsymbol{X}]P_0(S=\text{SW}|\boldsymbol{X}) \\
        &+\mathbb{E}_{P_0}[Y(D=1)-Y(D=0)|S=\text{ACO},\boldsymbol{X}]P_0(S=\text{ACO}|\boldsymbol{X}).
    \end{align*}    
Furthermore,
\begin{align*}
&\mathbb{E}_{P_0}[Y|Z=1_a,\boldsymbol{X}]-\mathbb{E}_{P_0}[Y|Z=0_a,\boldsymbol{X}]-h^{Y,a}(\boldsymbol{X})P(G=b|\boldsymbol{X})\\
=&\mathbb{E}_{P_0}[Y(Z=1_a)-Y(Z=0_a)|\boldsymbol{X}]\\
        =&\mathbb{E}_{P_0}[Y(D=1)-Y(D=0)|S=\text{ACO},\boldsymbol{X}]P_0(S=\text{ACO}|\boldsymbol{X}).
    \end{align*}    

    Therefore,
    \begin{align*}
        &\delta_{b,P_0}(\boldsymbol{X})-\delta_{a,P_0}(\boldsymbol{X})+(h^{Y,b}(\boldsymbol{X})P_0(G=a|\boldsymbol{X})+h^{Y,a}(\boldsymbol{X})P_0(G=b|\boldsymbol{X}))\\
        = &\mathbb{E}_{P_0}[Y(D=1)-Y(D=0)|S=\text{SW},\boldsymbol{X}]P_0(S=\text{SW}|\boldsymbol{X}).
    \end{align*}

    We can similarly derive the following equalities for $D$:
    \begin{align*}
        &P_0(S=\text{SW}|\boldsymbol{X})+P_0(S=\text{ACO}|\boldsymbol{X})\\
        =&\mathbbm{E}_{P_0}[D(Z=1_b)-D(Z=0_b)|\boldsymbol{X}]\\
        =&\mathbb{E}_{P_0}[D|Z=1_b,\boldsymbol{X}]-\mathbb{E}_{P_0}[D|Z=0_b,\boldsymbol{X}]+h^{D,b}(\boldsymbol{X})P_0(G=a|\boldsymbol{X})
    \end{align*}
    and
    \begin{align*}
        &P_0(S=\text{ACO}|\boldsymbol{X})\\
        =&\mathbbm{E}_{P_0}[D(Z=1_a)-D(Z=0_a)|\boldsymbol{X}]\\
        =&\mathbb{E}_{P_0}[D|Z=1_a,\boldsymbol{X}]-\mathbb{E}_{P_0}[D|Z=0_a,\boldsymbol{X}]-h^{D,a}(\boldsymbol{X})P_0(G=b|\boldsymbol{X}).
    \end{align*}
    Therefore,
    \begin{align*}
        &P_0(S=\text{SW}|\boldsymbol{X})\\
        =&\eta_{b,P_0}(\boldsymbol{X})-\eta_{a,P_0}(\boldsymbol{X})+(h^{D,b}(\boldsymbol{X})P_0(G=a|\boldsymbol{X})+h^{D,a}(\boldsymbol{X})P_0(G=b|\boldsymbol{X})).
    \end{align*}

    To sum up, we have
    \begin{align*}
        &\text{ACOATE}_{P_0}(\boldsymbol{X}) =\frac{\delta_{a,P_0}(\boldsymbol{X})-h^{Y,a}(\boldsymbol{X})P_0(G=b|\boldsymbol{X})}{\eta_{a,P_0}(\boldsymbol{X})-h^{D,a}(\boldsymbol{X})P_0(G=b|\boldsymbol{X})};\\
        &\text{SWATE}_{P_0}(\boldsymbol{X}) =\frac{\delta_{b,P_0}(\boldsymbol{X})-\delta_{a,P_0}(\boldsymbol{X})+(h^{Y,b}(\boldsymbol{X})P_0(G=a|\boldsymbol{X})+h^{Y,a}(\boldsymbol{X})P_0(G=b|{X}))}{\eta_{b,P_0}(\boldsymbol{X})-\eta_{a,P_0}(\boldsymbol{X})+(h^{D,b}(\boldsymbol{X})P_0(G=a|\boldsymbol{X})+h^{D,a}(\boldsymbol{X})P_0(G=b|\boldsymbol{X}))}.
    \end{align*}
    Furthermore,
    \begin{align*}
        \text{SWATE}_{P_0}&= \int  \text{SWATE}_{P_0}(\boldsymbol{X})f_{P_0}(\boldsymbol{x}|S=SW)d \mu(\boldsymbol{x})\\
        &= \int \frac{\delta_{b,P_0}(\boldsymbol{x})-\delta_{a,P_0}(\boldsymbol{x})+(h^{Y,b}(\boldsymbol{x})P(G=a|\boldsymbol{x})+h^{Y,a}(\boldsymbol{X})P_0(G=b|\boldsymbol{x}))}{\eta_{b,P_0}(\boldsymbol{X})-\eta_{a,P_0}(\boldsymbol{x})+(h^{D,b}(\boldsymbol{X})P_0(G=a|\boldsymbol{x})+h^{D,a}(\boldsymbol{X})P_0(G=b|\boldsymbol{x}))}\\
        &\times\frac{\eta_{b,P_0}(\boldsymbol{x})-\eta_{a,P_0}(\boldsymbol{x})+(h^{D,b}(\boldsymbol{x})P_0(G=a|\boldsymbol{X})+h^{D,a}(\boldsymbol{x})P_0(G=b|\boldsymbol{x}))}{P_0(S=\text{SW})}f_{P_0}(\boldsymbol{x})d \mu(\boldsymbol{x})\\
        & = \frac{\mathbbm{E}_{P_0}[\delta_{b,P_0}(\boldsymbol{X})-\delta_{a,P_0}(\boldsymbol{X})+(h^{Y,b}(\boldsymbol{X})P_0(G=a|\boldsymbol{X})+h^{Y,a}(\boldsymbol{X})P_0(G=b|\boldsymbol{X}))]}{\mathbbm{E}_{P_0}[\eta_{b,P_0}(\boldsymbol{X})-\eta_{a,P_0}(\boldsymbol{X})+(h^{D,b}(\boldsymbol{X})P_0(G=a|\boldsymbol{X})+h^{D,a}(\boldsymbol{X})P_0(G=b|\boldsymbol{X}))]}.
    \end{align*}
    Similarly we can prove the result for ACOATE.
\end{proof}

\subsection{Proof of Proposition \ref{prop: violation of nested IV pair assumption}}
\begin{proof}
Direct calculations yield
    \begin{align*}
        &\mathbbm{E}_{P_0}[Y|Z=1_b,\boldsymbol{X}]-\mathbbm{E}_{P_0}[Y|Z=0_b,\boldsymbol{X}]\\
        =&\mathbbm{E}_{P_0}[Y(D=1)-Y(D=0)|S=\text{SW},\boldsymbol{X}]P_0(S=\text{SW}|\boldsymbol{X})\\
        &+\mathbbm{E}_{P_0}[Y(D=1)-Y(D=0)|S=\text{ACO},\boldsymbol{X}]P_0(S=\text{ACO}|\boldsymbol{X}).
    \end{align*}
    Similarly,
     \begin{align*}
        &\mathbbm{E}_{P_0}[Y|Z=1_a,\boldsymbol{X}]-\mathbbm{E}_{P_0}[Y|Z=0_a,\boldsymbol{X}]\\
        =&\mathbbm{E}_{P_0}[Y(D=1)-Y(D=0)|S=\text{ACO},\boldsymbol{X}]P_0(S=\text{ACO}|\boldsymbol{X})\\
        &-\mathbbm{E}_{P_0}[Y(D=1)-Y(D=0)|S=\text{DFN},\boldsymbol{X}]P_0(S=\text{DFN}|\boldsymbol{X}).
    \end{align*}
    Therefore,
    \begin{align*}
        &(\mathbbm{E}_{P_0}[Y|Z=1_b,\boldsymbol{X}]-\mathbbm{E}_{P_0}[Y|Z=0_b,\boldsymbol{X}])-(\mathbbm{E}_{P_0}[Y|Z=1_a,\boldsymbol{X}]-\mathbbm{E}_{P_0}[Y|Z=0_a,\boldsymbol{X}])\\
        =&\mathbbm{E}_{P_0}[Y(D=1)-Y(D=0)|S=\text{SW},\boldsymbol{X}]P_0(S=\text{SW}|\boldsymbol{X})\\
        &+\mathbbm{E}_{P_0}[Y(D=1)-Y(D=0)|S=\text{DFN},\boldsymbol{X}]P_0(S=\text{DFN}|\boldsymbol{X})
    \end{align*}
Similarly,
\begin{align*}
        &(\mathbbm{E}_{P_0}[D|Z=1_b,\boldsymbol{X}]-\mathbbm{E}_{P_0}[D|Z=0_b,\boldsymbol{X}])-(\mathbbm{E}_{P_0}[D|Z=1_a,\boldsymbol{X}]-\mathbbm{E}_{P_0}[D|Z=0_a,\boldsymbol{X}])\\
=&P_0(S=SW|\boldsymbol{X})+P_0(S=DFN|\boldsymbol{X})
    \end{align*}
    Therefore,
    \begin{align}
      \theta_{\text{SW},P_0}(\boldsymbol{x}) &=  \lambda_{P_0}(\boldsymbol{x})\mathbbm{E}_{P_0}[Y(D=1)-Y(D=0)\mid S=\text{SW},\boldsymbol{X}=\boldsymbol{x}] + \nonumber\\
      &\{1-\lambda_{P_0}(\boldsymbol{x})\}\mathbbm{E}_{P_0}[Y(D=1)-Y(D=0)\mid S=\text{DFN}, \boldsymbol{X}=\boldsymbol{x}], 
    \end{align}
    where
    \begin{align*}
        \lambda_{P_0}(\boldsymbol{x}) = \frac{P_0(S=\text{SW}\mid \boldsymbol{X}=\boldsymbol{x})}{P_0(S=\text{SW}\mid \boldsymbol{X}=\boldsymbol{x})+P_0(S=\text{DFN}\mid \boldsymbol{X}=\boldsymbol{x})}.
    \end{align*}
\end{proof}

\subsection{Wald-type estimator for SWATE}
\label{subsec: wald estimator theorem, supp}

We propose the Wald-type estimator only when the identification assumptions hold without conditioning on $\boldsymbol{X}$.

A classical Wald estimator for ACOATE is $ \widehat{\psi}^{\text{Wald}}_{\text{ACO}} = \widehat{\delta}_a/\widehat{\eta}_a$, where
\begin{align*}
\widehat{\delta}_a = \frac{\sum_{Z_i = 1_a} Y_i}{\sum_i \mathbbm{1}\{{Z_i = 1_a}\}} - \frac{\sum_{Z_i = 0_a} Y_i}{\sum_i\mathbbm{1}\{{Z_i = 0_a}\}},\quad
    \widehat{\eta}_a = \frac{\sum_{Z_i = 1_a} D_i}{\sum_i \mathbbm{1}\{{Z_i = 1_a}\}} - \frac{\sum_{Z_i = 0_a} D_i}{\sum_i\mathbbm{1}\{{Z_i = 0_a}\}}.
\end{align*}

\noindent An analogous Wald-type estimator for SWATE can be formulated as follows:
\begin{equation*}
\widehat{\psi}^{\text{Wald}}_{\text{SW}} = \frac{\widehat{\delta}_b-\widehat{\delta}_a}{\widehat{\eta}_b-\widehat{\eta}_a},
\end{equation*}
where
\begin{align*}
    \widehat{\delta}_b = \frac{\sum_{Z_i = 1_b} Y_i}{\sum_i \mathbbm{1}\{{Z_i = 1_b}\}} - \frac{\sum_{Z_i = 0_b} Y_i}{\sum_i\mathbbm{1}\{{Z_i = 0_b}\}},\quad 
    \widehat{\eta}_b = \frac{\sum_{Z_i = 1_b} D_i}{\sum_i \mathbbm{1}\{{Z_i = 1_b}\}} - \frac{\sum_{Z_i = 0_b} D_i}{\sum_i\mathbbm{1}\{{Z_i = 0_b}\}}.
\end{align*}

\noindent Theorem \ref{thm_wald} summarizes the property of $\widehat{\psi}^{\text{Wald}}_{\text{SW}}$.
\begin{theorem}\label{thm_wald}
    Assume the random variables $Y$ has finite second moment. We have
    \begin{equation*}
        \sqrt{n}\left(\widehat{\psi}^{\text{Wald}}_{\text{SW}}-\Psi_{\text{SW},P_0}\right)\rightsquigarrow \mathcal{N}\left(0,\sigma_{\text{Wald}}^2\right),
    \end{equation*}
    where $\sigma^2_{\text{Wald}} = \mathbb{E}_{P_0}[\varphi_{\text{Wald,SW}}^2]$, and $\varphi_{\text{Wald,SW}}$ is the influence function of $\widehat{\psi}^{\text{Wald}}_{\text{SW}}$. An estimator $\widehat{\sigma}^2_{\text{Wald}}$ of $\sigma^2_{\text{Wald}}$ can be obtained using nonparametric bootstrap or via the delta method.
\end{theorem}

\begin{proof}
    Let
\begin{align*}
    &\widehat{\mu}_{Y,m} = \frac{1}{n}\sum_{i=1}^n Y_i1\{Z_i = m\}, \quad {\mu}_{Y,m} = \mathbb{E}_{P_0} [Y_i1\{Z_i = m\}] \\
    &\widehat{\mu}_{D,m} = \frac{1}{n}\sum_{i=1}^n D_i1\{Z_i = m\}, \quad {\mu}_{D,m} = \mathbb{E}_{P_0} [D_i1\{Z_i = m\}] \\
    &\widehat{\mu}_{Z,m} = \frac{1}{n}\sum_{i=1}^n 1\{Z_i = m\}, \quad {\mu}_{Z,m} = \mathbb{E}_{P_0} [1\{Z_i = m\}] 
\end{align*}
for $m\in \{a_1,a_0,b_1,b_0\}$. Then it's clear that $\hat{\mu}_{Y,m},\hat{\mu}_{D,m},\hat{\mu}_{Z,m}, m\in \{a_1,a_0,b_1,b_0\}$ are asymptotic linear estimators for ${\mu}_{Y,s},{\mu}_{D,s},{\mu}_{Z,s}, s\in\{a_1,a_0,b_1,b_0\}$, with influence functions
\begin{align*}
    &\varphi_{Y,m}= o\mapsto y\mathbbm{1}\{z=m\}-\mu_{Y,m}\\
    &\varphi_{D,m}= o\mapsto d\mathbbm{1}\{z=m\}-\mu_{D,m}\\
    &\varphi_{Z,m}= o\mapsto\mathbbm{1}\{z=m\}-\mu_{Z,m}
\end{align*}

Let
\begin{equation*}
f(\boldsymbol{x}_1,x_2,x_3,x_4,x_5,x_6,x_7,x_8,x_9,x_{10},x_{11},x_{12}) = \frac{\frac{x_1}{x_2}-\frac{x_3}{x_4}-\frac{x_5}{x_6}+\frac{x_7}{x_8}}{\frac{x_9}{x_2}-\frac{x_{10}}{x_4}-\frac{x_{11}}{x_6}+\frac{x_{12}}{x_8}}
\end{equation*}

Then
\begin{align*}
    \widehat{\psi}_{\text{SW}}^{\text{Wald}} = f(\widehat{\mu}_{Y,1_b},\widehat{\mu}_{Z,1_b},\widehat{\mu}_{Y,0_b},\widehat{\mu}_{Z,0_b},\widehat{\mu}_{Y,1_a},\widehat{\mu}_{Z,1_a},\widehat{\mu}_{Y,0_a},\widehat{\mu}_{Z,0_a},\widehat{\mu}_{D,1_b},\widehat{\mu}_{D,0_b},\widehat{\mu}_{D,1_a},\widehat{\mu}_{D,0_a})\\
    {\Psi}_{\text{SW},P_0}=f({\mu}_{Y,1_b},{\mu}_{Z,1_b},{\mu}_{Y,0_b},{\mu}_{Z,0_b},{\mu}_{Y,1_a},{\mu}_{Z,1_a},{\mu}_{Y,0_a},{\mu}_{Z,0_a},{\mu}_{D,1_b},{\mu}_{D,0_b},{\mu}_{D,1_a},{\mu}_{D,0_a})
\end{align*}
By delta method \citep{van2000asymptotic}, we know $\hat{\psi}_{\text{SW}}^{\text{Wald}}$ is asymptotically linear estimator of $\Psi_{SW,P_0}$ with influence function

\begin{equation*}
    \varphi_{\text{Wald,SW}}:o\mapsto [\nabla  f(\mu_{Y,D,Z})]^T{\varphi}_{Y,D,Z}(o)
\end{equation*}
where
\begin{align*}
    \mu_{Y,D,Z} = ({\mu}_{Y,1_b},{\mu}_{Z,1_b},{\mu}_{Y,0_b},{\mu}_{Z,0_b},{\mu}_{Y,1_a},{\mu}_{Z,1_a},{\mu}_{Y,0_a},{\mu}_{Z,0_a},{\mu}_{D,1_b},{\mu}_{D,0_b},{\mu}_{D,1_a},{\mu}_{D,0_a})^T,\\
    {\varphi}_{Y,D,Z}(o) = ({\varphi}_{Y,1_b},{\varphi}_{Z,1_b},{\varphi}_{Y,0_b},{\varphi}_{Z,0_b},{\varphi}_{Y,1_a},{\varphi}_{Z,1_a},{\varphi}_{Y,0_a},{\varphi}_{Z,0_a},{\varphi}_{D,1_b},{\varphi}_{D,0_b},{\varphi}_{D,1_a},{\varphi}_{D,0_a})^T.
\end{align*}
Therefore,
\begin{equation*}
\sqrt{n}\left(\hat{\psi}^{\text{Wald}}_{\text{SW}}-\Psi_{\text{SW},P_0}\right)\rightsquigarrow N\left(0,\sigma_{\text{Wald}}^2\right),
\end{equation*}
where $\sigma_{\text{Wald}}^2 = P_0\varphi_{\text{Wald},\text{SW}}^2$.
\end{proof}
\subsection{Proof of Theorem \ref{thm_gradient}}
\begin{proof}
Let $\{P_t, t \in [0,\delta)\}$ denote a one-dimensional regular submodel in $\mathcal{P}$, with $P_{t=0}=P_0$. Recall that $\mathcal{P}$ is a locally nonparametric model. Then the canonical gradient $D(P)$ at $P_0$ is the unique function which belongs to $L_0^2(P_0)$ that satisfies

\begin{equation}
    \frac{d \Psi(P_t)}{dt}\Big|_{t=0} = P_0(D(P_0)g),\label{pathwise_diff}
\end{equation}
where $g$ is the score function of submodel $\{P_t, t \in [0,\delta)\}$ at $t=0$. We observe that
\begin{align*}
    &\frac{d \Psi(P_t)}{dt}\Big|_{t=0}\\
   =& \frac{\frac{d}{dt}\big|_{t=0} \mathbb{E}_{P_t}[\mathbb{E}_{P_t}[Y|Z=b_1,\boldsymbol{X}]-\mathbb{E}_{P_t}[Y|Z=b_0,\boldsymbol{X}]-\mathbb{E}_{P_t}[Y|Z=a_1,\boldsymbol{X}]+\mathbb{E}_{P_t}[Y|Z=a_0,\boldsymbol{X}]]}{\mathbb{E}_{P_0}[\eta_{b,P_0}(\boldsymbol{X})-\eta_{a,P_0}(\boldsymbol{X})]} \\
   & - \Psi(P_0)\frac{\frac{d}{dt}\big|_{t=0} \mathbb{E}_{P_t}[\mathbb{E}_{P_t}[D|Z=b_1,\boldsymbol{X}]-\mathbb{E}_{P_t}[D|Z=b_0,\boldsymbol{X}]-\mathbb{E}_{P_t}[D|Z=a_1,\boldsymbol{X}]+\mathbb{E}_{P_t}[D|Z=a_0,\boldsymbol{X}]]}{\mathbb{E}_{P_0}[\eta_{b,P_0}(\boldsymbol{X})-\eta_{a,P_0}(\boldsymbol{X})]}
\end{align*}

It's well-known that the canonical gradients for $\mathbb{E}_{P_0}\big[\mathbb{E}_{P_0}[Y|Z=z,\boldsymbol{X}]\big]$ and $\mathbb{E}_{P_0}\big[\mathbb{E}_{P_0}[D|Z=z,\boldsymbol{X}]\big]$ are 
\begin{align*}
    &T_1(P_0)= E_{P_0}[Y|Z=z,\boldsymbol{X}] +\frac{\mathbbm{1}(Z=z)}{P_0(Z=z|\boldsymbol{X})}\big[Y-\mathbb{E}_{P_0}[Y|Z=z,\boldsymbol{X}]\big]-\mathbb{E}_{P_0}\big[\mathbb{E}_{P_0}[Y|Z=z,\boldsymbol{X}]\big]\\
    &\text{and}\\
    &T_2(P_0)= \mathbb{E}_{P_0}[D|Z=z,\boldsymbol{X}] +\frac{\mathbbm{1}(Z=z)}{P_0(Z=z|\boldsymbol{X})}\big[D-\mathbb{E}_{P_0}[D|Z=z,\boldsymbol{X}]\big]-\mathbb{E}_{P_0}\big[\mathbb{E}_{P_0}[D|Z=z,\boldsymbol{X}]\big]
\end{align*}
respectively, therefore
\begin{align*}
    \frac{d}{dt} \mathbb{E}_{P_t}[\mathbb{E}_{P_t}[Y|Z=z,\boldsymbol{X}]\Big|_{t=0} = P_0(T_1(P_0)g) \\
    \frac{d}{dt} \mathbb{E}_{P_t}[\mathbb{E}_{P_t}[Y|Z=z,\boldsymbol{X}]\Big|_{t=0} = P_0(T_2(P_0)g)
\end{align*}
therefore
\small\begin{equation*}
    \begin{split}
        D_{P}&=
        (z,x,d,y)\mapsto\frac{1}{\mathbb{E}_P[\eta_{b,P}(\boldsymbol{X})-\eta_{a,P}(\boldsymbol{X})]}\Bigg\{\frac{\mathbbm{1}\{z=1_b\}}{P(Z=1_b\mid \boldsymbol{X}=\boldsymbol{x})}\Big[y-\mathbb{E}_P[Y\mid Z=1_b,\boldsymbol{X}=\boldsymbol{x}]\Big]\\ 
        &-\frac{\mathbbm{1}\{z=0_b\}}{P(Z=0_b\mid \boldsymbol{X}=\boldsymbol{x})}\Big[y-\mathbb{E}_P[Y\mid Z=0_b,\boldsymbol{X}=\boldsymbol{x}]\Big]\\
        &-\frac{\mathbbm{1}\{z=1_a\}}{P(Z=1_a\mid \boldsymbol{X}=\boldsymbol{x})}\Big[y-\mathbb{E}_P[Y\mid Z=1_a,\boldsymbol{X}=\boldsymbol{x}]\Big] +\frac{\mathbbm{1}\{z=0_a\}}{P(Z=0_a\mid \boldsymbol{X}=\boldsymbol{x})}\Big[y-\mathbb{E}_P[Y\mid Z=0_a, \boldsymbol{X}=\boldsymbol{x}]\Big]\\
        &+\delta_{b,P}(\boldsymbol{x})-\delta_{a,P}(\boldsymbol{x})\Bigg\}- \frac{\Psi_{\text{SW},P}}{\mathbb{E}_P[\eta_{b,P}(\boldsymbol{X})-\eta_{a,P}(\boldsymbol{X})]}\Bigg\{\frac{\mathbbm{1}\{z=1_b\}}{P(Z=1_b\mid \boldsymbol{X}=\boldsymbol{x})}\Big[d-\mathbb{E}_P(D\mid Z=1_b, \boldsymbol{X}=\boldsymbol{x})\Big] \\
        &-\frac{\mathbbm{1}\{z=0_b\}}{P(Z=0_b\mid \boldsymbol{X}=\boldsymbol{x})}\Big[d-\mathbb{E}_P(D\mid Z=0_b,\boldsymbol{X}=\boldsymbol{x})\Big]-\frac{\mathbbm{1}\{Z=1_a\}}{P(Z=1_a\mid \boldsymbol{X}=\boldsymbol{x})}\Big[d-\mathbb{E}_P(D\mid Z=1_a, \boldsymbol{X}=\boldsymbol{x})\Big] \\
        &+\frac{\mathbbm{1}\{Z=0_a\}}{P(Z=0_a \mid \boldsymbol{X}=\boldsymbol{x})}\Big[d-\mathbb{E}_P(D\mid Z=0_a,\boldsymbol{X}=\boldsymbol{x})\Big]+\eta_{b,P}(\boldsymbol{x})-\eta_{a,P}(\boldsymbol{x})\Bigg\}.
        \end{split}
    \end{equation*}\normalsize
satisfies \eqref{pathwise_diff}.
    
\end{proof}

\subsection{Additional information for estimation part}
\label{subsec: additional information for estimation, supp}
\subsubsection{One-step estimator}
Similar to $\widehat{\psi}_{ee}$, we can also construct a one-step estimator based on the efficient influence function and cross-fitting.

A one-step estimator $\widehat{\psi}_{os}$ can be constructed as follows:
\begin{description}
    \item[Step I:]  For each $k\in[K]$, construct the one-step estimator $\widehat{\psi}_{os,k} = \Psi_{\widehat{P}_{n,k}} + \mathbb{P}_{n,k} D_{\widehat{P}_{n,k}}$ for $\Psi_{SW,P_0}$;

\item[Step II:] Construct the final cross-fitted one-step estimator as $\widehat{\psi}_{os} = \frac{1}{K}\sum_{k=1}^n \widehat{\psi}_{os,k}$.
\end{description}

\subsubsection{Regularity conditions for asymptotic normality of $\widehat{\psi}_{ee}$ and $\widehat{\psi}_{os}$}

\label{subsubsec: regularity conditions for ee and os, supp}
Suppose that
\begin{enumerate}
    \item There exists $0<\epsilon<0.5$ such that $\epsilon<P_0(Z|\boldsymbol{X})<1-\epsilon$, $\epsilon<P_0(D|Z,X)<1-\epsilon$,  $\epsilon<\mathbb{E}_{P_0}[\eta_{b,P_0}(\boldsymbol{X})-\eta_{a,P_0}(\boldsymbol{X})]< 1-\epsilon$,  $\epsilon<\widehat{P}_{n,k}(Z|\boldsymbol{X})<1-\epsilon$, $\epsilon<\widehat{P}_{n,k}(D|Z,X)<1-\epsilon$ 
 ($k=1,...K$),  there is a universal constant $C$ such that $|\mathbbm{E}_{P_0}[Y|Z,\boldsymbol{X}]|\leq C$ with probability one.
    \item The nuisance functions estimators are consistent:
    \begin{align*}
        &\Big\Vert \pi_{P_0}(z,\cdot)-\pi_{\widehat{P}_{n,k}}(z,\cdot)\Big\Vert_{L^2(P_0)} = o_p(1)\\
        &\Big\Vert \mu_{Y,P_0}(z,\cdot)-\mu_{Y,\widehat{P}_{n,k}}(z,\cdot)\Big\Vert_{L^2(P_0)} = o_p(1)\\
        &\Big\Vert \mu_{D,P_0}(z,\cdot)-\mu_{D,\widehat{P}_{n,k}}(z,\cdot)\Big\Vert_{L^2(P_0)} = o_p(1)
    \end{align*}
    for $z\in \{0_a,1_a,0_b,1_b\}$ and $k \in \{1,...,K\}$
    \item Rate conditions for nuisance functions estimation:\begin{enumerate}
        \item Rate double-robustness:
        \begin{align*}
            \Big\Vert \pi_{P_0}(z,\cdot)-\pi_{\widehat{P}_{n,k}}(z,\cdot)\Big\Vert_{L^2(P_0)}\Big\Vert \mu_{B,P_0}(z,\cdot)-\mu_{B,\widehat{P}_{n,k}}(z,\cdot)\Big\Vert_{L^2(P_0)}=o_p(1/\sqrt{n})
        \end{align*}
        for $z\in \{0_a,1_a,0_b,1_b\}$, $k \in \{1,...,K\}$, $B=Y$ or $D$.
        \item The convergence rates for all nuisance function estimators are faster than $n^{-1/4}$:
        \begin{align*}
             &\Big\Vert \pi_{P_0}(z,\cdot)-\pi_{\widehat{P}_{n,k}}(z,\cdot)\Big\Vert_{L^2(P_0)} = o_p(n^{-1/4})\\
        &\Big\Vert \mu_{Y,P_0}(z,\cdot)-\mu_{Y,\widehat{P}_{n,k}}(z,\cdot)\Big\Vert_{L^2(P_0)} = o_p(n^{-1/4})\\
        &\Big\Vert \mu_{D,P_0}(z,\cdot)-\mu_{D,\widehat{P}_{n,k}}(z,\cdot)\Big\Vert_{L^2(P_0)} = o_p(n^{-1/4})
        \end{align*}
        for $z\in \{0_a,1_a,0_b,1_b\}$, $k \in \{1,...,K\}$.
    \end{enumerate}
    
\end{enumerate}

For asymptotic linearity of $\hat{\psi}_{ee}$, we need to assume conditions 1, 2 and 3(a) hold, for asymptotic linearity of $\hat{\psi}_{os}$, we need to assume conditions 1, 2 and 3(b) hold.
\begin{proof}
    We firstly prove the results for $\hat{\psi}_{os}$. For probability measures $P$ and $\overline{P}$, define
    \begin{equation*}
        R(\overline{P},P) = \Psi_{\text{SW},\overline{P}}-\Psi_{\text{SW},{P}}+PD_{\overline{P}}.
    \end{equation*}
    We will use the notations with overbar to denote functionals defined by $\overline{P}$, and use the notations without overbar to denote functionals defined by ${P}$. The simplied notations are listed below:
    \begin{align*}
        &\overline{\pi}_{z} = \overline{P}(Z=z|\boldsymbol{X}=\boldsymbol{x}), {\pi}_{z} = {P}(Z=z|\boldsymbol{X}=\boldsymbol{x}),\\
        &\overline{\mu}_{Y,z} = \mathbbm{E}_{\overline{P}}[Y|Z=z,\boldsymbol{X}=\boldsymbol{x}],{\mu}_{Y,z} = \mathbbm{E}_{{P}}[Y|Z=z,\boldsymbol{X}=\boldsymbol{x}],\\
        &\overline{\mu}_{D,z} = \mathbbm{E}_{\overline{P}}[D|Z=z,\boldsymbol{X}=\boldsymbol{x}],{\mu}_{D,z} = \mathbbm{E}_{{P}}[D|Z=z,\boldsymbol{X}=\boldsymbol{x}],\\
        & \overline{\omega}_1 = \mathbbm{E}_{\overline{P}}[\delta_{b,\overline{P}}(\boldsymbol{X})-\delta_{a,\overline{P}}(\boldsymbol{X})],{\omega}_1 = \mathbbm{E}_{{P}}[\delta_{b,{P}}(\boldsymbol{X})-\delta_{a,{P}}(\boldsymbol{X})],\\
        & \overline{\omega}_2 = \mathbbm{E}_{\overline{P}}[\eta_{b,\overline{P}}(\boldsymbol{X})-\eta_{a,\overline{P}}(\boldsymbol{X})],{\omega}_2 = \mathbbm{E}_{{P}}[\eta_{b,{P}}(\boldsymbol{X})-\eta_{a,{P}}(\boldsymbol{X})]
    \end{align*}

    For $z\in \{1_b,0_b,1_a,0_a\}$, and $L$ is equal to either $Y$ or $D$, we have (we will use lower case to denote the realization of random variables),
    \begin{align*}
        &P\Bigg\{\frac{\mathbbm{1}\{z=1_b\}}{\overline{\pi}_{z}}\Big[l-\overline{\mu}_{L,z}\Big]+\overline{\mu}_{L,z}\Bigg\}\\
        =&P\Bigg\{\mathbbm{1}\{z=1_b\}\Big(\frac{1}{\overline{\pi}_{z}}-\frac{1}{{\pi}_{z}}+\frac{1}{{\pi}_{z}}\Big)\Big[(l-{\mu}_{L,z})+({\mu}_{L,z}-\overline{\mu}_{L,z})\Big]+(\overline{\mu}_{L,z}-{\mu}_{L,z})+{\mu}_{L,z}\Bigg\}\\
        =&P\Bigg\{\mathbbm{1}\{z=1_b\}\frac{1}{\pi_z}\Big[l-\mu_{L,z}\Big]+\mu_{L,z}\Bigg\}+P\Bigg\{\mathbbm{1}\{z=1_b\}\Big(\frac{1}{\overline{\pi}_z}-\frac{1}{{\pi}_z}\Big)\Big(\mu_{L,z}-\overline{\mu}_{L,z}\Big)\Bigg\}\\
        &+P\Bigg\{\mathbbm{1}\{z=1_b\}\frac{1}{\pi_{z}}(\overline{\mu}_{L,z}-{\mu}_{L,z})-(\overline{\mu}_{L,z}-{\mu}_{L,z})\Bigg\}+P\Bigg\{\mathbbm{1}\{z=1_b\}\Big(\frac{1}{\overline{\pi}_z}-\frac{1}{{\pi}_z}\Big)\Big(y-{\mu}_{L,z}\Big)\Bigg\}\\
        =&P\mu_{L,z}+P\Bigg\{\mathbbm{1}\{z=1_b\}\Big(\frac{1}{\overline{\pi}_z}-\frac{1}{{\pi}_z}\Big)\Big(\mu_{L,z}-\overline{\mu}_{L,z}\Big)\Bigg\}
    \end{align*}

Plug this into $R(\overline{P},P)$, we have
\begin{align*}
    &R(\overline{P},P)\\
    =&\Psi_{SW,\overline{P}}-\Psi_{SW,{P}}+PD_{\overline{P}}\\
    =&\frac{1}{\overline{\omega}_2}\Bigg\{P\Big[\mathbbm{1}\{z=1_b\}\Big(\frac{1}{\overline{\pi}_{1_b}}-\frac{1}{{\pi}_{1_b}}\Big)(\mu_{Y,1_b}-\overline{\mu}_{Y,1_b})\Big]-P\Big[\mathbbm{1}\{z=0_b\}\Big(\frac{1}{\overline{\pi}_{0_b}}-\frac{1}{{\pi}_{0_b}}\Big)(\mu_{Y,0_b}-\overline{\mu}_{Y,0_b})\Big] \\
    &-P\Big[\mathbbm{1}\{z=1_a\}\Big(\frac{1}{\overline{\pi}_{1_a}}-\frac{1}{{\pi}_{1_a}}\Big)(\mu_{Y,1_a}-\overline{\mu}_{Y,1_a})\Big]+P\Big[\mathbbm{1}\{z=0_a\}\Big(\frac{1}{\overline{\pi}_{0_a}}-\frac{1}{{\pi}_{0_a}}\Big)(\mu_{Y,0_a}-\overline{\mu}_{Y,0_a})\Big]\Bigg\}\\
    &-\frac{\overline{\omega}_1}{\overline{\omega}^2_2}\Bigg\{P\Big[\mathbbm{1}\{z=1_b\}\Big(\frac{1}{\overline{\pi}_{1_b}}-\frac{1}{{\pi}_{1_b}}\Big)(\mu_{D,1_b}-\overline{\mu}_{D,1_b})\Big]-P\Big[\mathbbm{1}\{z=0_b\}\Big(\frac{1}{\overline{\pi}_{0_b}}-\frac{1}{{\pi}_{0_b}}\Big)(\mu_{D,0_b}-\overline{\mu}_{D,0_b})\Big] \\
    &-P\Big[\mathbbm{1}\{z=1_a\}\Big(\frac{1}{\overline{\pi}_{1_a}}-\frac{1}{{\pi}_{1_a}}\Big)(\mu_{D,1_a}-\overline{\mu}_{D,1_a})\Big]+P\Big[\mathbbm{1}\{z=0_a\}\Big(\frac{1}{\overline{\pi}_{0_a}}-\frac{1}{{\pi}_{0_a}}\Big)(\mu_{D,0_a}-\overline{\mu}_{D,0_a})\Big]\Bigg\}\\
    &+\frac{1}{\overline{\omega}_2^2}(\omega_2-\overline{\omega}_2)(\overline{\omega}_1-\omega_1)+\frac{\omega_1}{\overline{\omega}_2^2\omega_2} (\omega_2-\overline{\omega}_2)^2
\end{align*}

Then for $k = 1,...,K$,
\begin{align*}
    &\hat{\psi}_{os,k}-\Psi_{SW,P_0}\\
=&\Psi_{SW,\hat{P}_{n,k}}+\mathbbm{P}_{n,k}D_{\hat{P}_n}-\Psi_{SW,P_0}\\
    =&(\mathbbm{P}_{n,k}-P_0)D_{P_0}+(\mathbbm{P}_{n,k}-P_0)(D_{\hat{P}_{n,k}}-D_{P_0})+R(\hat{P}_{n,k},P_0)
\end{align*}
We will show that $\sqrt{n}R(\hat{P}_{n,k},P_0)=o_p(1)$. Now we define new notations
\begin{align*}
        &\hat{\pi}_{z,k}(\boldsymbol{X}) = \hat{P}_{n,k}(Z=z|\boldsymbol{X}=\boldsymbol{x}), {\pi}_{0,z}(\boldsymbol{X}) = {P_0}(Z=z|\boldsymbol{X}=\boldsymbol{x}),\\
        &\hat{\mu}_{Y,z,k} = \mathbbm{E}_{\hat{P}_{n,k}}[Y|Z=z,\boldsymbol{X}=\boldsymbol{x}],{\mu}_{0,Y,z} = \mathbbm{E}_{{P}_0}[Y|Z=z,\boldsymbol{X}=\boldsymbol{x}],\\
        &\hat{\mu}_{D,z,k} = \mathbbm{E}_{\hat{P}_{n,k}}[D|Z=z,\boldsymbol{X}=\boldsymbol{x}],{\mu}_{0,D,z} = \mathbbm{E}_{{P}_0}[D|Z=z,\boldsymbol{X}=\boldsymbol{x}],\\
        & \hat{\omega}_{1,k} = \mathbbm{E}_{\hat{P}_{n,k}}[\delta_{b,\hat{P}_{n,k}}(\boldsymbol{X})-\delta_{a,\hat{P}_{n,k}}(\boldsymbol{X})],{\omega}_{0,1} = \mathbbm{E}_{{P}_0}[\delta_{b,{P}_0}(\boldsymbol{X})-\delta_{a,{P}_0}(\boldsymbol{X})],\\
        & \hat{\omega}_{2,k} = \mathbbm{E}_{\hat{P}_{n,k}}[\eta_{b,\hat{P}_{n,k}}(\boldsymbol{X})-\eta_{a,\hat{P}_{n,k}}(\boldsymbol{X})],{\omega}_{0,2} = \mathbbm{E}_{{P}_0}[\eta_{b,{P}_0}(\boldsymbol{X})-\eta_{a,{P}_0}(\boldsymbol{X})]
    \end{align*}

Then with probability approaching one, we know there exists a constant $C$ such that
\begin{align*}
    \Big|\frac{1}{\hat{\omega}_{2,k}}\Big|\leq C,\Big|\frac{1}{\hat{\omega}^2_{2,k}}\Big|\leq C,\Big|\frac{\hat{\omega}_{1,k}}{\hat{\omega}^2_{2,k}}\Big|\leq C,\Big|\frac{\hat{\omega}_{1,k}}{\hat{\omega}^2_{2,k}\omega_{0,2}}\Big|\leq C,
\end{align*}
Under our assumptions for nuisance parameter estimation, we have
\begin{align*}
    P_0\Big[\mathbbm{1}\{z=z\}\Big(\frac{1}{\hat{\pi}_{z,k}}-\frac{1}{{\pi}_{z}}\Big)(\mu_{B,z}-\hat{\mu}_{B,z,k})\Big]\leq C\Big\Vert  \frac{1}{\hat{\pi}_{z,k}}-\frac{1}{{\pi}_{z}}\Big\Vert_{L^2(P_0)}\Big\Vert  \mu_{B,z}-\hat{\mu}_{B,z,k} \Big\Vert_{L^2(P_0)} = o_p(n^{-1/2})
\end{align*}
for $z\in \{1_b,0_b,1_a,0_a\}$. Similarly, by the fact we used cross-fitting and Holder's inequality, we have 
\begin{align*}
    &|\omega_{0,2}-\hat{\omega}_{2,k}||\omega_{0,1}-\hat{\omega}_{1,k}|=o_p(n^{-1/2}),\\
    &|\omega_{0,2}-\hat{\omega}_{2,k}|^2=o_p(n^{-1/2})
\end{align*}
Therefore, we have proved that $R(\hat{P}_{n,k},P_0)=o_p(n^{-1/2})$. 

Now we discuss how to handle $(\mathbbm{P}_{n,k}-P_0)(D_{\hat{P}_{n,k}}-D_{P_0})$. Since we used cross-fitting, we have for fixed $t>0$,
\begin{align*}
    &P_0\Bigg\{ \Big| [\mathbbm{P}_{n,k}-P_0][D_{\hat{P}_{n,k}}-D_{P_0}]  \Big|> (n/K)^{-1/2}t \bigg|\text{ $k$-th subsample}\Bigg\} \\
    \leq &\min\Bigg\{ 1,\frac{\text{Var}\{[D_{\hat{P}_{n,k}}-D_{P_0}]|\text{ $k$-th subsample}\}}{t^2}   \Bigg\} \\
    = &\min\Bigg\{ 1,\frac{P_0\{[D_{\hat{P}_{n,k}}-D_{P_0}]^2|\text{ $k$-th subsample}\}}{t^2}   \Bigg\} \\
    = &\min \Bigg\{ 1,\frac{P_0[D_{\hat{P}_{n,k}}-D_{P_0}]^2}{t^2}   \Bigg\} \\
    = &\min \{1,o_p(1)\}
\end{align*}

Taking an expectation of both sides and applying the dominated convergence theorem shows that
\begin{align*}
    &P_0\Bigg\{ \Big| [\mathbbm{P}_{n,k}-P_0][D_{\hat{P}_{n,k}}-D_{P_0}]  \Big|> (n/K)^{-1/2}t \Bigg\} \\
    =& E[\min{1,o_p(1)}]\\
    \rightarrow & 0
\end{align*}
As $t>0$ was arbitrary, we have
\begin{equation*}
    (\mathbbm{P}_{n,k}-P_0)(D_{\hat{P}_{n,k}}-D_{P_0}) = o_p(n^{-1/2}).
\end{equation*}

To sum up,
\begin{align*}
    &\sqrt{n}(\hat{\psi}_{os}-\Psi_{SW,P_0})\\
    =&\sqrt{n}(\frac{1}{K}\sum_{k=1}^K (\hat{\psi}_{os,k}-\Psi_{SW,P_0}))\\
    =&\sqrt{n}(\frac{1}{K}\sum_{k=1}^K(\mathbbm{P}_{n,k}-P_0)D_{P_0})+o_p(1)\\
    =&\sqrt{n}(\mathbbm{P}_{n}-P_0)D_{P_0}\\
    \rightsquigarrow& N(0,P_0D^2_{P_0})
\end{align*}
where the last line holds if we further assume $P_0D^2_{P_0}<\infty$.

For $\widehat{\psi}_{ee}$, we notice that
\begin{equation*}
    \begin{split}
        \widehat{\psi}_{ee,k} = \frac{\mathbbm{P}_{n,k}A_{n,k}}{\mathbbm{P}_{n,k}B_{n,k}}
    \end{split}
\end{equation*}
where 
\begin{equation*}
\small
    \begin{split}
        A_{n,k} &= \frac{\mathbbm{1}\{z=1_b\}}{\widehat{P}_{n,k}(Z=1_b\mid \boldsymbol{X}=\boldsymbol{x})}\Big[y-\mathbb{E}_{\widehat{P}_{n,k}}[Y\mid Z=1_b,\boldsymbol{X}=\boldsymbol{x}]\Big]\\
        &-\frac{\mathbbm{1}\{z=0_b\}}{\widehat{P}_{n,k}(Z=0_b\mid \boldsymbol{X}=\boldsymbol{x})}\Big[y-\mathbb{E}_{\widehat{P}_{n,k}}[Y\mid Z=0_b,\boldsymbol{X}=\boldsymbol{x}]\Big]\\
        &-\frac{\mathbbm{1}\{z=1_a\}}{\widehat{P}_{n,k}(Z=1_a\mid \boldsymbol{X}=\boldsymbol{x})}\Big[y-\mathbb{E}_{\widehat{P}_{n,k}}[Y\mid Z=1_a,\boldsymbol{X}=\boldsymbol{x}]\Big] \\
        &+\frac{\mathbbm{1}\{z=0_a\}}{\widehat{P}_{n,k}(Z=0_a\mid \boldsymbol{X}=\boldsymbol{x})}\Big[y-\mathbb{E}_{\widehat{P}_{n,k}}[Y\mid Z=0_a, \boldsymbol{X}=\boldsymbol{x}]\Big]\\
        &+\delta_{b,\widehat{P}_{n,k}}(\boldsymbol{x})-\delta_{a,\widehat{P}_{n,k}}(\boldsymbol{x})
    \end{split}
\end{equation*}

\begin{equation*}
\small
    \begin{split}
        B_{n,k} &= \frac{\mathbbm{1}\{z=1_b\}}{\widehat{P}_{n,k}(Z=1_b\mid \boldsymbol{X}=\boldsymbol{x})}\Big[d-\mathbb{E}_{\widehat{P}_{n,k}}[D\mid Z=1_b,\boldsymbol{X}=\boldsymbol{x}]\Big]\\
        &-\frac{\mathbbm{1}\{z=0_b\}}{\widehat{P}_{n,k}(Z=0_b\mid \boldsymbol{X}=\boldsymbol{x})}\Big[d-\mathbb{E}_{\widehat{P}_{n,k}}[D\mid Z=0_b,\boldsymbol{X}=\boldsymbol{x}]\Big]\\
        &-\frac{\mathbbm{1}\{z=1_a\}}{\widehat{P}_{n,k}(Z=1_a\mid \boldsymbol{X}=\boldsymbol{x})}\Big[d-\mathbb{E}_{\widehat{P}_{n,k}}[D\mid Z=1_a,\boldsymbol{X}=\boldsymbol{x}]\Big] \\
        &+\frac{\mathbbm{1}\{z=0_a\}}{\widehat{P}_{n,k}(Z=0_a\mid \boldsymbol{X}=\boldsymbol{x})}\Big[d-\mathbb{E}_{\widehat{P}_{n,k}}[D\mid Z=0_a, \boldsymbol{X}=\boldsymbol{x}]\Big]\\
        &+\eta_{b,\widehat{P}_{n,k}}(\boldsymbol{x})-\eta_{a,\widehat{P}_{n,k}}(\boldsymbol{x})
    \end{split}
\end{equation*}

Since under the given conditions, $\mathbbm{P}_{n,k}A_{n,k}$ and $\mathbbm{P}_{n,k}B_{b,k}$ are asymptotically linear estimators for $\mathbb{E}_{P_0}[\delta_{b,P_0}-\delta_{a,P_0}]$ and $\mathbb{E}_{P_0}[\eta_{b,P_0}-\eta_{a,P_0}]$, respectively, the influence functions (denoted as $D_{A,P_0}$ and $D_{B,P_0}$) at $P_0$ are
\begin{align*}
    D_{A,P_0}: o \mapsto &\mathbb{E}_{P_0}[Y|Z=1_b,\boldsymbol{X}=\boldsymbol{x}] +\frac{\mathbbm{1}(z=1_b)}{P_0(Z=1_b|\boldsymbol{X}=\boldsymbol{x})}\big[y-\mathbb{E}_{P_0}[Y|Z=z,\boldsymbol{X}=\boldsymbol{x}]\big]\\
    &-\mathbb{E}_{P_0}[Y|Z=0_b,\boldsymbol{X}=\boldsymbol{x}] -\frac{\mathbbm{1}(z=0_b)}{P_0(Z=0_b|\boldsymbol{X}=\boldsymbol{x})}\big[y-\mathbb{E}_{P_0}[Y|Z=z,\boldsymbol{X}=\boldsymbol{x}]\big]\\
    &- \mathbb{E}_{P_0}[Y|Z=1_a,\boldsymbol{X}=\boldsymbol{x}] -\frac{\mathbbm{1}(z=1_a)}{P_0(Z=1_a|\boldsymbol{X}=\boldsymbol{x})}\big[y-\mathbb{E}_{P_0}[Y|Z=z,\boldsymbol{X}=\boldsymbol{x}]\big]\\
    &+  \mathbb{E}_{P_0}[Y|Z=o_a,\boldsymbol{X}=\boldsymbol{x}] +\frac{\mathbbm{1}(z=0_a)}{P_0(Z=0_a|\boldsymbol{X}=\boldsymbol{x})}\big[y-\mathbb{E}_{P_0}[Y|Z=z,\boldsymbol{X}=\boldsymbol{x}]\big]\\
    &-\mathbb{E}_{P_0}[\delta_{b,P_0}(\boldsymbol{X})-\delta_{a,P_0}(\boldsymbol{X})]
\end{align*}
and
\begin{align*}
    D_{B,P_0}: o \mapsto &\mathbb{E}_{P_0}[D|Z=1_b,\boldsymbol{X}=\boldsymbol{x}] +\frac{\mathbbm{1}(z=1_b)}{P_0(Z=1_b|\boldsymbol{X}=\boldsymbol{x})}\big[d-\mathbb{E}_{P_0}[D|Z=z,\boldsymbol{X}=\boldsymbol{x}]\big]\\
    &-\mathbb{E}_{P_0}[D|Z=0_b,\boldsymbol{X}=\boldsymbol{x}] -\frac{\mathbbm{1}(z=0_b)}{P_0(Z=0_b|\boldsymbol{X}=\boldsymbol{x})}\big[d-\mathbb{E}_{P_0}[D|Z=z,\boldsymbol{X}=\boldsymbol{x}]\big]\\
    &- \mathbb{E}_{P_0}[D|Z=1_a,\boldsymbol{X}=\boldsymbol{x}] -\frac{\mathbbm{1}(z=1_a)}{P_0(Z=1_a|\boldsymbol{X}=\boldsymbol{x})}\big[d-\mathbb{E}_{P_0}[D|Z=z,\boldsymbol{X}=\boldsymbol{x}]\big]\\
    &+  \mathbb{E}_{P_0}[D|Z=o_a,\boldsymbol{X}=\boldsymbol{x}] +\frac{\mathbbm{1}(z=0_a)}{P_0(Z=0_a|\boldsymbol{X}=\boldsymbol{x})}\big[d-\mathbb{E}_{P_0}[D|Z=z,\boldsymbol{X}=\boldsymbol{x}]\big]\\
    &-\mathbb{E}_{P_0}[\eta_{b,P_0}(\boldsymbol{X})-\eta_{a,P_0}(\boldsymbol{X})].
\end{align*}

Let $f(\boldsymbol{x}_1,x_2) = x_1/x_2$, then by delta method \citet{van2000asymptotic}, 
$\widehat{\psi}_{ee,k}$ should also be asymptotically linear, where the influence function should be
\begin{equation*}
    o \mapsto (\nabla f(\mathbb{E}_{P_0}[\delta_{b,P_0}-\delta_{a,P_0}],\mathbb{E}_{P_0}[\eta_{b,P_0}-\eta_{a,P_0}])^T(D_{A,P_0}(o),D_{B,P_0}(o))
\end{equation*}
which is exactly $D_{P_0}$. Therefore, we also have
\begin{equation*}
    \sqrt{n/K}(\widehat{\psi}_{ee,k}-\Psi_{SW,P_0}) = \sqrt{n/K}(\mathbbm{P}_{n,k}-P_0)D_{P_0} +o_p(1)
\end{equation*}
Therefore, 
\begin{equation*}
    \sqrt{n}(\hat{\psi}_{ee}-\Psi_{SW,P_0}) = \sqrt{n}(\mathbbm{P}_n-P_0)D_{P_0} +o_p(1).
\end{equation*}
\end{proof}
\subsection{Proof of Proposition \ref{prop_homo}}
\begin{proof}
    Suppose principal ignorability holds, then by definition of principal ignorability, we have 
    \begin{equation*}
        \text{ATE}_{P_0}(\boldsymbol{X}) = \text{SWATE}_{P_0}(\boldsymbol{X}) =\text{ACOATE}_{P_0}(\boldsymbol{X}).
    \end{equation*}
    Suppose no unmeasured common effect modifier holds, then
    \begin{align*}
        &(\mathbbm{E}_{P_0}[Y|\boldsymbol{X},Z=1_a]- \mathbbm{E}_{P_0}[Y|\boldsymbol{X},Z=0_a])-(\mathbbm{E}_{P_0}[Y|\boldsymbol{X},Z=1_b]- \mathbbm{E}_{P_0}[Y|\boldsymbol{X},Z=0_b])\\
        =&(\mathbbm{E}_{P_0}[Y(Z=1_a)|\boldsymbol{X}]- \mathbbm{E}_{P_0}[Y(Z=0_a)|\boldsymbol{X}])-(\mathbbm{E}_{P_0}[Y(Z=1_b)|\boldsymbol{X}]- \mathbbm{E}_{P_0}[Y(Z=0_b)|\boldsymbol{X}])
    \end{align*}
    Notice that 
    \begin{align*}
        &\mathbbm{E}_{P_0}[Y(Z=1_a)|\boldsymbol{X}]- \mathbbm{E}_{P_0}[Y(Z=0_a)|\boldsymbol{X}]\\
        =&\mathbbm{E}_{P_0}[D(Z=1_a)Y(D=1)+(1-D(Z=1_a))Y(D=0)|\boldsymbol{X}]- \\&\mathbbm{E}_{P_0}[D(Z=0_a)Y(D=1)+(1-D(Z=0_a))Y(D=0)|\boldsymbol{X}]\\
        =&\mathbbm{E}_{P_0}[(D(Z=1_a)-D(Z=0_a))(Y(D=1)-Y(D=0))|\boldsymbol{X}]\\
        =&\mathbbm{E}_{P_0}[D(Z=1_a)-D(Z=0_a)|\boldsymbol{X}]\mathbbm{E}_{P_0}[Y(D=1)-Y(D=0)|\boldsymbol{X}]
    \end{align*}
    similarly we have
    \begin{equation*}
        \mathbbm{E}_{P_0}[Y(Z=1_b)|\boldsymbol{X}]- \mathbbm{E}_{P_0}[Y(Z=0_b)|\boldsymbol{X}] = \mathbbm{E}_{P_0}[D(Z=1_b)-D(Z=0_b)|\boldsymbol{X}]\mathbbm{E}_{P_0}[Y(D=1)-Y(D=0)|\boldsymbol{X}]
    \end{equation*}
    Therefore,
    \begin{align*}
        &(\mathbbm{E}_{P_0}[Y|\boldsymbol{X},Z=1_a]- \mathbbm{E}_{P_0}[Y|\boldsymbol{X},Z=0_a])-(\mathbbm{E}_{P_0}[Y|\boldsymbol{X},Z=1_b]- \mathbbm{E}_{P_0}[Y|\boldsymbol{X},Z=0_b])\\
        =&(\mathbbm{E}_{P_0}[D(Z=1_a)-D(Z=0_a)|\boldsymbol{X}]-\mathbbm{E}_{P_0}[D(Z=1_b)-D(Z=0_b)|\boldsymbol{X}])\mathbbm{E}_{P_0}[Y(D=1)-Y(D=0)|\boldsymbol{X}]\\
        =&[(\mathbbm{E}_{P_0}[D|\boldsymbol{X},Z=1_a]-\mathbbm{E}_{P_0}[D|\boldsymbol{X},Z=0_a])-(\mathbbm{E}_{P_0}[D|\boldsymbol{X},Z=1_b]-\mathbbm{E}_{P_0}[D|\boldsymbol{X},Z=0_b])]\text{ATE}_{P_0}(\boldsymbol{X})
    \end{align*}
    Therefore, $\text{ATE}_{P_0}(\boldsymbol{X})=\text{SWATE}_{P_0}(\boldsymbol{X})$.
    Similarly, we have $\text{ATE}_{P_0}(\boldsymbol{X})=\text{ACOATE}_{P_0}(\boldsymbol{X})$.
\end{proof}

\subsection{Additional information for Theorem \ref{thm: gradient for beta}}
\label{subsec: additional information, projection test, supp}
\subsubsection{Explicit form of the canonical gradients}
The canonical gradients for $\beta^{(j)}_{P}$, $j=1,2,3$ at $P_0$, are $\phi^{(j)}_{P_0}$, $j=1,2,3$. $\phi^{(j)}_{P}$, $j=1,2,3$ are defined as
    \small\begin{align*}
        \phi^{(1)}_{P}:(z,\boldsymbol{x},d,y)\mapsto  C^{(1)}_{P}\frac{\partial\gamma(\boldsymbol{x};\beta)}{\partial \beta}\Big|_{\beta=\beta^{(1)}_P}\Bigg[D^{(1)}_{P}(z,\boldsymbol{x},d,y)-D^{(2)}_{P}(z,\boldsymbol{x},d,y)+\theta^{(1)}_{P}(\boldsymbol{x})-\gamma(\boldsymbol{x};\beta^{(1)}_P)\Bigg],\\
        \phi^{(2)}_{P}: (z,\boldsymbol{x},d,y)\mapsto C^{(2)}_{P}\frac{\partial\gamma(\boldsymbol{x};\beta)}{\partial \beta}\Big|_{\beta=\beta^{(2)}_P}\Bigg[D^{(1)}_{P}(z,\boldsymbol{x},d,y)-D^{(3)}_{P}(z,\boldsymbol{x},d,y)+\theta^{(2)}_{P}(\boldsymbol{x})-\gamma(\boldsymbol{x};\beta^{(2)}_P)\Bigg], \\
        \phi^{(3)}_{P}:(z,\boldsymbol{x},d,y)\mapsto  C^{(3)}_{P}\frac{\partial\gamma(\boldsymbol{x};\beta)}{\partial \beta}\Big|_{\beta=\beta^{(3)}_P}\Bigg[D^{(2)}_{P}(z,\boldsymbol{x},d,y)-D^{(3)}_{P}(z,\boldsymbol{x},d,y)+\theta^{(3)}_{P}(\boldsymbol{x})-\gamma(\boldsymbol{x};\beta^{(3)}_P)\Bigg].
    \end{align*} \normalsize
    where $C^{(j)}_{P}$ are constants, and  the expressions of $D^{(1)}_{P}$, $D^{(2)}_{P}$, and $D^{(3)}_{P}$ are

    \begingroup
\allowdisplaybreaks
    \small\begin{align*}
        D^{(1)}_{P}&=
        (z,\boldsymbol{x},d,y)\mapsto\frac{1}{\eta_{a,P}(\boldsymbol{x})}\Bigg\{\frac{1\{z=1_a\}}{P(Z=1_a\mid\boldsymbol{X} = \boldsymbol{x})}\Big[y-\mathbb{E}_P[Y\mid Z=1_a,\boldsymbol{X} = \boldsymbol{x}]\Big] \\
        &-\frac{1\{z=0_a\}}{P(Z=0_a\mid\boldsymbol{X} = \boldsymbol{x})}\Big[y-\mathbb{E}_P[Y\mid Z=0_a,\boldsymbol{X} = \boldsymbol{x}]\Big]\Bigg\}\\
        &- \frac{\delta_{a,P}(\boldsymbol{x})}{[\eta_{a,P}(\boldsymbol{x})]^2}\Bigg\{\frac{1\{Z=1_a\}}{P(Z=1_a\mid\boldsymbol{X} = \boldsymbol{x})}\Big[d-\mathbb{E}_P[D\mid Z=1_a,\boldsymbol{X} = \boldsymbol{x}]\Big] \\
        &-\frac{1\{Z=0_a\}}{P(Z=0_a\mid\boldsymbol{X} = \boldsymbol{x})}\Big[d-\mathbb{E}_P[D\mid Z=0_a,\boldsymbol{X} = \boldsymbol{x}]\Big]\Bigg\},\\ 
        D^{(2)}_{P}&=
        (z,x,d,y)\mapsto\frac{1}{\eta_{b,P}(\boldsymbol{x})-\eta_{a,P}(\boldsymbol{x})}\Bigg\{\frac{1\{z=1_b\}}{P(Z=1_b\mid\boldsymbol{X} = \boldsymbol{x})}\Big[y-\mathbb{E}_P[Y\mid Z=1_b,\boldsymbol{X} = \boldsymbol{x}]\Big]\\ 
        &-\frac{1\{z=0_b\}}{P(Z=0_b\mid\boldsymbol{X} = \boldsymbol{x})}\Big[y-\mathbb{E}_P[Y\mid Z=0_b,\boldsymbol{X} = \boldsymbol{x}]\Big]\\
        &-\frac{1\{z=1_a\}}{P(Z=1_a\mid\boldsymbol{X} = \boldsymbol{x})}\Big[y-\mathbb{E}_P[Y\mid Z=1_a,\boldsymbol{X} = \boldsymbol{x}]\Big] +\frac{1\{z=0_a\}}{P(Z=0_a\mid\boldsymbol{X} = \boldsymbol{x})}\Big[y-\mathbb{E}_P[Y\mid Z=0_a,\boldsymbol{X} = \boldsymbol{x}]\Big]\Bigg\}\\
        &- \frac{\delta_{b,P}(\boldsymbol{x})-\delta_{a,P}(\boldsymbol{x})}{[\eta_{b,P}(\boldsymbol{x})-\eta_{a,P}(\boldsymbol{x})]^2}\Bigg\{\frac{1\{z=1_b\}}{P(Z=1_b\mid\boldsymbol{X} = \boldsymbol{x})}\Big[d-\mathbb{E}_P[D\mid Z=1_b,\boldsymbol{X} = \boldsymbol{x}]\Big] \\
        &-\frac{1\{z=0_b\}}{P(Z=0_b\mid\boldsymbol{X} = \boldsymbol{x})}\Big[d-\mathbb{E}_P[D\mid Z=0_b,\boldsymbol{X} = \boldsymbol{x}]\Big]-\frac{1\{Z=1_a\}}{P(Z=1_a\mid\boldsymbol{X} = \boldsymbol{x})}\Big[d-\mathbb{E}_P[D\mid Z=1_a,\boldsymbol{X} = \boldsymbol{x}]\Big] \\
       &+\frac{1\{Z=0_a\}}{P(Z=0_a\mid\boldsymbol{X} = \boldsymbol{x})}\Big[d-\mathbb{E}_P[D\mid Z=0_a,\boldsymbol{X} = \boldsymbol{x}]\Big]\Bigg\},\\
         D^{(3)}_{P}&=
 (z,\boldsymbol{x},d,y)\mapsto\frac{1}{\eta_{b,P}(\boldsymbol{x})}\Bigg\{\frac{1\{z=1_b\}}{P(Z=1_b\mid\boldsymbol{X} = \boldsymbol{x})}\Big[y-\mathbb{E}_P[Y\mid Z=1_b,\boldsymbol{X} = \boldsymbol{x}]\Big] \\
         &-\frac{1\{z=0_b\}}{P(Z=0_b\mid\boldsymbol{X} = \boldsymbol{x})}\Big[y-\mathbb{E}_P[Y\mid Z=0_b,\boldsymbol{X} = \boldsymbol{x}]\Big]\Bigg\}\\
         &- \frac{\delta_{b,P}(\boldsymbol{x})}{[\eta_{b,P}(\boldsymbol{x})]^2}\Bigg\{\frac{1\{Z=1_b\}}{P(Z=1_b\mid\boldsymbol{X} = \boldsymbol{x})}\Big[d-\mathbb{E}_P[D\mid Z=1_b,\boldsymbol{X} = \boldsymbol{x}]\Big] \\
         &-\frac{1\{Z=0_b\}}{P(Z=0_b\mid\boldsymbol{X} = \boldsymbol{x})}\Big[d-\mathbb{E}_P[D\mid Z=0_b,\boldsymbol{X} = \boldsymbol{x}]\Big]\Bigg\}.
    \end{align*}
    
\endgroup
\normalsize When $\gamma(\boldsymbol{x};\beta)=\boldsymbol{x}^T\beta$, we have $C^{(1)}_{P} = C^{(2)}_{P} =C^{(3)}_{P}= \mathbb{E}_{P}[\boldsymbol{X}\boldsymbol{X}^T]^{-1}$.
\subsubsection{Proof}
\begin{proof}

$\beta_P$ satisfies the following equation:
\begin{align*}
\int q(\boldsymbol{x};\beta_{P})(\theta_{P}(\boldsymbol{X})-\gamma(\boldsymbol{x};\beta_{P}))dP(\boldsymbol{X}) = 0,
\end{align*}
where $q(\boldsymbol{x};\beta_P) = \frac{\gamma(\boldsymbol{x};\beta)}{\partial \beta}|_{\beta = \beta_P}$

Now we pick a one-dimensional parametric submodel $\{P_t:t\in[0,\epsilon)\}$, then we have
\begin{align*}
    \int q(\boldsymbol{x};\beta_{P_t})(\theta_{P_t}(\boldsymbol{X})-\gamma(\boldsymbol{x};\beta_{P_t}))dP_t(\boldsymbol{X}) = 0
\end{align*}
Take derivative with respect to $t$, then evaluate the derivative at $t=0$, we have

\begin{align*}
    &\int \frac{\partial q(\boldsymbol{x};\beta)}{\partial \beta}\Big|_{\beta = \beta_{P_0}}\frac{d \beta_{P_t}}{dt}\Big|_{t=0}[\theta_{P_0}(\boldsymbol{X})-\gamma(\boldsymbol{x};\beta_{P_0})]dP_0(\boldsymbol{X})+\\
    & \int q(\boldsymbol{x};\beta_{P_0})\Big[\frac{\theta_{P_t}(\boldsymbol{X})}{\partial t}\Big|_{t=0}-q^T(\boldsymbol{x};\beta_{P_0})\frac{\partial \beta_{P_t}}{\partial t}\Big|_{t=0}\Big] dP_0(\boldsymbol{X})\\
    & \int q(\boldsymbol{x};\beta_{P_0})[\theta_{P_0}(\boldsymbol{X})-\gamma(\boldsymbol{x};\beta_{P_0})]s_0(\boldsymbol{X})dP_0(\boldsymbol{X}) = 0
\end{align*}

Therefore
\begin{align*}
    C\frac{\partial \beta_{P_t}}{\partial t} = \int q(\boldsymbol{x};\beta_{P_0})\frac{\partial\theta_{P_t}(\boldsymbol{X})}{\partial t}\Big|_{t=0}dP_0(\boldsymbol{X}) +\mathbb{E}_{P_0}\Big[q(\boldsymbol{x};\beta_{P_0})[\theta_0(\boldsymbol{X})-\gamma(\boldsymbol{x};\beta_{P_0})]s_0(O)\Big],
\end{align*}
where $C$ is a constant, when $\gamma(\boldsymbol{x};\beta)=\boldsymbol{x}^T\beta$, we have $\frac{\partial q(\boldsymbol{x};\beta)}{\partial \beta}=0$, $q(\boldsymbol{x};\beta) = \boldsymbol{x}$ and therefore $C = \int \boldsymbol{x}^T\boldsymbol{x} dP_0$. Note that $q(\boldsymbol{x};\beta_{P_0})[\theta_0(\boldsymbol{X})-\gamma(\boldsymbol{x};\beta_{P_0})]$ is already mean zero, therefore we only need to arrange the first term. Since $\theta_{P_t}(\boldsymbol{X}) = \theta_{1, P_t}(\boldsymbol{X})-\theta_{2, P_t}(\boldsymbol{X})$. We first handle $\frac{\partial \theta_{1, P_t}(\boldsymbol{X})}{\partial t}$:
\begin{align*}
    &\frac{\partial \theta_{1, P_t}(\boldsymbol{X})}{\partial t}\Big|_{t=0} \\
    = & \frac{\frac{\partial \delta_{a,P_t}(\boldsymbol{X})}{\partial t}|_{t=0} \eta_{a,P_0}(\boldsymbol{X})-\delta_{a,P_0}(\boldsymbol{X}) \frac{\partial \eta_{a,P_t}(\boldsymbol{X})}{\partial t}|_{t=0}}{[\eta_{a,P_0}(\boldsymbol{X})]^2} \\
    =&\frac{1}{\eta_{a,P_0}(\boldsymbol{X})}\Big[\frac{\partial \delta_{a,P_t}(\boldsymbol{X})}{\partial t}\Big|_{t=0}  -\theta_{1,P_0}(\boldsymbol{X}) \frac{\partial \eta_{a,P_t}(\boldsymbol{X})}{\partial t}\Big|_{t=0}\Big]
\end{align*}
It remains to arrange
\begin{align*}
    \int q(\boldsymbol{x};\beta_{P_0})\frac{1}{\eta_{a,P_0}(\boldsymbol{X})}\frac{\delta_{a,P_t}(\boldsymbol{X})}{\partial t}\Big|_{t=0}dP_0(\boldsymbol{X}) - \int q(\boldsymbol{x};\beta_{P_0})\frac{\theta_{1,P_0}(\boldsymbol{X})}{\eta_{a,P_0}(\boldsymbol{X})}\frac{\partial \eta_{a,P_t}(\boldsymbol{X})}{\partial t}\Big|_{t=0} dP_0(\boldsymbol{X})
\end{align*}

For any $h (\boldsymbol{X})$, we calculate $\mathbbm{E}_{P_0}[h(\boldsymbol{X})\frac{\partial}{\partial t}\mathbb{E}_{P_t}[C|Z=z,\boldsymbol{X}]|_{t=0}]$, where $C$ is equal to either $Y$ or $D$.
\begin{align*}
    &\mathbbm{E}_{P_0}\Big[h(\boldsymbol{X})\frac{\partial}{\partial t}\mathbbm{E}_{P_t}[C|Z=z,\boldsymbol{X}]\Big|_{t=0}\Big]\\
    =& \mathbbm{E}_{P_0}\Big[h(\boldsymbol{X})\mathbbm{E}_{P_0}[CS_{C|Z=z,\boldsymbol{X}}(C,\boldsymbol{X})|Z=z,\boldsymbol{X}]\Big]\\
    =&\mathbbm{E}_{P_0}\Big[h(\boldsymbol{X})\mathbbm{E}_{P_0}[(C-\mathbbm{E}_{P_0}(C|Z=z,\boldsymbol{X})S_{C|Z=z,\boldsymbol{X}}(C,\boldsymbol{X})|Z=z,\boldsymbol{X}]\Big] \\
    =& \mathbbm{E}_{P_0}\Big[\frac{h(\boldsymbol{X})}{P_0(Z=z|\boldsymbol{X})}\mathbbm{E}_{P_0}[1\{Z=z\}(C-E_{P_0}(C|Z,X)S_{C|Z=z,X}(C,X)|\boldsymbol{X}]\Big]\\
    =&\mathbbm{E}_{P_0}\Bigg[\frac{h(\boldsymbol{X})\mathbbm{E}_{P_0}[1\{Z=z\}(C-\mathbbm{E}_{P_0}[C|Z,\boldsymbol{X}])S_{C|Z,\boldsymbol{X}}(C,Z,\boldsymbol{X})|\boldsymbol{X}]}{P_0(Z=z|\boldsymbol{X})}\Bigg] \\
    =&\mathbbm{E}_{P_0}\Bigg[\frac{h(\boldsymbol{X}) [1\{Z=z\}(C-\mathbbm{E}_{P_0}[C|Z,\boldsymbol{X}]) }{P_0(Z=z|\boldsymbol{X})}S_{C|Z,\boldsymbol{X}}(C,Z,\boldsymbol{X})\Bigg] 
\end{align*}
Furthermore,
\begin{align*}
    \mathbbm{E}_{P_0}\Bigg[\frac{h(\boldsymbol{X}) [1\{Z=z\}(C-\mathbbm{E}_{P_0}[C|Z,\boldsymbol{X}]) }{P_0(Z=z|\boldsymbol{X})}S_{Z,\boldsymbol{X}}(Z,\boldsymbol{X})\Bigg] =0
\end{align*}
Therefore we have
\begin{align*}
    \mathbbm{E}_{P_0}\Bigg[\frac{h(\boldsymbol{X}) [1\{Z=z\}(C-\mathbbm{E}_{P_0}[C|Z,\boldsymbol{X}]) }{P_0(Z=z|\boldsymbol{X})}S_{O}(O)\Bigg] =0
\end{align*}
Therefore,
\begin{align*}
    & \int q(\boldsymbol{x};\beta_{P_0})\frac{1}{\eta_{a,P_0}(\boldsymbol{X})}\frac{\delta_{a,P_t}(\boldsymbol{X})}{\partial t}\Big|_{t=0}dP_0(\boldsymbol{X})\\
     = & \mathbbm{E}_{P_0}\Bigg[\frac{q(\boldsymbol{x};\beta_0)}{\eta_{a,P_0}(\boldsymbol{X})}\Bigg[\frac{1\{Z=1_a\}}{P_0(Z=1_a|\boldsymbol{X})}(Y-\mathbbm{E}_{P_0}(Y|Z=1_a,\boldsymbol{X}))\\
     &-\frac{1\{Z=0_a\}}{P_0(Z=0_a|\boldsymbol{X})}(Y-\mathbbm{E}_{P_0}(Y|Z=0_a,\boldsymbol{X}))\Bigg]S_O(O)\Bigg]
\end{align*}
Similarly,
\begin{align*}
    & \int q(\boldsymbol{x};\beta_{P_0})\frac{\theta_{1,P_0}(\boldsymbol{X})}{\eta_{a,P_0}(\boldsymbol{X})}\frac{\eta_{a,P_t}(\boldsymbol{X})}{\partial t}\Big|_{t=0}dP_0(\boldsymbol{X})\\
     = & \mathbbm{E}_{P_0}\Bigg[\frac{q(\boldsymbol{x};\beta_0)\theta_{1,P_0}(\boldsymbol{X})}{\eta_{a,P_0}(\boldsymbol{X})}\Bigg[\frac{1\{Z=1_a\}}{P_0(Z=1_a|\boldsymbol{X})}(D-\mathbbm{E}_{P_0}(D|Z=1_a,\boldsymbol{X}))\\
     &-\frac{1\{Z=0_a\}}{P_0(Z=0_a|\boldsymbol{X})}(D-\mathbbm{E}_{P_0}(D|Z=0_a,\boldsymbol{X}))\Bigg]S_O(O)\Bigg]
\end{align*}
We repeat the similar calculation for the term involving $\theta_{2,P_t}$, then we can obtain the canonical gradient as stated.

\end{proof}

\subsection{Proof of Theorem \ref{thm_one_step_test}}
\label{subsec: additional information, np test, supp}
Suppose that
\begin{enumerate}
    \item There exists $0<\epsilon<0.5$ such that $\epsilon<P_0(Z|\boldsymbol{X})<1-\epsilon$, $\epsilon<P_0(D|Z,\boldsymbol{X})<1-\epsilon$, $\epsilon<\eta_{a,P_0}<1-\epsilon$, $\epsilon<\eta_{a,P_0}-\eta_{b,P_0}<1-\epsilon$, $\epsilon<\widehat{P}_{n,k}(Z|\boldsymbol{X})<1-\epsilon$, $\epsilon<\widehat{P}_{n,k}(D|Z,\boldsymbol{X})<1-\epsilon$,  $\epsilon<\eta_{a,\widehat{P}_{n,k}}<1-\epsilon$, $\epsilon<\eta_{a,\widehat{P}_{n,k}}-\eta_{b,\widehat{P}_{n,k}}<1-\epsilon$, ($k=1,...K$), there is a universal constant $C$ such that $|\mathbbm{E}_{P_0}[Y|Z,\boldsymbol{X}]|\leq C$ with probability one. 
    \item $\Vert \boldsymbol{X}\Vert_{\infty}<C$,$\lambda_{\min}(\frac{1}{n}\sum_{i=1}^n \boldsymbol{X}_i^T\boldsymbol{X}_i)>0$ with probability one. $\lambda_{\min}\mathbbm{E}[\boldsymbol{X}_i^T\boldsymbol{X}_i]>0$. Here for any matrix $H$, $\lambda_{\min}H$ denoted the smallest eigenvalue of $H$.
    \item Rate conditions for nuisance functions estimation: The convergence rates for all nuisance function estimators are faster then $n^{-1/4}$:
        \begin{align*}
             &\Big\Vert \pi_{P_0}(z,\cdot)-\pi_{\widehat{P}_{n,k}}(z,\cdot)\Big\Vert_{L^2(P_0)} = o_p(n^{-1/4})\\
        &\Big\Vert \mu_{Y,P_0}(z,\cdot)-\mu_{Y,\widehat{P}_{n,k}}(z,\cdot)\Big\Vert_{L^2(P_0)} = o_p(n^{-1/4})\\
        &\Big\Vert \mu_{D,P_0}(z,\cdot)-\mu_{D,\widehat{P}_{n,k}}(z,\cdot)\Big\Vert_{L^2(P_0)} = o_p(n^{-1/4})
        \end{align*}
        for $z\in \{0_a,1_a,0_b,1_b\}$, $k \in \{1,...,K\}$.
\end{enumerate}

\begin{proof}
    The proof is very similar to the proof of Theorem 4. We will only state the proof for test $T^{(1)}_{\text{Wald},\alpha}$. Since one-step estimator and estimating equation estimator are equivalent when our tests are based on best linear projection, we state the proof for one-step estimator. We define
    \begin{equation*}
        R(\overline{P},P) = \beta^{(1)}_{\overline{P}}-\beta^{(1)}_{{P}}+P\phi^{(1)}_{\overline{P}}.
    \end{equation*}
    Similar to the proof of Theorem 4, we define the following notations:
    \begin{align*}
        &\overline{\pi}_{z}(\boldsymbol{X}) = \overline{P}(Z=z|\boldsymbol{X}=\boldsymbol{x}), {\pi}_{z}(\boldsymbol{X}) = {P}(Z=z|\boldsymbol{X}=\boldsymbol{x}),\\
        &\overline{\mu}_{Y,z} = \mathbbm{E}_{\overline{P}}[Y|Z=z,\boldsymbol{X}=\boldsymbol{x}],{\mu}_{Y,z} = \mathbbm{E}_{{P}}[Y|Z=z,\boldsymbol{X}=\boldsymbol{x}],\\
        &\overline{\mu}_{D,z} = \mathbbm{E}_{\overline{P}}[D|Z=z,\boldsymbol{X}=\boldsymbol{x}],{\mu}_{D,z} = \mathbbm{E}_{{P}}[D|Z=z,\boldsymbol{X}=\boldsymbol{x}],\\
        & \bar{\delta}_{g} = \delta_{g,\overline{P}},{\delta}_{g} = \delta_{g,P},\bar{\eta}_{g} = \eta_{g,\overline{P}},{\eta}_{g} = \eta_{g,P}
    \end{align*}
    for $z\in \{1_b,0_b,1_a,0_a\}$ and $g\in \{a,b\}$.

    We first study for fixed $z'\in\{1_b,0_b,1_a,0_a\}$, direct calculation yields
    \begin{align*}
        &P\Bigg[\frac{1}{\bar{\eta}_g}\Bigg(\frac{\mathbbm{1}\{z=z'\}}{\bar{\pi}_{z'}}(y-\overline{\mu}_{Y,z'})\Bigg)\Bigg]\\
        =&P\Bigg[\Bigg(\frac{1}{\bar{\eta}_g}-\frac{1}{{\eta}_g}\Bigg)\Bigg(\frac{\mathbbm{1}\{z=z'\}}{{\pi}_{z'}}(y-{\mu}_{Y,{z'}})\Bigg)\Bigg]+P\Bigg[\frac{1}{{\eta}_g}\Bigg(\mathbbm{1}\{z={z'}\}\Bigg(\frac{1}{\overline{\pi}_{z'}}-\frac{1}{{\pi}_{z'}}\Bigg)(y-{\mu}_{Y,{z'}})\Bigg)\Bigg]\\
        &+P\Bigg[\frac{1}{{\eta}_g}\Bigg(\mathbbm{1}\{z={z'}\}\frac{1}{{\pi}_{z'}}(\overline{\mu}_{Y,{z'}}-{\mu}_{Y,z'})\Bigg)\Bigg]+P\Bigg[\frac{1}{{\eta}_g}\frac{\mathbbm{1}\{z={z'}\}}{{\pi}_{z'}}(y-{\mu}_{Y,z'})\Bigg]\\
        &+P\Bigg[\Bigg(\frac{1}{\overline{\eta}_g}-\frac{1}{{\eta}_g}\Bigg)\mathbbm{1}\{z=z'\}\frac{1}{\pi_{z'}}(\mu_{Y,z'}-\overline{\mu}_{Y,z'})\Bigg] + P\Bigg[\Bigg(\frac{1}{\overline{\eta}_g}-\frac{1}{{\eta}_g}\Bigg)\mathbbm{1}\{z=z'\}\Bigg(\frac{1}{\overline{\pi}_{z'}}-\frac{1}{{\pi}_{z'}}\Bigg)(y-\mu_{Y,z'})\Bigg]\\
        &+P\Bigg[\frac{1}{{\eta}_g}\mathbbm{1}\{z=z'\}\Bigg(\frac{1}{\overline{\pi}_{z'}}-\frac{1}{{\pi}_{z'}}\Bigg)(\mu_{Y,z'}-\overline{\mu}_{Y,z'})\Bigg]\\
        &+P\Bigg[\Bigg(\frac{1}{\overline{\eta}_g}-\frac{1}{{\eta}_g}\Bigg)\mathbbm{1}\{z=z'\}\Bigg(\frac{1}{\overline{\pi}_{z'}}-\frac{1}{{\pi}_{z'}}\Bigg)(\mu_{Y,z'}-\overline{\mu}_{Y,z'})\Bigg]\\
        =&P\Bigg[\frac{1}{\eta_g}(\mu_{Y,z'}-\overline{\mu}_{Y,z'})\Bigg]+P\Bigg[\Bigg(\frac{1}{\overline{\eta}_g}-\frac{1}{{\eta}_g}\Bigg)(\mu_{Y,z'}-\overline{\mu}_{Y,z'})\Bigg] +P\Bigg[\frac{1}{{\eta}_g}\mathbbm{1}\{z=z'\}\Bigg(\frac{1}{\overline{\pi}_{z'}}-\frac{1}{{\pi}_{z'}}\Bigg)(\mu_{Y,z'}-\overline{\mu}_{Y,z'})\Bigg]\\
        &+P\Bigg[\Bigg(\frac{1}{\overline{\eta}_g}-\frac{1}{{\eta}_g}\Bigg)\mathbbm{1}\{z=z'\}\Bigg(\frac{1}{\overline{\pi}_{z'}}-\frac{1}{{\pi}_{z'}}\Bigg)(\mu_{Y,z'}-\overline{\mu}_{Y,z'})\Bigg]\\
    \end{align*}

Then
\begin{align*}
    &\beta^{(1)}_{\overline{P}}-\beta^{(1)}_{{P}}+P\phi^{(1)}_{\overline{P}}\\
    =&(\overline{P}\boldsymbol{x}\boldsymbol{x}^T)^{-1}\overline{P}\Bigg[\boldsymbol{x}\Bigg(\frac{\overline{\delta}_a}{\overline{\eta}_a}-\frac{\overline{\delta}_b-\overline{\delta}_a}{\overline{\eta}_b-\overline{\eta}_a}\Bigg)\Bigg]-(P\boldsymbol{x}\boldsymbol{x}^T)^{-1}P\Bigg[\boldsymbol{x}\Bigg(\frac{{\delta}_a}{{\eta}_a}-\frac{{\delta}_b-{\delta}_a}{{\eta}_b-{\eta}_a}\Bigg)\Bigg]\\
    &+(\overline{P}\boldsymbol{x}\boldsymbol{x}^T)^{-1}P\Bigg[\boldsymbol{x}\Bigg(D_{\overline{P}}^1-D_{\overline{P}}^2+\frac{\overline{\delta}_a}{\overline{\eta}_a}-\frac{\overline{\delta}_b-\overline{\delta}_a}{\overline{\eta}_b-\overline{\eta}_a}-x^T(\overline{P}\boldsymbol{x}\boldsymbol{x}^T)^{-1}\overline{P}\Bigg[\boldsymbol{x}\Bigg(\frac{\overline{\delta}_a}{\overline{\eta}_a}-\frac{\overline{\delta}_b-\overline{\delta}_a}{\overline{\eta}_b-\overline{\eta}_a}\Bigg)\Bigg]\Bigg)\Bigg]\\
    =&\Bigg\{(\overline{P}\boldsymbol{x}\boldsymbol{x}^T)^{-1}\overline{P}\Bigg[\boldsymbol{x}\frac{\overline{\delta}_a}{\overline{\eta}_a}\Bigg]-(P\boldsymbol{x}\boldsymbol{x}^T)^{-1}P\Bigg[\boldsymbol{x}\frac{{\delta}_a}{{\eta}_a}\Bigg]+(\overline{P}\boldsymbol{x}\boldsymbol{x}^T)^{-1}P\Bigg[\boldsymbol{x}\Bigg(D_{\overline{P}}^1+\frac{\overline{\delta}_a}{\overline{\eta}_a}-x^T(\overline{P}\boldsymbol{x}\boldsymbol{x}^T)^{-1}\overline{P}\Bigg[\boldsymbol{x}\frac{\overline{\delta}_a}{\overline{\eta}_a}\Bigg]\Bigg)\Bigg]\Bigg\}\\
    &-\Bigg\{(\overline{P}\boldsymbol{x}\boldsymbol{x}^T)^{-1}\overline{P}\Bigg[\boldsymbol{x}\frac{\overline{\delta}_b-\overline{\delta}_a}{\overline{\eta}_b-\overline{\eta}_a}\Bigg]-(P\boldsymbol{x}\boldsymbol{x}^T)^{-1}P\Bigg[\boldsymbol{x}\frac{{\delta}_b-{\delta}_a}{{\eta}_b-{\eta}_a}\Bigg]\\
    &+(\overline{P}\boldsymbol{x}\boldsymbol{x}^T)^{-1}P\Bigg[\boldsymbol{x}\Bigg(D_{\overline{P}}^2+\frac{\overline{\delta}_b-\overline{\delta}_a}{\overline{\eta}_b-\overline{\eta}_a}-x^T(\overline{P}\boldsymbol{x}\boldsymbol{x}^T)^{-1}\overline{P}\Bigg[\boldsymbol{x}\frac{\overline{\delta}_b-\overline{\delta}_a}{\overline{\eta}_b-\overline{\eta}_a}\Bigg]\Bigg)\Bigg]\Bigg\}
\end{align*}
Let
\begin{align*}
    A:&=(\overline{P}\boldsymbol{x}\boldsymbol{x}^T)^{-1}\overline{P}\Bigg[\boldsymbol{x}\frac{\overline{\delta}_a}{\overline{\eta}_a}\Bigg]-(P\boldsymbol{x}\boldsymbol{x}^T)^{-1}P\Bigg[\boldsymbol{x}\frac{{\delta}_a}{{\eta}_a}\Bigg]+(\overline{P}\boldsymbol{x}\boldsymbol{x}^T)^{-1}P\Bigg[\boldsymbol{x}\Bigg(D_{\overline{P}}^1+\frac{\overline{\delta}_a}{\overline{\eta}_a}-x^T(\overline{P}\boldsymbol{x}\boldsymbol{x}^T)^{-1}\overline{P}\Bigg[\boldsymbol{x}\frac{\overline{\delta}_a}{\overline{\eta}_a}\Bigg]\Bigg)\Bigg]\\
    B:&=(\overline{P}\boldsymbol{x}\boldsymbol{x}^T)^{-1}\overline{P}\Bigg[\boldsymbol{x}\frac{\overline{\delta}_b-\overline{\delta}_a}{\overline{\eta}_b-\overline{\eta}_a}\Bigg]-(P\boldsymbol{x}\boldsymbol{x}^T)^{-1}P\Bigg[\boldsymbol{x}\frac{{\delta}_b-{\delta}_a}{{\eta}_b-{\eta}_a}\Bigg]\\
    &+(\overline{P}\boldsymbol{x}\boldsymbol{x}^T)^{-1}P\Bigg[\boldsymbol{x}\Bigg(D_{\overline{P}}^2+\frac{\overline{\delta}_b-\overline{\delta}_a}{\overline{\eta}_b-\overline{\eta}_a}-x^T(\overline{P}\boldsymbol{x}\boldsymbol{x}^T)^{-1}\overline{P}\Bigg[\boldsymbol{x}\frac{\overline{\delta}_b-\overline{\delta}_a}{\overline{\eta}_b-\overline{\eta}_a}\Bigg]\Bigg)\Bigg]
\end{align*}

Direct calculation yields
\begin{align*}
    A=&(\overline{P}\boldsymbol{x}\boldsymbol{x}^T)^{-1}P(\boldsymbol{x}x^T)\Big(P(\boldsymbol{x}x^T)^{-1}-\overline{P}(\boldsymbol{x}x^T)^{-1}\Big)\Bigg(\overline{P}x\frac{\overline{\delta}_a}{\overline{\eta}_a}-{P}x\frac{{\delta}_a}{{\eta}_a}\Bigg)\\
    &+((\overline{P}\boldsymbol{x}\boldsymbol{x}^T)^{-1}P(\boldsymbol{x}x^T)-I)\Big((P\boldsymbol{x}\boldsymbol{x}^T)^{-1}-(\overline{P}\boldsymbol{x}\boldsymbol{x}^T)^{-1}\Big)P\Big(\boldsymbol{x}\frac{\delta_a}{\eta_a}\Big)\\
    &+P\Bigg[\boldsymbol{x}\Bigg(\frac{1}{\overline{\eta}_a}-\frac{1}{{\eta}_a}\Bigg)(\mu_{Y,1_a}-\overline{\mu}_{Y,1_a})\Bigg] +P\Bigg[\boldsymbol{x}\frac{1}{{\eta}_a}\mathbbm{1}\{z=1_a\}\Bigg(\frac{1}{\overline{\pi}_{1_a}}-\frac{1}{{\pi}_{1_a}}\Bigg)(\mu_{Y,1_a}-\overline{\mu}_{Y,1_a})\Bigg]\\
    &-P\Bigg[\boldsymbol{x}\Bigg(\frac{1}{\overline{\eta}_a}-\frac{1}{{\eta}_a}\Bigg)(\mu_{Y,0_a}-\overline{\mu}_{Y,0_a})\Bigg] -P\Bigg[\boldsymbol{x}\frac{1}{{\eta}_a}\mathbbm{1}\{z=0_a\}\Bigg(\frac{1}{\overline{\pi}_{0_a}}-\frac{1}{{\pi}_{0_a}}\Bigg)(\mu_{Y,0_a}-\overline{\mu}_{Y,0_a})\Bigg]\\
    &-P\Bigg[\boldsymbol{x}\Bigg(\frac{\overline{\delta}_a}{\overline{\eta}^2_a}-\frac{{\delta}_a}{{\eta}^2_a}\Bigg)(\mu_{D,1_a}-\overline{\mu}_{D,1_a})\Bigg] +P\Bigg[\boldsymbol{x}\frac{{\delta}_a}{{\eta}^2_a}\mathbbm{1}\{z=1_a\}\Bigg(\frac{1}{\overline{\pi}_{1_a}}-\frac{1}{{\pi}_{1_a}}\Bigg)(\mu_{D,1_a}-\overline{\mu}_{D,1_a})\Bigg]\\
    &+P\Bigg[\boldsymbol{x}\Bigg(\frac{\overline{\delta}_a}{\overline{\eta}^2_a}-\frac{{\delta}_a}{{\eta}^2_a}\Bigg)(\mu_{D,0_a}-\overline{\mu}_{D,0_a})\Bigg] +P\Bigg[\boldsymbol{x}\frac{{\delta}_a}{{\eta}^2_a}\mathbbm{1}\{z=0_a\}\Bigg(\frac{1}{\overline{\pi}_{0_a}}-\frac{1}{{\pi}_{1_a}}\Bigg)(\mu_{D,0_a}-\overline{\mu}_{D,0_a})\Bigg]\\
    &+P\Bigg[\boldsymbol{x}\Bigg(\frac{1}{\overline{\eta}_a}-\frac{1}{{\eta}_a}\Bigg)\mathbbm{1}\{z=1_a\}\Bigg(\frac{1}{\overline{\pi}_{1_a}}-\frac{1}{{\pi}_{1_a}}\Bigg)(\mu_{Y,1_a}-\overline{\mu}_{Y,1_a})\Bigg]\\
    &-P\Bigg[\boldsymbol{x}\Bigg(\frac{1}{\overline{\eta}_a}-\frac{1}{{\eta}_a}\Bigg)\mathbbm{1}\{z=0_a\}\Bigg(\frac{1}{\overline{\pi}_{0_a}}-\frac{1}{{\pi}_{0_a}}\Bigg)(\mu_{Y,0_a}-\overline{\mu}_{Y,0_a})\Bigg]\\
    &- P\Bigg[\boldsymbol{x}\Bigg(\frac{\overline{\delta}_a}{\overline{\eta}^2_a}-\frac{{\delta}_a}{{\eta}^2_a}\Bigg)\mathbbm{1}\{z=0_a\}\Bigg(\frac{1}{\overline{\pi}_{1_a}}-\frac{1}{{\pi}_{1_a}}\Bigg)(\mu_{D,1_a}-\overline{\mu}_{D,1_a})\Bigg]\\
    &+ P\Bigg[\boldsymbol{x}\Bigg(\frac{\overline{\delta}_a}{\overline{\eta}^2_a}-\frac{{\delta}_a}{{\eta}^2_a}\Bigg)\mathbbm{1}\{z=0_a\}\Bigg(\frac{1}{\overline{\pi}_{0_a}}-\frac{1}{{\pi}_{0_a}}\Bigg)(\mu_{D,0_a}-\overline{\mu}_{D,0_a})\Bigg]\\
    &+P\Bigg[\frac{\boldsymbol{x}\delta_a}{\eta_a^2\overline{\eta}_a}\Big[\delta_a(\overline{\eta}_a-{\eta}_a)^2+\eta_a(\overline{\eta}_a-\eta_a)(\delta_a-\overline{\delta}_a)\Big]\Bigg]
\end{align*}

Similarly, for $B$, direct calculation yields
\begin{equation*}\small
    \begin{split}
    B=&(\overline{P}\boldsymbol{x}\boldsymbol{x}^T)^{-1}P(\boldsymbol{x}x^T)\Big(P(\boldsymbol{x}x^T)^{-1}-\overline{P}(\boldsymbol{x}x^T)^{-1}\Big)\Bigg(\overline{P}x\frac{\overline{\delta}_b-\overline{\delta}_a}{\overline{\eta}_b-\overline{\eta}_a}-{P}x\frac{{\delta}_b-{\delta}_a}{{\eta}_b-{\eta}_a}\Bigg)\\
    &+((\overline{P}\boldsymbol{x}\boldsymbol{x}^T)^{-1}P(\boldsymbol{x}x^T)-I)\Big((P\boldsymbol{x}\boldsymbol{x}^T)^{-1}-(\overline{P}\boldsymbol{x}\boldsymbol{x}^T)^{-1}\Big)P\Big(\boldsymbol{x}\frac{\delta_b-\delta_a}{\eta_b-\eta_a}\Big)\\
    &+P\Bigg[\boldsymbol{x}\Bigg(\frac{1}{\overline{\eta}_b-\overline{\eta}_a}-\frac{1}{{\eta}_b-{\eta}_a}\Bigg)(\mu_{Y,1_b}-\overline{\mu}_{Y,1_b})\Bigg] +P\Bigg[\boldsymbol{x}\frac{1}{{\eta}_b-{\eta}_a}\mathbbm{1}\{z=1_b\}\Bigg(\frac{1}{\overline{\pi}_{1_b}}-\frac{1}{{\pi}_{1_b}}\Bigg)(\mu_{Y,1_b}-\overline{\mu}_{Y,1_b})\Bigg]\\
    &-P\Bigg[\boldsymbol{x}\Bigg(\frac{1}{\overline{\eta}_b-\overline{\eta}_a}-\frac{1}{{\eta}_b-{\eta}_a}\Bigg)(\mu_{Y,0_b}-\overline{\mu}_{Y,0_b})\Bigg] -P\Bigg[\boldsymbol{x}\frac{1}{{\eta}_b-{\eta}_a}\mathbbm{1}\{z=0_b\}\Bigg(\frac{1}{\overline{\pi}_{0_b}}-\frac{1}{{\pi}_{0_b}}\Bigg)(\mu_{Y,0_b}-\overline{\mu}_{Y,0_b})\Bigg]\\
    &-P\Bigg[\boldsymbol{x}\Bigg(\frac{1}{\overline{\eta}_b-\overline{\eta}_a}-\frac{1}{{\eta}_b-{\eta}_a}\Bigg)(\mu_{Y,1_a}-\overline{\mu}_{Y,1_a})\Bigg] -P\Bigg[\boldsymbol{x}\frac{1}{{\eta}_b-{\eta}_a}\mathbbm{1}\{z=1_a\}\Bigg(\frac{1}{\overline{\pi}_{1_a}}-\frac{1}{{\pi}_{1_a}}\Bigg)(\mu_{Y,1_a}-\overline{\mu}_{Y,1_a})\Bigg]\\
    &+P\Bigg[\boldsymbol{x}\Bigg(\frac{1}{\overline{\eta}_b-\overline{\eta}_a}-\frac{1}{{\eta}_b-{\eta}_a}\Bigg)(\mu_{Y,0_a}-\overline{\mu}_{Y,0_a})\Bigg] +P\Bigg[\boldsymbol{x}\frac{1}{{\eta}_b-{\eta}_a}\mathbbm{1}\{z=0_a\}\Bigg(\frac{1}{\overline{\pi}_{0_a}}-\frac{1}{{\pi}_{0_a}}\Bigg)(\mu_{Y,0_a}-\overline{\mu}_{Y,0_a})\Bigg]\\
    &-P\Bigg[\boldsymbol{x}\Bigg(\frac{\overline{\delta}_b-\overline{\delta}_a}{\overline{\eta}_b-\overline{\eta}_a}-\frac{{\delta}_b-{\delta}_a}{{\eta}_b-{\eta}_a}\Bigg)(\mu_{D,1_b}-\overline{\mu}_{D,1_b})\Bigg] -P\Bigg[\boldsymbol{x}\frac{{\delta}_b-{\delta}_a}{{\eta}_b-{\eta}_a}\mathbbm{1}\{z=1_b\}\Bigg(\frac{1}{\overline{\pi}_{1_b}}-\frac{1}{{\pi}_{1_b}}\Bigg)(\mu_{D,1_b}-\overline{\mu}_{D,1_b})\Bigg]\\
    &+P\Bigg[\boldsymbol{x}\Bigg(\frac{\overline{\delta}_b-\overline{\delta}_a}{\overline{\eta}_b-\overline{\eta}_a}-\frac{{\delta}_b-{\delta}_a}{{\eta}_b-{\eta}_a}\Bigg)(\mu_{D,0_b}-\overline{\mu}_{D,0_b})\Bigg] +P\Bigg[\boldsymbol{x}\frac{{\delta}_b-{\delta}_a}{{\eta}_b-{\eta}_a}\mathbbm{1}\{z=0_b\}\Bigg(\frac{1}{\overline{\pi}_{0_b}}-\frac{1}{{\pi}_{0_b}}\Bigg)(\mu_{D,0_b}-\overline{\mu}_{D,0_b})\Bigg]
    \\
    &+P\Bigg[\boldsymbol{x}\Bigg(\frac{\overline{\delta}_b-\overline{\delta}_a}{\overline{\eta}_b-\overline{\eta}_a}-\frac{{\delta}_b-{\delta}_a}{{\eta}_b-{\eta}_a}\Bigg)(\mu_{D,1_a}-\overline{\mu}_{D,1_a})\Bigg] +P\Bigg[\boldsymbol{x}\frac{{\delta}_b-{\delta}_a}{{\eta}_b-{\eta}_a}\mathbbm{1}\{z=1_a\}\Bigg(\frac{1}{\overline{\pi}_{1_a}}-\frac{1}{{\pi}_{1_a}}\Bigg)(\mu_{D,1_a}-\overline{\mu}_{D,1_a})\Bigg]\\
    &-P\Bigg[\boldsymbol{x}\Bigg(\frac{\overline{\delta}_b-\overline{\delta}_a}{\overline{\eta}_b-\overline{\eta}_a}-\frac{{\delta}_b-{\delta}_a}{{\eta}_b-{\eta}_a}\Bigg)(\mu_{D,0_a}-\overline{\mu}_{D,0_a})\Bigg] -P\Bigg[\boldsymbol{x}\frac{{\delta}_b-{\delta}_a}{{\eta}_b-{\eta}_a}\mathbbm{1}\{z=0_a\}\Bigg(\frac{1}{\overline{\pi}_{0_a}}-\frac{1}{{\pi}_{0_a}}\Bigg)(\mu_{D,0_a}-\overline{\mu}_{D,0_a})\Bigg]\\
    &+P\Bigg[\boldsymbol{x}\Bigg(\frac{1}{\overline{\eta}_b-\overline{\eta}_a}-\frac{1}{{\eta}_b-{\eta}_a}\Bigg)\mathbbm{1}\{z=1_b\}\Bigg(\frac{1}{\overline{\pi}_{1_b}}-\frac{1}{{\pi}_{1_b}}\Bigg)(\mu_{Y,1_b}-\overline{\mu}_{Y,1_b})\Bigg]\\
    &-P\Bigg[\boldsymbol{x}\Bigg(\frac{1}{\overline{\eta}_b-\overline{\eta}_a}-\frac{1}{{\eta}_b-{\eta}_a}\Bigg)\mathbbm{1}\{z=0_b\}\Bigg(\frac{1}{\overline{\pi}_{0_b}}-\frac{1}{{\pi}_{0_b}}\Bigg)(\mu_{Y,0_b}-\overline{\mu}_{Y,0_b})\Bigg]\\
    &-P\Bigg[\boldsymbol{x}\Bigg(\frac{1}{\overline{\eta}_b-\overline{\eta}_a}-\frac{1}{{\eta}_b-{\eta}_a}\Bigg)\mathbbm{1}\{z=1_a\}\Bigg(\frac{1}{\overline{\pi}_{1_a}}-\frac{1}{{\pi}_{1_a}}\Bigg)(\mu_{Y,1_a}-\overline{\mu}_{Y,1_a})\Bigg]\\
    &+P\Bigg[\boldsymbol{x}\Bigg(\frac{1}{\overline{\eta}_b-\overline{\eta}_a}-\frac{1}{{\eta}_b-{\eta}_a}\Bigg)\mathbbm{1}\{z=0_b\}\Bigg(\frac{1}{\overline{\pi}_{0_a}}-\frac{1}{{\pi}_{0_a}}\Bigg)(\mu_{Y,0_a}-\overline{\mu}_{Y,0_a})\Bigg]\\
    &-P\Bigg[\boldsymbol{x}\Bigg(\frac{\overline{\delta}_b-\overline{\delta}_a}{\overline{\eta}_b-\overline{\eta}_a}-\frac{{\delta}_b-\delta_a}{{\eta}_b-{\eta}_a}\Bigg)\mathbbm{1}\{z=1_b\}\Bigg(\frac{1}{\overline{\pi}_{1_b}}-\frac{1}{{\pi}_{1_b}}\Bigg)(\mu_{D,1_b}-\overline{\mu}_{D,1_b})\Bigg]\\
    &+P\Bigg[\boldsymbol{x}\Bigg(\frac{\overline{\delta}_b-\overline{\delta}_a}{\overline{\eta}_b-\overline{\eta}_a}-\frac{{\delta}_b-\delta_a}{{\eta}_b-{\eta}_a}\Bigg)\mathbbm{1}\{z=0_b\}\Bigg(\frac{1}{\overline{\pi}_{0_b}}-\frac{1}{{\pi}_{0_b}}\Bigg)(\mu_{D,0_b}-\overline{\mu}_{D,0_b})\Bigg]\\
    &+P\Bigg[\boldsymbol{x}\Bigg(\frac{\overline{\delta}_b-\overline{\delta}_a}{\overline{\eta}_b-\overline{\eta}_a}-\frac{{\delta}_b-\delta_a}{{\eta}_b-{\eta}_a}\Bigg)\mathbbm{1}\{z=1_a\}\Bigg(\frac{1}{\overline{\pi}_{1_a}}-\frac{1}{{\pi}_{1_a}}\Bigg)(\mu_{D,1_a}-\overline{\mu}_{D,1_a})\Bigg]\\
    &-P\Bigg[\boldsymbol{x}\Bigg(\frac{\overline{\delta}_b-\overline{\delta}_a}{\overline{\eta}_b-\overline{\eta}_a}-\frac{{\delta}_b-\delta_a}{{\eta}_b-{\eta}_a}\Bigg)\mathbbm{1}\{z=0_b\}\Bigg(\frac{1}{\overline{\pi}_{0_a}}-\frac{1}{{\pi}_{0_a}}\Bigg)(\mu_{D,0_a}-\overline{\mu}_{D,0_a})\Bigg]
\end{split}
\end{equation*}
Now we show how to bound $R(\widehat{P}_{n,k},P_0)$. 

\begin{align*}
&(\mathbbm{P}_{n,k} \boldsymbol{x}\boldsymbol{x}^T)^{-1}P_0(\boldsymbol{x}x^T)\Big((P_0\boldsymbol{x}\boldsymbol{x}^T)^{-1}-\mathbbm{P}_{n,k}(\boldsymbol{x}x^T)^{-1}\Big)\Bigg(\mathbbm{P}_{n,k}\boldsymbol{x}\frac{\widehat{\delta}_{a,k}}{\widehat{\eta}_{a,k}}-{P}_0x\frac{{\delta}_{a,0}}{{\eta}_{a,0}}\Bigg)\\
\leq & \Vert (\mathbbm{P}_{n,k} \boldsymbol{x}\boldsymbol{x}^T)^{-1}\Vert \Vert P_0(\boldsymbol{x}x^T)\Vert \Big \Vert\Big((P_0\boldsymbol{x}\boldsymbol{x}^T)^{-1}-(\mathbbm{P}_{n,k}\boldsymbol{x}\boldsymbol{x}^T)^{-1}\Big)\Big\Vert \Bigg\Vert\Bigg(\mathbbm{P}_{n,k}\boldsymbol{x}\frac{\widehat{\delta}_{a,k}}{\widehat{\eta}_{a,k}}-{P}_0x\frac{{\delta}_{a,0}}{{\eta}_{a,0}}\Bigg)\Bigg\Vert\\
\lesssim & \Big\Vert\Big((P\boldsymbol{x}\boldsymbol{x}^T)^{-1}-(\mathbbm{P}_{n,k}\boldsymbol{x}\boldsymbol{x}^T)^{-1}\Big\Vert_{F} \Bigg\Vert\Bigg(\mathbbm{P}_{n,k}\boldsymbol{x}\frac{\widehat{\delta}_{a,k}}{\widehat{\eta}_{a,k}}-{P}_0x\frac{{\delta}_{a,0}}{{\eta}_{a,0}}\Bigg)\Bigg\Vert\\
\lesssim & O_p(1/\sqrt{n})o_p(1)\\
=& o_p(1/\sqrt{n})
\end{align*}

Similarly, we can show that
\begin{align*}
    ((\mathbbm{P}_{n,k}\boldsymbol{x}\boldsymbol{x}^T)^{-1}P_0(\boldsymbol{x}x^T)-I)\Big((P_0\boldsymbol{x}\boldsymbol{x}^T)^{-1}-(\mathbbm{P}_{n,k}\boldsymbol{x}\boldsymbol{x}^T)^{-1}\Big)P_0\Big(\boldsymbol{x}\frac{\delta_{a,0}}{\eta_{a,0}}\Big) = o_p(1/\sqrt{n})
\end{align*}

Also,
\begin{align*}
&P_0\Bigg[\boldsymbol{x}\Bigg(\frac{\widehat{\delta}_{a,k}}{\widehat{\eta}^2_{a,k}}-\frac{{\delta}_{a,0}}{{\eta}^2_{a,0}}\Bigg)(\mu_{D,z,0}-\widehat{\mu}_{D,z,k})\Bigg]\\
\lesssim & P_0\Bigg[\Bigg\vert\Bigg(\frac{\widehat{\delta}_{a,k}}{\widehat{\eta}^2_{a,k}}-\frac{{\delta}_{a,0}}{{\eta}^2_{a,0}}\Bigg)(\mu_{D,z,0}-\widehat{\mu}_{D,z,k})\Bigg\vert\Bigg] \\
\lesssim & P_0\Big[\vert\hat{\delta}_{a,k}-\delta_{a,0}\vert \vert\mu_{D,z,0}-\widehat{\mu}_{D,z,k}\vert\Big]+P_0\Big[\vert\hat{\eta}_{a,k}-\eta_{a,0}\vert \vert\mu_{D,z,0}-\widehat{\mu}_{D,z,k}\vert\Big]\\
\leq & \Bigg(\Big\Vert\hat{\delta}_{a,k}-\delta_{a,0}\Big\Vert_{L^2(P_0)}+\Big\Vert\hat{\eta}_{a,k}-\eta_{a,0}\Big\Vert_{L^2(P_0)}\Bigg)\Big\Vert\mu_{D,z,0}-\widehat{\mu}_{D,z,k}\Big\Vert\\
=&o_p(1/\sqrt{n})
\end{align*}
holds for $z\in \{0_a,1_a,0_b,1_b\}$. Similarly, we can show
\begin{align*}
    &P_0\Bigg[\boldsymbol{x}\Bigg(\frac{1}{\widehat{\eta}_{a,k}}-\frac{1}{{\eta}_{a,0}}\Bigg)(\mu_{Y,z,0}-\widehat{\mu}_{Y,z,k})\Bigg]=o_p(1/\sqrt{n})\\
    &P_0\Bigg[\boldsymbol{x}\frac{1}{{\eta}_{a,0}}\mathbbm{1}\{z=z'\}\Bigg(\frac{1}{\widehat{\pi}_{z,k}}-\frac{1}{{\pi}_{z',0}}\Bigg)(\mu_{Y,z',0}-\widehat{\mu}_{Y,z',k})\Bigg] = o_p(1/\sqrt{n}) \\
    & P_0\Bigg[\boldsymbol{x}\frac{{\delta}_{a,0}}{{\eta}^2_{a,0}}\mathbbm{1}\{z=z'\}\Bigg(\frac{1}{\widehat{\pi}_{z',k}}-\frac{1}{{\pi}_{z',0}}\Bigg)(\mu_{D,z',0}-\widehat{\mu}_{D,z',0})\Bigg] = o_p(1/\sqrt{n})\\
    & P_0\Bigg[\boldsymbol{x}\Bigg(\frac{1}{\widehat{\eta}_{a,k}}-\frac{1}{{\eta}_{a,0}}\Bigg)\mathbbm{1}\{z=z'\}\Bigg(\frac{1}{\widehat{\pi}_{z',0}}-\frac{1}{{\pi}_{z',0}}\Bigg)(\mu_{Y,z',0}-\overline{\mu}_{Y,z',0})\Bigg] = o_p(1/\sqrt{n})\\
    &P_0\Bigg[\boldsymbol{x}\Bigg(\frac{\widehat{\delta}_{a,k}}{\widehat{\eta}^2_{a,k}}-\frac{{\delta}_{a,0}}{{\eta}^2_{a,0}}\Bigg)\mathbbm{1}\{z=z'\}\Bigg(\frac{1}{\widehat{\pi}_{z',k}}-\frac{1}{{\pi}_{z',0}}\Bigg)(\mu_{D,z',0}-\widehat{\mu}_{D,z',0})\Bigg]=o_p(1/\sqrt{n})\\
    & P_0\Bigg[\frac{x\delta_{a,0}}{\eta_{a,0}^2\widehat{\eta}_{a,k}}\Big[\delta_{a,0}(\widehat{\eta}_{a,k}-{\eta}_{a,0})^2+\eta_{a,0}(\widehat{\eta}_{a,k}-\eta_{a,0})(\delta_{a,0}-\widehat{\delta}_{a,k})\Big]\Bigg] = o_p(1/\sqrt{n})
\end{align*}

To sum up, $A = o_p(1/\sqrt{n})$. Similarly, we can prove $B = o_p(1/\sqrt{n})$. Therefore, $R(\hat{P}_{n,k},P_0) = o_p(1/\sqrt{n})$.

Following the same steps in  Theorem 4, we know

\begin{equation*}
    \sqrt{n}(\widehat{\beta}^{(1)}_{os}-\beta^{(1)}_{P_0}) = \mathbbm{P}_n\phi^{(1)}_{P_0}+o_p(1)
\end{equation*}
Therefore,
\begin{equation*}
        \sqrt{n}(\widehat{\beta}^{(1)}_{os} - \beta_{P_0}) \rightsquigarrow \mathcal{N}\left(0,~ \mathbb{E}_{P_0}\left[\phi^{(j)T}_{P_0}\phi^{(j)}_{P_0}\right]\right),
 \end{equation*}
 and
 \begin{align*}
     W^{(j)}\rightsquigarrow \chi^2(d_x)
 \end{align*}
 Therefore, $T^{(j)}_{Wald}$ is an asymptotically size-$\alpha$ test.
\end{proof}

\subsection{A technical lemma}
\label{subsec: the lemma, supp}

\begin{lemma}
The following statements are equivalent: (\romannumeral 1). $\theta_{P_0}(\boldsymbol{X}) = 0$ for all $x\in \mathcal{X}_1\times \mathcal{X}_2$. (\romannumeral 2). $\Omega_{P_0}(c):=\int \xi_{P_0}(\boldsymbol{X})1_c(\boldsymbol{X})dP_0(\boldsymbol{X})=0 $ for all $c\in \mathcal{X}$. (\romannumeral 3). $\Vert \Omega_{P_0} \Vert_{P_0,p}=0$ for $p\geq 1$ or $p=\infty$.
\end{lemma}

\begin{proof}
We will show (\romannumeral 1) $\Leftrightarrow$ (\romannumeral 2). 
    $\Rightarrow$ is obvious. We only need to show $\Leftarrow$. 

    For $\Leftarrow$, it is equivalent to show $A$ $\Rightarrow$ $B$, where $A$ is 
    \begin{equation*}
        \text{There exists an $x^*$, such that $\xi_{P_0}(\boldsymbol{x}^*)>0$}.
    \end{equation*}
    and $B$ is
    \begin{equation*}
        \text{There exists an $c$, such that $\Omega_{P_0}(c)>0$}.
    \end{equation*}
    We prove this by contradiction. That is, the statement that there exists an $x^*$, such that $\xi_{P_0}(\boldsymbol{x}^*)>0$ contradicts with the statement that $\Omega_{P_0}(c)=0$ for all $c$.

Let $x^*=(\boldsymbol{x}^*_{1},...,x^{*}_{d_1},...,x^{*}_{d_1+d_2})$ such that $x^*\in \mathcal{X}_1 \times \mathcal{X}_2$. Then there exists a set, and $C:= (\boldsymbol{x}^*_1,...,x^*_{d_1},...,(a_{d_1+1},b_{d_1+1})$,$...$,$(a_{d_1+d_2},b_{d_1+d_2}))$ such that 
\begin{equation*}
    P_0(C)>0
\end{equation*}
$|a_i<x^*_i<b_i|$, $|b_i-a_i|$ small enough such that $\xi_{P_0}(\boldsymbol{X})>0$ for $x\in C$. We assume $(-\infty,x_1^*]\times ... \times (-\infty,x_{d_1}^*]\cap \mathcal{X}_1$ is not empty, the case when it is empty can be handled similarly. Then there must exist a point $(a_1,...,a_{d_1})\in \mathcal{X}_1$ such that $\mathcal{X}_1\cap (a_1,x_1^*]\times ... (a_{d_1},x_{d_1}^*] = (\boldsymbol{x}_1^*,...,x_{d_1}^*)$. Then

\begin{equation*}
    \int \xi_{P_0}(\boldsymbol{X})1\{x\in (a_1,x_1^*]\times ...\times (a_{d_1},x_{d_1}^*] \times (a_{d_1+1},b_{d_1+1}) \times... \times (a_{d_2},b_{d_2})\}dP_0(\boldsymbol{X})>0
\end{equation*}
Since $X_2$ is absolutely continuous, then 

    \begin{equation}
    \int \theta_{P_0}(\boldsymbol{X})1\{x\in (a_1,x_1^*]\times ...\times(a_{d_1},x_{d_1}^*] \times (a_{d_1+1},b_{d_1+1}] \times... \times (a_{d_2},b_{d_2}]\}dP_0(\boldsymbol{X})>0 \label{eq: lemma1}
\end{equation}
Let $b_i = x_i^*$ for $i\leq d_1$, define 
    \begin{equation*}
        C = \{c = (c_1,...,c_n):c_i \in \{a_i,b_i\}, i = 1,...,d_1+d_2\}
    \end{equation*}
    and  let $n(c) = \#\{i:c_i = a_i\}$. Then if $\Omega_{P_0}(c) = 0$ for all $c\in\mathbb{R}^{d_x}$,
    \begin{align*}
        0 & = \sum_{c\in C}(-1)^{n(c)}\Omega_{P_0}(c)\\
        & = \int \xi_{P_0} (\boldsymbol{X})1_{(a_1,b_1](\boldsymbol{X})\times \cdots (a_{d_1+d_2},b_{d_1+d_2}]}d{P_0}(\boldsymbol{X}),
    \end{align*}
    which contradicts with \eqref{eq: lemma1}.

    (\romannumeral 2) $\Rightarrow$ (\romannumeral 3) is obvious. To show (\romannumeral 3) $\Rightarrow$ (\romannumeral 2), we just need to notice (\romannumeral 3) implies $\Omega_{P_0}$ is 0 almost everywhere and the covariates are either continuous or discrete.
\end{proof}

\subsection{Proof of Proposition \ref{prop: gradient Omega(c)}}
\begin{proof}
     We will pick $\Omega_{P}^{(j)}(c)$ as example. We pick a one-dimensional parametric submodel $\{P_t:t\in[0,\epsilon)\}$, our goal is to calculate
    \begin{align*}
        \frac{d\Omega_{P_t}^{(1)}(c)}{dt}\Big|_{t=0} = \frac{d \mathbbm{E}_{P_t}[\theta_{ACO,P_t}(\boldsymbol{X})\mathbbm{1}_c(\boldsymbol{X})]}{dt}\Big|_{t=0} - \frac{d \mathbbm{E}_{P_t}[\theta_{SW,P_t}(\boldsymbol{X})\mathbbm{1}_c(\boldsymbol{X})]}{dt}\Big|_{t=0}
    \end{align*}
    Similar to the proof of theorem 5, one can show that
    \begin{align*}
       & \frac{d \mathbbm{E}_{P_t}[\theta_{ACO,P_t}(\boldsymbol{X})-\mathbbm{1}_c(\boldsymbol{X})]}{dt}\Big|_{t=0}\\
        =& \mathbbm{E}_{P_0}\Bigg[(\theta_{ACO,P_t}(\boldsymbol{X})-\mathbbm{E}_{P_0}[\theta_{ACO,P_t}(\boldsymbol{X})]+D_P^{(1)})\mathbbm{1}_c(\boldsymbol{X})S(O)\Bigg]
    \end{align*}
    and
    \begin{align*}
       & \frac{d \mathbbm{E}_{P_t}[\theta_{SW,P_t}(\boldsymbol{X})-\mathbbm{1}_c(\boldsymbol{X})]}{dt}\Big|_{t=0}\\
        =& \mathbbm{E}_{P_0}\Bigg[(\theta_{SW,P_t}(\boldsymbol{X})-\mathbbm{E}_{P_0}[\theta_{SW,P_t}(\boldsymbol{X})]+D_P^{(2)})\mathbbm{1}_c(\boldsymbol{X})S(O)\Bigg]
    \end{align*}
    so that
    \begin{align*}
        D^{(1)*}_P(c)=o\mapsto D^{(1)}_{P}(z,x,d,y)\mathbbm{1}_{c}(\boldsymbol{X})-D^{(2)}_{P}(g,z,x,d,y)\mathbbm{1}_{c}(\boldsymbol{X})+\theta^{(1)}_{P}(\boldsymbol{X})\mathbbm{1}_{c}(\boldsymbol{X})-\Omega^{(1)}_P(c)
    \end{align*}
\end{proof}

\subsection{Proof of Theorem \ref{thm: T_np properties}}
\label{subsec: NP test proof, supp}

Suppose that
\begin{enumerate}
    \item There exists $0<\epsilon<0.5$ such that $\epsilon<P_0(Z|\boldsymbol{X})<1-\epsilon$, $\epsilon<P_0(D|Z,X)<1-\epsilon$, $\epsilon<\eta_{a,P_0}<1-\epsilon$, $\epsilon<\eta_{a,P_0}-\eta_{b,P_0}<1-\epsilon$, $\epsilon<\widehat{P}_{n,k}(Z|\boldsymbol{X})<1-\epsilon$, $\epsilon<\widehat{P}_{n,k}(D|Z,X)<1-\epsilon$,  $\epsilon<\eta_{a,\widehat{P}_{n,k}}<1-\epsilon$, $\epsilon<\eta_{a,\widehat{P}_{n,k}}-\eta_{b,\widehat{P}_{n,k}}<1-\epsilon$, ($k=1,...K$), there is a universal constant $C$ such that $|\mathbbm{E}_{P_0}[Y|Z,\boldsymbol{X}]|\leq C$, $|\mathbbm{E}_{\widehat{P}_{n,k}}[Y|Z,\boldsymbol{X}]|\leq C$ ($k=1,...K$) with probability one. 
    \item $\{D^{(j)*}_{P_0}(c):c\in \mathcal{X}\}$ is $P_0$-Donsker for $j\in \{1,2,3\}$.
    \item Rate conditions for nuisance functions estimation: The convergence rates for all nuisance function estimators are faster then $n^{-1/4}$:
        \begin{align*}
             &\Big\Vert \pi_{P_0}(z,\cdot)-\pi_{\widehat{P}_{n,k}}(z,\cdot)\Big\Vert_{L^2(P_0)} = o_p(n^{-1/4})\\
        &\Big\Vert \mu_{Y,P_0}(z,\cdot)-\mu_{Y,\widehat{P}_{n,k}}(z,\cdot)\Big\Vert_{L^2(P_0)} = o_p(n^{-1/4})\\
        &\Big\Vert \mu_{D,P_0}(z,\cdot)-\mu_{D,\widehat{P}_{n,k}}(z,\cdot)\Big\Vert_{L^2(P_0)} = o_p(n^{-1/4})
        \end{align*}
        for $z\in \{0_a,1_a,0_b,1_b\}$, $k \in \{1,...,K\}$.
        \item Let $\mathbb{G}^{(j)}_0$ be the limiting process of $\{\sqrt{n}[\widehat{\Omega}^{(j)}(c)-{\Omega}^{(j)}_{P_0}(c)],c\in \mathcal{X}\}$, the distribution function of $\Vert \mathbb{G}^{(j)}_0\Vert_{P_0,p}$ is strictly increasing and continuous in a neighborhood of its $\alpha$ quantile.
\end{enumerate}
\begin{proof}
    First we will establish asymptotic linearity of $\widehat{\Omega}^{(1)}_{ee,k}(c)$ for a given $c$, so that we can establish asymptotic linearity of $\widehat{\Omega}^{(1)}_{ee}(c)$.

    We firstly define \begin{equation*}
        R(\overline{P},P;c) = \Omega^{(1)}_{\overline{P}}(c)-\Omega^{(1)}_{{P}}(c)+PD^{(1)*}_{\overline{P}}(c).
    \end{equation*}

    Similar to the proof of theorem 6, $R(\overline{P},P;c) = A+B$, where 
    \begin{align*}
    A=
    &P\Bigg[\mathbbm{1}_c\Bigg(\frac{1}{\overline{\eta}_a}-\frac{1}{{\eta}_a}\Bigg)(\mu_{Y,1_a}-\overline{\mu}_{Y,1_a})\Bigg] +P\Bigg[\mathbbm{1}_c\frac{1}{{\eta}_a}\mathbbm{1}\{z=1_a\}\Bigg(\frac{1}{\overline{\pi}_{1_a}}-\frac{1}{{\pi}_{1_a}}\Bigg)(\mu_{Y,1_a}-\overline{\mu}_{Y,1_a})\Bigg]\\
    &-P\Bigg[\mathbbm{1}_c\Bigg(\frac{1}{\overline{\eta}_a}-\frac{1}{{\eta}_a}\Bigg)(\mu_{Y,0_a}-\overline{\mu}_{Y,0_a})\Bigg] -P\Bigg[\mathbbm{1}_c\frac{1}{{\eta}_a}\mathbbm{1}\{z=0_a\}\Bigg(\frac{1}{\overline{\pi}_{0_a}}-\frac{1}{{\pi}_{0_a}}\Bigg)(\mu_{Y,0_a}-\overline{\mu}_{Y,0_a})\Bigg]\\
    &-P\Bigg[\mathbbm{1}_c\Bigg(\frac{\overline{\delta}_a}{\overline{\eta}^2_a}-\frac{{\delta}_a}{{\eta}^2_a}\Bigg)(\mu_{D,1_a}-\overline{\mu}_{D,1_a})\Bigg] +P\Bigg[\mathbbm{1}_c\frac{{\delta}_a}{{\eta}^2_a}\mathbbm{1}\{z=1_a\}\Bigg(\frac{1}{\overline{\pi}_{1_a}}-\frac{1}{{\pi}_{1_a}}\Bigg)(\mu_{D,1_a}-\overline{\mu}_{D,1_a})\Bigg]\\
    &+P\Bigg[\mathbbm{1}_c\Bigg(\frac{\overline{\delta}_a}{\overline{\eta}^2_a}-\frac{{\delta}_a}{{\eta}^2_a}\Bigg)(\mu_{D,0_a}-\overline{\mu}_{D,0_a})\Bigg] +P\Bigg[\mathbbm{1}_c\frac{{\delta}_a}{{\eta}^2_a}\mathbbm{1}\{z=0_a\}\Bigg(\frac{1}{\overline{\pi}_{0_a}}-\frac{1}{{\pi}_{1_a}}\Bigg)(\mu_{D,0_a}-\overline{\mu}_{D,0_a})\Bigg]\\
    &+P\Bigg[\mathbbm{1}_c\Bigg(\frac{1}{\overline{\eta}_a}-\frac{1}{{\eta}_a}\Bigg)\mathbbm{1}\{z=1_a\}\Bigg(\frac{1}{\overline{\pi}_{1_a}}-\frac{1}{{\pi}_{1_a}}\Bigg)(\mu_{Y,1_a}-\overline{\mu}_{Y,1_a})\Bigg]\\
    &-P\Bigg[\mathbbm{1}_c\Bigg(\frac{1}{\overline{\eta}_a}-\frac{1}{{\eta}_a}\Bigg)\mathbbm{1}\{z=0_a\}\Bigg(\frac{1}{\overline{\pi}_{0_a}}-\frac{1}{{\pi}_{0_a}}\Bigg)(\mu_{Y,0_a}-\overline{\mu}_{Y,0_a})\Bigg]\\
    &- P\Bigg[\mathbbm{1}_c\Bigg(\frac{\overline{\delta}_a}{\overline{\eta}^2_a}-\frac{{\delta}_a}{{\eta}^2_a}\Bigg)\mathbbm{1}\{z=0_a\}\Bigg(\frac{1}{\overline{\pi}_{1_a}}-\frac{1}{{\pi}_{1_a}}\Bigg)(\mu_{D,1_a}-\overline{\mu}_{D,1_a})\Bigg]\\
    &+ P\Bigg[\mathbbm{1}_c\Bigg(\frac{\overline{\delta}_a}{\overline{\eta}^2_a}-\frac{{\delta}_a}{{\eta}^2_a}\Bigg)\mathbbm{1}\{z=0_a\}\Bigg(\frac{1}{\overline{\pi}_{0_a}}-\frac{1}{{\pi}_{0_a}}\Bigg)(\mu_{D,0_a}-\overline{\mu}_{D,0_a})\Bigg]\\
    &+P\Bigg[\frac{\mathbbm{1}_c\delta_a}{\eta_a^2\overline{\eta}_a}\Big[\delta_a(\overline{\eta}_a-{\eta}_a)^2+\eta_a(\overline{\eta}_a-\eta_a)(\delta_a-\overline{\delta}_a)\Big]\Bigg]
\end{align*}

Similarly, for $B$, direct calculation yields
\begin{equation*}\small
    \begin{split}
    B=&P\Bigg[\mathbbm{1}_c\Bigg(\frac{1}{\overline{\eta}_b-\overline{\eta}_a}-\frac{1}{{\eta}_b-{\eta}_a}\Bigg)(\mu_{Y,1_b}-\overline{\mu}_{Y,1_b})\Bigg] +P\Bigg[\mathbbm{1}_c\frac{1}{{\eta}_b-{\eta}_a}\mathbbm{1}\{z=1_b\}\Bigg(\frac{1}{\overline{\pi}_{1_b}}-\frac{1}{{\pi}_{1_b}}\Bigg)(\mu_{Y,1_b}-\overline{\mu}_{Y,1_b})\Bigg]\\
    &-P\Bigg[\mathbbm{1}_c\Bigg(\frac{1}{\overline{\eta}_b-\overline{\eta}_a}-\frac{1}{{\eta}_b-{\eta}_a}\Bigg)(\mu_{Y,0_b}-\overline{\mu}_{Y,0_b})\Bigg] -P\Bigg[\mathbbm{1}_c\frac{1}{{\eta}_b-{\eta}_a}\mathbbm{1}\{z=0_b\}\Bigg(\frac{1}{\overline{\pi}_{0_b}}-\frac{1}{{\pi}_{0_b}}\Bigg)(\mu_{Y,0_b}-\overline{\mu}_{Y,0_b})\Bigg]\\
    &-P\Bigg[\mathbbm{1}_c\Bigg(\frac{1}{\overline{\eta}_b-\overline{\eta}_a}-\frac{1}{{\eta}_b-{\eta}_a}\Bigg)(\mu_{Y,1_a}-\overline{\mu}_{Y,1_a})\Bigg] -P\Bigg[\mathbbm{1}_c\frac{1}{{\eta}_b-{\eta}_a}\mathbbm{1}\{z=1_a\}\Bigg(\frac{1}{\overline{\pi}_{1_a}}-\frac{1}{{\pi}_{1_a}}\Bigg)(\mu_{Y,1_a}-\overline{\mu}_{Y,1_a})\Bigg]\\
    &+P\Bigg[\mathbbm{1}_c\Bigg(\frac{1}{\overline{\eta}_b-\overline{\eta}_a}-\frac{1}{{\eta}_b-{\eta}_a}\Bigg)(\mu_{Y,0_a}-\overline{\mu}_{Y,0_a})\Bigg] +P\Bigg[\mathbbm{1}_c\frac{1}{{\eta}_b-{\eta}_a}\mathbbm{1}\{z=0_a\}\Bigg(\frac{1}{\overline{\pi}_{0_a}}-\frac{1}{{\pi}_{0_a}}\Bigg)(\mu_{Y,0_a}-\overline{\mu}_{Y,0_a})\Bigg]\\
    &-P\Bigg[\mathbbm{1}_c\Bigg(\frac{\overline{\delta}_b-\overline{\delta}_a}{\overline{\eta}_b-\overline{\eta}_a}-\frac{{\delta}_b-{\delta}_a}{{\eta}_b-{\eta}_a}\Bigg)(\mu_{D,1_b}-\overline{\mu}_{D,1_b})\Bigg] -P\Bigg[\mathbbm{1}_c\frac{{\delta}_b-{\delta}_a}{{\eta}_b-{\eta}_a}\mathbbm{1}\{z=1_b\}\Bigg(\frac{1}{\overline{\pi}_{1_b}}-\frac{1}{{\pi}_{1_b}}\Bigg)(\mu_{D,1_b}-\overline{\mu}_{D,1_b})\Bigg]\\
    &+P\Bigg[\mathbbm{1}_c\Bigg(\frac{\overline{\delta}_b-\overline{\delta}_a}{\overline{\eta}_b-\overline{\eta}_a}-\frac{{\delta}_b-{\delta}_a}{{\eta}_b-{\eta}_a}\Bigg)(\mu_{D,0_b}-\overline{\mu}_{D,0_b})\Bigg] +P\Bigg[\mathbbm{1}_c\frac{{\delta}_b-{\delta}_a}{{\eta}_b-{\eta}_a}\mathbbm{1}\{z=0_b\}\Bigg(\frac{1}{\overline{\pi}_{0_b}}-\frac{1}{{\pi}_{0_b}}\Bigg)(\mu_{D,0_b}-\overline{\mu}_{D,0_b})\Bigg]
    \\
    &+P\Bigg[\mathbbm{1}_c\Bigg(\frac{\overline{\delta}_b-\overline{\delta}_a}{\overline{\eta}_b-\overline{\eta}_a}-\frac{{\delta}_b-{\delta}_a}{{\eta}_b-{\eta}_a}\Bigg)(\mu_{D,1_a}-\overline{\mu}_{D,1_a})\Bigg] +P\Bigg[\mathbbm{1}_c\frac{{\delta}_b-{\delta}_a}{{\eta}_b-{\eta}_a}\mathbbm{1}\{z=1_a\}\Bigg(\frac{1}{\overline{\pi}_{1_a}}-\frac{1}{{\pi}_{1_a}}\Bigg)(\mu_{D,1_a}-\overline{\mu}_{D,1_a})\Bigg]\\
    &-P\Bigg[\mathbbm{1}_c\Bigg(\frac{\overline{\delta}_b-\overline{\delta}_a}{\overline{\eta}_b-\overline{\eta}_a}-\frac{{\delta}_b-{\delta}_a}{{\eta}_b-{\eta}_a}\Bigg)(\mu_{D,0_a}-\overline{\mu}_{D,0_a})\Bigg] -P\Bigg[\mathbbm{1}_c\frac{{\delta}_b-{\delta}_a}{{\eta}_b-{\eta}_a}\mathbbm{1}\{z=0_a\}\Bigg(\frac{1}{\overline{\pi}_{0_a}}-\frac{1}{{\pi}_{0_a}}\Bigg)(\mu_{D,0_a}-\overline{\mu}_{D,0_a})\Bigg]\\
    &+P\Bigg[\mathbbm{1}_c\Bigg(\frac{1}{\overline{\eta}_b-\overline{\eta}_a}-\frac{1}{{\eta}_b-{\eta}_a}\Bigg)\mathbbm{1}\{z=1_b\}\Bigg(\frac{1}{\overline{\pi}_{1_b}}-\frac{1}{{\pi}_{1_b}}\Bigg)(\mu_{Y,1_b}-\overline{\mu}_{Y,1_b})\Bigg]\\
    &-P\Bigg[\mathbbm{1}_c\Bigg(\frac{1}{\overline{\eta}_b-\overline{\eta}_a}-\frac{1}{{\eta}_b-{\eta}_a}\Bigg)\mathbbm{1}\{z=0_b\}\Bigg(\frac{1}{\overline{\pi}_{0_b}}-\frac{1}{{\pi}_{0_b}}\Bigg)(\mu_{Y,0_b}-\overline{\mu}_{Y,0_b})\Bigg]\\
    &-P\Bigg[\mathbbm{1}_c\Bigg(\frac{1}{\overline{\eta}_b-\overline{\eta}_a}-\frac{1}{{\eta}_b-{\eta}_a}\Bigg)\mathbbm{1}\{z=1_a\}\Bigg(\frac{1}{\overline{\pi}_{1_a}}-\frac{1}{{\pi}_{1_a}}\Bigg)(\mu_{Y,1_a}-\overline{\mu}_{Y,1_a})\Bigg]\\
    &+P\Bigg[\mathbbm{1}_c\Bigg(\frac{1}{\overline{\eta}_b-\overline{\eta}_a}-\frac{1}{{\eta}_b-{\eta}_a}\Bigg)\mathbbm{1}\{z=0_b\}\Bigg(\frac{1}{\overline{\pi}_{0_a}}-\frac{1}{{\pi}_{0_a}}\Bigg)(\mu_{Y,0_a}-\overline{\mu}_{Y,0_a})\Bigg]\\
    &-P\Bigg[\mathbbm{1}_c\Bigg(\frac{\overline{\delta}_b-\overline{\delta}_a}{\overline{\eta}_b-\overline{\eta}_a}-\frac{{\delta}_b-\delta_a}{{\eta}_b-{\eta}_a}\Bigg)\mathbbm{1}\{z=1_b\}\Bigg(\frac{1}{\overline{\pi}_{1_b}}-\frac{1}{{\pi}_{1_b}}\Bigg)(\mu_{D,1_b}-\overline{\mu}_{D,1_b})\Bigg]\\
    &+P\Bigg[\mathbbm{1}_c\Bigg(\frac{\overline{\delta}_b-\overline{\delta}_a}{\overline{\eta}_b-\overline{\eta}_a}-\frac{{\delta}_b-\delta_a}{{\eta}_b-{\eta}_a}\Bigg)\mathbbm{1}\{z=0_b\}\Bigg(\frac{1}{\overline{\pi}_{0_b}}-\frac{1}{{\pi}_{0_b}}\Bigg)(\mu_{D,0_b}-\overline{\mu}_{D,0_b})\Bigg]\\
    &+P\Bigg[\mathbbm{1}_c\Bigg(\frac{\overline{\delta}_b-\overline{\delta}_a}{\overline{\eta}_b-\overline{\eta}_a}-\frac{{\delta}_b-\delta_a}{{\eta}_b-{\eta}_a}\Bigg)\mathbbm{1}\{z=1_a\}\Bigg(\frac{1}{\overline{\pi}_{1_a}}-\frac{1}{{\pi}_{1_a}}\Bigg)(\mu_{D,1_a}-\overline{\mu}_{D,1_a})\Bigg]\\
    &-P\Bigg[\mathbbm{1}_c\Bigg(\frac{\overline{\delta}_b-\overline{\delta}_a}{\overline{\eta}_b-\overline{\eta}_a}-\frac{{\delta}_b-\delta_a}{{\eta}_b-{\eta}_a}\Bigg)\mathbbm{1}\{z=0_b\}\Bigg(\frac{1}{\overline{\pi}_{0_a}}-\frac{1}{{\pi}_{0_a}}\Bigg)(\mu_{D,0_a}-\overline{\mu}_{D,0_a})\Bigg]
\end{split}
\end{equation*}
By similar argument in Theorem 6 and $\mathbbm{1}_c\leq 1$, we know $\sup_{c}|R(\widehat{P}_{n,k},P_0)(c)| = o_p(1/\sqrt{n})$. Now we consider the empirical process term
\begin{equation*}
    \mathbb{G}_{n,k}(D^{(1)*}_{\widehat{P}_{n,k}}(c)-D^{(1)*}_{P_0}(c))=\sqrt{n_k}(\mathbbm{P}_{n,k}-P_0)(D^{(1)*}_{\widehat{P}_{n,k}}(c)-D^{(1)*}_{P_0}(c))
\end{equation*}
Define $\mathcal{F} :=\{D^{(1)*}_{\widehat{P}_{n,k}}(c)-D^{(1)*}_{P_0}(c):c\in \mathcal{X}\}$, then $\sup_{c}\vert \mathbb{G}_{n,k}(D^{(1)*}_{\widehat{P}_{n,k}}(c)-D^{(1)*}_{P_0}(c)) \vert$ can be written as $\sup_{f\in \mathcal{F}}\vert \mathbb{G}_{n,k}f \vert$. To show the supreme of this empirical process is $o_p(1)$, we will bound
\begin{align*}
    \mathbbm{E}_{P_0}\Bigg[\sup_{f\in \mathcal{F}_{n,k}} \Big\vert \mathbb{G}_{n,k}f \Big\vert \Bigg] = \mathbbm{E}_{P_0}\Bigg[\mathbbm{E}_{P_0}\Big[\sup_{f\in \mathcal{F}_{n,k}} \vert \mathbb{G}_{n,k}f \vert \Big|O_{i}, i \in I_k \Big]\Bigg]. 
\end{align*}
After condition on $\{O_{i}, i \in I^c_k\}$, the estimated nuisance functions can be treated as fixed. Let $F_{n,k}(o)$ be an envelop function of $\mathcal{F}_{n,k}$. Then by Theorem 2.14.1 of \citet{van2023weak}. It holds that
\begin{align*}
    \mathbbm{E}_{P_0}\Big[\sup_{f\in \mathcal{F}_{n,k}} \vert \mathbb{G}_{n,k}f \vert \Big|O_{i}, i \in I_k \Big] \lesssim \sqrt{\mathbbm{E}_{P_0}\Bigg[F_{n,k}(O)^2\Bigg|O_i, i\in I^c_k\Bigg]} J(1,\mathcal{F}_{n,k}|F_{n,k},L^2)
\end{align*}
where $J(1,\mathcal{F}_{n,k}|F_{n,k},L^2)$ is the uniform entropy integral. Our goal is to show that
\begin{itemize}
    \item $J(1,\mathcal{F}_{n,k}|F_{n,k},L^2) = O(1)$
    \item There exists an $F_{n,k}(o)$ such that $\sqrt{\mathbbm{E}_{P_0}\Bigg[F_{n,k}(O)^2\Bigg|O_i, i\in I^c_k\Bigg]}=o(1)$.
\end{itemize}
To show $J(1,\mathcal{F}_{n,k}|F_{n,k},L^2) = O(1)$, we notice that $\mathcal{F}_{n,k}$ is a linear combination of elements from the following three class
\begin{align*}
    & \mathcal{A}_1 :=\{[D^{(1)*}_{\widehat{P}_{n,k}}(o)-D^{(1)*}_{{P}_{0}}(o)]\mathbbm{1}_c:c\in \mathcal{C}\}\\
    & \mathcal{A}_2 :=\Bigg\{\Big[\frac{\delta_{a,\widehat{P}_{n,k}}}{\eta_{a,\widehat{P}_{n,k}}}-\frac{\delta_{a,{P}_{0}}}{\eta_{a,{P}_{0}}}\Big]\mathbbm{1}_c:c\in \mathcal{C}\Bigg\}\\
    & \mathcal{A}_3 :=\Bigg\{\Big[\mathbbm{P}_{n,k}\frac{\delta_{a,\widehat{P}_{n,k}}}{\eta_{a,\widehat{P}_{n,k}}}\mathbbm{1}_c-P_0\frac{\delta_{a,{P}_{0}}}{\eta_{a,{P}_{0}}}\mathbbm{1}_c\Big]:c\in \mathcal{C}\Bigg\}
\end{align*}
Since $\mathcal{A}_4:=\{\mathbbm{1}_c:c\in \mathcal{X}\}$ is a VC class with VC index $d_x+1$, by Theorem 2.6.7 of \citet{van2023weak}, the covering number satisfies
\begin{equation*}
    N(\epsilon \Vert F \Vert_{Q,2},\mathcal{A}_4, L^2(Q))\leq K \Big(\frac{1}{\epsilon}\Big)^{2d_x}
\end{equation*}
By the similar reasoning from the proof of Lemma 2 in \citet{westling2022nonparametric}, We know $J(1,\mathcal{F}_{n,k}|F_{n,k},L^2) = O(1)$.

Next we derive an envelope function $F_{n,k}$, and verify it satisfies $\sqrt{\mathbbm{E}_{P_0}\Bigg[F_{n,k}(O)^2\Bigg|O_i, i\in I^c_k\Bigg]}=o(1)$.

\begin{align*}
    &D_{\widehat{P}_{n,k}}^{(1)*}\mathbbm{1}_c-D_{P_0}^{(1)*}(c)\\
    =&\underbrace{\Big[D^{(1)}_{\widehat{P}_{n,k}}\mathbbm{1}_c+\theta_{ACO,\widehat{P}_{n,k}}\mathbbm{1}_c-\int \mathbbm{1}_c\theta_{ACO,\widehat{P}_{n,k}}d{P}_{n,k}\Big]-\Big[D^{(1)}_{P_0}\mathbbm{1}_c+\theta_{ACO,P_0}-\int \mathbbm{1}_c\theta_{ACO,P_0}dP_0\Big]}_{A}\\
    &+\underbrace{\Big[D^{(2)}_{\widehat{P}_{n,k}}\mathbbm{1}_c +\theta_{SW,\widehat{P}_{n,k}}-\int \mathbbm{1}_c\theta_{SW,\widehat{P}_{n,k}}d{P}_{n,k}\Big]-\Big[D^{(2)}_{P_0}\mathbbm{1}_c +\theta_{SW,P_0}-\int \mathbbm{1}_c\theta_{SW,P_0}dP_0\Big]}_{B}
\end{align*}
We firstly consider term $A$. For $z'\in \{o_a,1_a,0_b,1_b\}$,
\begin{align*}
    &\Bigg\vert\frac{\mathbbm{1}_c}{\eta_{a,\widehat{P}_{n,k}}}\Bigg[\frac{\mathbbm{1}\{z=z'\}}{{\widehat{\pi}_{z',k}}}[y-\widehat{\mu}_{Y,z',k}]\Bigg]-\frac{\mathbbm{1}_c}{\eta_{a,P_0}}\Bigg[\frac{\mathbbm{1}\{z=z'\}}{{{\pi}_{z',0}}}[y-{\mu}_{Y,z',0}]\Bigg]\Bigg\vert\\
    \leq&\Bigg\vert\Bigg(\frac{1}{\eta_{a,\widehat{P}_{n,k}}}-\frac{1}{\eta_{a,P_0}}\Bigg)\Bigg[\frac{\mathbbm{1}\{z=z'\}}{{\widehat{\pi}_{z',k}}}[y-\widehat{\mu}_{Y,z',k}]\Bigg]+\frac{1}{\eta_{a,P_0}}\Bigg[\mathbbm{1}\{z=z'\}\Bigg(\frac{1}{{\widehat{\pi}_{z',k}}}-\frac{1}{{\pi}_{z',0}}\Bigg)[y-\widehat{\mu}_{Y,z',k}]\Bigg]\\
    &+\frac{1}{\eta_{a,P_0}}\Bigg[\mathbbm{1}\{z=z'\}\frac{1}{{\pi}_{z',0}}[({\mu}_{Y,z',0}-\widehat{\mu}_{Y,z',k})]\Bigg]\Bigg\vert\\
    \leq & C\Big\vert \eta_{a,\widehat{P}_{n,k}}-\eta_{a,{P}_{0}}\Big\vert(|y|+C)+C\Big\vert \widehat{\pi}_{z',k}-{\pi}_{z',0}  \Big\vert(|y|+C) +C(\vert \widehat{\mu}_{Y,z',k}-{\mu}_{Y,z',0} \vert) \\
    \leq&  C\max_{z\in \{0_a,1_a\}}\Big\vert \widehat{\mu}_{D,z,k}-{\mu}_{D,z,0}\Big\vert(|y|+C)+C\Big\vert \widehat{\pi}_{z',k}-{\pi}_{z',0}  \Big\vert(|y|+C) +C\vert \widehat{\mu}_{Y,z',k}-{\mu}_{Y,z',0} \vert
\end{align*}
Similarly,
\begin{align*}
    &\Bigg\vert\frac{\mathbbm{1}_c\delta_{a,\widehat{P}_{n,k}}}{\eta^2_{a,\widehat{P}_{n,k}}}\Bigg[\frac{\mathbbm{1}\{z=z'\}}{{\widehat{\pi}_{z',k}}}[d-\widehat{\mu}_{D,z',k}]\Bigg]-\frac{\mathbbm{1}_c\delta_{a,P_0}}{\eta^2_{a,P_0}}\Bigg[\frac{\mathbbm{1}\{z=z'\}}{{{\pi}_{z',0}}}[d-{\mu}_{D,z',0}]\Bigg]\Bigg\vert\\
    \leq&  C\max_{z\in \{0_a,1_a\},B\in \{Y,D\}}\Big\vert \widehat{\mu}_{B,z,k}-{\mu}_{B,z,0}\Big\vert+C\Big\vert \widehat{\pi}_{z',k}-{\pi}_{z',0}  \Big\vert +C\vert \widehat{\mu}_{D,z',k}-{\mu}_{D,z',0} \vert
\end{align*}

Furthermore, we have
\begin{align*}
    &\vert \theta_{ACO,\widehat{P}_{n,k}}\mathbbm{1}_c - \theta_{ACO,{P}_{0}}\mathbbm{1}_c\vert \leq C\max_{z\in \{0_a,1_a\},B\in \{Y,D\}}\Big\vert \widehat{\mu}_{B,z,k}-{\mu}_{B,z,0}\Big\vert\\
    & \int \mathbbm{1}_c\theta_{ACO,\widehat{P}_{n,k}}d{P}_{n,k} - \int \mathbbm{1}_c\theta_{ACO,{P}_{0}}d{P}_{0}\\
    &\leq \sup_{c\in \mathcal{C}}\vert (\mathbbm{P}_{n,k}-P_0)\mathbbm{1}_c\theta_{ACO,\widehat{P}_{n,k}}\vert +C\max_{z\in \{0_a,1_a\},B\in \{Y,D\}}P_0\Big\vert \widehat{\mu}_{B,z,k}-{\mu}_{B,z,0}\Big\vert
\end{align*}
Therefore,
\begin{align*}
    |A|\leq &\Big\vert D^{(1)}_{\widehat{P}_{n,k}}\mathbbm{1}_c -D^{(1)}_{P_0}\mathbbm{1}_c\Big\vert+\Big\vert \theta_{ACO,\widehat{P}_{n,k}}\mathbbm{1}_c-\theta_{ACO,P_0}\mathbbm{1}_c\Big \vert +\Big \vert \int \mathbbm{1}_c\theta_{ACO,\widehat{P}_{n,k}}d\mathbbm{P}_{n,k}-\int \mathbbm{1}_c\theta_{ACO,P_0}dP_0\Big]\\
    \leq& C\max_{z\in \{0_a,1_a\},B\in \{Y,D\}}\Big\vert \widehat{\mu}_{B,z,k}-{\mu}_{B,z,0}\Big\vert(|y|+C)+C\sup_{z\in \{0_a,1_a\}}\Big\vert \widehat{\pi}_{z,k}-{\pi}_{z,0}  \Big\vert(|y|+C)\\
    &+\sup_{c\in \mathcal{C}}\vert (\mathbbm{P}_{n,k}-P_0)\mathbbm{1}_c\theta_{ACO,\widehat{P}_{n,k}}\vert +C\max_{z\in \{0_a,1_a\},B\in \{Y,D\}}P_0\Big\vert \widehat{\mu}_{B,z,k}-{\mu}_{B,z,0}\Big\vert
\end{align*}
We can obtain a similar bound for $B$, to sum up, we can take
\begin{align*}
    &F_{n,k}\\
    =&C\max_{z\in \{0_a,1_a,0_b,1_b\},B\in \{Y,D\}}\Big\vert \widehat{\mu}_{B,z,k}-{\mu}_{B,z,0}\Big\vert(|y|+C)+C\sup_{z\in \{0_a,1_a,0_b,1_b\}}\Big\vert \widehat{\pi}_{z,k}-{\pi}_{z,0}  \Big\vert(|y|+C)\\
    &+\sup_{c\in \mathcal{C}}\vert (\mathbbm{P}_{n,k}-P_0)\mathbbm{1}_c\theta_{SW,\widehat{P}_{n,k}}\vert +\sup_{c\in \mathcal{C}}\vert (\mathbbm{P}_{n,k}-P_0)\mathbbm{1}_c\theta_{ACO,\widehat{P}_{n,k}}\vert\\
    &+C\max_{z\in \{0_a,1_a,0_b,1_b\},B\in \{Y,D\}}P_0\Big\vert \widehat{\mu}_{B,z,k}-{\mu}_{B,z,0}\Big\vert
\end{align*}
and our assumption equipped with triangle inequality implies that $\Vert F_{n,k} \Vert_{P_0,2}\xrightarrow[]{P_0} 0$, uniform boundedness of $\Vert F_{n,k} \Vert_{P_0,2}$ impplies $E[\Vert F_{n,k} \Vert_{P_0,2}]\rightarrow 0$ therefore,
\begin{align*}
   &\mathbbm{E}_{P_0}\Big[\sup_{f\in \mathcal{F}_{n,k}} \vert \mathbb{G}_{n,k}f \vert \Big]\\
   = &\mathbbm{E}_{P_0}\Bigg[\mathbbm{E}_{P_0}\Big[\sup_{f\in \mathcal{F}_{n,k}} \vert \mathbb{G}_{n,k}f \vert \Big|O_{i}, i \in I_k \Big]\Bigg] \\
   \lesssim & \mathbbm{E}_{P_0}\Bigg\{\sqrt{\mathbbm{E}_{P_0}\Bigg[F_{n,k}(O)^2\Bigg|O_i, i\in I^c_k\Bigg]} \Bigg\}J(1,\mathcal{F}_{n,k}|F_{n,k},L^2)\\
   =&o(1)O(1) = o(1)
\end{align*}
To sum up, we have already shown that
\begin{equation*}
    \max_{k}\sup_{c}\vert \mathbb{G}_{n,k}(D^{(1)*}_{\widehat{P}_{n,k}}(c)-D^{(1)*}_{P_0}(c)) \vert = o_p(1)
\end{equation*}
and the one-step estimator has the following expansion:
\begin{align*}
        \widehat{\Omega}^{(1)}_{ee}(c)-\Omega^{(1)}_{P_0}(c) = \frac{1}{{n}}\sum_{i=1}^n D^{(1)*}_{P_0}(c)(O_i)+r_n(c).
    \end{align*}
with $\sup_{c\in \mathcal{X}} |r_n(c)| = o_p(1/\sqrt{n})$.

The rest of the proof is similar to that of \citet{westling2022nonparametric}.
\end{proof}

\subsection{Proofs for Section \ref{sec: additional identification details, supp}}

\subsubsection{Proof for Theorem \ref{thm: G direct effect,1,supp}}

\begin{proof}
    For SWATE, Since we have
    \begin{align*}
    &\mathbb{E}_{P_0}[Y\mid Z=1_b,\boldsymbol{X}]-\mathbb{E}_{P_0}[Y\mid Z=0_b,\boldsymbol{X}]\\
     =&\mathbb{E}_{P_0}[Y(Z=1_b)\mid Z=1_b,G=b,\boldsymbol{X}]-\mathbb{E}_{P_0}[Y(Z=0_b)|Z=0_b,G=b, X]\\
        =&\mathbb{E}_{P_0}[Y(Z=1_b)-Y(Z=0_b)|G=b,\boldsymbol{X}]\\
        =&\mathbb{E}_{P_0}[Y(Z=1_b)-Y(Z=0_b)|\boldsymbol{X}]\\
        =&\mathbb{E}_{P_0}[Y(D=1,G=b)-Y(D=0,G=b)|S=\text{SW},\boldsymbol{X}]P_0(S=\text{SW}|\boldsymbol{X}) \\
        &+\mathbb{E}_{P_0}[Y(D=1,G=b)-Y(D=0,G=b)|S=\text{ACO},\boldsymbol{X}]P_0(S=\text{ACO}|\boldsymbol{X})\\
        =&\mathbb{E}_{P_0}[Y(D=1)-Y(D=0)|S=\text{SW},\boldsymbol{X}]P_0(S=\text{SW}|\boldsymbol{X}) \\
        &+\mathbb{E}_{P_0}[Y(D=1)-Y(D=0)|S=\text{ACO},\boldsymbol{X}]P_0(S=\text{ACO}|\boldsymbol{X})
    \end{align*}
    and
    \begin{align*}
        &\mathbb{E}_{P_0}[Y|Z=1_a,\boldsymbol{X}]-\mathbb{E}_{P_0}[Y|Z=0_a,\boldsymbol{X}]\\
        =&\mathbb{E}_{P_0}[Y(Z=1_a)|Z=1_a,G=a,\boldsymbol{X}=\boldsymbol{x}]-\mathbb{E}_{P_0}[Y(Z=0_a)|Z=0_a,G=a,\boldsymbol{X}=\boldsymbol{x}]\\
        =&\mathbb{E}_{P_0}[Y(Z=1_a)-Y(Z=0_a)|G=a,\boldsymbol{X}=\boldsymbol{x}]\\
        =&\mathbb{E}_{P_0}[Y(Z=1_a)-Y(Z=0_a)|\boldsymbol{X}=\boldsymbol{x}]\\
        =&\mathbb{E}_{P_0}[Y(D=1, G=a)-Y(D=0, G=a)|S=\text{ACO},\boldsymbol{X}]P_0(S=\text{ACO}|\boldsymbol{X})\\
        =&\mathbb{E}_{P_0}[Y(D=1)-Y(D=0)|S=\text{ACO},\boldsymbol{X}]P_0(S=\text{ACO}|\boldsymbol{X})
    \end{align*}
    therefore
    \begin{align*}
    &\mathbb{E}_{P_0}[Y(D=1)-Y(D=0)|S=\text{SW},\boldsymbol{X}]P_0(S=\text{SW}|\boldsymbol{X})\\
        =&(\mathbb{E}_{P_0}[Y|Z=1_b,\boldsymbol{X}]-\mathbb{E}_{P_0}[Y|Z=0_b,\boldsymbol{X}])-(\mathbb{E}_{P_0}[Y|Z=1_a,\boldsymbol{X}]-\mathbb{E}_{P_0}[Y|Z=0_a,\boldsymbol{X}])
    \end{align*}
Similarly, we have
\begin{align*}
    &\mathbb{E}_{P_0}[D|Z=1_b,\boldsymbol{X}]-\mathbb{E}_{P_0}[D|Z=0_b,\boldsymbol{X}]\\
    =&\mathbb{E}_{P_0}[D|Z=1_b,G=b,\boldsymbol{X}]-\mathbb{E}_{P_0}[D|Z=0_b,G=b, X]\\
    =&\mathbb{E}_{P_0}[D(Z=1_b)-D(Z=0_b)|G=b,\boldsymbol{X}]\\
    =&\mathbb{E}_{P_0}[D(Z=1_b)-D(Z=0_b)|\boldsymbol{X}]\\
    =&P_0(S=\text{ACO}|\boldsymbol{X})+P_0(S=\text{SW}|\boldsymbol{X})
\end{align*}
and
\begin{align*}
&\mathbb{E}_{P_0}[D|Z=1_a,\boldsymbol{X}]-E[D|Z=0_a,\boldsymbol{X}]\\
    =&\mathbb{E}_{P_0}[D|Z=1_a,G=a,\boldsymbol{X}]-\mathbb{E}_{P_0}[D|Z=0_a,G=a, X]\\
    =&\mathbb{E}_{P_0}[D(Z=1_a)-D(Z=0_a)|G=a,\boldsymbol{X}]\\
    =&\mathbb{E}_{P_0}[D(Z=1_a)-D(Z=0_a)|\boldsymbol{X}]\\
    =&P_0(S=\text{ACO}|\boldsymbol{X})
\end{align*}
    Therefore we have
    \begin{align*}
       (\mathbb{E}_{P_0}[D|Z=1_b,\boldsymbol{X}]-\mathbb{E}_{P_0}[D|Z=0_b,\boldsymbol{X}])- (\mathbb{E}_{P_0}[D|Z=1_a,\boldsymbol{X}]-\mathbb{E}_{P_0}[D|Z=0_a,\boldsymbol{X}]) = P_0(S=\text{SW}|\boldsymbol{X})
    \end{align*}
    Therefore,
    \begin{align*}
        SWATE_{P_0}(\boldsymbol{X})&:=\mathbb{E}_{P_0}[Y(D=1)-Y(D=0)|S=\text{SW},\boldsymbol{X}]\\
        & = \frac{(\mathbb{E}_{P_0}[Y|Z=1_b,\boldsymbol{X}]-\mathbb{E}_{P_0}[Y|Z=0_b,\boldsymbol{X}])-(\mathbb{E}_{P_0}[Y|Z=1_a,\boldsymbol{X}]-\mathbb{E}_{P_0}[Y|Z=0_a,\boldsymbol{X}])}{(\mathbb{E}_{P_0}[D|Z=1_b,\boldsymbol{X}]-\mathbb{E}_{P_0}[D|Z=0_b,\boldsymbol{X}])- (\mathbb{E}_{P_0}[D|Z=1_a,\boldsymbol{X}]-\mathbb{E}_{P_0}[D|Z=0_a,\boldsymbol{X}])}
    \end{align*}
    and
    \begin{align*}
        \text{SWATE}_{P_0}&= \int  \text{SWATE}_{P_0}(\boldsymbol{X})f_{P_0}(\boldsymbol{x}|S=SW)d \mu(\boldsymbol{X})\\
        &= \int \frac{\delta_{b,P_0}(\boldsymbol{X})-\delta_{a,P_0}(\boldsymbol{X})}{\eta_{b,P_0}(\boldsymbol{X})-\eta_{a,P_0}(\boldsymbol{X})}\frac{\eta_{b,P_0}(\boldsymbol{X})-\eta_{a,P_0}(\boldsymbol{X})}{P_0(S=\text{SW})}f_{P_0}(\boldsymbol{X})d \mu(\boldsymbol{X})\\
        & = \frac{\mathbb{E}_{P_0}[\delta_{b,P_0}(\boldsymbol{X})-\delta_{a,P_0}(\boldsymbol{X})]}{P_0(S=\text{SW})}\\
        & = \frac{\mathbb{E}_{P_0}[\delta_{b,P_0}(\boldsymbol{X})-\delta_{a,P_0}(\boldsymbol{X})]}{\mathbb{E}_{P_0}[\eta_{b,P_0}(\boldsymbol{X})-\eta_{a,P_0}(\boldsymbol{X})]}
    \end{align*}
where $f_{P_0}(\boldsymbol{x}|S=SW)$ and $f_{P_0}(\boldsymbol{X})$ are conditional density of and density functions of $X$ implied by $P_0$.
    Similarly,  for $\text{ACOATE}$, we have
    \begin{align*}
        \text{ACOATE}_{P_0}(\boldsymbol{X})&:=\mathbb{E}_{P_0}[Y(D=1)-Y(D=0)|S=\text{ACO},\boldsymbol{X}]\\
        & = \frac{\mathbb{E}_{P_0}[Y|Z=1_a,\boldsymbol{X}]-\mathbb{E}_{P_0}[Y|Z=0_a,\boldsymbol{X}]}{\mathbb{E}_{P_0}[D|Z=1_a,\boldsymbol{X}]-\mathbb{E}_{P_0}[D|Z=0_a,\boldsymbol{X}]}
    \end{align*}
    and
    \begin{equation*}
        \text{ACOATE}_{P_0} = \frac{\mathbb{E}_{P_0}[\delta_{a,P_0}(\boldsymbol{X})]}{\mathbb{E}_{P_0}[\eta_{a,P_0}(\boldsymbol{X})]}
    \end{equation*}
\end{proof}
\subsection{Proof for Theorem \ref{thm: G direct effect,2,supp}}
\begin{proof}
    Direct calculation gives
    \begin{align*}
    &\mathbb{E}_{P_0}[Y\mid Z=1_b,\boldsymbol{X}]-\mathbb{E}_{P_0}[Y\mid Z=0_b,\boldsymbol{X}]\\
     =&\mathbb{E}_{P_0}[Y(D=0)+(Y(D=1)-Y(D=0))D\mid Z=1_b,G=b,\boldsymbol{X}]\\
     &-\mathbb{E}_{P_0}[Y(D=0)+(Y(D=1)-Y(D=0))D|Z=0_b,G=b,\boldsymbol{X}]\\
     =&\mathbb{E}_{P_0}[Y(D=0,Z=1_b)+(Y(D=1,Z=1_b)-Y(D=0,Z=1_b))D(Z=1_b)\mid Z=1_b,G=b,\boldsymbol{X}]\\
     &-\mathbb{E}_{P_0}[Y(D=0,Z=0_b)+(Y(D=1,Z=0_b)-Y(D=0,Z=0_b))D(Z=0_b)|Z=0_b,G=b,\boldsymbol{X}]\\
     =&\mathbb{E}_{P_0}[Y(D=0,Z=1_b)+(Y(D=1,Z=1_b)-Y(D=0,Z=1_b))D(Z=1_b)\mid G=b,\boldsymbol{X}]\\
     &-\mathbb{E}_{P_0}[Y(D=0,Z=0_b)+(Y(D=1,Z=0_b)-Y(D=0,Z=0_b))D(Z=0_b)|G=b,\boldsymbol{X}]\\
        =&\mathbb{E}_{P_0}[(Y(D=1,Z=1_b)-Y(D=0,Z=1_b))(D(Z=1_b)-D(Z=0_b))|G=b,\boldsymbol{X}]\\
        =&\mathbb{E}_{P_0}[(Y(D=1,G=b)-Y(D=0,G=b))(D(Z=1_b)-D(Z=0_b))|G=b,\boldsymbol{X}]\\
        =&\mathbb{E}_{P_0}[(Y(D=1)-Y(D=0))(D(Z=1_b)-D(Z=0_b))|G=b,\boldsymbol{X}]\\
        =&\mathbb{E}_{P_0}[Y(D=1)-Y(D=0)|S=\text{SW},G=b,\boldsymbol{X}]P_0(S=\text{SW}|G=b,X) \\
        &+\mathbb{E}_{P_0}[Y(D=1)-Y(D=0)|S=\text{ACO},G=b,\boldsymbol{X}]P_0(S=\text{ACO}|G=b,X)\\
        =&\mathbb{E}_{P_0}[Y(D=1)-Y(D=0)|S=\text{SW},\boldsymbol{X}]P_0(S=\text{SW}|\boldsymbol{X}) \\
        &+\mathbb{E}_{P_0}[Y(D=1)-Y(D=0)|S=\text{ACO},\boldsymbol{X}]P_0(S=\text{ACO}|\boldsymbol{X}).
    \end{align*}

    Following the similar derivation, we have 

    \begin{align*}
        &\mathbb{E}_{P_0}[Y|Z=1_a,\boldsymbol{X}]-\mathbb{E}_{P_0}[Y|Z=0_a,\boldsymbol{X}] = \mathbb{E}_{P_0}[Y(D=1)-Y(D=0)|S=\text{ACO},\boldsymbol{X}]P_0(S=\text{ACO}|\boldsymbol{X}),\\
        &\mathbb{E}_{P_0}[D|Z=1_b,\boldsymbol{X}]-\mathbb{E}_{P_0}[D|Z=0_b,\boldsymbol{X}] = P_0(S=\text{ACO}|\boldsymbol{X})+P_0(S=\text{SW}|\boldsymbol{X}),\\
        &\mathbb{E}_{P_0}[D|Z=1_a,\boldsymbol{X}]-E[D|Z=0_a,\boldsymbol{X}] = P_0(S=\text{ACO}|\boldsymbol{X}).
    \end{align*}
    Therefore we end up with the same identification formula.
\end{proof}

\clearpage
\section{Additional simulation results}
\subsection{Data-generating process for simulation studies of SWATE estimation}
\label{subsec: main simu details, supp}

We generate datasets according to the following factorial design:

\noindent \textbf{Sample sizes:} $N=1000$, $2000$, $5000$ and $10000$.

\noindent  \textbf{Baseline covariates:} We consider a eight-dimensional covariate $\boldsymbol{X}=(\boldsymbol{X}_1, \boldsymbol{X}_2,\dots,\boldsymbol{X}_8)$, where $(\boldsymbol{X}_1,\boldsymbol{X}_2,\boldsymbol{X}_3)\sim\mathcal{MVN}(\mathbf{\mu},\mathbf{\Sigma})$, with $\mathbf{\mu} = (0,1,-0.5)$, $\mathbf{\Sigma} = {\scriptsize\begin{pmatrix}
    1 & 0.2 &-0.3 \\
    0.2 & 1 & 0.1 \\
    -0.3& 0.1 &1 \\
\end{pmatrix}},$ and truncated at $-4$ and $4$,
$\boldsymbol{X}_4,\boldsymbol{X}_5,\boldsymbol{X}_6 \sim \text{Bernoulli}(0.5)$, $\boldsymbol{X}_7\sim \text{Uniform}(-3,3)$, and $\boldsymbol{X}_8\sim \text{Binomial}(4,0.5)$.

\noindent \textbf{Instrumental variable:} We generate a binary variable $G \in \{a, b\}$ according to $P(G = a) = 1 - \text{expit}(1+0.2X_1-0.1X_2+0.3X_3)$ and $P(G = b) = 1 - P(G = a)$, where $\text{expit}(\boldsymbol{X}) = \exp(\boldsymbol{X})/(1+\exp(\boldsymbol{X}))$ is the inverse of the logit function. Conditioning on $G =a$ or $b$, we then generate a binary IV $Z \in \{0_a, 1_a\}$ when $G = a$ and $Z \in \{0_b, 1_b\}$ when $G = b$:
\begin{equation*}
\begin{split}\vspace{-6pt}
    &P(Z=1_a \mid G=a) = \text{expit}(1+0.5X_1-X_2+0.7X_3), \\
    &P(Z=1_b \mid G=b) = \text{expit}(0.5+0.6X_1+0.3X_2+0.4X_3).
\end{split}
\end{equation*}\vspace{-6pt}
\noindent According to this data generating process, the IV assignment is dependent on observed covariates and different for $G = a$ or $b$.

\noindent\textbf{Unmeasured confounder:} We generate an unmeasured confounder $U \sim \mathcal{N}(0,0.6)$. Scalar representation of unobserved confounding is widely adopted in the literature and referred to as a ``principal unobserved covariate" \citep{rosenbaum2023propensity}.

\noindent\textbf{Principal stratum:} The principal stratum status of participant $i$, denoted as $S_i$, is generated via a multinomial distribution: $P(S_i = j) = \frac{g_j}{\sum_{j}g_j}$, $j \in \{\text{ANT}, \text{SW1}, \text{SW2},$ $ \text{AT-NT}, \text{ACO}, \text{NT-AT}, \text{AAT}\}$, where $SW1$ and $SW2$ correspond to participants who convert from always-takers or never-takers to compliers, respectively. We specify $g_j$ as follows:
{\small \begin{equation*}
\begin{split}
    &g_{ANT} = \exp(1-X_2+0.7X_3+0.3U),
    \qquad\qquad~ g_{SW1} = \exp(\alpha_1+\alpha_2(\boldsymbol{x}_1+2X_2-X_3)+\alpha_3U),\\
    &g_{SW2} = \exp(\alpha_1+\alpha_2(-X_1-2X_2)+\alpha_3U),
    \quad~~~ g_{AT-NT} = \exp(1+0.5X_1+X_2+0.5X_3+0.5U),\\
    &g_{ACO} = \exp(1+0.8X_1-2X_2-2X_3+0.5U),
    \quad g_{NT-AT} = \exp(1-0.5X_1-1X_2-0.5X_3-0.5U),\\
    &g_{AAT} = \exp(1+2X_1+2X_3-U),
    \end{split}
\end{equation*}}
\noindent where parameters $(
\alpha_1,
\alpha_2,
\alpha_3)$ control the proportions of switchers. We consider the following $4$ choices of $(\alpha_1, \alpha_2,
\alpha_3)$: $(-0.2,0.1,0.0005)$, $(0.5,0.2,0.05)$, $(0.3,0.5,0.1)$, and $(1,1,1)$, corresponding to $11\%$, $22\%$, $32\%$, and $66\%$ switchers in the population, respectively. 

\noindent\textbf{Treatment received:} Treatment received $D$ is determined by $Z$ and $S$.

\noindent\textbf{Outcome:} We consider two sets of data generating processes for potential outcomes.
\noindent For a continuous $Y$, we generate $Y(0) = 1+X_1+X_2+X_3+X_4+U+\epsilon$ and $Y(1)$ as follows: 
{\small\begin{equation*}
\begin{split}
    Y(1) = &\mathbbm{1}\{S= \text{ANT} \text{ or } \text{AAT}\}\cdot(1+X_1+2X_2+2X_3+X_4+U)\\
    &+\mathbbm{1}\{S= \text{AT-NT} \text{ or } \text{NT-AT}\}\cdot(1+X_1+X_2+2X_3+X_4+U)\\
    &+\mathbbm{1}\{S= \text{SWI}\}\cdot(\beta_1+\beta_2X_1+2X_2+\beta_3X_3+X_4+U)\\
    & +\mathbbm{1}\{S= \text{ACO}\}\cdot(1+X_1+2X_2+0.2X_2^2+X_3+X_4+U)+\epsilon,
\end{split}
\end{equation*}}
where $(\beta_1,\beta_2,\beta_3) = (2,2,2)$ or $(4,4,4)$ control the effect size of SWATE, and $\epsilon \sim \mathcal{N}(0,1)$.

\noindent For a binary outcome, potential outcomes are generated as follows:
{\small \begin{equation*}
\begin{split}
    P(Y(0)=1) &=
    \mathbbm{1}\{S= \text{ANT} \text{ or } \text{AAT}\}\cdot\text{expit}(1+X_1+2X_2+2X_3+X_4+X_5-X_6-X_7+X_8+U)\\
    &+\mathbbm{1}\{S= \text{AT-NT} \text{ or } \text{NT-AT}\}\cdot\text{expit}\{1+X_1+X_2+2X_3+X_4+X_5-X_6-X_7+X_8+U\}\\
    &+\mathbbm{1}\{S= \text{SWI}\}\cdot\text{expit}(\beta_1+\beta_2X_1+2X_2+X_3+\beta_3X_6-X_7+2X_8+U)\\
    & +\mathbbm{1}\{S= \text{ACO}\}\cdot\text{expit}(1+X_1+2X_2+0.2X_2^2+X_3+X_4+X_5-X_6-X_7+X_8+U),\\
    P(Y(1)=1) &= \text{expit}(1+X_1+X_2+X_3+U),\\
\end{split}
\end{equation*}}
where $(\beta_1,\beta_2,\beta_3) = (0,1,-1)$ or $(0,2,-3)$ control the size of SWATE. \textcolor{black}{For a binary outcome, SWATE is the causal risk difference (RD) among switchers. The true SWATE was calculated using Monte Carlo methods based on a simulated dataset of size 1 million.}

\subsection{Simulation results for testing implications of homogeneity}
\label{subsec: simu test homogeneity, np test, supp}
We evaluate and compare the level and power of the best linear projection-based test, $L^2$ norm-based tests (Cramer-von Mises-type), and $L^\infty$ norm-based tests (Kolmogorov-Smirnov-type) under various data generating processes in this section. Datasets were generated according to the following factorial design:

\noindent \textbf{Sample sizes:} $N = 2000$, $5000$, and $10000$ for the projection-based tests. $N=2000$ and $5000$ for non-parametric tests ($L^2$ and $L^\infty$ norm-based). We did not pursue $N = 10000$ for the non-parametric tests because of the large computation cost.

\noindent  \textbf{Baseline covariates:} We consider $\boldsymbol{X}=(\boldsymbol{X}_1, \boldsymbol{X}_2)$ from a bivariate normal distribution with $\mathbf{\mu} = (0,0)$ and $\mathbf{\Sigma} =I_2$.

\noindent \textbf{Instrumental variable:} We generate a  variable $G \in \{a, b\}$ as follows: 
{\begin{equation*}
\begin{split}
    &P(G = a) = \text{expit}(0.1\mathbbm{1}\{X_1>0\}-0.1\mathbbm{1}\{X_2>0\}).
\end{split}
\end{equation*}}

\noindent We generate a binary IV for $G = a$ or $b$ as follows:
{\begin{equation*}
\begin{split}
    &P(Z=1_a \mid G=a) = 0.5+0.1X_1^* - 0.1X_2^*, \\
    &P(Z=1_b \mid G=b) = 0.5 - 0.1X_1^* + 0.1X_2^*,
\end{split}
\end{equation*}}
where $X^*_1 = 1\{X_1>0\}, X^*_2 = 1\{X_2>0\}$

\noindent\textbf{Unmeasured confounder:} We generate an unmeasured confounder $U \sim \mathcal{N}(-0.3,0.3)$.

\noindent\textbf{Principal stratum:} The principal stratum status of participant $i$, denoted as $S_i$, is generated through a multinomial distribution. To this end, we define $g_j$ as follows:
{\small \begin{equation*}
\begin{split}
    &g_{\text{ANT}} = \exp(1-X^*_2+0.3\mathbbm{1}\{U>0\}),
    \qquad\qquad\quad g_{\text{SW1}} = \exp(3.5+0.5X^*_1+X^*_2+0.1\mathbbm{1}\{U>0\}),\\
    &g_{\text{SW2}} = \exp(3.5+0.5X^*_1+X^*_2+0.1\mathbbm{1}\{U>0\}),
    ~g_{\text{AT-NT}} = \exp(1+0.5X^*_1+X^*_2+0.5\mathbbm{1}\{U>0\}),\\
    &g_{\text{ACO}} = \exp(1+0.8X^*_1-2X^*_2+0.5\mathbbm{1}\{U>0\}),
    ~~g_{\text{NT-AT}} = \exp(1-0.5X^*_1-1X^*_2-0.5\mathbbm{1}\{U>0\}),\\
    &g_{\text{AAT}} = \exp(1+2X^*_1-\mathbbm{1}\{U>0\}),
    \end{split}
\end{equation*}}
and let $P(S_i = j) = \frac{g_j}{\sum_{j}g_j}$ for $j\in \{\text{ANT}, \text{AT-NT}, \text{NT-AT}, \text{AAT}\}$, $P(S=\text{SW1}) = P(S=\text{SW2}) = \frac{(1-\alpha)\cdot(g_{\text{SW1}}+g_{\text{SW2}}+g_{\text{ACO}})}{2\sum_{j}g_j}$, and $P(S=\text{ACO}) =\frac{\alpha\cdot(g_{\text{SW1}}+g_{\text{SW2}}+g_{\text{ACO}})}{\sum_{j}g_j}$,
where $\alpha$ controls the proportion of switchers. We let $\alpha = 0.1$, $0.2$, $0.4$, $0.6$, $0.8$, and $0.9$, which correspond to $9\%$, $18\%$, $36\%$, $54\%$, $72\%$, and $81\%$ switchers, respectively. Under this data generating process, the proportion of switchers and always-compliers, i.e., $P(S=\text{SW})+P(S=\text{ACO})$, is fixed at $90\%$.

\noindent\textbf{Treatment received:} Treatment received $D$ is determined by $Z$ and $S$.

\noindent\textbf{Outcome:} We consider the following data generating process for potential outcomes:
{\small\begin{equation*}
\begin{split}
Y(0) = &1+X_1+X_2+U+\epsilon, \\
    Y(1) = &\mathbbm{1}\{S= \text{ANT} \text{ or } \text{AAT}\}\cdot(1+X_1+X_2+U)
    +\mathbbm{1}\{S= \text{AT-NT} \text{ or } \text{NT-AT}\}\cdot(1+X_1+X_2+U)\\
    &+\mathbbm{1}\{S= \text{SW}\}\cdot(\beta_1+\beta_2X_1+\beta_3X_2+U)
     +\mathbbm{1}\{S= \text{ACO}\}\cdot(1+2X_1+2X_2+U)+\epsilon,
\end{split}
\end{equation*}}
\noindent where parameters $(\beta_1,\beta_2,\beta_3)$ control the effect size of SWATE, and $\epsilon \sim N(0,1)$. We consider $(\beta_1,\beta_2,\beta_3)= (1,2,2)$, $(1.5,2.5,2.5)$, and $(2,3,3)$. When $\beta = (1,2,2)$, the null hypothesis $H^{(j)}_0$ holds, i.e., $\theta^{(j)}_{P_0}=0$. Conversely, $\beta = (1,2.5,2.5)$ and $(2,3,3)$ correspond to $\theta^{(j)}_{P_0}\neq 0$. For each null hypothesis $H^{(j)}_{0}$, $j = 1, 2, 3$, we construct the projection-based, $L^2$ norm-based, and $L^\infty$ norm-based tests and repeated the simulation $1000$ times.

\begin{figure}[H]
    \centering
\includegraphics[width=0.99\textwidth]{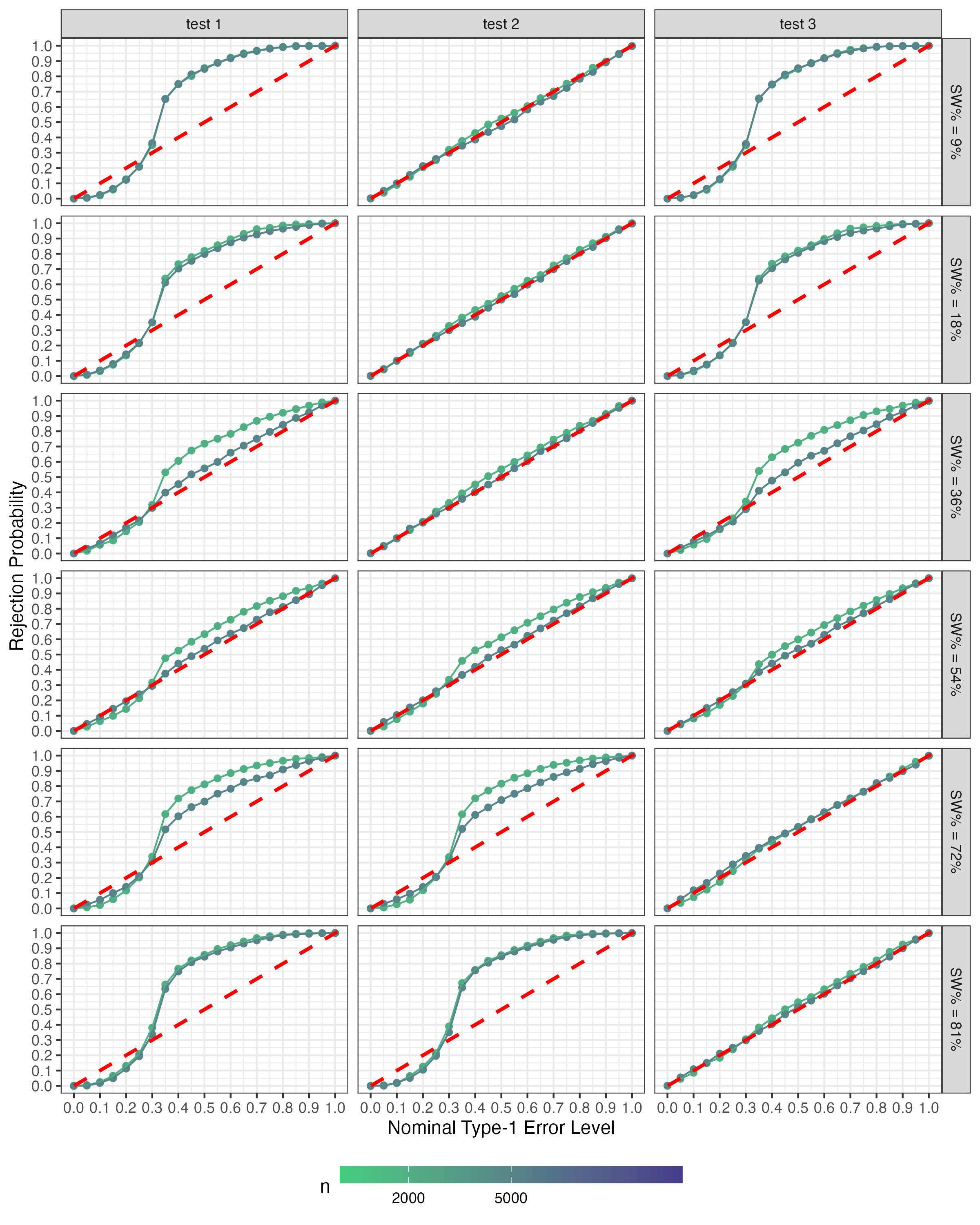}
    \caption{Simulation results for type-one-error control for Kolmogorov-Smirnov-type test.} The red dash line is the theoretical asymptotic size for each setting under different levels.
    \label{fig: test_size_KS}
\end{figure}

\begin{figure}[H]
    \centering
\includegraphics[width=0.99\textwidth]{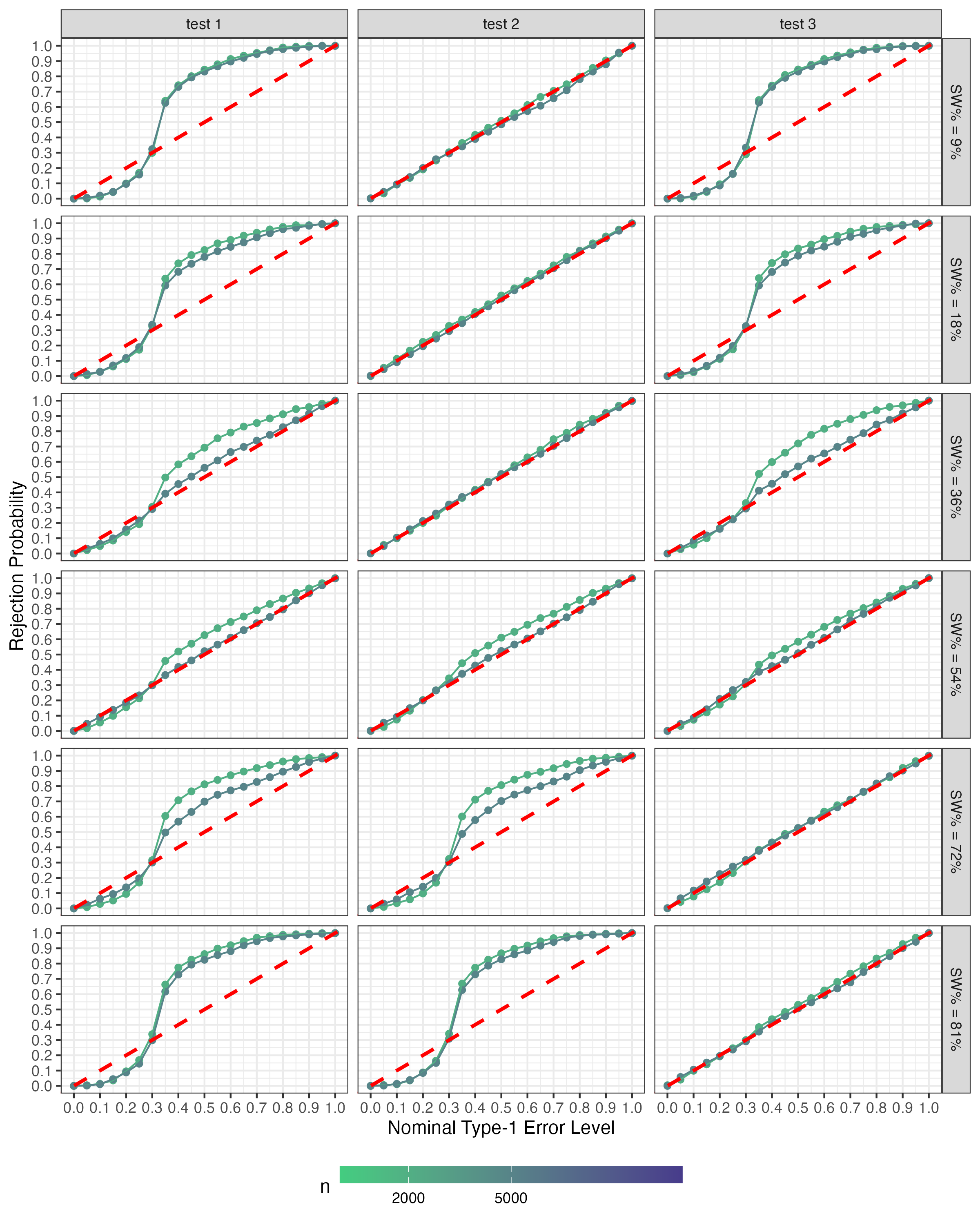}
    \caption{Simulation results for type-one-error control for Cramér–von Mises-type test.} The red dash line is the theoretical asymptotic size for each setting under different levels.
    \label{fig: test_size_CM}
\end{figure}

Figure \ref{fig: test_power} in the Supplemental Material plots the nominal Type-I error rate versus the empirical rejection proportion for the projection-based test. For ${T}^{(1)}_{\text{Wald},\alpha}$ that compares $\text{SWATE}_{P_0}(\boldsymbol{X})$ to $\text{ACOATE}_{P_0}(\boldsymbol{X})$, the rejection fraction aligns well with the nominal level when the proportion of switchers is not too high or too low. On the other hand, the approximation of the null distribution is not accurate when the proportion of switchers is small and hence $\text{SWATE}_{P_0}(\boldsymbol{X})$ is not accurately estimated, or when the proportion of always-compliers is small and hence $\text{ACOATE}_{P_0}(\boldsymbol{X})$ is not well estimated. Analogously, the null distribution of ${T}^{(2)}_{\text{Wald},\alpha}$ is best approximated when the proportion of switchers is small and that of ${T}^{(3)}_{\text{Wald},\alpha}$ is best approximated when the proportion of switchers is large. Importantly, we did not observe an inflation of type-I error for a level-$0.05$ test in any setting. \textcolor{black}{We observed similar patterns for the Type-I error control for the $L^2$ and $L^\infty$ norm-based tests (Figure \ref{fig: test_size_KS} and \ref{fig: test_size_CM}).} 

Figure \ref{fig: test_power} further compares the power of each test under different settings. In all settings, when the proportion of switchers is small (e.g., $9\%$ and $18\%$), ${T}^{(2)}_{\text{Wald},0.05}$ often outperforms ${T}^{(1)}_{\text{Wald},0.05}$ and ${T}^{(3)}_{\text{Wald},0.05}$, and ${T}^{(3)}_{\text{Wald},0.05}$ is the least powerful. When the proportion of switchers is high (e.g., $72\%$ and $81\%$), ${T}^{(3)}_{\text{Wald},0.05}$ is the most powerful among three tests. Lastly, when the proportion of switchers is moderate (e.g., $36\%$ and $54\%$), ${T}^{(2)}_{\text{Wald},0.05}$ outperforms the other two tests although the three tests have similar performance when the sample size is large. As the sample size increases, the power of all three tests increases. In empirical studies, we would recommend ${T}^{(2)}_{\text{Wald},0.05}$ when the proportion of always-compliers is high and ${T}^{(3)}_{\text{Wald},0.05}$ when the proportion of switchers is high. 

\begin{figure}[H]
    \centering
\includegraphics[width=0.99\textwidth]{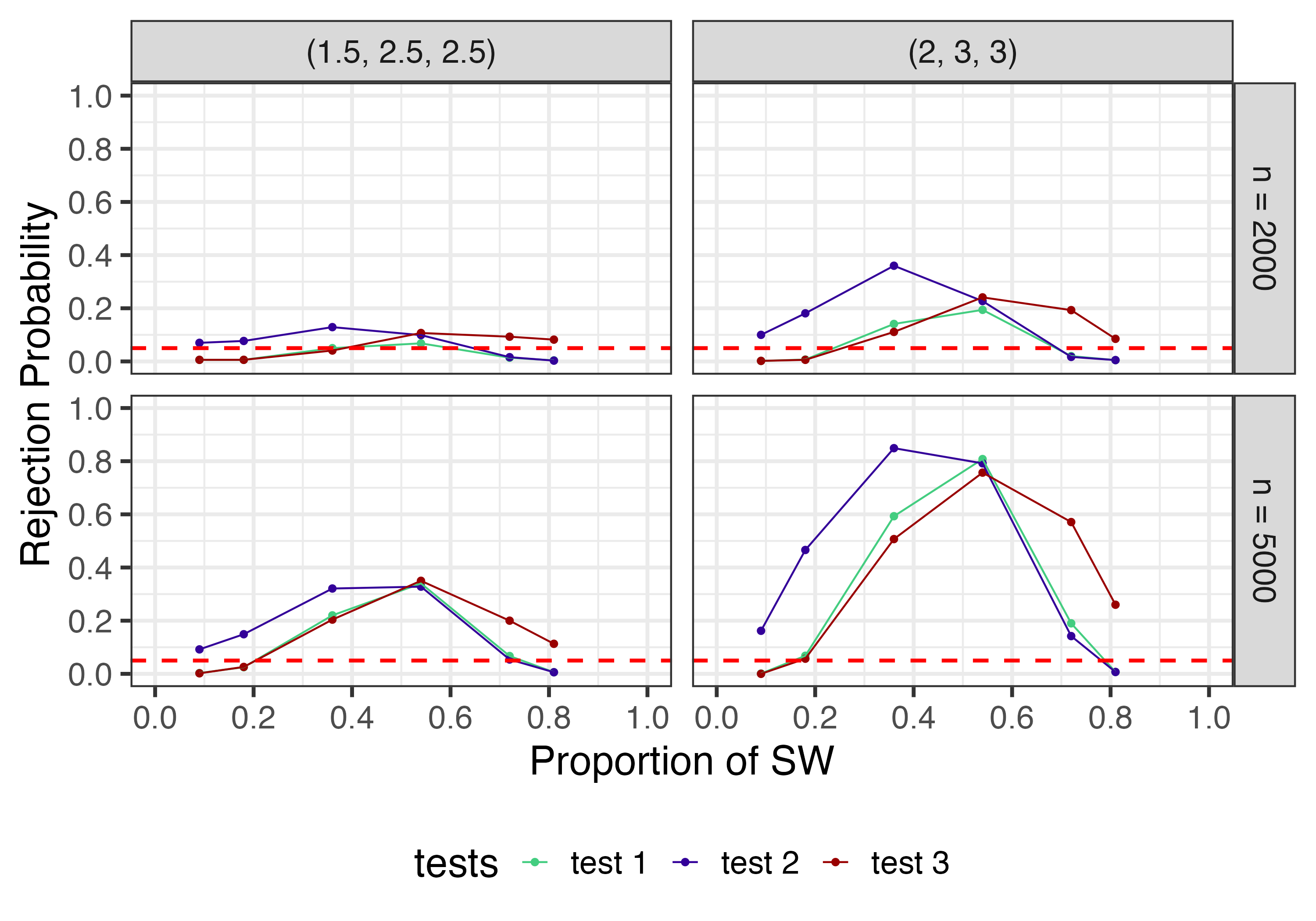}
    \caption{Comparison of power for Kolmogorov-Smirnov-type test. The red dashed lines correspond to the nominal level. Test 1 compares $\text{SWATE}_{P_0}(\boldsymbol{X})$ to $\text{ACOATE}_{P_0}(\boldsymbol{X})$; test 2 compares $\text{COATE}_{P_0}(\boldsymbol{X})$ to $\text{ACOATE}_{P_0}(\boldsymbol{X})$; test 3 compares $\text{COATE}_{P_0}(\boldsymbol{X})$ to $\text{SWATE}_{P_0}(\boldsymbol{X})$.}
    \label{fig: test_power_KS}
\end{figure}

\begin{figure}[H]
    \centering
\includegraphics[width=0.99\textwidth]{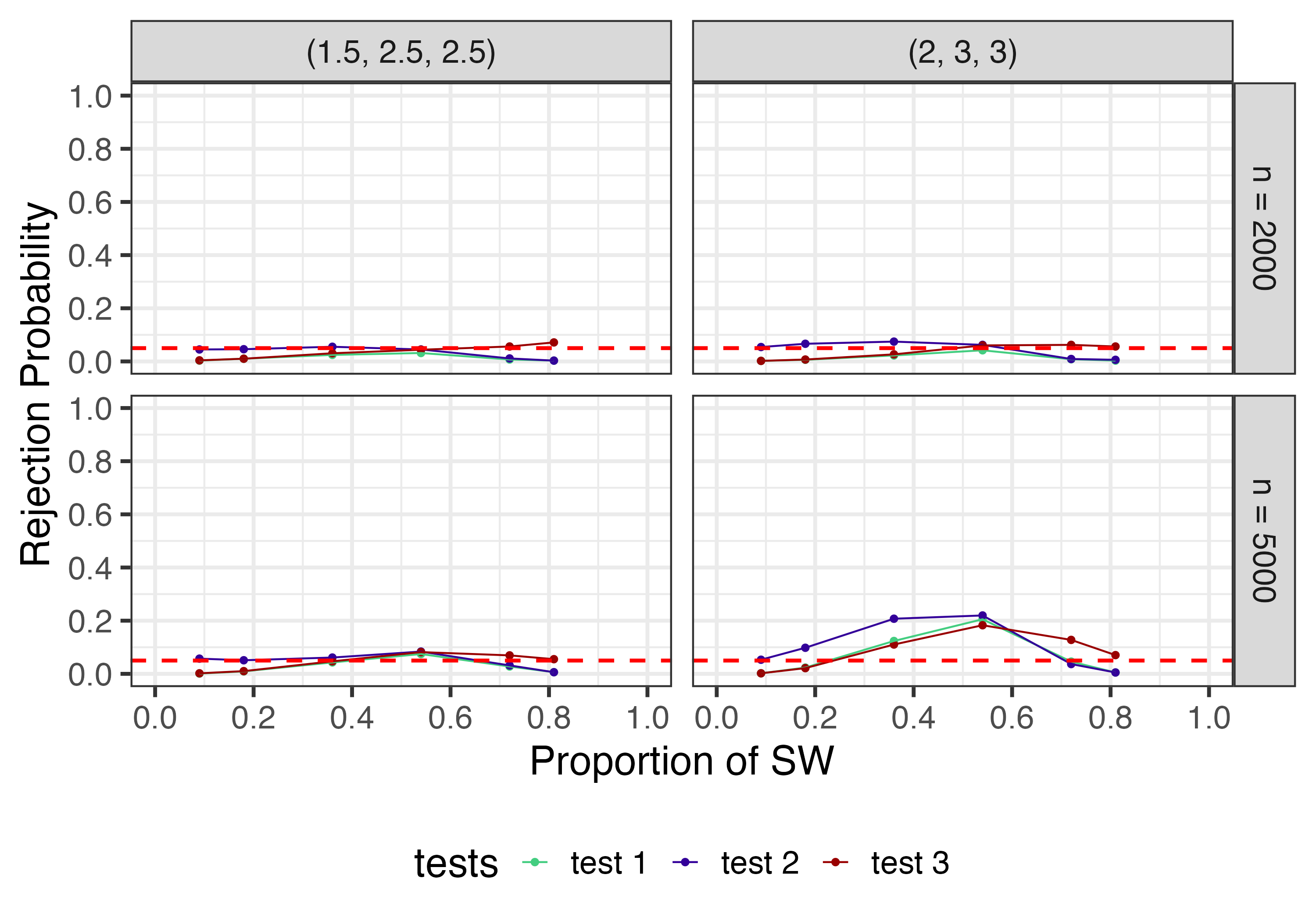}
    \caption{Comparison of power for Cramér–von Mises-type test. The red dashed lines correspond to the nominal level. Test 1 compares $\text{SWATE}_{P_0}(\boldsymbol{X})$ to $\text{ACOATE}_{P_0}(\boldsymbol{X})$; test 2 compares $\text{COATE}_{P_0}(\boldsymbol{X})$ to $\text{ACOATE}_{P_0}(\boldsymbol{X})$; test 3 compares $\text{COATE}_{P_0}(\boldsymbol{X})$ to $\text{SWATE}_{P_0}(\boldsymbol{X})$.}
    \label{fig: test_power_CM}
\end{figure}

\begin{figure}[H]
    \centering
\includegraphics[width=0.99\textwidth]{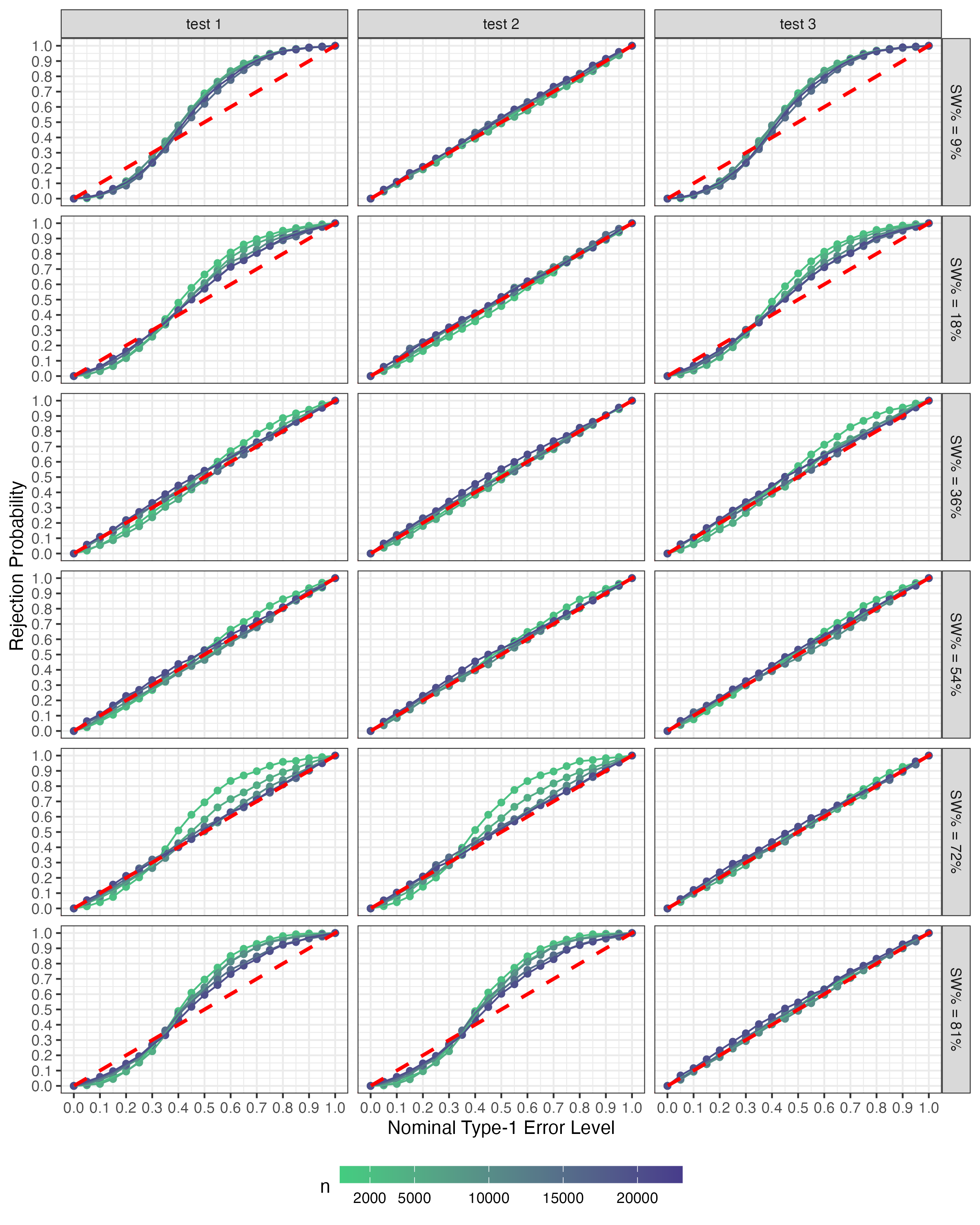}
    \caption{Simulation results for type-one-error control for the projection-based tests with nonlinear conditional treatment effects. The red dash line is the theoretical asymptotic size for each setting under different levels.} 
    \label{fig: project_test_size_nonlinear}
\end{figure}

\begin{figure}[ht]
    \centering
\includegraphics[width=0.85\textwidth]{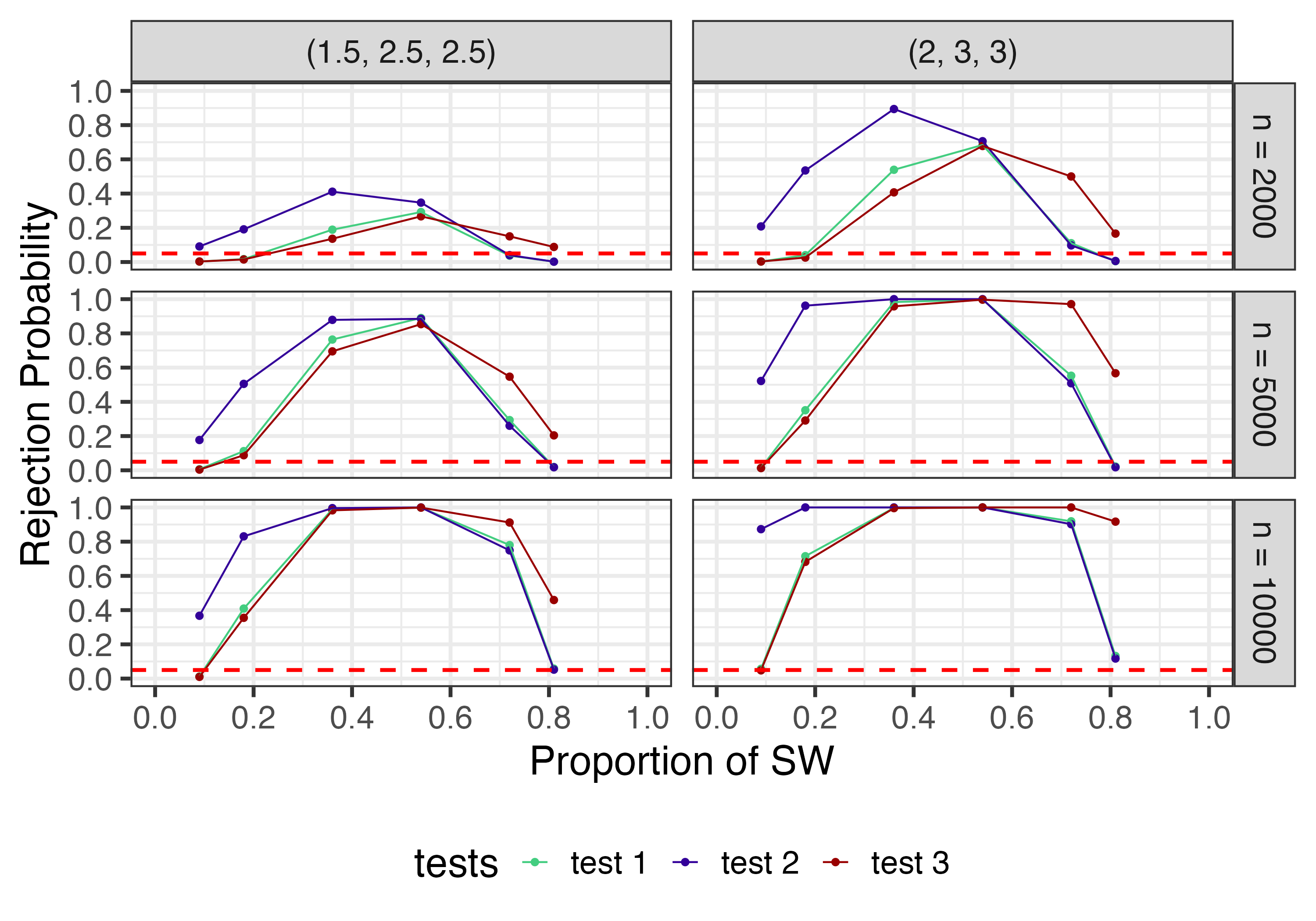}
    \caption{Comparison of power for the projection-based tests with nonlinear conditional treatment effects. The red dashed lines correspond to the nominal level. Test 1 compares $\text{SWATE}_{P_0}(\boldsymbol{X})$ to $\text{ACOATE}_{P_0}(\boldsymbol{X})$; test 2 compares $\text{COATE}_{P_0}(\boldsymbol{X})$ to $\text{ACOATE}_{P_0}(\boldsymbol{X})$; test 3 compares $\text{COATE}_{P_0}(\boldsymbol{X})$ to $\text{SWATE}_{P_0}(\boldsymbol{X})$.}
    \label{fig: project_test_power_nonlinear}
\end{figure}

\begin{figure}[ht]
    \centering
\includegraphics[width=0.99\textwidth]{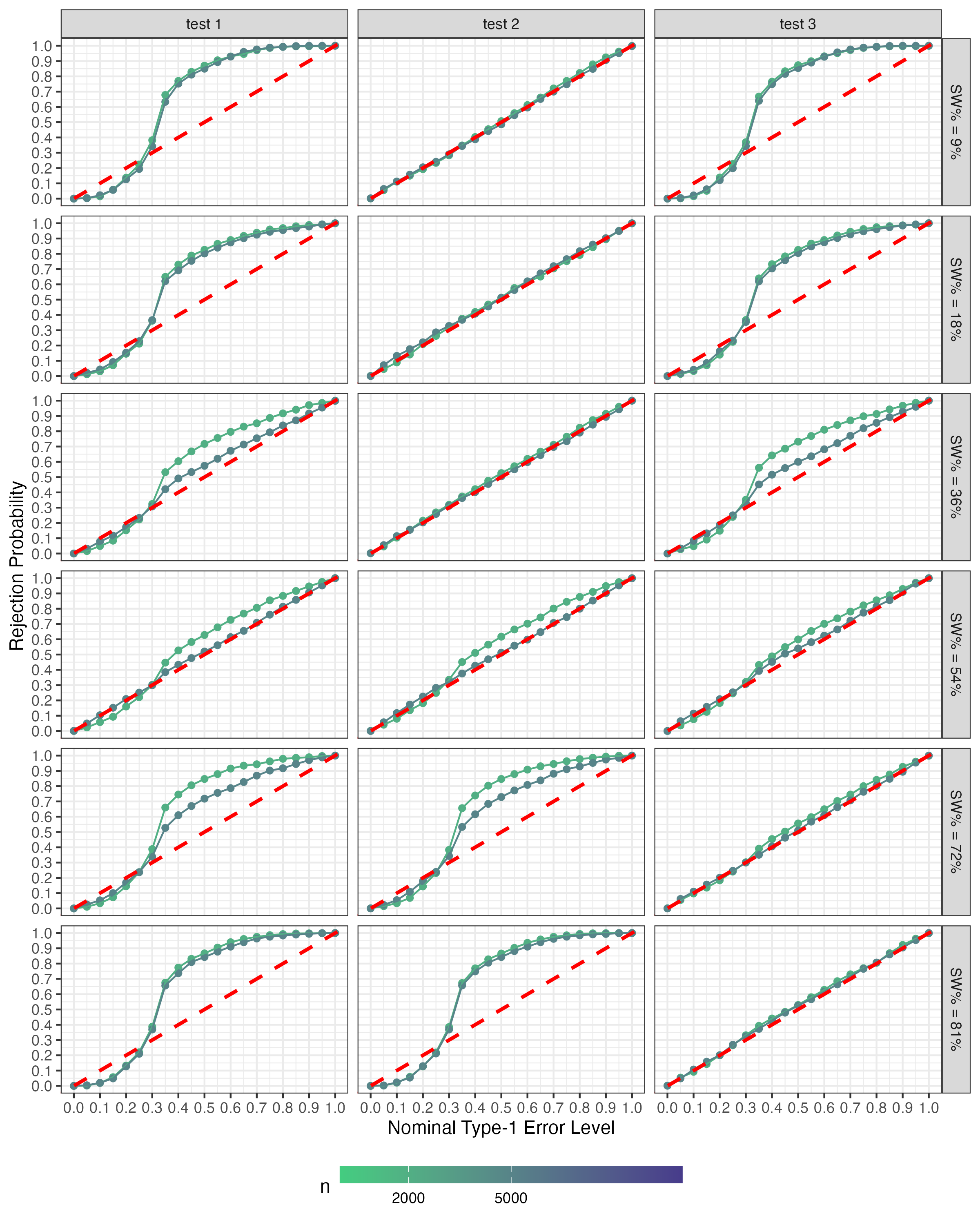}
    \caption{Simulation results for type-one-error control for Kolmogorov-Smirnov-type test with nonlinear conditional treatment effects. The red dash line is the theoretical asymptotic size for each setting under different levels.} 
    \label{fig: test_size_KM_nonlinear}
\end{figure}

\begin{figure}[ht]
    \centering
\includegraphics[width=0.99\textwidth]{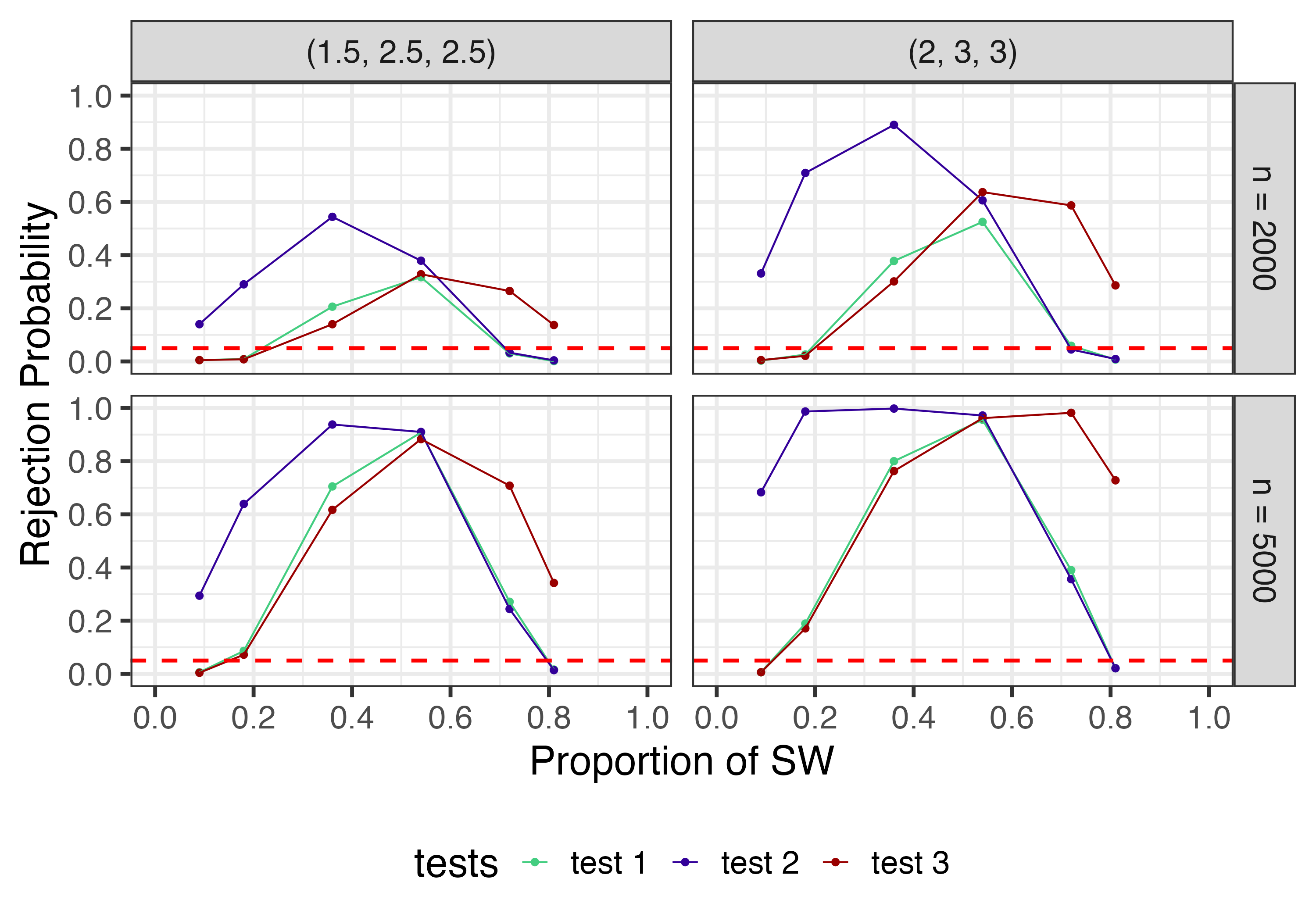}
    \caption{Comparison of power for Kolmogorov-Smirnov-type test with nonlinear conditional treatment effects. The red dashed lines correspond to the nominal level. Test 1 compares $\text{SWATE}_{P_0}(\boldsymbol{X})$ to $\text{ACOATE}_{P_0}(\boldsymbol{X})$; test 2 compares $\text{COATE}_{P_0}(\boldsymbol{X})$ to $\text{ACOATE}_{P_0}(\boldsymbol{X})$; test 3 compares $\text{COATE}_{P_0}(\boldsymbol{X})$ to $\text{SWATE}_{P_0}(\boldsymbol{X})$.}
    \label{fig: test_power_KS_nonlinear}
\end{figure}

\begin{figure}[ht]
    \centering
\includegraphics[width=0.99\textwidth]{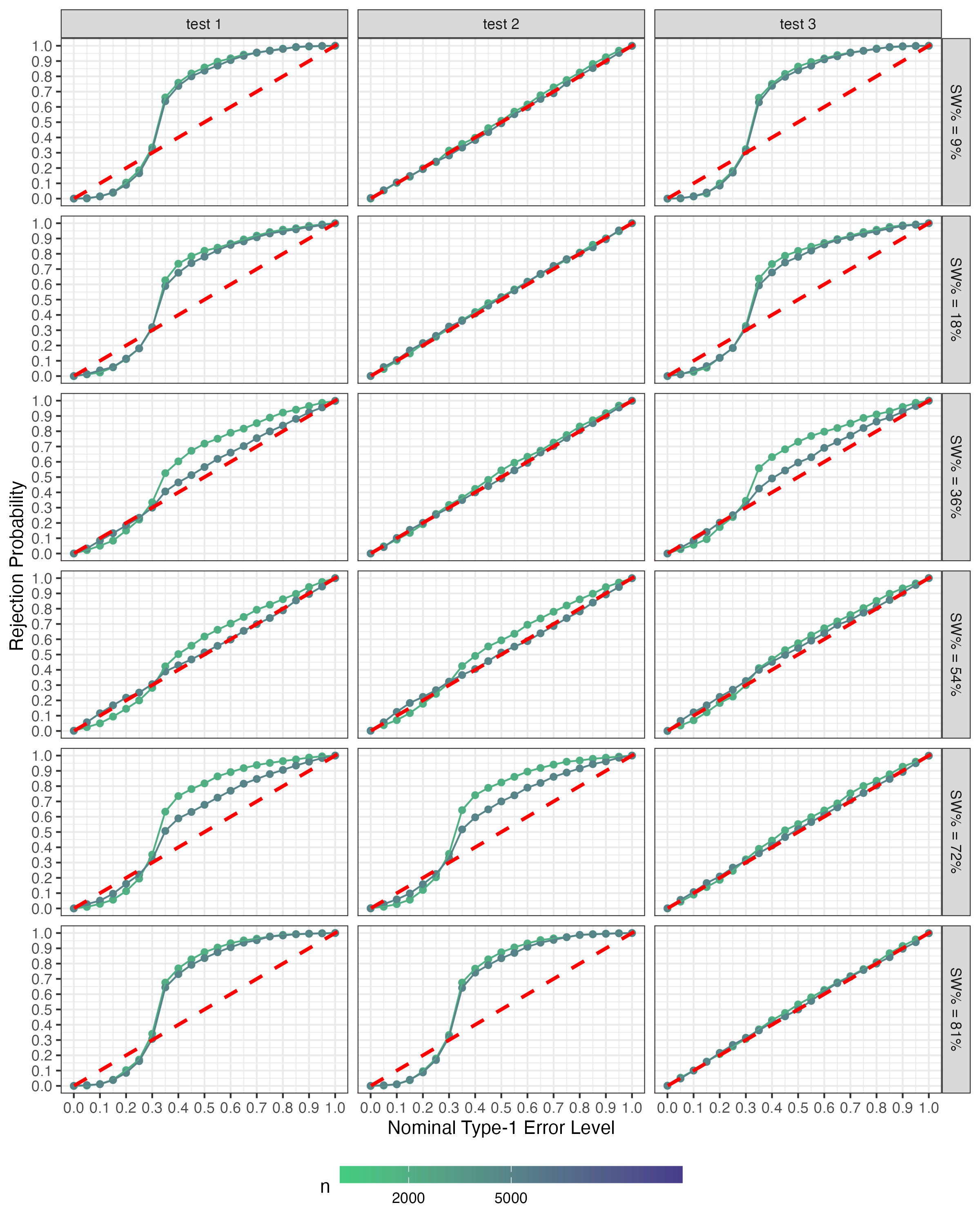}
    \caption{Simulation results for type-one-error control for Cramér–von Mises-type test with nonlinear conditional treatment effects. The red dash line is the theoretical asymptotic size for each setting under different levels.} 
    \label{fig: test_size_CM_nonlinear}
\end{figure}

\begin{figure}[ht]
    \centering
\includegraphics[width=0.99\textwidth]{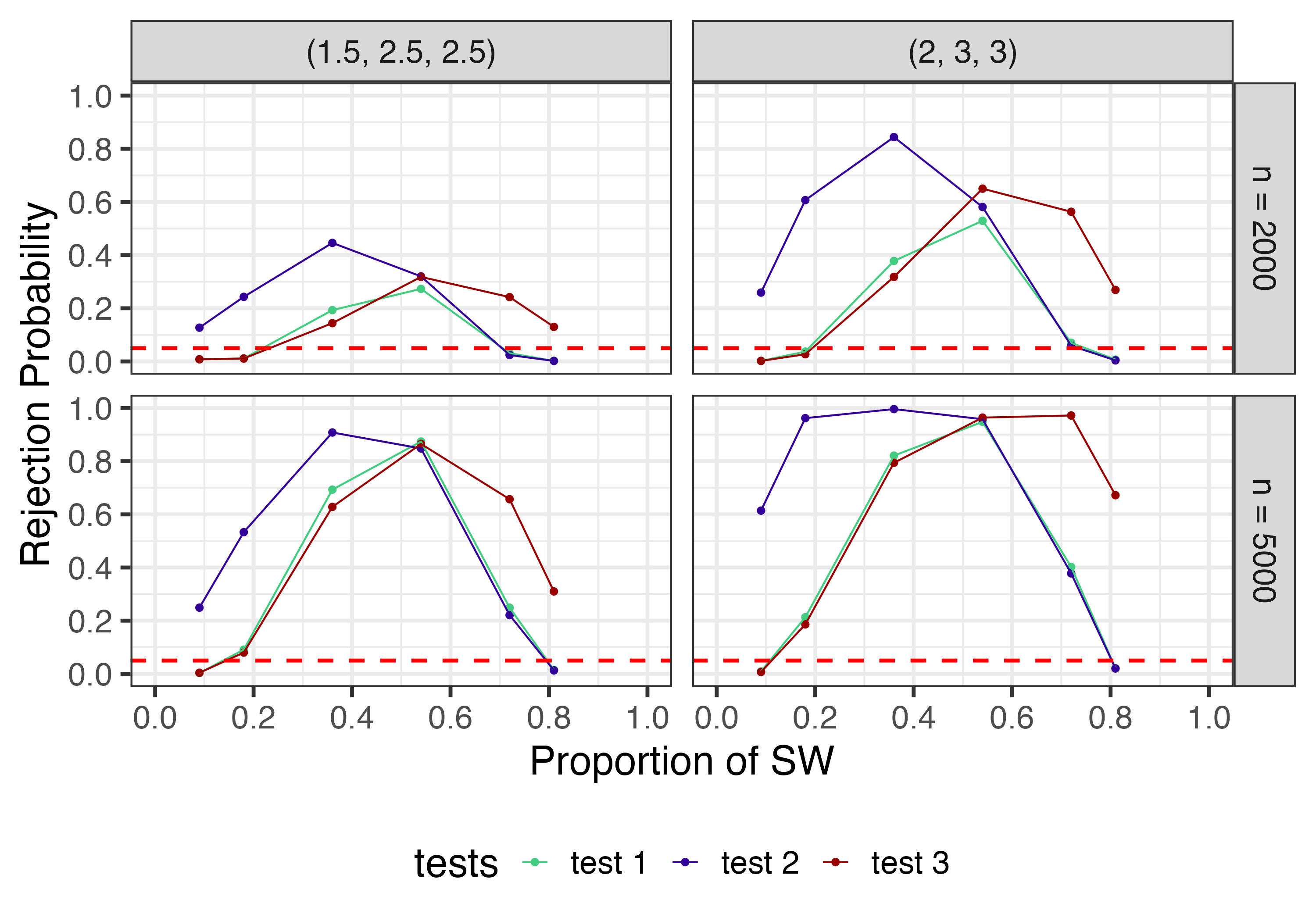}
    \caption{Comparison of power for Cramér–von Mises-type test with nonlinear conditional treatment effects. The red dashed lines correspond to the nominal level. Test 1 compares $\text{SWATE}_{P_0}(\boldsymbol{X})$ to $\text{ACOATE}_{P_0}(\boldsymbol{X})$; test 2 compares $\text{COATE}_{P_0}(\boldsymbol{X})$ to $\text{ACOATE}_{P_0}(\boldsymbol{X})$; test 3 compares $\text{COATE}_{P_0}(\boldsymbol{X})$ to $\text{SWATE}_{P_0}(\boldsymbol{X})$.}
    \label{fig: test_power_CM_nonlinear}
\end{figure}

{\color{black}
Figure \ref{fig: test_power_KS} and \ref{fig: test_power_CM} are analogous to Figure \ref{fig: test_power} and exhibit the power of each non-parametric test. We found the power of the projection-based test considerably higher than that of either non-parametric test when the conditional treatment effects are linear in covariates. The power of the Cramer-von Mises-type tests almost equaled its nominal Type-I error rate except when both the effect size and sample size were large (e.g. when $\beta = (2,3,3)$ and $N=5000$).

We then considered additional data generating processes where the conditional treatment effects are nonlinear in covariates; see Supplemental Material \ref{subsec: addtional simulation, nonlinear, supp}. Under these scenarios, we observed a significant improvement in the power of both non-parametric tests. Nevertheless, we found that projection-based tests still had more favorable or comparable power compared to two non-parametric alternatives; see Figures \ref{fig: project_test_size_nonlinear}---\ref{fig: test_power_CM_nonlinear} for details.




Finally, we compared the computation cost of the projection-based and $L^p$ norm-based, non-parametric tests. When $n = 5000$, it took roughly one minute to conduct a projection-based test, with cross-fitting using machine learning methods taking the most computation time. On the other hand, it took more than $15$ minutes to conduct a $L^p$ norm-based, non-parametric test, as simulating a Gaussian process could be computationally intensive.}





\begin{figure}[H]
    \centering
\includegraphics[width=0.99\textwidth]{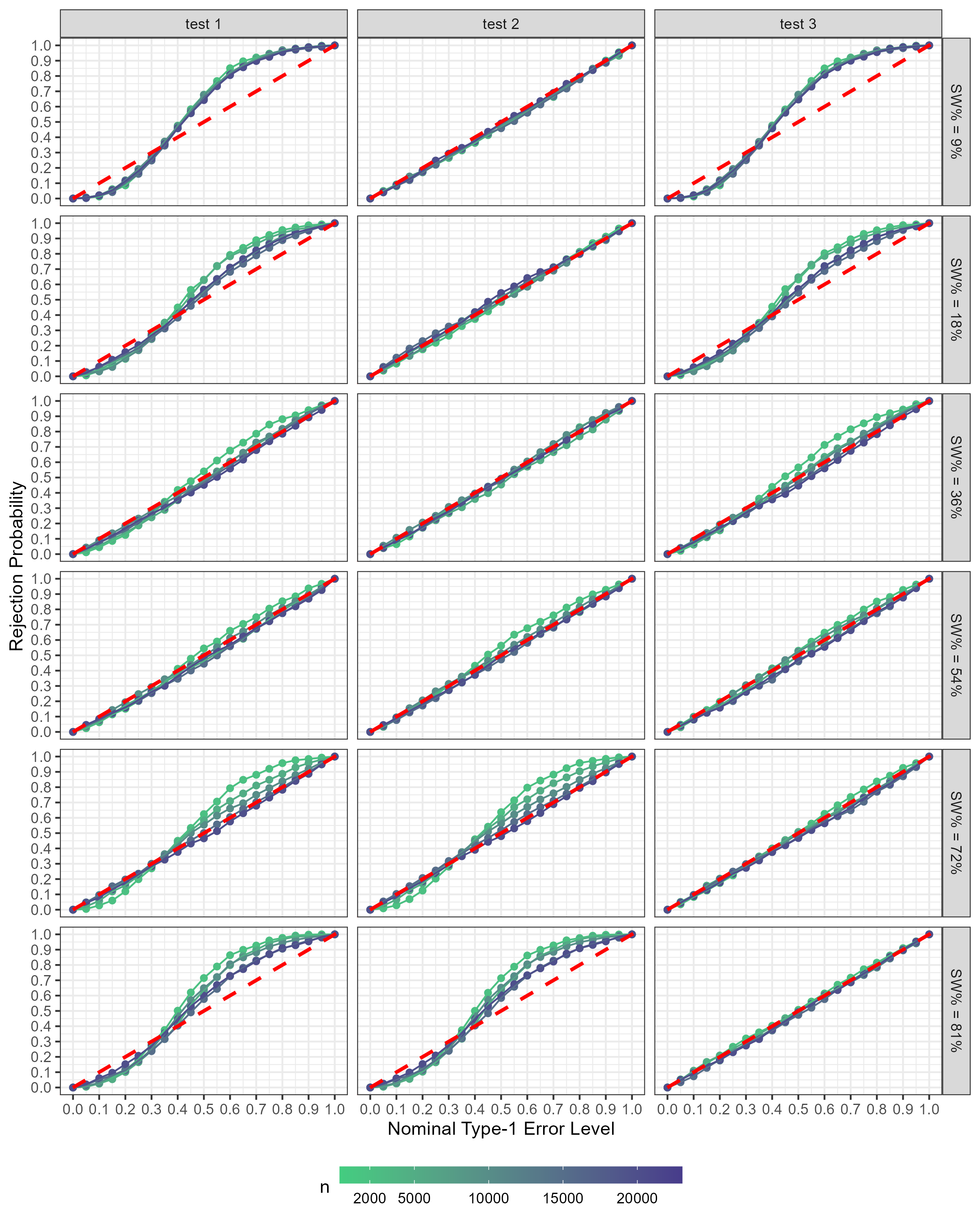}
    \caption{Simulation results for type-one-error control for the projection-based tests.} The red dash line is the theoretical asymptotic size for each setting under different levels.
    \label{fig: test_size_proj}
\end{figure}
\newpage
\begin{figure}[ht]
    \centering
\includegraphics[width=0.99\textwidth]{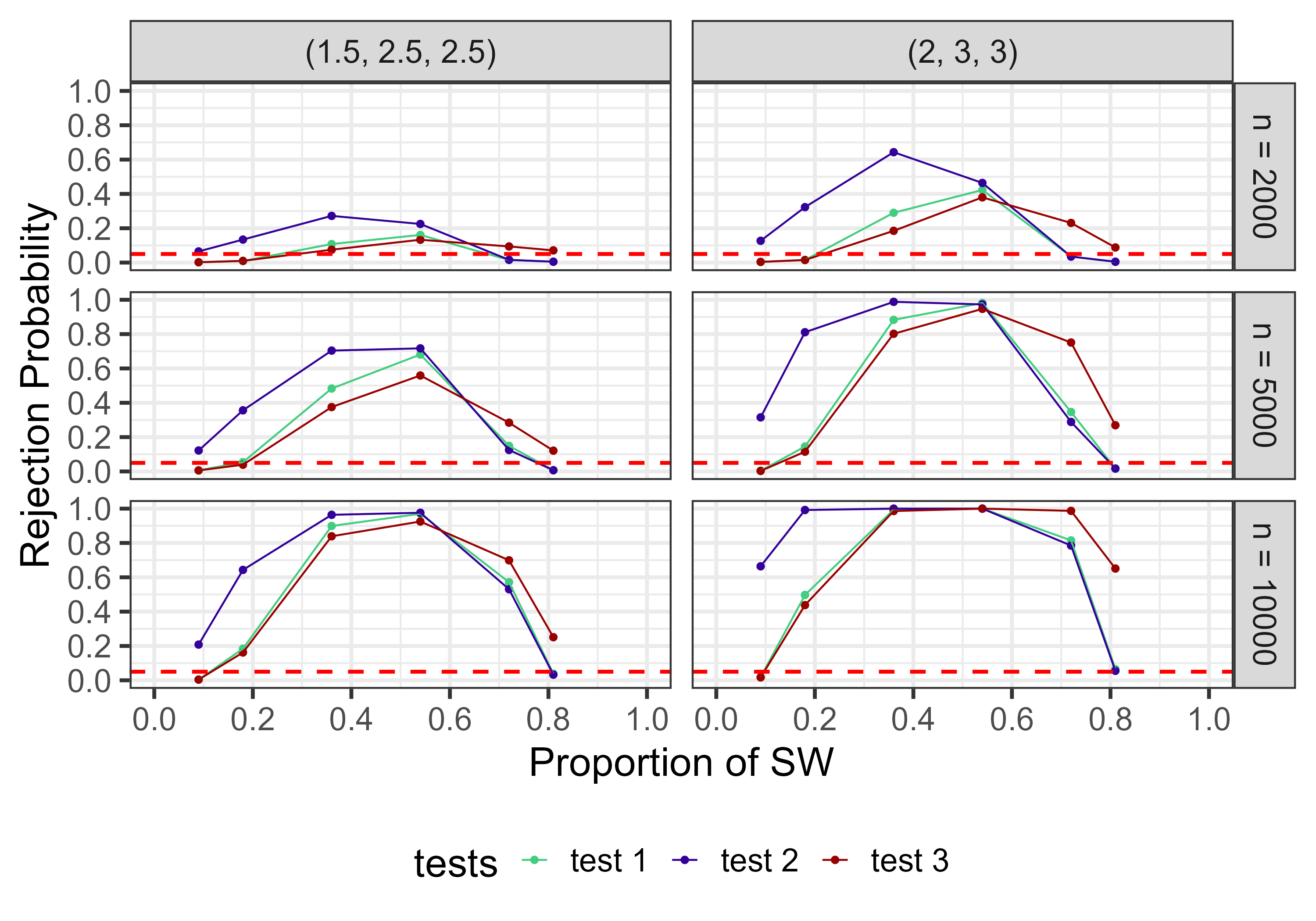}
    \caption{Comparison of power for the projection-based tests. The red dashed lines correspond to the nominal level. Test 1 compares $\text{SWATE}_{P_0}(\boldsymbol{X})$ to $\text{ACOATE}_{P_0}(\boldsymbol{X})$; test 2 compares $\text{COATE}_{P_0}(\boldsymbol{X})$ to $\text{ACOATE}_{P_0}(\boldsymbol{X})$; test 3 compares $\text{COATE}_{P_0}(\boldsymbol{X})$ to $\text{SWATE}_{P_0}(\boldsymbol{X})$.}
    \label{fig: test_power}
\end{figure}


\subsection{Additional simulation results for proposed estimator}
\label{subsec: simulation, estimators, supp}
\newpage
\begin{figure}[ht]
    \centering
\includegraphics[width=0.85\textwidth]{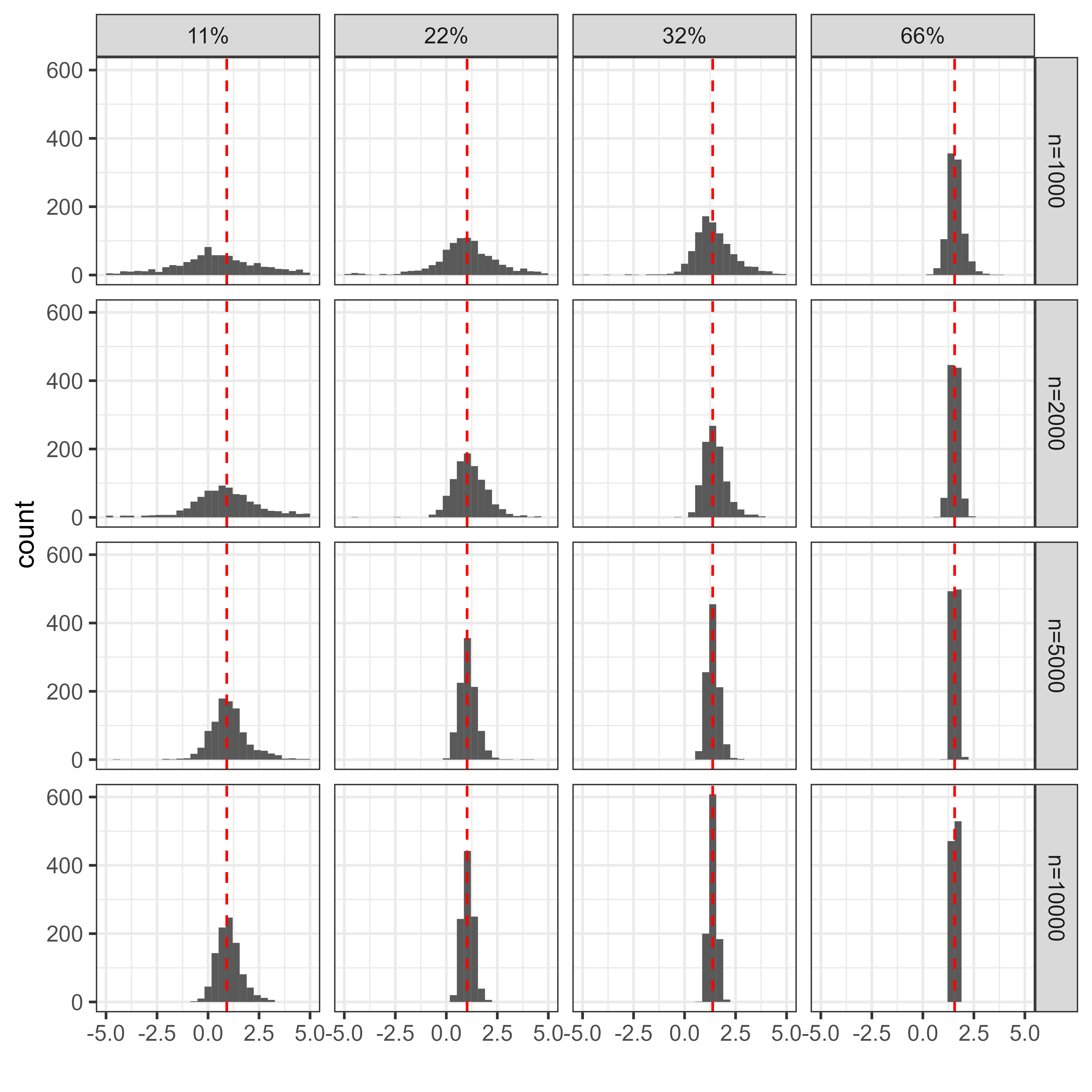}
    \caption{Sampling distribution of the estimating equation-based estimator when the outcome is continuous. Simulations are repeated 1000 times. Datasets are generating with $\beta=(2,2,2)$. Four  proportions of SW are considered. The dashed red lines represents the ground truth SWATE in each setting.}
    \label{fig: SWATE_est_continuous}
\end{figure}

\begin{figure}[H]
    \centering
\includegraphics[width=\textwidth]{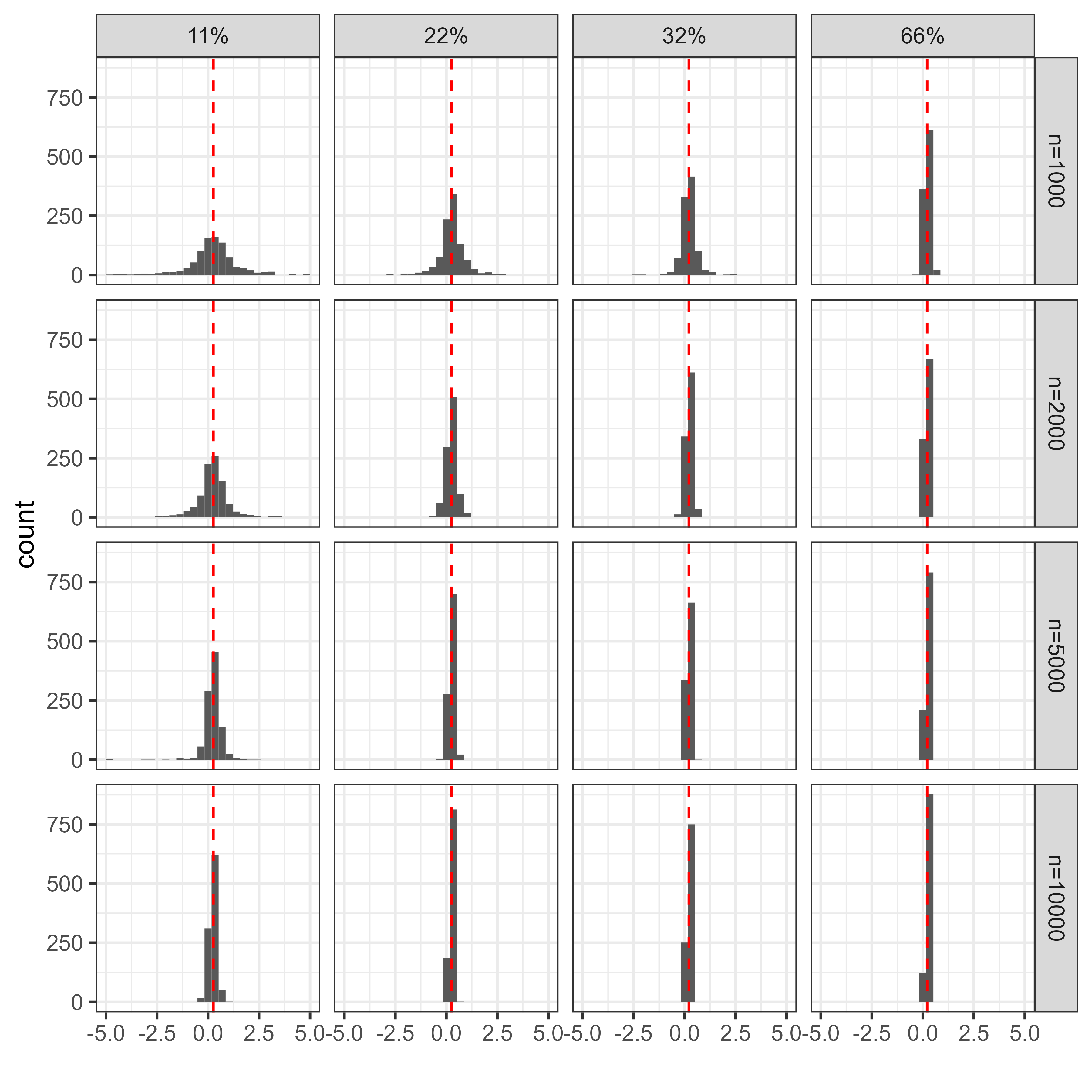}
    \caption{Sampling distribution of estimating equation estimator when the outcome is binary. Simulations are repeated 1000 times. Datasets are generating with $\beta=(0,1,-1)$. Four proportions of SW are considered. The dashed red lines represents the ground truth SWATE in each setting.}
    \label{fig: SWATE_est_binary_ee}
\end{figure}

\begin{figure}[H]
    \centering
\includegraphics[width=\textwidth]{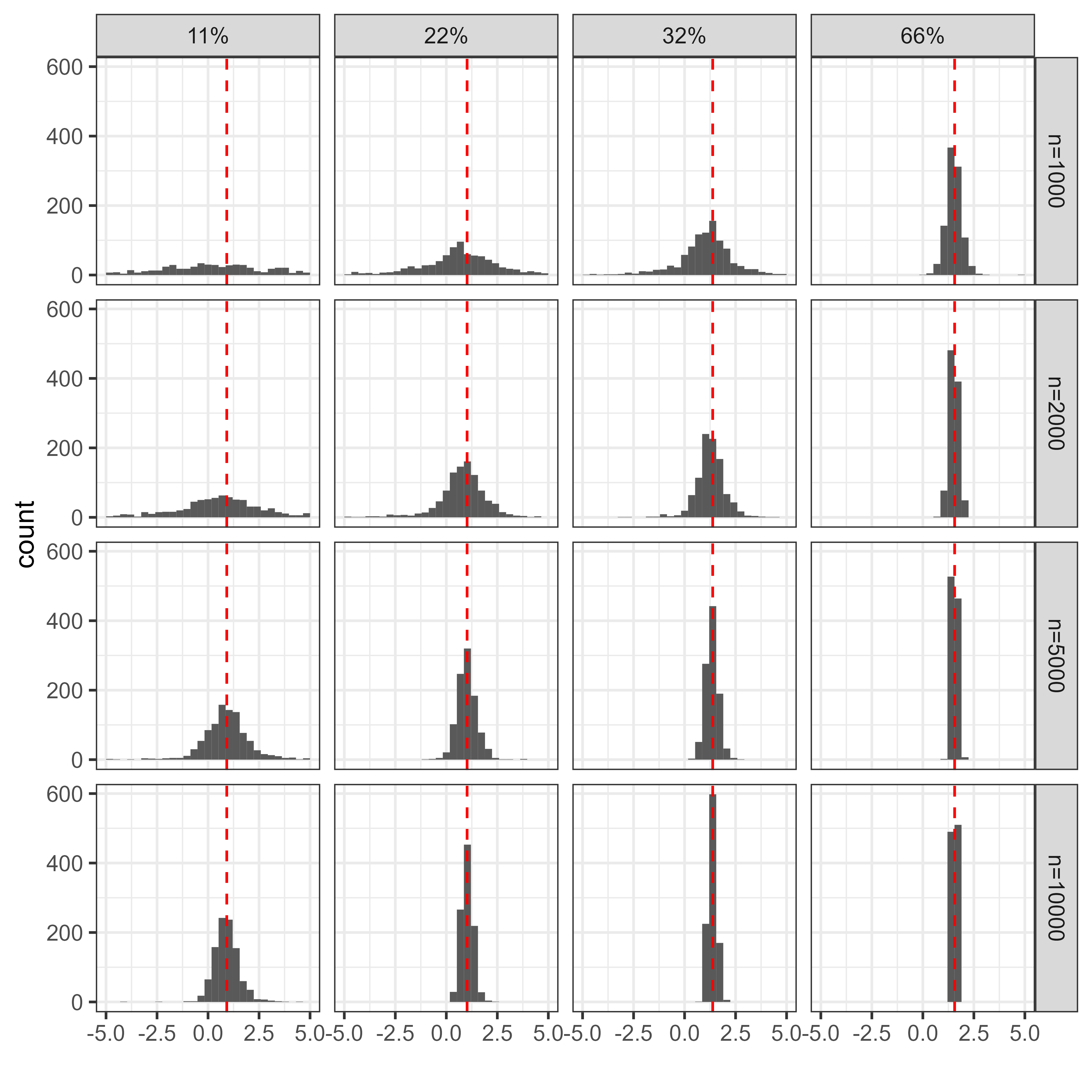}
    \caption{Sampling distribution of one-step estimator when the outcome is continuous. Simulations are repeated 1000 times. Datasets are generating with $\beta=(0,1,-1)$. Four proportions of SW are considered. The dashed red lines represents the ground truth SWATE in each setting.}
    \label{fig: SWATE_est_continuous_os}
\end{figure}

\begin{figure}[H]
    \centering
\includegraphics[width=\textwidth]{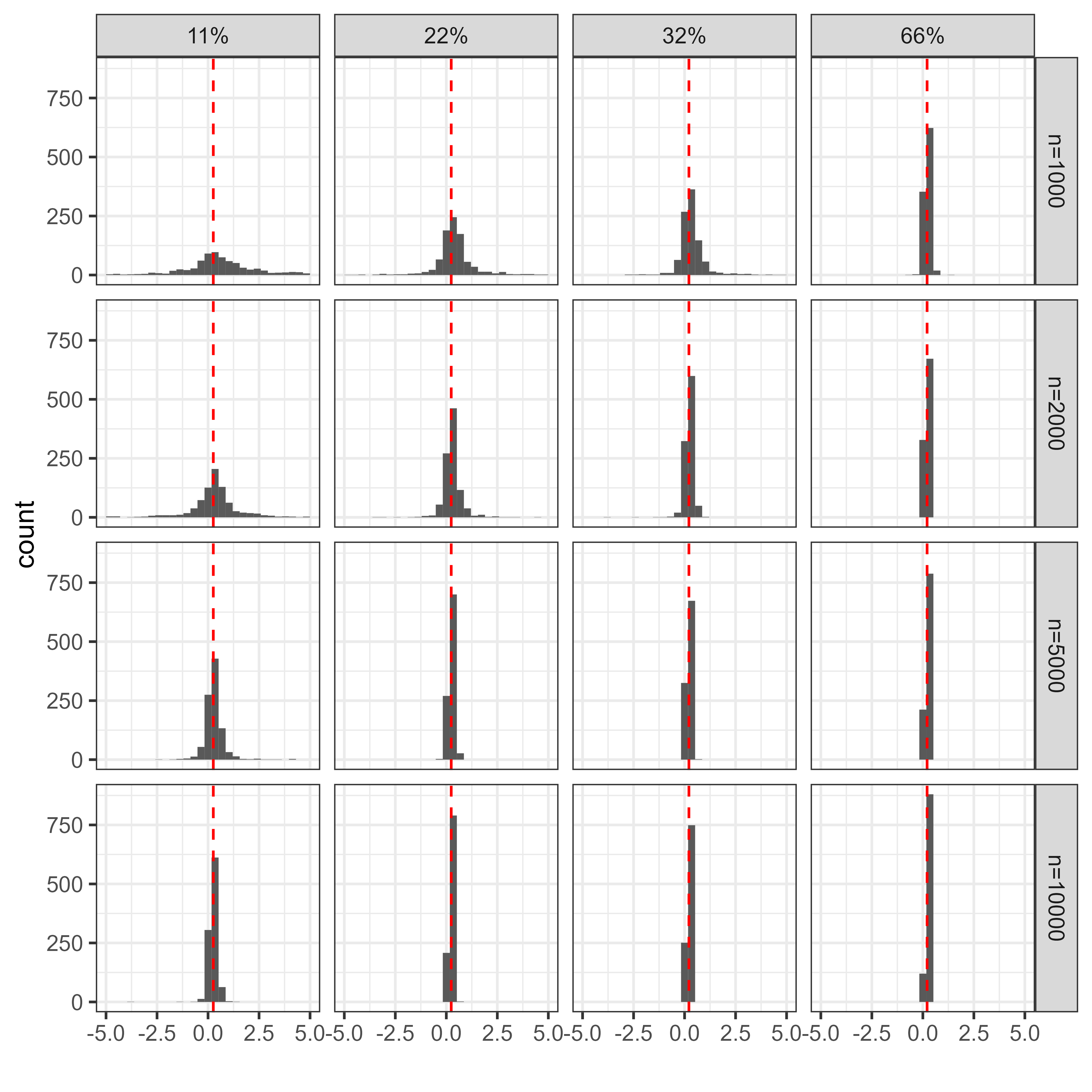}
    \caption{Sampling distribution of one-step estimator when the outcome is binary. Simulations are repeated 1000 times. Datasets are generating with $\beta=(0,1,-1)$. Four proportions of SW are considered. The dashed red lines represents the ground truth SWATE in each setting.}
    \label{fig: SWATE_est_binary_os}
\end{figure}
\begin{table}[htbp]
  \centering
  \footnotesize
  \begin{threeparttable}[b]
  \caption{Simulation results  for $\widehat{\psi}_{ee}$ when the outcome is binary}
    \begin{tabular}{cccccccccc}
    \toprule
    $\beta$  & SWI $\%$   & SWATE & Sample size & Estimate & Relative bias & Bias  & Se est & Coverage & Accep. rate \\
    \midrule
    \multirow{16}[8]{*}{(0,1-1)} & \multirow{4}[2]{*}{11\%} & \multirow{4}[2]{*}{0.257} & 1000  & 0.383 & 0.955 & 3.704 & 17.015 & 0.912 & 0.999 \\
          &       &       & 2000  & -0.070 & -0.328 & -1.272 & 4.664 & 0.956 & 1.000 \\
          &       &       & 5000  & 0.246 & -0.012 & -0.045 & 0.337 & 0.966 & 1.000 \\
          &       &       & 10000 & 0.247 & -0.011 & -0.043 & 0.171 & 0.949 & 1.000 \\
\cmidrule{2-10}          & \multirow{4}[2]{*}{22\%} & \multirow{4}[2]{*}{0.241} & 1000  & 0.192 & -0.048 & -0.200 & 1.232 & 0.937 & 1.000 \\
          &       &       & 2000  & 0.236 & -0.005 & -0.021 & 0.304 & 0.963 & 1.000 \\
          &       &       & 5000  & 0.248 & 0.008 & 0.032 & 0.136 & 0.962 & 1.000 \\
          &       &       & 10000 & 0.247 & 0.006 & 0.026 & 0.087 & 0.944 & 1.000 \\
\cmidrule{2-10}          & \multirow{4}[2]{*}{32\%} & \multirow{4}[2]{*}{0.212} & 1000  & 0.279 & 0.067 & 0.313 & 0.395 & 0.939 & 1.000 \\
          &       &       & 2000  & 0.226 & 0.014 & 0.064 & 0.178 & 0.976 & 1.000 \\
          &       &       & 5000  & 0.210 & -0.003 & -0.012 & 0.095 & 0.963 & 1.000 \\
          &       &       & 10000 & 0.213 & 0.001 & 0.003 & 0.063 & 0.964 & 1.000 \\
\cmidrule{2-10}          & \multirow{4}[2]{*}{66\%} & \multirow{4}[2]{*}{0.208} & 1000  & 0.224 & 0.016 & 0.075 & 0.121 & 0.934 & 1.000 \\
          &       &       & 2000  & 0.204 & -0.004 & -0.018 & 0.073 & 0.951 & 1.000 \\
          &       &       & 5000  & 0.206 & -0.003 & -0.012 & 0.043 & 0.956 & 1.000 \\
          &       &       & 10000 & 0.205 & -0.003 & -0.016 & 0.029 & 0.956 & 1.000 \\
    \midrule
    \multirow{16}[8]{*}{(0,2,-3)} & \multirow{4}[2]{*}{11\%} & \multirow{4}[2]{*}{0.524} & 1000  & 1.091 & 0.531 & 1.013 & 20.669 & 0.922 & 0.998 \\
          &       &       & 2000  & 0.576 & 0.051 & 0.097 & 2.751 & 0.935 & 1.000 \\
          &       &       & 5000  & 0.581 & 0.057 & 0.109 & 0.321 & 0.961 & 1.000 \\
          &       &       & 10000 & 0.541 & 0.016 & 0.031 & 0.174 & 0.952 & 1.000 \\
\cmidrule{2-10}          & \multirow{4}[2]{*}{22\%} & \multirow{4}[2]{*}{0.518} & 1000  & 0.805 & 0.287 & 0.555 & 1.293 & 0.924 & 1.000 \\
          &       &       & 2000  & 0.555 & -1.261 & -2.434 & 0.311 & 0.971 & 0.999 \\
          &       &       & 5000  & 0.530 & 0.012 & 0.023 & 0.140 & 0.968 & 1.000 \\
          &       &       & 10000 & 0.521 & 0.003 & 0.006 & 0.088 & 0.940 & 1.000 \\
\cmidrule{2-10}          & \multirow{4}[2]{*}{32\%} & \multirow{4}[2]{*}{0.511} & 1000  & 0.510 & -0.002 & -0.003 & 0.376 & 0.960 & 1.000 \\
          &       &       & 2000  & 0.520 & 0.009 & 0.018 & 0.188 & 0.970 & 1.000 \\
          &       &       & 5000  & 0.512 & 0.001 & 0.002 & 0.098 & 0.974 & 1.000 \\
          &       &       & 10000 & 0.512 & 0.001 & 0.001 & 0.064 & 0.968 & 1.000 \\
\cmidrule{2-10}          & \multirow{4}[2]{*}{66\%} & \multirow{4}[2]{*}{0.510} & 1000  & 0.522 & 0.012 & 0.023 & 0.126 & 0.935 & 1.000 \\
          &       &       & 2000  & 0.513 & 0.003 & 0.006 & 0.077 & 0.950 & 1.000 \\
          &       &       & 5000  & 0.512 & 0.001 & 0.003 & 0.044 & 0.955 & 1.000 \\
          &       &       & 10000 & 0.510 & 0.000 & -0.001 & 0.030 & 0.954 & 1.000 \\
    \bottomrule
    \label{tab:simu_SWATE_ee_bin}
    \end{tabular}%
    \begin{tablenotes}
       \item [1] True values of SWATE under each setting.
       \item [2] Truncated means of point estimates. Point estimates are truncated if absolute value are bigger than $500$.
       \item [3] $90\%$ Winsorized mean of standard error estimates.
       \item [4] Proportion of point estimates which are not truncated.
     \end{tablenotes}
  \end{threeparttable}
  
\end{table}%

\begin{table}[htbp]
  \centering
  \footnotesize
  \begin{threeparttable}[b]
  \caption{Simulation results  for $\widehat{\psi}_{os}$ when the outcome is continuous}
    \begin{tabular}{cccccccccc}
    \toprule
    $\beta$  & SWI $\%$   & SWATE & Sample size & Estimate & Relative bias & Bias  & Se est & Coverage & Accep. rate \\
    \midrule
    \multirow{16}[8]{*}{(2,2,2)} & \multirow{4}[2]{*}{11\%} & \multirow{4}[2]{*}{0.917} & 1000  & -3.122 & -4.040 & -4.402 & 43.614 & 0.863 & 0.947 \\
          &       &       & 2000  & -5.087 & -6.004 & -6.543 & 10.774 & 0.864 & 0.976 \\
          &       &       & 5000  & 0.744 & -0.174 & -0.190 & 1.018 & 0.928 & 0.997 \\
          &       &       & 10000 & 0.902 & -0.015 & -0.017 & 0.546 & 0.932 & 1.000 \\
\cmidrule{2-10}          & \multirow{4}[2]{*}{22\%} & \multirow{4}[2]{*}{1.019} & 1000  & -7.244 & -8.263 & -8.109 & 4.852 & 0.849 & 0.987 \\
          &       &       & 2000  & 0.597 & -0.422 & -0.415 & 0.966 & 0.914 & 0.998 \\
          &       &       & 5000  & 1.008 & -0.011 & -0.011 & 0.430 & 0.943 & 1.000 \\
          &       &       & 10000 & 1.025 & 0.006 & 0.006 & 0.273 & 0.953 & 1.000 \\
\cmidrule{2-10}          & \multirow{4}[2]{*}{32\%} & \multirow{4}[2]{*}{1.377} & 1000  & -0.404 & -1.781 & -1.293 & 1.259 & 0.885 & 0.996 \\
          &       &       & 2000  & 1.232 & -0.146 & -0.106 & 0.606 & 0.931 & 1.000 \\
          &       &       & 5000  & 1.334 & -0.044 & -0.032 & 0.304 & 0.936 & 1.000 \\
          &       &       & 10000 & 1.364 & -0.014 & -0.010 & 0.202 & 0.946 & 1.000 \\
\cmidrule{2-10}          & \multirow{4}[2]{*}{66\%} & \multirow{4}[2]{*}{1.557} & 1000  & 1.531 & -0.026 & -0.017 & 0.341 & 0.938 & 1.000 \\
          &       &       & 2000  & 1.522 & -0.036 & -0.023 & 0.210 & 0.934 & 1.000 \\
          &       &       & 5000  & 1.550 & -0.007 & -0.005 & 0.122 & 0.938 & 1.000 \\
          &       &       & 10000 & 1.557 & -0.001 & -0.001 & 0.083 & 0.951 & 1.000 \\
    \midrule
    \multirow{16}[8]{*}{(4,4,4)} & \multirow{4}[2]{*}{11\%} & \multirow{4}[2]{*}{0.906} & 1000  & -4.529 & -5.436 & -5.994 & 51.315 & 0.859 & 0.946 \\
          &       &       & 2000  & -1.545 & -2.452 & -2.704 & 9.969 & 0.856 & 0.971 \\
          &       &       & 5000  & 0.090 & -0.817 & -0.901 & 1.216 & 0.910 & 0.992 \\
          &       &       & 10000 & 0.941 & 0.035 & 0.038 & 0.607 & 0.935 & 1.000 \\
\cmidrule{2-10}          & \multirow{4}[2]{*}{22\%} & \multirow{4}[2]{*}{1.146} & 1000  & -2.536 & -3.682 & -3.211 & 5.562 & 0.865 & 0.978 \\
          &       &       & 2000  & -0.637 & -1.783 & -1.555 & 1.298 & 0.906 & 0.995 \\
          &       &       & 5000  & 1.080 & -0.067 & -0.058 & 0.546 & 0.953 & 1.000 \\
          &       &       & 10000 & 1.108 & -0.038 & -0.033 & 0.340 & 0.942 & 1.000 \\
\cmidrule{2-10}          & \multirow{4}[2]{*}{32\%} & \multirow{4}[2]{*}{1.699} & 1000  & -0.306 & -2.006 & -1.180 & 1.721 & 0.907 & 0.998 \\
          &       &       & 2000  & 1.412 & -0.287 & -0.169 & 0.853 & 0.945 & 1.000 \\
          &       &       & 5000  & 1.615 & -0.084 & -0.050 & 0.425 & 0.955 & 1.000 \\
          &       &       & 10000 & 1.688 & -0.011 & -0.006 & 0.278 & 0.953 & 1.000 \\
\cmidrule{2-10}          & \multirow{4}[2]{*}{66\%} & \multirow{4}[2]{*}{2.315} & 1000  & 2.199 & -0.117 & -0.050 & 0.574 & 0.943 & 1.000 \\
          &       &       & 2000  & 2.286 & -0.029 & -0.013 & 0.350 & 0.952 & 1.000 \\
          &       &       & 5000  & 2.284 & -0.032 & -0.014 & 0.202 & 0.954 & 1.000 \\
          &       &       & 10000 & 2.298 & -0.018 & -0.008 & 0.138 & 0.967 & 1.000 \\
    \bottomrule
    \label{tab:simu_SWATE_os_con}
    \end{tabular}%
    \begin{tablenotes}
       \item [1] True values of SWATE under each setting.
       \item [2] Truncated means of point estimates. Point estimates are truncated if absolute value are bigger than $500$.
       \item [3] $90\%$ Winsorized mean of standard error estimates.
       \item [4] Proportion of point estimates which are not truncated.
     \end{tablenotes}
  \end{threeparttable}
  
\end{table}%

\begin{table}[htbp]
  \centering
  \footnotesize
  \begin{threeparttable}[b]
  \caption{Simulation results  for $\widehat{\psi}_{os}$ when the outcome is binary}
    \begin{tabular}{cccccccccc}
    \toprule
    $\beta$  & SWI $\%$   & SWATE & Sample size & Estimate & Relative bias & Bias  & Se est & Coverage & Accep. rate \\
    \midrule
    \multirow{16}[8]{*}{(0,1-1)} & \multirow{4}[2]{*}{11\%} & \multirow{4}[2]{*}{0.257} & 1000  & 0.769 & 0.512 & 1.984 & 2.679 & 0.914 & 0.790 \\
          &       &       & 2000  & 0.404 & 0.146 & 0.568 & 1.090 & 0.917 & 0.884 \\
          &       &       & 5000  & 0.284 & 0.026 & 0.100 & 0.370 & 0.922 & 0.988 \\
          &       &       & 10000 & 0.255 & -0.003 & -0.012 & 0.172 & 0.937 & 1.000 \\
\cmidrule{2-10}          & \multirow{4}[2]{*}{22\%} & \multirow{4}[2]{*}{0.241} & 1000  & 0.444 & 0.204 & 0.846 & 0.859 & 0.915 & 0.931 \\
          &       &       & 2000  & 0.219 & -0.022 & -0.090 & 0.324 & 0.924 & 0.998 \\
          &       &       & 5000  & 0.242 & 0.002 & 0.007 & 0.136 & 0.953 & 1.000 \\
          &       &       & 10000 & 0.243 & 0.002 & 0.010 & 0.087 & 0.956 & 1.000 \\
\cmidrule{2-10}          & \multirow{4}[2]{*}{32\%} & \multirow{4}[2]{*}{0.212} & 1000  & 0.381 & 0.168 & 0.792 & 0.414 & 0.927 & 0.984 \\
          &       &       & 2000  & 0.233 & 0.021 & 0.097 & 0.186 & 0.955 & 1.000 \\
          &       &       & 5000  & 0.215 & 0.002 & 0.011 & 0.095 & 0.944 & 1.000 \\
          &       &       & 10000 & 0.210 & -0.002 & -0.011 & 0.063 & 0.956 & 1.000 \\
\cmidrule{2-10}          & \multirow{4}[2]{*}{66\%} & \multirow{4}[2]{*}{0.208} & 1000  & 0.214 & 0.006 & 0.029 & 0.120 & 0.954 & 1.000 \\
          &       &       & 2000  & 0.209 & 0.000 & 0.002 & 0.073 & 0.945 & 1.000 \\
          &       &       & 5000  & 0.205 & -0.003 & -0.013 & 0.043 & 0.951 & 1.000 \\
          &       &       & 10000 & 0.205 & -0.003 & -0.015 & 0.029 & 0.952 & 1.000 \\
    \midrule
    \multirow{16}[8]{*}{(0,2,-3)} & \multirow{4}[2]{*}{11\%} & \multirow{4}[2]{*}{0.524} & 1000  & 0.506 & -0.019 & -0.036 & 2.304 & 0.917 & 0.796 \\
          &       &       & 2000  & 0.495 & -0.029 & -0.056 & 1.725 & 0.919 & 0.894 \\
          &       &       & 5000  & 0.461 & -0.064 & -0.122 & 0.363 & 0.930 & 0.994 \\
          &       &       & 10000 & 0.513 & -0.011 & -0.022 & 0.179 & 0.931 & 0.999 \\
\cmidrule{2-10}          & \multirow{4}[2]{*}{22\%} & \multirow{4}[2]{*}{0.518} & 1000  & 0.505 & -0.013 & -0.026 & 0.853 & 0.904 & 0.923 \\
          &       &       & 2000  & 0.440 & -0.078 & -0.151 & 0.331 & 0.930 & 0.987 \\
          &       &       & 5000  & 0.508 & -0.010 & -0.020 & 0.139 & 0.963 & 1.000 \\
          &       &       & 10000 & 0.521 & 0.003 & 0.005 & 0.089 & 0.953 & 1.000 \\
\cmidrule{2-10}          & \multirow{4}[2]{*}{32\%} & \multirow{4}[2]{*}{0.511} & 1000  & 0.445 & -0.067 & -0.131 & 0.397 & 0.911 & 0.989 \\
          &       &       & 2000  & 0.491 & -0.020 & -0.040 & 0.197 & 0.953 & 1.000 \\
          &       &       & 5000  & 0.504 & -0.007 & -0.014 & 0.098 & 0.950 & 1.000 \\
          &       &       & 10000 & 0.509 & -0.002 & -0.004 & 0.065 & 0.965 & 1.000 \\
\cmidrule{2-10}          & \multirow{4}[2]{*}{66\%} & \multirow{4}[2]{*}{0.51} & 1000  & 0.514 & 0.004 & 0.008 & 0.130 & 0.939 & 1.000 \\
          &       &       & 2000  & 0.508 & -0.003 & -0.005 & 0.077 & 0.951 & 1.000 \\
          &       &       & 5000  & 0.507 & -0.004 & -0.007 & 0.045 & 0.956 & 1.000 \\
          &       &       & 10000 & 0.507 & -0.003 & -0.006 & 0.030 & 0.950 & 1.000 \\
    \bottomrule
    \label{tab:simu_SWATE_os_bin}
    \end{tabular}%
    \begin{tablenotes}
       \item [1] True values of SWATE under each setting.
       \item [2] Truncated means of point estimates. Point estimates are truncated if absolute value are bigger than $500$.
       \item [3] $90\%$ Winsorized mean of standard error estimates.
       \item [4] Proportion of point estimates which are not truncated.
     \end{tablenotes}
  \end{threeparttable}
  
\end{table}%


\subsection{Additional simulation results when the conditional treatment effects are not linear in covariates} 
\label{subsec: addtional simulation, nonlinear, supp}
\noindent\textbf{Outcome:} We consider the following new data generating process for potential outcomes to allow for nonlinear conditional treatment effects:
{\small\begin{equation*}
\begin{split}
Y(0) = &1+X_1+X_2+U+\epsilon, \\
    Y(1) = &\mathbbm{1}\{S= \text{ANT} \text{ or } \text{AAT}\}\cdot(1+X_1+X_2+U)
    +\mathbbm{1}\{S= \text{AT-NT} \text{ or } \text{NT-AT}\}\cdot(1+X_1+X_2+U)\\
    &+\mathbbm{1}\{S= \text{SW}\}\cdot(\beta_1+2X_1+2X_2+\beta_2\cos(\boldsymbol{x}_1)+\beta_3\text{expit}(\boldsymbol{x}_2)+U)\\
     &+\mathbbm{1}\{S= \text{ACO}\}\cdot(\beta_1+2X_1+2X_2+2\cos(\boldsymbol{x}_1)+2\text{expit}(\boldsymbol{x}_2)+U)+\epsilon,
\end{split}
\end{equation*}}

The data generating distributions for other variables stay the same. We consider $(\beta_1,\beta_2,\beta_3)= (1,2,2)$, $(1,2.5,2.5)$, and $(2,3,3)$. When $\beta = (1,2,2)$, the null hypothesis $H^{(j)}_0$ holds, i.e., $\theta^{(j)}_{P_0}=0$. Conversely, $\beta = (1,2.5,2.5)$ and $(2,3,3)$ correspond to $\theta^{(j)}_{P_0}\neq 0$. For each null hypothesis $H^{(j)}_{0}$, $j = 1, 2, 3$, we construct the projection-based, $L^2$ norm-based, and $L^\infty$ norm-based tests and repeated the simulation $1000$ times.


\subsection{Additional simulation results when IV assumptions are violated}
\label{subsec: simulation when IV assumption fail, supp}
{ 
The proposed tests can be viewed as joint tests for the identification assumptions. In this session, we consider the case when the exclusion restriction assumption is violated, that is, $Z$ has direct effect on the outcome $Y$. Let $Z^* = 1\{Z\in \{1_a,1_b\}\}$. We modify only the outcome models and leave all other models unchanged.

\noindent\textbf{Outcome:} We consider the following data generating process for potential outcomes:
{\small\begin{equation*}
\begin{split}
Y(0) = &1+X_1+X_2+U+Z^*+\epsilon, \\
    Y(1) = &\mathbbm{1}\{S= \text{ANT} \text{ or } \text{AAT}\}\cdot(1+X_1+X_2+U)
    +\mathbbm{1}\{S= \text{AT-NT} \text{ or } \text{NT-AT}\}\cdot(1+X_1+X_2+U)\\
    &+\mathbbm{1}\{S= \text{SW}\}\cdot(\beta_1+\beta_2X_1+\beta_3X_2+U)
     +\mathbbm{1}\{S= \text{ACO}\}\cdot(1+2X_1+2X_2+U)+Z^*+\epsilon,
\end{split}
\end{equation*}}
\noindent where parameters $(\beta_1,\beta_2,\beta_3)$ control the effect size of SWATE, and $\epsilon \sim N(0,1)$. We consider $(\beta_1,\beta_2,\beta_3)= (1,2,2)$, $(1,2.5,2.5)$, and $(2,3,3)$. In our previous simulation settings in Section \ref{subsec: simu test homogeneity}, $(\beta_1,\beta_2,\beta_3)= (1,2,2)$ corresponds to the null hypothesis. Under the new outcome model the conditional SWATE still equals the conditional ACOATE. However, the core IV assumption is violated, thus the identification formulas we proposed are no longer valid. We expect the our proposed tests should reject the null hypothesis under all of those three settings. Figure \ref{fig: project_test_power_mis} shows the power of the proposed projection tests when the exclusion restriction assumption is violated. In contrast to the results in  Section \ref{subsec: simu test homogeneity}, when $(\beta_1,\beta_2,\beta_3)= (1,2,2)$ our proposed test still have non-trivial power.

\begin{figure}
    \centering
    \includegraphics[width=1\linewidth]{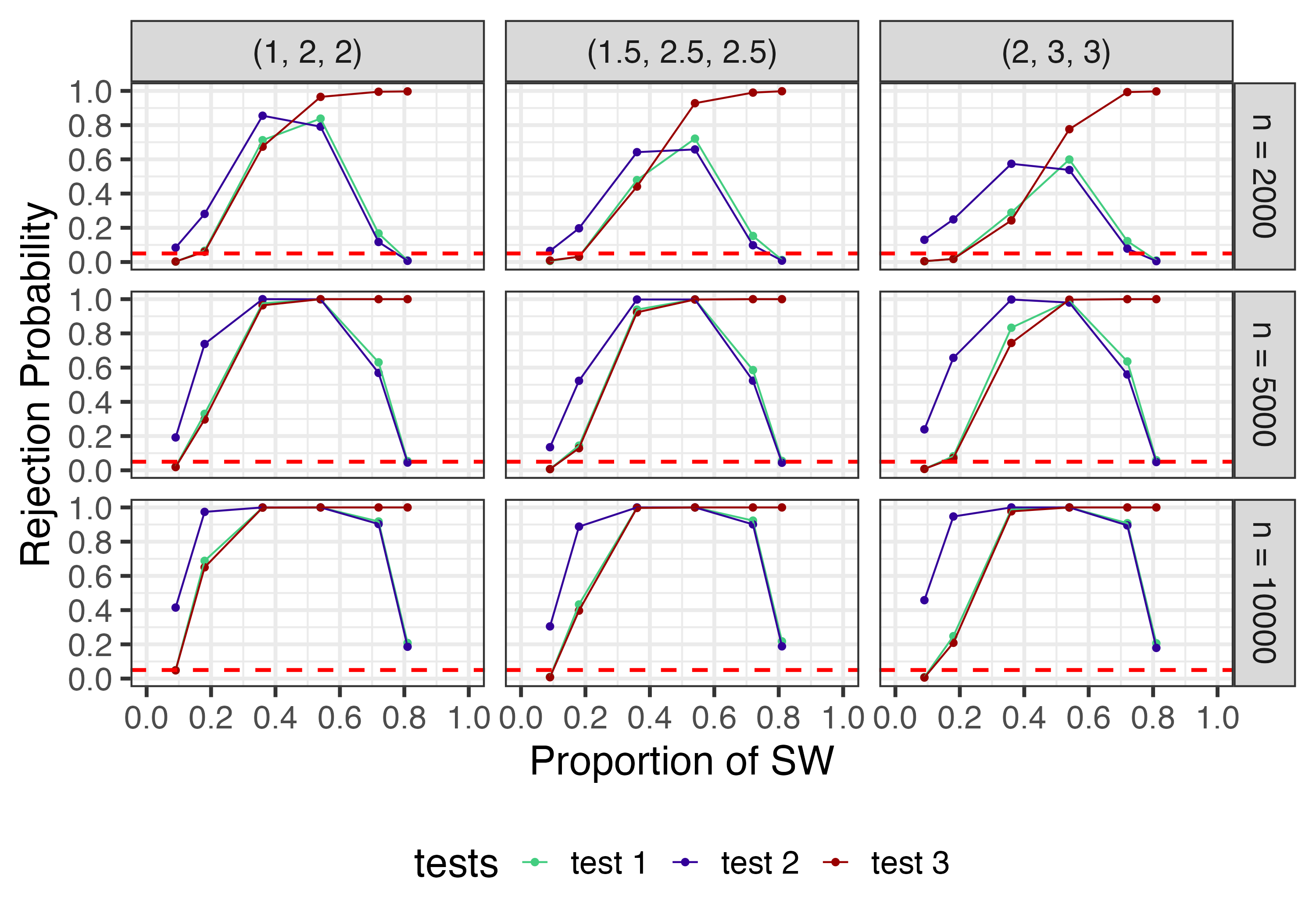}
    \caption{Comparison of power for the projection-based tests when the exclusion restriction assumption is violated. The red dashed lines correspond to the nominal level. Test 1 compares $\text{SWATE}_{P_0}(\boldsymbol{X})$ to $\text{ACOATE}_{P_0}(\boldsymbol{X})$; test 2 compares $\text{COATE}_{P_0}(\boldsymbol{X})$ to $\text{ACOATE}_{P_0}(\boldsymbol{X})$; test 3 compares $\text{COATE}_{P_0}(\boldsymbol{X})$ to $\text{SWATE}_{P_0}(\boldsymbol{X})$.}
    \label{fig: project_test_power_mis}
\end{figure}
}

\clearpage
\section{Additional real data analysis results}
\label{sec: additional real data, supp}
\subsection{Descriptive statistics}
\begin{table}[htbp]
  \centering
  \caption{Baseline characteristics by study stages and arms. Continuous variables are presented as mean (sd) and categorical variables are presented as count (proportion).}
  \resizebox{\textwidth}{!}{\begin{tabular}{lcccc}
    \toprule
          & \multicolumn{2}{c}{Prior to 1997 ($G=a$)} & \multicolumn{2}{c}{After 1997 ($G=b$)} \\
    \midrule
          & Control arm & Treatment arm & Control arm & Treatment arm \\
    \midrule
    Samlpe size & 4,210 & 4,204 & 4,970 & 4,978 \\
    Treatment uptake (\%) &       &       &       &  \\
      \qquad No screening &  4210 (100.0)  &  2063 (49.1)  &  4970 (100.0)  &   989 (19.9)  \\
      \qquad Screening &     0 (  0.0)  &  2141 (50.9)  &     0 (  0.0)  &  3989 (80.1)  \\
    Age   & 64.75 (5.09) & 64.69 (5.08) & 60.40 (5.26) & 60.42 (5.31) \\
    Sex (\%) &       &       &       &  \\
     \qquad  Female &  2642 ( 62.8)  &  2638 (62.7)  &  2810 ( 56.5)  &  2798 (56.2)  \\
     \qquad  Male &  1568 ( 37.2)  &  1566 (37.3)  &  2160 ( 43.5)  &  2180 (43.8)  \\
    Race (\%) &       &       &       &  \\
     \qquad  White &  3346 ( 79.5)  &  3289 (78.2)  &  4266 ( 85.8)  &  4278 (85.9)  \\
     \qquad  Minority &   864 ( 20.5)  &   915 (21.8)  &   704 ( 14.2)  &   700 (14.1)  \\
    Education (\%) &       &       &       &  \\
     \qquad  No high school &   585 ( 13.9)  &   616 (14.7)  &   351 (  7.1)  &   369 ( 7.4)  \\
     \qquad  High school &  1775 ( 42.2)  &  1639 (39.0)  &  1586 ( 31.9)  &  1589 (31.9)  \\
     \qquad  College or above &  1850 ( 43.9)  &  1949 (46.4)  &  3033 ( 61.0)  &  3020 (60.7)  \\
    Smoking Status (\%) &       &       &       &  \\
     \qquad  Non-smoker &  1803 ( 42.8)  &  1798 (42.8)  &  2101 ( 42.3)  &  2076 (41.7)  \\
     \qquad  Current smoker &   562 ( 13.3)  &   573 (13.6)  &   700 ( 14.1)  &   729 (14.6)  \\
     \qquad  Former smoker &  1845 ( 43.8)  &  1833 (43.6)  &  2169 ( 43.6)  &  2173 (43.7)  \\
    BMI (\%) &       &       &       &  \\
     \qquad  $<25$ &  1439 ( 34.2)  &  1446 (34.4)  &  1438 ( 28.9)  &  1486 (29.9)  \\
     \qquad  $>25$ &  2771 ( 65.8)  &  2758 (65.6)  &  3532 ( 71.1)  &  3492 (70.1)  \\
    Colorectal cancer (\%) &       &       &       &  \\
     \qquad  No Confirmed Cancer &  4128 ( 98.1)  &  4139 (98.5)  &  4905 ( 98.7)  &  4920 (98.8)  \\
     \qquad  Confirmed Cancer &    82 (  1.9)  &    65 ( 1.5)  &    65 (  1.3)  &    58 ( 1.2)  \\
    Following time & 12.08 (3.87) & 11.45 (4.47) &  9.64 (2.31) &  9.55 (2.47) \\
    Rate (per 1000 person-yrs) & 1.61 & 1.35 & 1.36 & 1.22 \\
    \bottomrule
    \end{tabular}}
  \label{tab:descriptive statistics, supp}%
\end{table}%

\begin{table}[htbp]
  \centering
  \caption{Estimated baseline characteristics of switchers (SW) and always-compliers (ACO)}
    \begin{tabular}{lcc}
    \toprule
          & Switchers    & Always-compliers \\
    \midrule
    Age   & 62.6  & 62.1 \\
    Sex (\%) &       &  \\
      \qquad Female & 71.6\% & 50.4\% \\
      \qquad Male & 28.4\%  & 49.6\%\\
    Race (\%) &       &  \\
      \qquad White & 88.0\% & 80.8\% \\
      \qquad Minority & 12.0\% & 19.2\% \\
    Education (\%) &       &  \\
      \qquad No high school & 9.3\% & 10.0\% \\
      \qquad High school & 40.0\% & 33.1\% \\
      \qquad College or above & 50.7\% & 56.9\% \\
    Smoking Status (\%) &       &  \\
      \qquad Non-smoker & 41.5\% & 42.8\% \\
      \qquad Current smoker & 15.3\% & 12.1\% \\
      \qquad Former smoker & 43.2\% & 45.1\% \\
    BMI (\%) &       &  \\
      \qquad $<25$ & 36.0\% & 28.7\% \\
      \qquad $>25$ & 64.0\% & 71.3\% \\
    \bottomrule
    \end{tabular}%
  \label{tab:characterize SW and AC}%
\end{table}%

\clearpage
\subsection{Integrated analysis of multiple clinical sites}
\label{subsec: case study analysis of multiple sites, supp}
We consider an integrated analysis of three clinical sites as a second application of the proposed method. We consider the Henry Ford Health system during the dual consent stage (denoted as $G=a$), University of Colorado (denoted as $G=b$), and University of Alabama at Birmingham (denoted as $G=c$). The compliance rates in these three sites are 50.9\%, 82.6\% and 96.0\%, respectively (see Table \ref{tab: descriptive statistics 2} in Supplementary Material). We consider the following nested IV assumption: IV pair $\mathcal{Z}_a = \{1_a,0_a\}$ is nested in $\mathcal{Z}_b = \{1_b,0_b\}$, and $\mathcal{Z}_c = \{1_c,0_c\}$. Using the notation introduced in Section \ref{subsec: multiple IV pairs, supp}, we can write  $\mathcal{Z}_a\preceq \mathcal{Z}_b$ and $\mathcal{Z}_a\preceq \mathcal{Z}_c$. Two switcher populations can be defined accordingly, one associated with $\mathcal{Z}_a$ and $\mathcal{Z}_b$ and the other with $\mathcal{Z}_a$ and $\mathcal{Z}_c$.

Table \ref{tab: charactrization 2} in the Supplemental Material characterizes the mean values of covariates among switchers and always-compliers associated with $\mathcal{Z}_a$ and $\mathcal{Z}_b$ and with $\mathcal{Z}_a$ and $\mathcal{Z}_c$. Interestingly, in both cases, we found the switcher population was more likely to be female, white, and having BMI $< 25$. 

We also implemented the proposed one-step estimator and estimating equation-based estimator for the switcher average treatment effects. For the dataset with Henry Fold Health system and University of Colorado, the risk difference was estimated to be $-0.008$ ($95\%$ CI: $-0.033$ to $0.017$) for switchers and $-0.011$ ($95\%$ CI: $-0.023$ to $0.001$) for always-compliers based on the one-step estimator. The results obtained from the estimating equation-based estimator were almost the same: $-0.008$ ($95\%$ CI: $-0.033$ to $0.017$) for switchers and $-0.011$ ($95\%$ CI: $-0.023$ to $0.001$) for always-compliers. The risk difference estimated by Wald estimator is $-0.012$ ($95\%$CI: $-0.040$ to $0.015$) or switchers and $-0.009$ ($95\%$ CI: $-0.021$ to $0.002$ for always-compliers). The test statistics based on best least-squares projection are 5.06, 3.31 and 5.92, respectively. Therefore there is no evidence to reject the null hypothesis that there is no treatment effect heterogeneity. For the dataset with Henry Fold Health system and University of Alabama Birmingham, the 15-year risk difference was estimated to be $-0.006$ ($95\%$ CI: $-0.017$ to $0.006$) for switchers and $-0.008$ ($95\%$ CI: $-0.017$ to $0.002$) for always-compliers based on the one-step estimator. The results obtained from the estimating equation-based estimator were also very similar: $-0.006$ ($95\%$ CI: $-0.017$ to $0.006$) for switchers and $-0.008$ ($95\%$ CI: $-0.017$ to $0.002$) for always-compliers. The risk difference estimated by Wald estimator is $0.002$ ($95\%$CI: $-0.018$ to $0.022$) for switchers and $-0.007$ ($95\%$ CI: $-0.018$ to $0.004$ for always-compliers). The test statistics based on best least-squares projection are 3.88, 4.56 and 4.11 respectively. Therefore, we have no evidence to reject the null hypothesis.

\begin{table}[htbp]
  \centering
  \tiny
  \caption{Baseline characteristics by study stages and arms. Continuous variables are presented as mean (sd) and categorical variables are presented as count (proportion).}
       \begin{tabular}{lcccccc}
    \toprule
          & \multicolumn{2}{c}{Henry Ford Health System} & \multicolumn{2}{c}{University of Colorado} & \multicolumn{2}{p{13.59em}}{University of Alabama\newline{}Birmingham} \\
    \midrule
          & Control arm & Treatment arm & Control arm & Treatment arm & Control arm & Treatment arm \\
    \midrule
    Samlpe size & 4,210 & 4,204 & 6,278 & 6,284 & 3,073 & 3,074 \\
    Treatment uptake (\%) &       &       &       &       &       &  \\
      \quad No screening &  4210 (100.0)  &  2063 (49.1)  &  6278 (100.0)  &  1094 (17.4)  &  3073 (100.0)  &   123 ( 4.0)  \\
      \quad screening &     0 (  0.0)  &  2141 (50.9)  &     0 (  0.0)  &  5190 (82.6)  &     0 (  0.0)  &  2951 (96.0)  \\
    Age   & 64.75 (5.09) & 64.69 (5.08) & 62.38 (5.11) & 62.33 (5.08) & 62.12 (5.21) & 62.09 (5.21) \\
    Sex (\%) &       &       &       &       &       &  \\
     \quad  Female &  2642 ( 62.8)  &  2638 (62.7)  &  2520 ( 40.1)  &  2529 (40.2)  &  1981 ( 64.5)  &  1978 (64.3)  \\
     \quad  Male &  1568 ( 37.2)  &  1566 (37.3)  &  3758 ( 59.9)  &  3755 (59.8)  &  1092 ( 35.5)  &  1096 (35.7)  \\
    Race (\%) &       &       &       &       &       &  \\
     \quad  White &  3346 ( 79.5)  &  3289 (78.2)  &  5411 ( 86.2)  &  5437 (86.5)  &  2075 ( 67.5)  &  2097 (68.2)  \\
     \quad  Minority &   864 ( 20.5)  &   915 (21.8)  &   867 ( 13.8)  &   847 (13.5)  &   998 ( 32.5)  &   977 (31.8)  \\
    Education (\%) &       &       &       &       &       &  \\
    \quad   Less than high school &   585 ( 13.9)  &   616 (14.7)  &   347 (  5.5)  &   313 ( 5.0)  &   478 ( 15.6)  &   444 (14.4)  \\
    \quad   High school graduate &  1775 ( 42.2)  &  1639 (39.0)  &  1473 ( 23.5)  &  1484 (23.6)  &  1037 ( 33.7)  &  1059 (34.5)  \\
     \quad  college or above &  1850 ( 43.9)  &  1949 (46.4)  &  4458 ( 71.0)  &  4487 (71.4)  &  1558 ( 50.7)  &  1571 (51.1)  \\
    Smoking Status (\%) &       &       &       &       &       &  \\
    \quad   Non-smoker &  1803 ( 42.8)  &  1798 (42.8)  &  2495 ( 39.7)  &  2556 (40.7)  &  1555 ( 50.6)  &  1590 (51.7)  \\
    \quad   Current smoker &   562 ( 13.3)  &   573 (13.6)  &   709 ( 11.3)  &   704 (11.2)  &   390 ( 12.7)  &   378 (12.3)  \\
    \quad   Former smoker &  1845 ( 43.8)  &  1833 (43.6)  &  3074 ( 49.0)  &  3024 (48.1)  &  1128 ( 36.7)  &  1106 (36.0)  \\
    BMI (\%) &       &       &       &       &       &  \\
    \quad   $<$25 &  1439 ( 34.2)  &  1446 (34.4)  &  2464 ( 39.2)  &  2518 (40.1)  &   916 ( 29.8)  &   955 (31.1)  \\
   \quad    $>$25 &  2771 ( 65.8)  &  2758 (65.6)  &  3814 ( 60.8)  &  3766 (59.9)  &  2157 ( 70.2)  &  2119 (68.9)  \\
    Colorectal cancer (\%) &       &       &       &       &       &  \\
    \quad   No Confirmed Cancer &  4128 ( 98.1)  &  4139 (98.5)  &  6165 ( 98.2)  &  6219 (99.0)  &  3023 ( 98.4)  &  3035 (98.7)  \\
   \quad    Confirmed Cancer &    82 (  1.9)  &    65 ( 1.5)  &   113 (  1.8)  &    65 ( 1.0)  &    50 (  1.6)  &    39 ( 1.3)  \\
    Following time & 12.08 (3.87) & 11.45 (4.47) & 11.68 (2.94) & 11.70 (3.00) &  9.56 (2.10) &  9.62 (2.07) \\
    Rate (per 1000 person-yrs) & 1.61  & 1.35  & 1.54  & 0.8   & 1.7   & 1.32 \\
    \bottomrule
    \end{tabular}%
  \label{tab: descriptive statistics 2}%
\end{table}%

\begin{table}[htbp]
  \centering
  \caption{Estimated baseline characteristics of switchers (SW) and always-compliers (ACO) for the dataset with Herry Fold Health system (HF) and University of Colorado (UC), and Herry Fold Health system (HF) and University of Alabama Birmingham (UAB)}
    \begin{tabular}{lcccc}
    \toprule
          & \multicolumn{2}{c}{HF and UC} & \multicolumn{2}{c}{HF and UAB} \\
    \midrule
          & Switchers & Always-compliers & Switchers & Always-compliers \\
    \midrule
    Age   & 63.7  & 63.1  & 63.8  & 63.4 \\
    Sex (\%) &       &       &       &  \\
    \quad  Female & 64.9\% & 39.9\% & 73.2\% & 54.9\% \\
    \quad    Male & 35.1\% & 60.1\% & 26.8\% & 45.1\% \\
    Race (\%) &       &       &       &  \\
    \quad    White & 89.0\% & 81.7\% & 78.8\% & 71.4\% \\
    \quad    Minority & 11.0\% & 18.3\% & 21.2\% & 28.6\% \\
    Education (\%) &       &       &       &  \\
    \quad    Less than high school & 6.9\% & 8.3\% & 14.5\% & 14.1\% \\
    \quad    High school graduate & 34.5\% & 27.5\% & 41.1\% & 35.2\% \\
    \quad    college or above & 58.6\% & 64.2\% & 44.4\% & 50.7\% \\
    Smoking Status (\%) &       &       &       &  \\
    \quad    Non-smoker & 42.2\% & 41.5\% & 47.2\% & 46.4\% \\
    \quad    Current smoker & 13.1\% & 10.3\% & 14.2\% & 11.2\% \\
    \quad    Former smoker & 44.7\% & 48.2\% & 38.6\% & 42.4\% \\
    BMI (\%) &       &       &       &  \\
    \quad    $<25$ & 44.4\% & 34.5\% & 36.1\% & 29.7\% \\
    \quad    $>25$ & 55.6\% & 65.5\% & 63.9\% & 70.3\% \\
    \bottomrule
    \end{tabular}%
  \label{tab: charactrization 2}%
\end{table}%